\documentclass[12pt,preprint,eqsecnum,floatfix,nofootinbib,amsmath,amssymb,tightenlines]{revtex4}
\usepackage{graphicx}
\usepackage{dcolumn}
\usepackage{bm}
\usepackage{epsfig,psfrag,amsmath,amssymb,float,lscape}
\input{epsf}

\begin{document}
\setcounter{footnote}{0}
\def\half{{\textstyle {1 \over 2}}}
\def\eg{{\it e.g.~}}
\def\cf{{\it cf.~}}
\def\vs#1{vs.~\(#1)}
\def\viz{{\it viz.~}}
\def\se{\raise.15ex\hbox{$/$}\kern-.56em\hbox{$\epsilon$}}
\def\ie{{\it i.e.,\ }}
\def\gtwid{\mathrel{\raise.3ex\hbox{$>$\kern-.75em\lower1ex\hbox{$\sim$}}}}
\def\ltwid{\mathrel{\raise.3ex\hbox{$<$\kern-.75em\lower1ex\hbox{$\sim$}}}}
\def\ed{{\it ed.~by\ }}

\title{SPACETIME QUANTUM MECHANICS AND THE\\
QUANTUM MECHANICS OF SPACETIME\footnote{Lectures
given at the 1992 Les Houches \'Ecole d'\'et\'e, {\sl Gravitation et
Quantifications},\hfill\break July 9 -- 17, 1992.}}
                                                                                
\author{\bf James B. Hartle}
\email{hartle@physics.ucsb.edu}
\affiliation{Department of Physics, University of California\\
Santa Barbara, CA 93106-9530 USA}
\date{\today}
                                                                                
\maketitle
                                                                                
\tableofcontents
                                                                                
\section{Introduction}
\setcounter{footnote}{0}

These lectures are not about the quantization of any particular theory
of gravitation.  Rather they are about how to formulate quantum
mechanics generally enough
 so that it can answer questions in any  quantum
theory of spacetime.  They are not concerned with any particular theory
of the dynamics of gravity but rather with the quantum
framework for prediction in such theories generally.

It is reasonable to ask why an elementary course of lectures on quantum
mechanics should be needed in a school on the quantization of gravity.
We have standard courses in quantum mechanics that are taught in every
graduate school.  Why aren't these sufficient?
They are not sufficient because the formulations of quantum
mechanics usually taught in these courses is insufficiently general for
constructing a quantum theory of gravity suitable for application to
all the domains in
which we would like to apply it.  There are at least two counts on which
the usual formulations of quantum mechanics are not general enough:
They do not discuss the quantum mechanics of closed systems such as the
universe as a whole, and they do not address the ``problem of time'' in
quantum gravity.

The $S$-matrix is one important question to which quantum gravity should
supply an answer.  We cannot expect to test its matrix-elements that involve
external, Planck-energy gravitons any time in the near future. However,
we might hope that, since gravity couples universally to all forms of
matter, we might see imprints of Planck scale physics in
testable scattering experiments at more accessible energies with more
familiar constituents.  For the calculation of $S$-matrix elements the
usual formulations of quantum mechanics are adequate.

Cosmology, however,
 provides questions of a very different character to which a
quantum theory of gravity should also supply answers.  In our past
there is  an
epoch of the early universe when quantum gravity was important.
The remnants of this early time are all about us.
In these remnants of the Planck era we may hope to find some of the most
direct tests of any quantum theory of gravity.  However, it is not an
$S$-matrix that is relevant for these predictions.  We live in the
middle of this particular experiment.

Beyond simply describing the quantum dynamics of the early universe we
have today a more ambitious aim.  We aim, in the subject that has come
to be called quantum cosmology, to provide a {\it theory} of the initial
condition of the universe that will predict testable correlations among
observations today.  There are no realistic predictions of any kind that
do not depend on this initial condition if only very weakly.
Predictions of certain observations may be testably sensitive to its
details.  These include the familiar large scale features of
the universe --- its the approximate homogeneity and isotropy,
its vast age when compared with the Planck scale, and the
spectrum of fluctuations that were the progenitors of the galaxies.
Features on familiar scales, such as the homogeneity of the
thermodynamic arrow of time and the existence of a domain of
applicability of classical physics, may also depend centrally on the nature of
this quantum initial condition.  It has even been suggested that such
microscopic features as the coupling constants of the effective
interactions of the elementary particles may depend in part on the
nature of this quantum initial condition \cite{Haw83, Col88, GS88}.
It is to explain such
phenomena that a theory of the initial condition of the universe is just
as necessary and just as fundamental as a unified
quantum theory of all interactions including gravity. There is no other
place to turn.\footnote{For a review of some current proposals for
theories of the initial condition see Halliwell \cite{Hal91}.}

Providing a theory of the universe's quantum initial condition appears to
be a different enterprise from providing a manageable theory of the
quantum gravitational dynamics.  Specifying the initial condition is
analogous to  specifying
the initial state   while specifying the dynamics is analogous to
specifying the Hamiltonian.  Certainly these two goals
are pursued in different ways today.  String theorists deal with a
deep and subtle theory but are not able to answer deep questions about
cosmology.  Quantum cosmologists are interested in predicting features
like the large scale structure but are limited to working with cutoff
versions of the low-energy effective theory of gravity --- general
relativity.  However, it is possible that these two fundamental
questions are related. That is suggested, for example, by the ``no
boundary'' theory of the initial condition \cite{HH83}
whose wave function of the
universe is {\it derived} from the fundamental action for gravity and matter.
Is there one compelling principle that will specify both a unified
theory of dynamics and an initial condition?

The usual,``Copenhagen'', formulations of the quantum mechanics of
measured subsystems are inadequate for quantum cosmology. These
formulations assumed a division of the universe into ``observer'' and
``observed''. But in cosmology there can be no such fundamental
division. They assumed that fundamentally quantum theory is about the
results of ``measurements''. But measurements and observers cannot be
fundamental notions in a theory which seeks to describe the early
universe where neither existed. These formulations posited the
existence of an external ``classical domain''. But in quantum mechanics
there are no variables that behave classically in all circumstances. For
these reasons ``Copenhagen'' quantum mechanics must be generalized for
application to closed systems --- most generally and correctly the
universe as a whole.

I shall describe in these lectures the so called post-Everett
formulation of the quantum mechanics of closed systems.
This has its origins in the work of Everett
\cite{Eve57}
and has been developed by
 many.\footnote{Some notable earlier papers in the Everett to
post-Everett
development of the quantum mechanics of closed systems are those of
Everett \cite{Eve57}, Wheeler \cite{Whe57}, Gell-Mann \cite{Gel63},
Cooper and VanVechten \cite{CV69}, DeWitt \cite{DeW70}, Geroch
\cite{Ger84},
Mukhanov \cite{Muk85}, Zeh \cite{Zeh71},
Zurek \cite{Zur81, Zur91, Zur92},
Joos and Zeh \cite{JZ85}, Griffiths
\cite{Gri84}, Omn\`es \cite{Omnsum}, and Gell-Mann
and Hartle \cite{GH90a}. Some of the earlier papers
are collected in the reprint
volume edited by DeWitt and Graham \cite{DG73}.}
The post-Everett framework stresses that the probabilities of
alternative, coarse-grained, time histories are the most
general object of quantum mechanical prediction.  It stresses the
consistency of probability sum rules as the primary criterion for
determining which sets of histories may be assigned probabilities
rather than any notion of ``measurement''.  It stresses the
absence of quantum mechanical interference between individual histories,
or decoherence, as a sufficient condition for the consistency of
probability sum rules.  It stresses the importance of the initial
condition of the closed system in determining which sets of histories
decohere and which do not.  It does not posit the existence of the
quasiclassical domain of everyday experience but seeks to explain it as
an emergent feature of the initial condition of the
universe.

The second count on which the familiar framework of quantum needs to be
generalized for quantum cosmology concerns the nature of the alternatives
to which a quantum theory that includes  gravitation
 assigns probabilities --- loosely speaking the
nature of its ``observables''.  The usual formulations of quantum
mechanics deal with alternatives defined at definite moments of time.
They are concerned, for example, with the probabilities of alternative
positions of a particle at  definite moments of time or alternative
field configurations on spacelike surfaces.  When a background spacetime
geometry is fixed, as in special relativistic field theory, that
geometry
gives an
unambiguous meaning to the notions of ``at a moment of time'' or ``on a
spacelike surface''. However, in quantum gravity spacetime geometry is
not fixed; it is quantum mechanically variable and generally without
definite value.  Given two points it is not in general meaningful to say
whether they are separated by a spacelike, timelike, or null interval
much less what the magnitude of that interval is.  In a covariant theory
of quantum spacetime it is, therefore, not possible to assign an
meaning to alternatives ``at a moment of time'' except in the
case of alternatives that are independent of time, that is, in the case
of constants of
the motion.  This is a very limited class of
observables!\footnote{Although it is argued by some to be
enough. See Rovelli \cite{Rov90a}.}

The problem of alternatives is one aspect of what is called
``problem of time'' in quantum gravity.\footnote{Classic
papers on the ``problem of time'' are those of Wheeler
\cite{Whe79} and Kucha\v r \cite{Kuc81a}.
For recent, lucid reviews see Kucha\v r \cite{Kuc92}, Isham \cite{Ish92},
\cite{Ishpp}, and Unruh \cite{Unr91}.}  Broadly speaking this is the
conflict between the requirement of usual Hamiltonian formulations of
quantum mechanics for privileged set of spacelike surfaces
and the requirements
of general covariance which mean no one set of spacelike surfaces can be
more privileged than any other.  There is already a nascent conflict in
special relativity where there are many sets of spacelike surfaces.
However, the causal structure provided by the fixed background spacetime
geometry provides a resolution.  The Hamiltonian quantum mechanics constructed
by utilizing
one set of spacelike surfaces is unitarily equivalent to that using any
other.  But in quantum gravity there is no fixed background spacetime, no
corresponding notion of causality and no corresponding unitary
equivalence
either.  For these reasons a generalization of familiar Hamiltonian
quantum mechanics is needed for quantum gravity.

Various resolutions of the problem of time in quantum gravity have been
proposed.  They range from breaking general covariance by singling out a
 particular privileged set of spacelike surfaces to abandoning
spacetime as a fundamental
variable.\footnote{As in the lectures of Ashtekar in
this volume.}
I will not review these proposals
and the serious difficulties from which they suffer.\footnote{Not
least because there exist comprehensive recent reviews by
Isham \cite{Ishpp},
Kucha\v r \cite{Kuc92}, and Unruh \cite{Unr91}.} Rather in these lectures,
I shall describe a
different approach.  This is to resolve the problem of time by
using the sum-over-histories approach to quantum mechanics to generalize
it and bring it to fully four-dimensional,
spacetime form so that it does not need a privileged notion of
time.\footnote{The
use of the sum-over-histories formulation of quantum mechanics to
resolve the problem of time has been advocated in various ways by 
C.~Teitelboim \cite{Tei83a}, by
R.~Sorkin \cite{Sor89} , and by the author
\cite{Har86b, Har88a, Har88b, Har89b, Har91a, Har91b, Har92}.
These lectures are a summary and, to a
certain extent, an attempt at sketching a completion of the program
begun in these latter papers. In particular,
Section VIII might be viewed as the
successor promised to \cite{Har88a} and \cite{Har88b}.}
The key
to this generalization will be generalizing the alternatives
that are potentially
assigned probabilities by quantum theory to a much larger class of
spacetime alternatives that are not defined on spacelike surfaces.

We do not have today a complete, manageable, agreed-upon quantum theory
of the dynamics of spacetime with which to illustrate the formulations of
quantum mechanics I shall discuss.  The search for such a theory is
mainly what this school is about! In the face of this difficulty we
shall proceed in a way time-honored  in physics.  We shall consider
models. Making virtue out of necessity, this will enable us to consider
the various aspects of the problems we expect to encounter in quantum
gravity in simplified contexts.

To understand the quantum mechanics of closed systems we shall consider
in Sections II and III a universe in a box neglecting gravitation all
together.  This will enable us to construct explicit models of
decoherence and the emergence of classical behavior.

To address the question of the alternatives in quantum gravity we shall
begin by introducing a very general framework for quantum theory called
{\it generalized quantum mechanics} in
Section IV.  Section V describes a generalized sum-over-histories quantum
mechanics for non-relativistic systems which is in fully spacetime form.
Dynamics are described by spacetime path integrals, but more importantly
a spacetime notion of alternative is introduced --- partitions
of the paths into exhaustive sets of exclusive classes.  In Section VI
these ideas are applied to gauge theories which are the most familiar
type of theory exhibiting a symmetry.  The general notion of alternative
here is a {\it gauge invariant} partition of spacetime histories of the
gauge potential.  In Section VII, we consider two models which,
like theories of spacetime, are invariant under reparametrizations of
the time.  These are parametrized non-relativistic mechanics
and the relativistic particle. The general notion of alternative is
a reparametrization invariant partition of the paths.

A generalized sum-over-histories quantum mechanics for Einstein's
general relativity is sketched in Section VIII.
The general notion of alternative is a
diffeomorphism invariant partition of four-dimensional spacetime
metrics and matter field configurations.
Of course, we have no certain evidence that general relativity makes sense as a
quantum theory.  One can, however, view general relativity as a kind of
formal model for the interpretative issues that will arise in any theory
of quantum gravity.  More fundamentally,  general relativity is
(under reasonable assumptions) the unique low energy limit of any
quantum theory of gravity \cite{Des70, BD75}.
Any quantum theory of gravity must therefore describe
the probabilities of alternatives for four-dimensional histories of
spacetime geometry no matter how distantly related are its fundamental
variables. Understanding the quantum mechanics of general relativity is
therefore a necessary approximation in any quantum theory of gravity and
for that reason we explore it here.

Any proposed generalization of usual quantum mechanics has the heavy
obligation to recover that familiar framework in suitable limiting
cases.  The ``Copenhagen'' quantum mechanics of measured subsystems is
not incorrect or in conflict with the quantum mechanics of closed
systems described here.  Copenhagen quantum mechanics is an
{\it approximation} to that more general framework that is appropriate when
certain approximate features of the universe such as the existence of
classically behaving measuring apparatus can be idealized as exact.  In
a similar way, as we shall describe in Section IX, how familiar Hamiltonian
quantum mechanics with its preferred notion of time is an approximation
to a more general sum-over-histories quantum mechanics of spacetime
geometry that is appropriate for those epochs and those scales when the
universe, as a consequence of its initial condition and dynamics, {\it does}
exhibit a classical spacetime geometry that can supply a notion of time.

\section{The Quantum Mechanics of Closed
Systems}\footnotemark{This section has been adapted from the
author's contribution to the Festshrift for C.W.~Misner \cite{Har93}}

\setcounter{footnote}{0}

\subsection{Quantum Mechanics and Cosmology}

As we mentioned in the Introduction, the Copenhagen frameworks for quantum
mechanics, as they were formulated
in the '30s and '40s and as they exist in most textbooks today, are
inadequate for quantum cosmology.
 Characteristically these formulations
assumed, as {\it external} to the framework of wave function and
Schr\"odinger equation, the classical domain we see all about us.  Bohr
\cite{Boh58}
spoke of phenomena which could be alternatively described in classical
language.  In their classic text, Landau and Lifschitz \cite{LL58}
 formulated
quantum mechanics in terms of a separate classical physics.  Heisenberg
and others stressed the central role of an external, essentially
classical, observer.\footnote{For a clear statement of
this point of view, see London and Bauer \cite{LB39}.}
Characteristically, these formulations assumed a
possible division of the
world into
``observer'' and ``observed'', assumed that ``measurements'' are the
primary
focus of
scientific statements and, in effect,  posited the existence of an
external
``classical domain''.  However, in a theory of the whole thing there can
be no
fundamental division into observer and observed.  Measurements and
observers
cannot be fundamental notions in a theory that seeks to describe the
early
universe when neither existed.  In a basic formulation of quantum
mechanics
there is no reason in general for there to be any variables that exhibit
classical behavior in all circumstances.
  Copenhagen quantum mechanics thus needs to be generalized to
provide a quantum framework for cosmology. In this section we shall give
a simplified introduction to that generalization.

It was Everett who, in 1957, first suggested how to generalize the
Copenhagen frameworks so as to apply quantum mechanics to
closed systems such as
cosmology.
Everett's idea was to take quantum mechanics seriously and apply it
to the
universe as a whole.  He showed how an observer could be considered part
of
this system and how its activities --- measuring, recording, calculating
probabilities, etc. --- could be described within quantum mechanics.
Yet the
Everett analysis was not complete.  It did not adequately describe
within
 quantum mechanics the origin of the ``quasiclassical domain'' of familiar
experience nor, in an observer independent way, the meaning of the
``branching''
that replaced the notion of measurement.  It did not distinguish from
among
the vast number of choices of quantum mechanical observables that are in
principle available to an observer, the particular choices that, in
fact,
describe the
quasiclassical domain.

In this section we shall give an introductory review of the basic ideas
of what has come to be called the ``post-Everett'' formulation of
quantum mechanics for closed systems.
This aims at a coherent formulation of quantum mechanics for the
universe as a whole that is a framework to explain rather than  posit the
classical domain of everyday experience.  It is an attempt at an
extension, clarification, and completion of the Everett interpretation.
The particular exposition follows the work of Murray Gell-Mann and the
author \cite{GH90a, GH90b}
that builds on the contributions of many others,
especially those of Zeh \cite{Zeh71}, Zurek \cite{Zur81},
Joos and Zeh \cite{JZ85}, Griffiths \cite{Gri84}, and Omn\`es (\eg
as reviewed in \cite{Omn92}).  The exposition we shall give in this
section will be informal and simplified. We will return to greater
precision and generality in Sections III and IV.

\subsection{Probabilities in General and Probabilities in Quantum
Mechanics}

Even apart from quantum mechanics, there is no certainty in this
world and
therefore physics deals in
probabilities.
It deals most generally with the probabilities for
alternative
time histories of the universe.  From these, conditional probabilities
can be constructed that are
appropriate when some features about our specific history are known
and further ones are to be predicted.

 To understand what probabilities mean for a single closed system,
it is best to understand
how they
are used.  We deal, first of all, with probabilities for {\sl single}
events
of the {\sl single} system.
When these
probabilities become sufficiently close to zero or one there is a
definite
prediction on which we may act.  How sufficiently close to zero or one the
probabilities must be depends on the circumstances in which they are
applied.
There is no certainty that the sun will come up tomorrow at the time
printed in
our daily newspapers.  The sun may be destroyed by a neutron star now
racing
across the galaxy at near light speed.  The earth's rotation rate could
undergo
a quantum fluctuation.  An error could have been made in the computer
that
extrapolates the motion of the earth.  The printer could have made a
mistake in
setting the type.  Our eyes may deceive us in reading the time.  Yet, we
watch the sunrise at the appointed time because we compute, however
imperfectly, that the probability of these alternatives is sufficiently
low.

A quantum mechanics of a single system such as the universe must
incorporate a theory of the system's initial condition and dynamics.
Probabilities for alternatives that differ from zero and one may be of
interest (as in predictions of the weather) but to test the theory we
must search
among the different possible alternatives to find those whose
probabilities are predicted to be near zero or one. Those are the
definite predictions with which we can test the theory.
 Various strategies can be employed to identify situations where
probabilities
are near zero or one.  Acquiring information and considering the
conditional
probabilities based on it is one such strategy.  Current theories of the
initial condition of the
universe predict almost no probabilities near zero or one without
further
conditions.  The ``no
boundary'' wave function of the universe, for example, does not predict
the
present position of the sun on the sky.  However, it will predict  that
the
conditional probability for the sun to be at the position predicted by
classical celestial mechanics given a few previous positions is a number
very
near unity.

Another strategy to isolate probabilities near zero or one is to consider
ensembles
of repeated observations of identical subsystems in the closed system.
There are no
genuinely
infinite ensembles in the world so we are necessarily concerned with the
probabilities for deviations of the behavior of a finite ensemble
from the expected behavior of an infinite one.  These are probabilities
for a
single feature (the deviation) of a single system (the whole
ensemble).\footnote{For a more quantitative discussion of the connection
between
statistical probabilities and the probabilities of a single system see
\cite{Har91a}, Section II.1.1 and the references therein.}

 The existence of large ensembles of repeated observations in identical
circumstances and their ubiquity in laboratory science should not,
therefore,
obscure the
fact that in the last analysis physics must predict probabilities for
the
single system that is the ensemble as a whole.  Whether it is the
probability
of a successful marriage, the probability of the present galaxy-galaxy
correlation function, or the probability of the fluctuations in an
ensemble of
repeated observations, we must deal with the probabilities of single
events in
single systems.
In geology, astronomy, history, and cosmology, most predictions of
interest
have this character.
The  goal of physical theory is, therefore, most generally to predict
the
probabilities of histories of single events of a single system.

 Probabilities need be assigned to histories by physical theory only
up to the
accuracy they are used.  Two theories that predict probabilities for the
sun
not rising tomorrow at its classically calculated time that are both
well
beneath the standard on which we act are equivalent for all practical
purposes
as far as this prediction is concerned. It is often convenient,
therefore, to deal with approximate probabilities which satisfy the
rules of probability theory up to the standard they are used.

The characteristic feature of a quantum mechanical theory is that not
every set of alternative
histories that may be described can be assigned  probabilities.  Nowhere is
this
more clearly illustrated than in the two-slit experiment illustrated in
Figure 1.  In the usual
``Copenhagen'' discussion if we have
 not measured which of the two slits
the
electron passed through on its way to being detected at the screen, then
we are not permitted to assign probabilities to these alternative
histories.
It would be inconsistent to do so since the correct probability sum rule
would not be satisfied.  Because of interference, the probability to
arrive
at $y$ is not the sum of the probabilities to arrive at $y$ going
through
the upper or lower slit:
\begin{equation}
p(y) \not= p_U (y) + p_L (y) 
\label{twoone}
\end{equation}
because
\begin{equation}
|\psi_L (y) + \psi_U (y) |^2 \not= |\psi_L (y) |^2 + |\psi_U (y) |^2
\, . 
\label{twotwo}
\end{equation}

\begin{figure}[t]
\begin{center}
\includegraphics[width=4in]{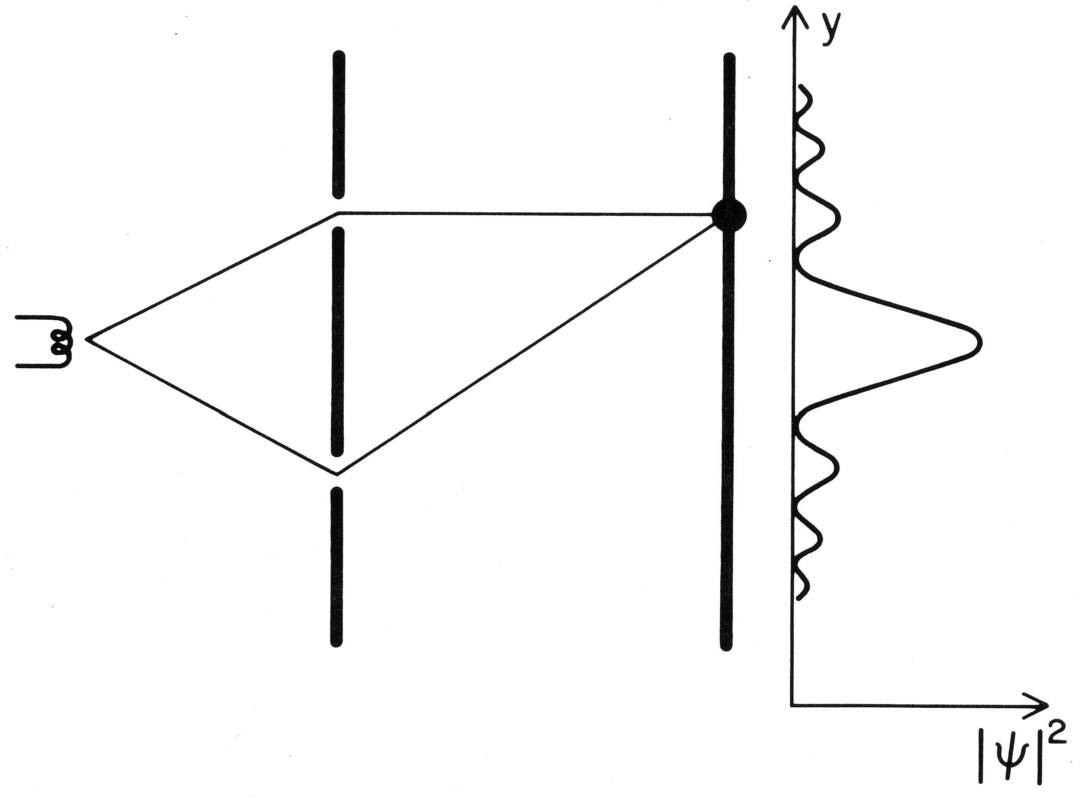}
\caption{The two-slit
experiment.  An
electron gun at right emits an electron traveling towards a screen with
two slits, its progress in space recapitulating its evolution in time.  When
precise detections are made of an ensemble of such electrons at the
screen it
is not possible, because of interference, to assign a probability to the
alternatives of whether an individual electron went through the upper
slit or
the lower slit.  However, if the electron interacts with apparatus that
measures which slit it passed through, then these alternatives decohere
and probabilities can be assigned.}
\end{center}
\end{figure}

If we {\it have} measured which slit the electron went through, then the
interference is destroyed, the sum rule obeyed, and we {\it can}
meaningfully
assign probabilities to these alternative histories.

A rule is thus needed in quantum theory to determine which sets of
alternative histories are assigned probabilities and which are not.
In Copenhagen quantum mechanics, the rule is that probabilities are
assigned to histories of alternatives of a subsystem
that are {\it measured} and not in
general otherwise. It is the generalization of this rule that we seek in
constructing a quantum mechanics of closed systems.

\subsection{Probabilities for a Time Sequence of Measurements}

To establish some notation, let us review in more detail the usual
``Copenhagen''
rules for the probabilities of time sequences of ideal measurements
of a subsystem using the two-slit experiment of Figure 1
 as an example.

Alternatives of the subsystem are represented by projection operators in
the Hilbert space which describes it.  Thus, in the two
slit experiment, the alternative that the electron passed through the
upper slit is represented by the projection operator
\begin{equation}
P_U = \Sigma_s \int_U d^3x \left|\vec x, s\rangle\langle\vec x,
s\right|
\label{threeone}
\end{equation}
where $|\vec x, s\rangle$ is a localized state of the electron with
spin component $s$, and the integral is over a volume around the upper
slit.  There is a similar projection operator $P_L$ for the alternative
that the electron goes through the lower slit.  These are exclusive
alternatives and they are exhaustive.  These properties, as well as the
requirements of being projections, are represented by the
relations
\begin{equation}
P_L^2 = P^2_U = 1\ ,\quad P_L P_U = 0\ , \quad P_U + P_L = I\, .
\label{threetwo}
\end{equation}
There is a similarly defined set of projection operators $\{P_{y_k}\}$
representing the alternative position intervals of arrival at the screen.

We can now state the rule for the joint probability that an electron
initially in a state $|\psi(t_0)\rangle$ at $t=t_0$ is determined by an
ideal measurement at time $t_1$ to have passed through the upper slit and
measured at time $t_2$ to arrive at point $y_k$ on the screen.  If one
likes, one can imagine the case when the electron is in a narrow wave packet
in the horizontal direction with a velocity defined as sharply as possible
consistent with the uncertainty principle.  The joint probability is
negligible unless $t_1$ and $t_2$ correspond to the times of flight to
the slits and to the screen respectively.

The first step in calculating the joint probability is to evolve the state of
the electron to the time $t_1$ of the first measurement
\begin{equation}
\bigl | \psi (t_1)\bigr\rangle = e^{-iH(t_1-t_0)/\hbar}\bigr|
\psi(t_0)\bigr\rangle\, . 
\label{threethree}
\end{equation}
The probability that the outcome of the measurement at time $t_1$
is that the electron passed through the upper slit is:
\begin{equation}
({\rm Probability\ of}\ U) = \left\Vert P_U\big |\psi(t_1)\big\rangle
\right\Vert^2 
\label{threefour}
\end{equation}
where $\Vert \cdot\Vert$ denotes the norm of a vector in Hilbert space.
If the outcome was the upper slit, and the measurement was an ``ideal''
one, that disturbed the electron as little as possible in making its
determination, then after the measurement the state vector is reduced to
\begin{equation}
\frac{P_U |\psi(t_1)\rangle}{\Vert P_U |\psi(t_1)\rangle\Vert}\, .
\label{threefive}
\end{equation}
This is evolved to the time of the next measurement
\begin{equation}
|\psi(t_2)\rangle =
e^{-iH(t_2-t_1)/\hbar}\frac{P_U|\psi(t_1)\rangle}{\Vert P_U
|\psi(t_1)\rangle\Vert}\, . 
\label{threesix}
\end{equation}
The probability of being detected at time $t_2$ in one of a set of position
intervals
on the screen centered at $y_k, k=1, 2, \cdots$
{\it given} that the electron passed through the upper slit is
\begin{equation}
({\rm Probability\ of}\ y_k\  {\rm given}\ U) = \left\Vert P_{y_k}
|\psi(t_2)\rangle\right\Vert^2\, .
\label{threeseven}
\end{equation}

The {\it joint} probability that the electron is measured to have gone
through
the upper slit {\it and} is detected at $y_k$ is the product of the
conditional probability \eqref{threeseven} with the probability 
\eqref{threefour}
that the electron passed through $U$.  The latter factor
cancels the denominator in
\eqref{threesix} so that combining all of the above equations in this
section, we have
\begin{equation}
({\rm Probability\ of}\ y_k\ {\rm and}\ U) = \left\Vert P_{y_k}
e^{-iH(t_2-t_1)/\hbar} P_U e^{-iH(t_1-t_0)/\hbar}\big |\psi(t_0)\big
\rangle\right\Vert^2\, . 
\label{threeeight}
\end{equation}
With Heisenberg picture projections this takes the even simpler
form
\begin{equation}
({\rm Probability\ of}\ y_k\ {\rm and}\ U)=\left\Vert P_{y_k} (t_2) P_U (t_1)
\ \big |\psi(t_0)\rangle\right\Vert^2\, . 
\label{threenine}
\end{equation}
where, for example,
\begin{equation}
P_U(t) = e^{iHt/\hbar} P_U e^{-iHt/\hbar}\, . 
\label{threeten}
\end{equation}
The formula \eqref{threenine} is a compact and unified expression of the two
laws of evolution that characterize the quantum mechanics of measured
subsystems --- unitary evolution in between measurements and reduction
of the wave packet at a measurement.\footnote{As has been noted by
many authors, \eg Groenewold \cite{Gro52} and Wigner \cite{Wig63} among the
earliest.}
  The important thing to remember
about the expression \eqref{threenine} is that everything in it ---
projections, state vectors, and Hamiltonian --- refer to the Hilbert space
of a subsystem, in this example the Hilbert space of the
electron that is measured.

Thus, in ``Copenhagen'' quantum mechanics, it is
measurement that determines which histories can be assigned
probabilities and formulae like \eqref{threenine} that determine what these
probabilities are.  As we mentioned, we cannot have such  rules in the quantum
mechanics
of closed systems because there is no fundamental division of a closed system
into measured
subsystem and measuring apparatus and no fundamental reason for
the closed system to contain classically behaving measuring apparatus in all
circumstances.
We need a more observer-independent,
measurement-independent, classical domain-independent rule for which
histories of a closed system can be assigned probabilities and what
these probabilities are.  The next section describes this rule.

\subsection{Post-Everett Quantum Mechanics}

It is easiest to introduce
the rules of post-Everett quantum mechanics, by first making a
simple assumption.  That is to neglect gross quantum fluctuations in
the geometry of spacetime, and assume a fixed background spacetime geometry
which
supplies a definite meaning to the notion of time.  This  is
an excellent approximation on accessible scales for times later than
$10^{-43}$ sec after the big bang.  The familiar apparatus of Hilbert
space, states, Hamiltonian, and other operators may then be applied to
process of prediction.  Indeed, in this context the quantum mechanics of
cosmology is in no way distinguished from the quantum mechanics of a
large isolated box, perhaps expanding, but containing both the observed
and its observers (if any).

\begin{figure}[t]
\begin{center}
\includegraphics[width=6in]{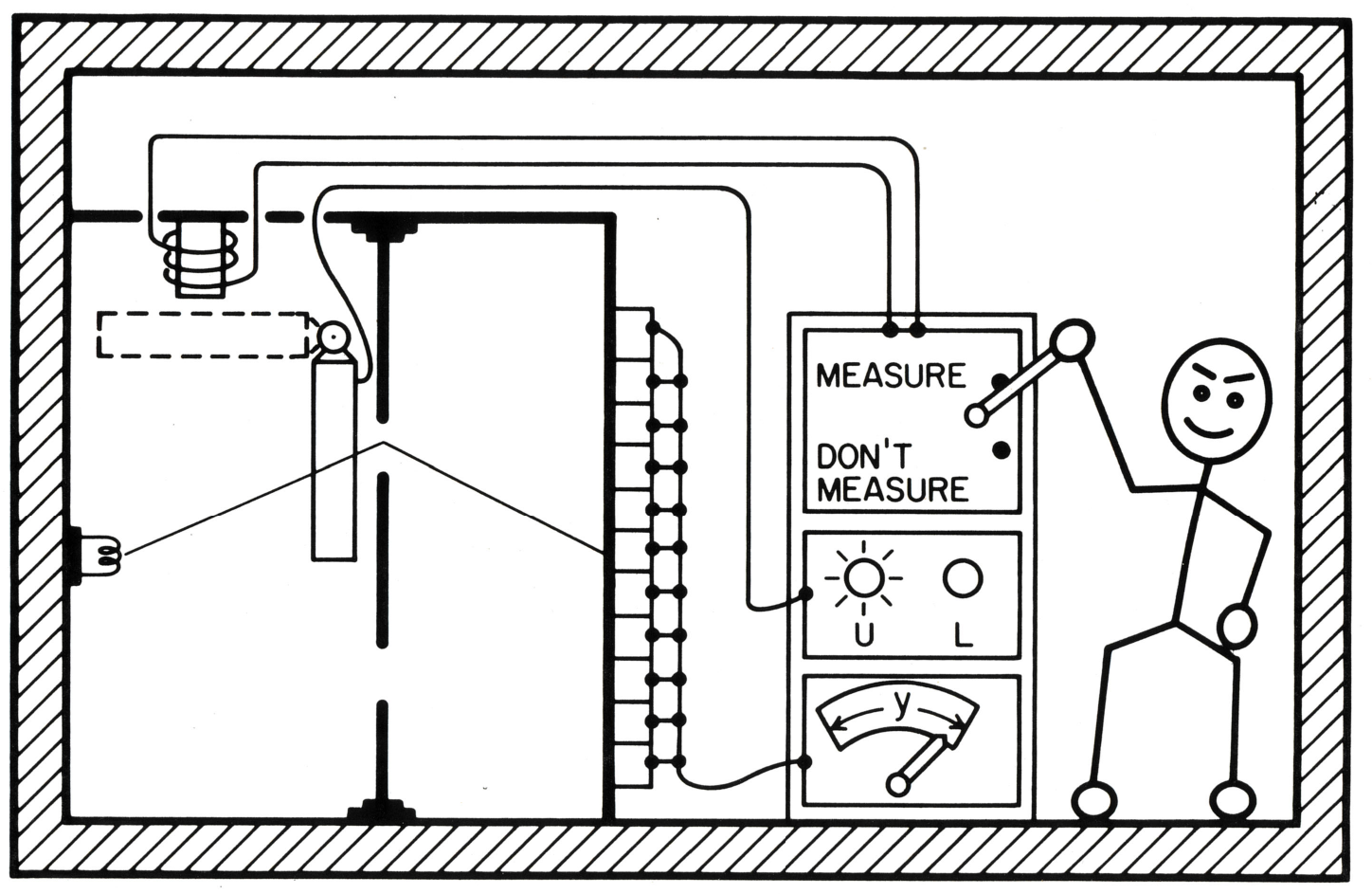}
\caption{A model closed quantum system.  At
one fundamental level of description this system consists of a large
number of electrons, nucleons, and excitations of the electromagnetic
field.  However, the initial state of the system is such that at a
coarser level description it contains an observer together with the
necessary apparatus for carrying out a two-slit experiment.
Alternatives for the system include whether the ``system'' contains a
two-slit experiment or not, whether it contains an observer or not,
whether the observer measured which slit the electron passed through or
did not, whether the electron passed through the upper or lower slit,
the alternative positions of arrival of the electron at the screen, the
alternative arrival positions registered by the apparatus, the
registration of these in the brain of the observer, etc., etc., etc.  Each
exhaustive set of exclusive alternatives is represented by an exhaustive
set of orthogonal projection operators on the Hilbert space of the
closed system.  Time sequences of such sets of alternatives describe
sets of alternative coarse-grained histories of the closed system.
Quantum theory assigns probabilities to the individual alternative
histories in such a set when there is negligible quantum mechanical
interference between them, that is, when the set of histories
decoheres.}
\end{center}
\end{figure}

A set of alternative histories for such a closed system is specified by
giving exhaustive sets of exclusive alternatives at a sequence of times.
Consider a model closed system with a quantity of matter initially in a
pure state that can be described as an observer and two-slit experiment,
with appropriate apparatus for producing the electrons, detecting which
slit they passed through, and measuring their position of arrival on the
screen (Figure 2).  Some alternatives for the whole system are:
\begin{enumerate}
\item Whether or not the observer decided to measure which slit the
electron went through.
\item Whether the electron went through the upper or lower slit.
\item The alternative positions, $y_1, \cdots, y_N$, that the
electron could have arrived at the screen.
\end{enumerate}
These sets of alternatives at a sequence of times define a set of
histories whose characteristic branching structure is shown in Figure 3.
An individual history in the set is specified by some particular
sequence of alternatives, \eg measured, upper, $y_9$.

\begin{figure}[t]
\begin{center}
\includegraphics[width=5in]{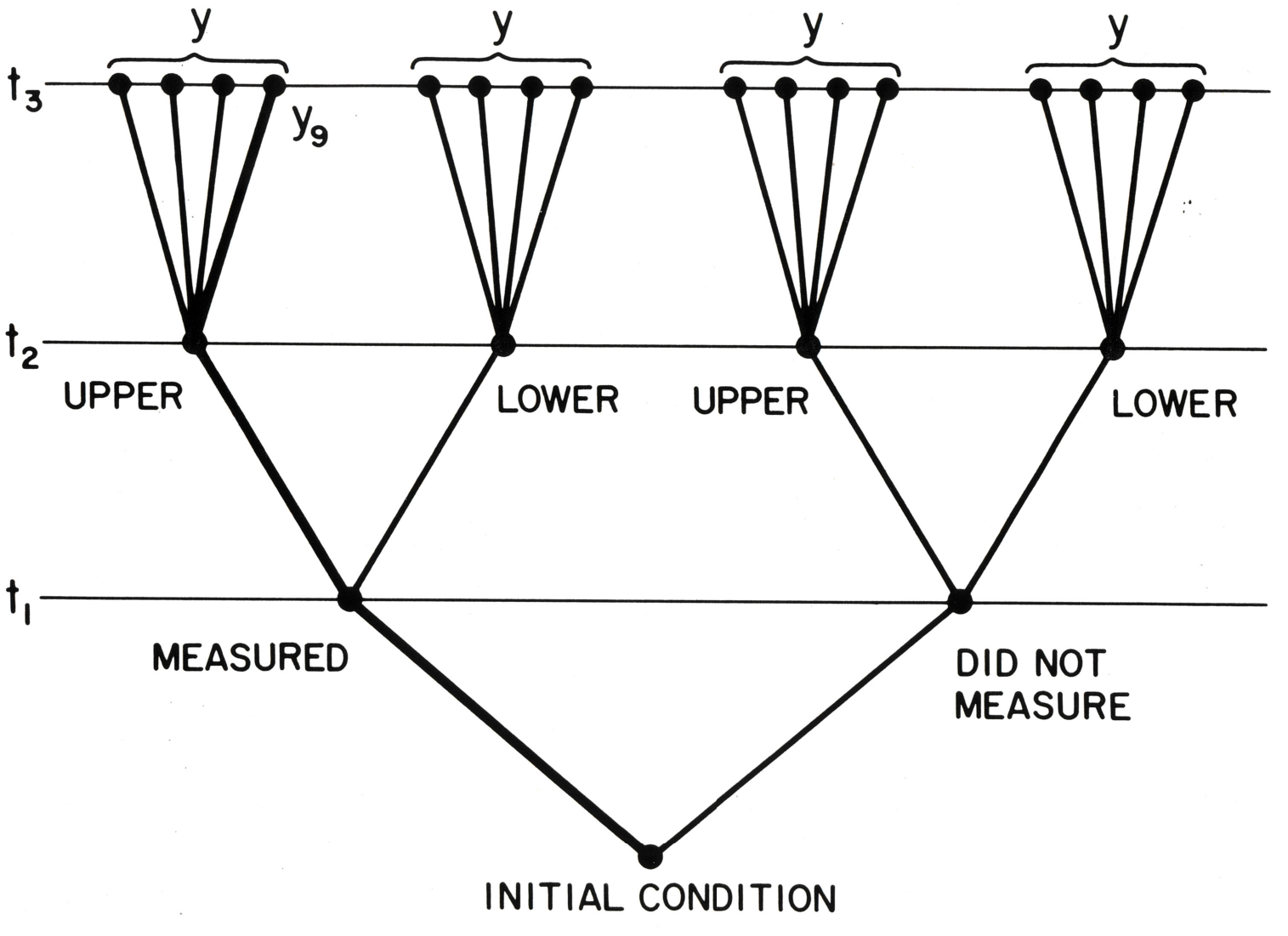}
\caption{Branching structure of a set of
alternative histories.  This figure illustrates the set of alternative
histories for the model closed system of Figure 2 defined by the
alternatives of whether the observer
decided to measure or did not decide to measure which slit the electron went
through at time $t_1$, whether the electron went through the upper slit
or through the lower slit at time $t_2$, and the alternative positions
of arrival at the screen at time  $t_3$.  A single branch corresponding to the
alternatives that the measurement was carried out, the electron went
through the upper slit, and arrived at point $y_9$ on the screen is
illustrated by the heavy line.\\
The illustrated set of histories does {\it not}
decohere because there is significant quantum mechanical interference
between the branch where no measurement was carried out and the electron
went through the upper slit and the similar branch where it went through
the lower slit.
A related set of histories that does decohere can be obtained by
replacing the alternatives at time $t_2$ by the following set of three
alternatives: (a record of the decision shows a measurement was
initiated and the electron went through
the upper slit); (a record of the decision shows a measurement was
initiated and the electron went through the lower slit); (a record of
the decision shows that the measurement was not initiated).  The
vanishing of the interference between the alternative values of the
record and the alternative configurations of apparatus ensures the
decoherence of this set of alternative histories.}
\end{center}
\end{figure}

Many other sets of alternative histories are possible for the closed
system.  For example, we could have included alternatives describing the
readouts of the apparatus that detects the position that the electron
arrived on the screen.  If the initial condition corresponded to a good
experiment there should be a high correlation between these alternatives
and the position that the electron arrives at the screen.  We could discuss
alternatives corresponding to thoughts in the observer's brain, or to
the individual positions of the atoms in the apparatus, or to the
possibilities that these atoms reassemble in some completely different
configuration. There are a vast number of possibilities.

Characteristically the alternatives that are of use to us as observers
are very {\it coarse grained},
distinguishing only very few of the degrees of freedom of a large closed
system and distinguishing these only at a small subset of the possible
times.  This is especially true if we recall that our box with observer
and two-slit experiment is only an idealized model.  The most general
closed system is the universe itself, and, as we shall show, the only
realistic closed systems are of cosmological dimensions.  Certainly, we
utilize only very, very coarse-grained descriptions of the universe as a
whole.

Let us now state the rules that determine which coarse-grained
sets of histories of a closed system may be assigned probabilities and what
those
probabilities are.  The essence of the rules can be
found in the work of Bob Griffiths \cite{Gri84}.
The general framework was extended
by Roland Omn\`es \cite{Omnsum}
and was independently, but later, arrived at by Murray
Gell-Mann and the author \cite{GH90a}.
The idea is simple: The obstacle to assigning probabilites is the
failure of the probability
sum rules due to quantum interference.
Probabilities can be therefore be assigned to just those sets of
alternative histories of a closed system for which there is negligible
interference between the individual histories in the set as a
consequence of the {\it particular} initial state the closed system has,
 and for which, therefore, all probability sum rules {\it are}
satisfied.  Let us now give this idea a precise expression.

Sets of alternatives at one moment of time, for example the set of
alternative position intervals $\{y_k\}$ at which the electron might
arrive at the screen,  are represented by exhaustive sets of
orthogonal projection operators.  Employing the Heisenberg picture these
can be denoted $\{P_{\alpha} (t)\}$ where $\alpha$ ranges over a set of
integers and $t$ denotes the time at which the alternatives are defined. A
particular alternative corresponds to a particular $\alpha$. For
example, in the two-slit experiment, $\alpha=9$ might be the alternative
that the electron arrives in the position interval $y_9$ at the screen.
$P_9(t)$ would be a projection on that interval at time $t$.
Sets of alternative histories are defined by giving sequences of sets of
alternatives at definite moments of time $t_1,\dots,t_n$ We denote the
sequence of such sets by
$\{P^1_{\alpha_1} (t_1)\}\ , \{P^2_{\alpha_2} (t_2)\}, \cdots,
\{P^n_{\alpha_n} (t_n)\}$. The sets are in general different at
different times. For example in the two-slit experiment
$\{P^2_{\alpha_2}(t_2)\}$ could be the set which distinguishes whether
the electron went through the upper slit or the lower slit at time
$t_2$, while $\{P^3_{\alpha_3}(t_3)\}$ might distinguish various positions
of arrival at the final screen at time $t_3$. More generally the
$\{P^k_{\alpha_k}(t_k)\}$ might be projections onto ranges of momentum or the
ranges of the eigenvalues of any other Hermitian operator at time $t_k$.
The superscript $k$ distinguishes these different sets in a sequence.
Each set of $P$'s satisfies
\begin{equation}
\sum\nolimits_{\alpha_k} P^k_{\alpha_k} (t_k) = I\ ,\quad P^k_{\alpha_k}
(t_k) P^k_{\alpha^\prime_k} (t_k) = \delta_{\alpha_k\alpha^\prime_k}
P^k_{\alpha_k} (t_k)\, ,
\label{fourone}
\end{equation}
showing that they represent an exhaustive set of exclusive
alternatives.
An individual history corresponds to a particular sequence  $(\alpha_1, \cdots,
\alpha_n)\equiv \alpha$ and,
for each history, there is a corresponding chain of {\it time ordered}
projection
operators
\begin{equation}
C_\alpha \equiv P^n_{\alpha_n} (t_n) \cdots P^1_{\alpha_1} (t_1)
\, .
\label{fourtwo}
\end{equation}
Such histories are said to be {\it coarse-grained} when, as is typically
the case, the $P$'s are not projections onto a basis (a complete set of
states) and when there is not a set of $P$'s at each and every time.

As an example, in the two-slit experiment illustrated in Figure 2
consider
the history in which the observer decided at time $t_1$ to measure which
slit the electron goes through, in which the electron goes through the
upper slit at time $t_2$, and arrives at the screen in position interval
$y_9$ at time $t_3$. This would be represented by the chain
\begin{equation}
P^3_{y_9} (t_3) P^2_U (t_2) P^1_{\rm meas} (t_1)
\label{fourthree}
\end{equation}
in an obvious notation. Evidently this is a very coarse-grained history,
involving only three times and ignoring most of the coordinates of the
particles that make up the apparatus in the closed system.
 As far as the description of histories is
concerned, the only difference between this situation and
that of the ``Copenhagen'' quantum mechanics of measured subsystems is
the following:  The sets of operators $\{P^k_{\alpha_k} (t_k)\}$
defining alternatives for the closed system act on the Hilbert space of
the closed system that includes the variables describing any apparatus,
observers, their constituent particles,
 and anything else.  The operators defining alternatives in
Copenhagen quantum mechanics act only on the Hilbert space of the
measured subsystem.

When the initial state is pure, it can be resolved into {\it branches}
corresponding to the individual members of any set of alternative
histories.  (The generalization to an impure initial density matrix is
not difficult and will be discussed in the next section.)
Denote the initial state by
$|\Psi\rangle$ in the Heisenberg picture.  Then
\begin{equation}
|\Psi\rangle = \sum\nolimits_\alpha C_\alpha |\Psi\rangle =
\sum\limits_{\alpha_1,\cdots, \alpha_n} P^n_{\alpha_n} (t_n) \cdots
P^1_{\alpha_1} (t_1) |\Psi\rangle\, . 
\label{fourfour}
\end{equation}
This identity follows by applying the first of \eqref{fourone} to all the
sums over $\alpha_k$ in turn.  The vector
\begin{equation}
C_\alpha|\Psi\rangle
\label{fourfive}
\end{equation}
is the {\it branch} of $|\Psi\rangle$
 corresponding to the individual history $\alpha$ and
\eqref{fourfour} is the resolution of the initial state into branches.

When the branches corresponding to a set of alternative histories are
sufficiently orthogonal, the set of histories is said to {\it decohere}.
More precisely a set of histories decoheres when
\begin{equation}
\langle \Psi | C^\dagger_{\alpha} C_{\alpha\prime} |\Psi \rangle \approx
0\ ,\quad {\rm for\ }\ \ \alpha \not= \alpha^\prime\, .
\label{foursix}
\end{equation}
Here, two histories $\alpha=(\alpha_1 \cdots \alpha_n)$ and
$\alpha'=(\alpha'_1 \cdots \alpha'_n)$ are equal when {\it all} the
$\alpha_k = \alpha'_k$ and are unequal when {\it any} $\alpha_k \ne
\alpha'_k$.
We shall return to the standard with which decoherence should be
enforced, but first let us examine its meaning and consequences.

Decoherence means the absence of quantum mechanical interference between
the individual histories of a coarse-grained set.
  Probabilities can be
assigned to the individual histories in a decoherent set of alternative
histories because decoherence implies the probability sum rules
necessary for a consistent assignment.  The probability of an individual
history $\alpha$ is
\begin{equation}
p(\alpha) = \left\Vert C_\alpha |\Psi\rangle\right\Vert^2\, . 
\label{fourseven}
\end{equation}

To see how decoherence implies the probability sum rules, let us
consider an example in which there are just three sets of alternatives
at times $t_1, t_2$, and $t_3$.  A typical sum rule might be
\begin{equation}
\sum\nolimits_{\alpha_2} p\left(\alpha_3, \alpha_2, \alpha_1\right) =
p\left(\alpha_3, \alpha_1\right)\, .
\label{foureight}
\end{equation}
We shall now show that \eqref{foursix} and \eqref{fourseven} imply 
\eqref{foureight}.  To do
that
write out the left hand side of \eqref{foureight} using \eqref{fourseven} and
suppress the time labels for compactness.
\begin{equation}
\sum\nolimits_{\alpha_2} p\left(\alpha_3, \alpha_2, \alpha_1\right) =
\sum\nolimits_{\alpha_2} \bigl\langle \Psi| P^1_{\alpha_1}
P^2_{\alpha_2} P^3_{\alpha_3} P^3_{\alpha_3} P^2_{\alpha_2}
P^1_{\alpha_1} |\Psi\bigr\rangle\, .
\label{fournine}
\end{equation}
Decoherence means that the sum on the right hand side of \eqref{fournine} can
be written with negligible error as
\begin{equation}
\sum\nolimits_{\alpha_2} p\left(\alpha_3, \alpha_2, \alpha_1\right)
\approx \sum\nolimits_{\alpha^\prime_2\alpha_2} \bigl\langle \Psi |
P^1_{\alpha_1}  P^2_{\alpha^\prime_2} P^3_{\alpha_3} P^3_{\alpha_3}
P^2_{\alpha_2} P^1_{\alpha_1} |\Psi\bigr\rangle\, . 
\label{fourten}
\end{equation}
the extra terms in the sum being vanishingly small.
But now, applying the first of \eqref{fourone} we see
\begin{equation}
\sum\nolimits_{\alpha_2} p\left(\alpha_3, \alpha_2, \alpha_1\right)
\approx \bigl\langle \Psi| P^1_{\alpha_1}
 P^3_{\alpha_3} P^3_{\alpha_3}
P^1_{\alpha_1} |\Psi\bigr\rangle = p\left(\alpha_3, \alpha_1\right)
\label{foureleven}
\end{equation}
so that the sum rule \eqref{foureight} is satisfied.

Given an initial state $|\Psi\rangle$ and a Hamiltonian $H$, one could,
in principle, identify all possible sets of decohering histories.  Among
these will be the exactly decohering sets where the orthogonality of the
branches is exact.  Indeed, trivial examples can be supplied by
resolving $|\Psi\rangle$ into a sum of orthogonal vectors
$|\Psi_{\alpha_1}\rangle$,
resolving these into vectors $|\Psi_{\alpha_2\alpha_1}\rangle$
such that the whole
set is orthogonal, and so on for $n$ steps.  The result is a resolution
of $|\Psi\rangle$ into exactly orthogonal branches
$|\Psi_{\alpha_n\cdots \alpha_1}\rangle$. By introducing suitable
projections and assigning them times $t_1, \cdots, t_n$, this set of
branches could be represented in the form \eqref{fourfour} giving an
exactly decoherent set of histories. Indeed, if the
$|\Psi_{\alpha_n\cdots\alpha_1}\rangle$ are not complete, there are
typically many different choices of projections that will do this.

Exactly  decoherent sets of histories are
 thus not difficult to achieve mathematically, but
such artifices
will not, in general, have a simple
description in terms of fundamental fields nor any connection, for
example, with the quasiclassical domain  of familiar experience.  For
this reason sets of histories that approximately decohere are also of
interest.  As we will argue in the next two sections, realistic
mechanisms lead to the decoherence of a set of histories describing a
quasiclassical domain that decohere
 to an excellent approximation as measured by
\cite{DH92}
\begin{equation}
\left|\left\langle\Psi\right| C^\dagger_\alpha C_{\alpha^\prime}
\left|\Psi\right\rangle\right| <<< \left\Vert C_\alpha \left| \Psi
\bigl\rangle \right\Vert \cdot \bigl\Vert C_{\alpha^\prime} \right| \Psi
\bigr\rangle \bigr\Vert,\quad {\rm for}\ \alpha^\prime \not= \alpha
\, .
\label{fourelevena}
\end{equation}

  When the
decoherence condition \eqref{foursix} is only approximately enforced, the
probability sum rules such as \eqref{foureight} will be only approximately
obeyed.  However, as discussed earlier, probabilities for single
systems are meaningful up to the standard they are used.
  Approximate probabilities for which
the sum rules are satisfied to a comparable standard may therefore also
be employed in the process of prediction.  When we speak of approximate
decoherence and approximate probabilities we mean decoherence achieved
and probability sum rules satisfied beyond any standard that might be
conceivably contemplated for the accuracy of prediction and the
comparison of theory with experiment.

We thus have a picture of the collection of all possible sets of
alternative coarse-grained histories of a closed system.  Within that
collection are the sets of histories that decohere and are assigned
approximate probabilities by quantum theory.  Within that collection
are the sets of histories describing the
quasiclassical domain of utility for everyday experience as we shall
describe in Section II.7.

Decoherent sets of alternative histories of the universe
are what can be utilized in the process of prediction
in quantum mechanics, for they
may be assigned probabilities.  Decoherence thus generalizes
and replaces the notion of ``measurement'', which served this role in
the Copenhagen interpretations.  Decoherence is a more precise, more
objective,
more observer-independent idea and gives a definite meaning to Everett's
branches.  For example, if their associated
histories
decohere, we may assign
probabilities to various values of reasonable scale density fluctuations
in the early universe whether or not anything like a ``measurement'' was
carried out on them and certainly whether or not there was an
``observer''
to do it.

\subsection{The Origins of Decoherence in Our Universe}

What are the features of coarse-grained sets of histories that
decohere in our universe?
In seeking to answer this question it is important to keep in mind the
basic
aspects of the theoretical framework on which decoherence depends.
Decoherence of a set of alternative histories is not a property of their
operators {\it alone}.  It depends  on the relations of those
operators to the initial state $|\Psi\rangle$, the Hamiltonian $H$, and the
fundamental
fields.  Given these, we could, in principle, {\it compute} which sets
of
alternative histories decohere.

We are not likely to carry out a computation of all decohering sets of
alternative histories for the universe, described in terms of the
fundamental
fields, any time in the near future, if ever. It is therefore important
to investigate specific mechanisms by which decoherence occurs.
Let us begin with a very simple model due
 to Joos and Zeh \cite{JZ85} in its essential features.
We consider the two-slit example again, but this time suppose that in the
neighborhood
of the slits there is a gas of photons or other light particles
colliding with
the electrons.  Physically it is easy to see what happens, the
random uncorrelated collisions carry away delicate phase
correlations
between the beams even if the trajectories of the
electrons are not affected much.
  The interference pattern is destroyed and it is
possible
to assign probabilities to whether the electron went through the upper
slit or
the lower slit.

\begin{figure}[t]
\begin{center}
\includegraphics[width=4in]{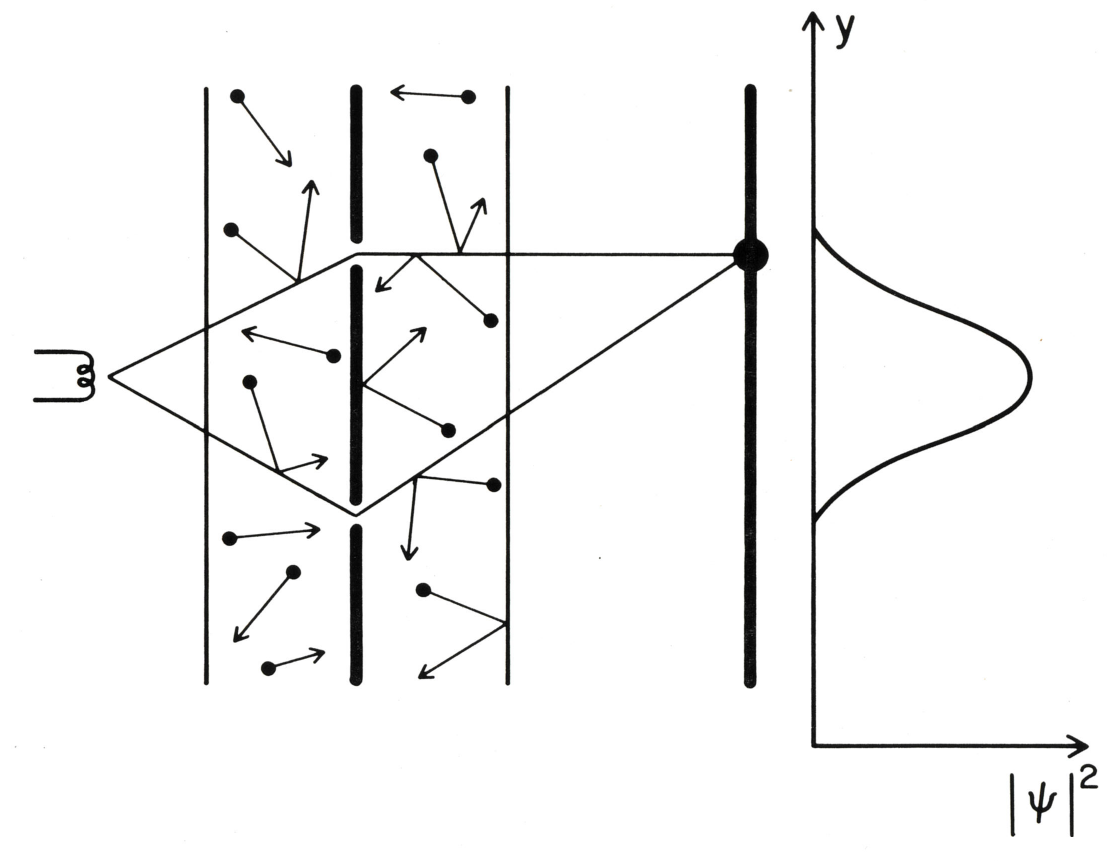}
\caption{The two-slit experiment
with an
interacting gas. Near the slits light particles of a gas collide with
the electrons.  Even if the collisions do not affect the trajectories of the
electrons very much they can still carry away the phase correlations
between
the histories in which the electron arrived at point $y_k$ on the screen
by passing through the upper slit and that in which it arrived at the same
point
by passing through the lower slit.  A coarse graining that consisted
only
of these two alternative histories of the electron would approximately
decohere as a consequence of the interactions with the gas given
adequate
density, cross-section, etc.  Interference is destroyed and
probabilities can be
assigned to these alternative histories of the electron in a way that
they
could not be if the gas were not present (\cf Fig.~1).  The lost phase
information is still available in correlations between states of the gas
and
states of the electron.  The alternative histories of the electron would
not
decohere in a coarse graining that included both the histories of the
electron
{\sl and} operators that were sensitive to the correlations between the
electrons and the gas.\\
This model illustrates a widely occurring mechanism by which certain
types of
coarse-grained sets of alternative histories decohere in the universe.}
\end{center}
\end{figure}

Let us see how this picture in words is given precise
meaning
in  mathematics.  Initially, suppose the state of the entire
system is a state of the electron $|\psi >$ and $N$ distinguishable
``photons''
in states $|\varphi_1 \rangle$, $|\varphi_2 \rangle$, etc., viz.
\begin{equation}
|\Psi \rangle = |\psi \rangle  |\varphi_1 \rangle |\varphi_2 >\cdots
|\varphi_N\rangle\, .
\label{fiveone}
\end{equation}
The electron state
$|\psi \rangle$ is a coherent superposition of a state in which the electron
passes
through the upper slit $|U \rangle$ and the lower slit $|L \rangle$.
Explicitly:
\begin{equation}
|\psi \rangle = \alpha |U \rangle + \beta |L \rangle\, .
\label{fivetwo}
\end{equation}
Both states are wave packets in horizontal position,
 $x$, so that position in $x$
recapitulates history in time.  We now ask whether the history where the
electron passes through the upper slit and arrives at a detector
defining an interval
$y_k$ on the screen, decoheres from that in which it passes through
the lower slit and arrives the interval $y_k$, as a consequence of the initial
condition of this ``universe''.  That is, as in Section 4, we ask
whether the two branches
\begin{equation}
P_{y_k}(t_2) P_U(t_1) |\Psi\rangle\quad , \quad P_{y_k}(t_2) P_L(t_1)
 |\Psi\rangle 
\label{fivethree}
\end{equation}
are nearly orthogonal, the times of the projections being those for the
nearly classical motion in $x$.  We work this out in the
Schr\"odinger picture where the initial state evolves, and the
projections on the electron's position are applied to it at the
appropriate times.

Collisions occur, but the states $|U\rangle$
 and  $|L \rangle$ are left more
or less undisturbed.  The states of the ``photons'' are, of course,
significantly affected.  If the photons are dilute enough to be
scattered only once
by the electron in its time to traverse the gas, the two branches 
\eqref{fivethree}
will be
approximately
\begin{subequations}
\label{fivefour}
\begin{equation}
\alpha P_{y_k} |U \rangle S_U |\varphi_1 \rangle S_U |\varphi_2 \rangle
\cdots S_U |\varphi_N\rangle\, ,
\label{fivefour a}
\end{equation}
and
\begin{equation}
\beta\ P_{y_k} |L \rangle S_L |\varphi_1 \rangle S_L |\varphi_2 \rangle
\cdots S_L |\varphi_N\rangle\, .
\label{fivefour b}
\end{equation}
\end{subequations}
Here, $S_U$ and $S_L$ are the scattering matrices from an electron in
the vicinity of the upper slit and the lower slit respectively.  The two
branches
 in \eqref{fivefour} decohere because the states of the ``photons'' are nearly
orthogonal.
The overlap of the branches is proportional to
\begin{equation}
\langle\varphi_1 |S^{\dagger}_U S_L |\varphi_1 \rangle\langle
\varphi_2 | S^{\dagger}_U
S_L |\varphi_2
\rangle\cdots \langle\varphi_N |S^\dagger_U S_L\ |\varphi_N \rangle
\, .\label{fivefive}
\end{equation}
Now, the $S$-matrices for scattering off an electron at the upper position or
the lower
position
can be connected to that of an electron at the origin by a translation
\begin{subequations}
\label{fivesix}
\begin{eqnarray}
S_U & = & \exp(-i\vec k \cdot{\vec x}_U) S\ \exp(+i {\vec k} \cdot  {\vec
x}_U)\, ,
\label{fivesix a}\\
S_L & = & \exp(-i{\vec k} \cdot{\vec x}_L) S\ \exp(+i {\vec k} \cdot  {\vec
x}_L)\, .
\label{fivesix b}
\end{eqnarray}
\end{subequations}
Here, $\hbar{\vec k}$ is the momentum of a photon, ${\vec x}_U$ and ${\vec
x}_L$
are the positions of the slits and $S$ is the  scattering matrix from
an electron at the origin.
\begin{equation}
\langle{\vec k}^\prime |S|{\vec k}\rangle = \delta^{(3)}\bigl({\vec k} - {\vec
k}^\prime
\bigr)
+ \frac{i}{2\pi\omega_k} f\bigl({\vec k}, {\vec k}^\prime\bigr)
\delta
\bigl({\omega_k} - \omega^\prime_k\bigr)\, ,
\label{fiveseven}
\end{equation}
where $f$ is the scattering amplitude and $\omega_k = |{\vec k}|$.

Consider the case where all the photons are in plane wave states in an
interaction volume $V$, all having the same energy $\hbar\omega$, but
with
random
orientations for their momenta.  Suppose further that the energy is low
so that
the electron is not much disturbed by a scattering
and low enough so the wavelength is
much longer than the separation between the slits, $k|{\vec  x}_U - {\vec
x}_L |<< 1$.
 It is then possible to work out the overlap.  The answer
according to
Joos and Zeh \cite{JZ85} is
\begin{equation}
\left(1-\frac{(k|{\vec x}_U - {\vec
x}_L|)^2}{8\pi^2V^{2/3}}\sigma\right)^N
\label{fiveeight}
\end{equation}
where $\sigma$ is the effective scattering cross section.  Even if
$\sigma$ is small, as $N$ becomes large this tends to zero.
In this way decoherence becomes a
quantitative phenomenon.

What such models convincingly show is that decoherence is frequent and
widespread in the universe.  Joos and Zeh  calculate that a
superposition of two positions of a grain of dust, 1mm apart, is
decohered simply by the scattering of the cosmic background radiation on
the time-scale of a nanosecond.  The existence of such mechanisms means
that the only realistic isolated systems are of cosmological dimensions.
So widespread is this kind of phenomena with the
initial condition and dynamics of our universe, that we may meaningfully
speak of habitually decohering variables such as the center of mass
positions of massive bodies.

\subsection{The Copenhagen Approximation}

What is the relation of the familiar Copenhagen quantum mechanics
described in Section II.3 to the more general ``post-Everett'' quantum
mechanics of closed systems described in Sections II.4 and II.5?
Copenhagen quantum mechanics predicts the probabilities of the histories
of measured subsystems.  Measurement situations may be described in
a closed system that contains both measured
subsystem and measuring apparatus.\footnote{For a more detailed model
of measurement situations in the quantum mechanics of closed systems see
{\it e.g.} \cite{Har91a}, Section II.10}.
measurement  In a typical measurement situation
the values of a variable not normally decohering become correlated with
alternatives of the apparatus that decohere because of {\it its}
 interactions
with the rest of the closed system.  The correlation means
that the measured alternatives decohere because the alternatives of the
apparatus decohere.

The recovery of the Copenhagen rule for when probabilities may be
assigned is immediate.  Measured quantities are correlated with
decohering histories.  Decohering histories can be assigned
probabilities.  Thus in the two-slit experiment (Figure 1), when the
electron interacts with an apparatus that determines which slit it
passed through, it is the decoherence of the alternative configurations
of the apparatus that register this determination
 that enables probabilities to be assigned to  the alternatives for
electron.

There is nothing incorrect about Copenhagen quantum
mechanics.  Neither is it, in any sense, opposed to the post-Everett
formulation of the quantum mechanics of closed systems.
 It is an {\it approximation} to the more general framework
appropriate in the special cases of measurement situations and when the
decoherence of alternative configurations of the apparatus  may be
idealized as exact and instantaneous.  However, while measurement
situations imply decoherence, they are only special cases of decohering
histories.  Probabilities may be assigned to alternative positions of the
moon and to  alternative values of
density fluctuations near the big bang in a universe in
which these alternatives decohere, whether or not they were participants in a
measurement situation and certainly whether or not there was an observer
registering their values.

\subsection{Quasiclassical Domains}

As observers of the universe, we deal with coarse-grained histories that
reflect our
own limited sensory perceptions, extended by instruments, communication
and
records  but in the end characterized by a large amount of ignorance.
Yet, we have the impression that the universe exhibits a much finer-grained
set of histories,
independent of us,
defining an always decohering ``quasiclassical domain'',
to which our senses are adapted, but deal with only a small part of it.
If we are preparing for a journey into a yet unseen part of the
universe, we do not believe that we need to equip ourselves with
spacesuits having detectors sensitive, say, to coherent superpositions of
position or other unfamiliar quantum variables.  We expect that the
familiar quasiclassical variables will decohere and be approximately
correlated in time by classical deterministic laws in any new part of
the universe we may visit just as they are here and now.

In a generalization of quantum mechanics which does not {\it posit} the
existence of a classical domain, the domain of
 applicability of classical physics
must be {\it explained}.
For a quantum mechanical system to exhibit classical behavior there must
be
some restriction on its state and some coarseness in how it is
described.
This is clearly illustrated in the quantum mechanics of a single
particle.  Ehrenfest's theorem shows that generally
\begin{equation}
M\frac{d^2\langle x\rangle}{dt^2} = \left \langle -\frac{\partial
V}{\partial
x}\right \rangle\, .
\label{oneone}
\end{equation}
However, only for special states, typically narrow wave packets, will
this
become
an equation of motion for $\langle x \rangle$ of the form
\begin{equation}
M\frac{d^2\langle x \rangle}{dt^2} = -\frac{\partial V(\langle x
\rangle)}{\partial x}\, .
\label{onetwo}
\end{equation}
For such special states, successive observations of position in time
will exhibit the classical correlations predicted by the equation
of motion \eqref{onetwo} {\sl provided} that
these observations are coarse enough so that the properties of the state
which
allow \eqref{onetwo} to replace  the general relation
\eqref{oneone} are not affected by these observations.  An {\sl exact}
determination of position, for example,  would yield a completely
delocalized
wave packet an
instant later and \eqref{onetwo} would no longer be a good approximation to
\eqref{oneone}.  Thus, even for large systems, and in particular for the 
universe as a
whole, we can expect classical behavior only for certain initial states
and then only when a sufficiently coarse grained description is used.

If classical behavior is {\it in general}  a consequence only of a certain
class
of states  in quantum
mechanics, then, as a particular case, we can expect to have classical
spacetime
only for certain states in quantum gravity.  The classical spacetime
geometry
we see all about us in the late universe is not a property of every state
in a
theory where geometry fluctuates quantum mechanically. Rather, it is
traceable
fundamentally to restrictions on the initial condition.
Such restrictions are likely to be generous in that, as in the single
particle
case, many different states will exhibit classical features. The
existence
of classical spacetime and the applicability of classical physics
 are thus not likely to be very restrictive conditions on constructing
a theory of the initial condition. Fundamentally, however, the existence
of one or more quasiclassical domains of the universe must be a
prediction of any successful theory of its initial condition and
dynamics, and thus an important problem for quantum cosmology.

Roughly speaking, a quasiclassical domain should be a set of alternative
histories that decoheres according to a realistic principle
of decoherence, that is maximally refined  consistent with that notion of
 decoherence, and
whose individual histories exhibit as much as possible patterns of
classical
correlation in time.
To make the question of the existence of one or more quasiclassical
domains into a {\it calculable} question in quantum cosmology
 we need measures of how close a set of histories comes to
constituting a ``quasiclassical domain''.  A quasiclassical domain
cannot be a
    {\it completely}
fine-grained description for then it would not decohere.  It cannot
consist
{\it entirely} of a few ``classical variables'' repeated over and over
because
sometimes
we may measure something highly quantum mechanical.  These variables
cannot be
{\it always} correlated in time by classical laws because sometimes
quantum
mechanical phenomena cause deviations from classical physics.  We need
measures
for maximality and classicality \cite{GH90a}.

It
is possible to give crude arguments for the type of
habitually decohering
operators we expect to occur over and over again in a set of
histories defining a quasiclassical domain \cite{GH90a}.
 Such habitually decohering
operators are called ``quasiclassical operators''.
In the earliest instants of the universe the operators defining
spacetime on
scales well above the Planck scale emerge from the quantum fog as
quasiclassical.  Any theory of the initial condition that does
not
imply this is simply inconsistent with observation in a manifest way.
A background spacetime is thus defined and conservation laws arising
from its symmetries have meaning. Then,
where there are suitable conditions of low temperature, density,
etc., various sorts of
hydrodynamic variables may emerge as quasiclassical operators.  These
are
integrals over suitably small volumes of densities of conserved or
nearly
conserved quantities.  Examples are densities of energy, momentum,
baryon
number, and, in later epochs, nuclei, and even chemical species.  The
sizes of
the volumes are limited above by maximality and are limited below by
classicality
because they require sufficient ``inertia'' resulting from their
approximate conservation  to enable them to resist
deviations
from predictability caused by their interactions with one another, by
quantum
spreading, and by the quantum and statistical fluctuations resulting
from
interactions with the rest of the universe that accomplish decoherence
\cite{GH90a}.
Suitable integrals of densities of
approximately conserved quantities are thus candidates for habitually
decohering quasiclassical operators.
These ``hydrodynamic variables'' {\it are} among the principle variables
of classical physics.

It would be in such ways that the classical domain of familiar
experience could
be an
emergent property of the fundamental description of the universe,
not generally in quantum
mechanics,
but as a consequence of our specific initial condition and the
Hamiltonian
describing evolution.  Whether a closed system exhibits a quasiclassical
domain, and, indeed, whether it exhibits more than one essentially
inequivalent domain, thus become calculable questions in the quantum
mechanics of closed systems.

The founders of quantum mechanics
were right in pointing out that something external to the framework of
wave function and the Schr\"odinger equation {\it is} needed to
interpret the theory.  But it is not a postulated classical domain to
which quantum mechanics does not apply.  Rather it is the initial
condition of the universe that, together with the action function of the
elementary particles and the throws of the quantum dice since the
beginning, is the likely origin of quasiclassical domain(s) within
quantum theory itself.

\section{Decoherence in General, Decoherence in Particular,\\
and the Emergence of Classical Behavior}
\setcounter{footnote}{0}

\subsection{A More General Formulation of the
Quantum Mechanics of Closed Systems}

The basic ideas of post-Everett quantum mechanics were introduced in the
preceding section.  We can briefly recapitulate these as follows: The most
general
predictions of quantum mechanics are the probabilities of alternative
coarse-grained histories of a closed system in an exhaustive set of such
histories.  Not every set of coarse-grained histories can be assigned
probabilities because of quantum mechanical interference and the
consequent failure of probability sum rules.  Rather, probabilities are
predicted only for those decohering sets of histories for which
interference between the individual members is negligible as a
consequence of the system's initial condition and Hamiltonian and the
probability sum rules therefore obeyed.    Among
the decohering sets implied by the initial condition of our universe are
those constituting the quasiclassical domain of familiar experience.

The discussion of Section II was oversimplified in several respects.
For example, we restricted attention to pure initial states, considered
only sets of alternatives at definite moments of time, considered only
sets of alternatives at any one moment that were independent of
alternatives at other moments of time, and assumed a fixed background
spacetime.
None of these restrictions is realistic.  In the rest of these lectures
we shall be pursuing the necessary generalizations needed for a more
realistic formulation.  In this section we develop a more general
framework still assuming a fixed spacetime geometry that supplies a
meaning to time and still restricting attention to alternatives at
definite moments of time.

\subsubsection{Fine-Grained and Coarse-Grained Histories}

We consider a closed quantum mechanical system described by a Hilbert
space ${\cal H}$.  As described in Section II, a set of alternatives at one
moment of time is
described by a set of orthogonal Heisenberg projection operators
$\{P^k_{\alpha_k}(t_k)\}$ satisfying \eqref{fourone}. The operators
corresponding to the same alternatives at different times are related by
unitary evolution
\begin{equation}
P^k_{\alpha_k} (t_k) = e^{iHt_k/\hbar}\, P^k_{\alpha_k} (0)\,
e^{-iHt_k/\hbar}\, .
\label{threeoneonea}
\end{equation}
 Sequences of such sets of
alternatives at, say, times $t_1, \cdots, t_n$ define a set of
alternative histories for the closed system.  The individual histories
in  such a set consist of particular chains of alternatives $\alpha =
(\alpha_1, \cdots, \alpha_n)$ and are represented by the corresponding
chains of projection operators, $C_\alpha$,  as in \eqref{fourtwo}.

Sets of histories described in this way are in general {\it
coarse-grained} because they do not define alternatives at each and
every time and because the projections specifying the
 alternatives are not onto {\it complete}
sets of states (one-dimensional projections onto a basis) at the times when
they are
defined.  The {\it fine-grained sets} of histories on a time interval
$[0, T]$ are defined by giving sets of {\it one-dimensional} projections at
each time  and so are represented by continuous products of one-dimensional
projections.  These are the most refined descriptions of the quantum
mechanical system possible. There are many different sets of
fine-grained histories. A simple example of fine- and coarse-grained
histories occurs when
${\cal H}$ is the space of square integrable functions on a
configuration space of generalized coordinates $\{q^i\}$ (for example,
modes of field configurations on a spacelike surface). Exhaustive sets
of exclusive coordinate ranges at a sequence of times define a set of
coarse-grained histories.  If the ranges are made smaller and smaller
and more and more dense in time, these increasingly fine-grained
histories come closer and closer to representing continuous paths
$q^i(t)$ on the interval $[0,T]$. These paths are the starting point for
a sum-over-histories formulation of quantum mechanics.  Operators
$C_\alpha$ corresponding to the individual paths themselves do not exist
because there are no exactly localized states in ${\cal H}$, but the
$C_\alpha$ on the finer- and finer-grained histories described above
represent them in the familiar way continuous spectra are handled
in quantum mechanics.

A set of alternatives at one moment of time may be further
coarse-grained by taking the union of alternatives corresponding to the
logical operation  ``or''.  If $P_a$ and $P_b$ are the projections
corresponding to alternatives ``$a$'' and ``$b$'' respectively,
then  $P_a + P_b $ is the projection corresponding
to the alternative
 ``$a$ or $b$''.  This is the simplest example of an operation of {\sl
coarse-graining}.
This operation ``or'' can be applied to histories. If $C_\alpha$ is the
operator representing one history in a coarse-grained set, and $C_\beta$
is another, then the coarser grained alternative in which the system
follows either history $\alpha$ {\it or} history $\beta$
is represented by
\begin{equation}
C_{\alpha\ {\rm or}\ \beta} = C_\alpha + C_\beta\,  .  
\label{threeoneone}
\end{equation}
Thus, if $\{c_\alpha\}$ is a set of alternative histories for the closed system
defined by sequences of alternatives at definite moments of time, then
the general notion of a coarse graining of this set of histories is
a partition of the $\{c_\alpha\}$ into exclusive classes
$\{c_{\bar\alpha}\}$. The classes are the individual histories in
the coarser grained set and are represented by operators, called {\it
class operators}, that are {\it
sums} of the chains of the constituent projections in the finer-grained set:
\begin{equation}
C_{\bar\alpha} = \sum\limits_{\alpha\epsilon\bar\alpha} C_\alpha\, .
\label{threeonetwo}
\end{equation}
When the $C_\alpha$ are chains of projections we have:
\begin{equation}
C_{\bar\alpha} =
\sum\limits_{(\alpha_1, \cdots, \alpha_n)\epsilon\bar\alpha}
P^n_{\alpha_n} (t_n) \cdots P^1_{\alpha_1} (t_1)\, .
\label{threeonetwoa}
\end{equation}
These $\{C_{\bar\alpha}\}$ may sometimes be representable as chains of
projections (as when the sum is over alternatives at just one time).
However, they will not {\it generally } be chains of projections. The
general operator corresponding to a coarse-grained history will thus be
a class operator of the form \eqref{threeonetwoa}.

In a similar manner one can define operations of fine-graining.  For
example, introducing a set of alternatives at a time when there was none
before is an operation of fine-graining as is splitting the projections of
an existing set at one time into more
 mutually orthogonal ones.  Continued fine-graining would eventually
result in a completely fine-grained set of histories. All coarse-grained
sets of histories are therefore coarse grainings of at least one
fine-grained set.

Sets of histories are partially ordered by the operations of coarse
graining and fine graining.  For any pair of sets of histories, the
least coarse-grained set, of which they are both fine grainings, can be
defined.  However, there is not, in general, a unique fine-grained set
of which they are both a coarse graining. There is an operation of
``join'' but not of ``meet''.

So far we have considered histories defined  by sets of alternatives at
sequences of times that are independent of one another. For realistic
situations we are interested in sets of histories in which (assuming
causality) the {\it set} of alternatives and their times
are dependent on the
{\it particular}
alternatives and {\it particular} times
that define the history at earlier times. Such sets of
histories are said to be {\it branch dependent}. A more complete notation
would be to write:
\begin{equation}
P^n_{\alpha_n} \left(t_n; \alpha_{n-1}, t_{n-1}, \cdots,
\alpha_1, t_1
\right)\, P^{n-1}_{\alpha_{n-1}} \left(t_{n-1}; \alpha_{n-2}, t_{n-2},
 \cdots,
\alpha_1, t_1\right) \cdots P^1_{\alpha_1}(t_1) \, ,
\label{threeonethree}
\end{equation}
for histories represented by chains of such projections. Here
$\{P^k_{\alpha_k}(t_k; \alpha_{k-1}, t_{k-1},
 \cdots$, 
 $\alpha_1, t_1)\}$
are an exhaustive set of orthogonal projection operators as $\alpha_k$
varies, keeping $\alpha_{k-1}, t_{k-1},
\cdots,\alpha_1, t_1$ fixed. Nothing more than
replacing chains in \eqref{threeonetwoa} by \eqref{threeonethree} is needed
to complete the generalization to branch dependent histories.

Branch dependence is important, for example, in describing realistic
quasiclassical domains because past events may determine what is a
suitable quasiclassical variable. For instance, if a quantum fluctuation
gets amplified so that a galaxy condenses in one branch and no such
condensation occurs in other branches, then what are suitable
quasiclassical variables in the region where the galaxy would form is
branch dependent. While branch dependent sets of histories are clearly
important for a description of realistic quasiclassical domains, we
shall not make much use of them in these lectures devoted to general
frameworks and frequently use the notation in \eqref{threeonetwoa} as an
abbreviation for the more precise \eqref{threeonethree}.

\subsubsection{The Decoherence Functional}

Quantum mechanical interference between individual histories in a
coarse-grained set is measured by a {\it decoherence functional}.
This is a
complex-valued functional on pairs of histories in a coarse-grained set
depending on the initial condition of the closed system.
If $c_{\alpha^\prime}$ and $c_\alpha$ are a pair of histories,
$C_{\alpha^\prime}, C_\alpha$ are the corresponding operators
as in \eqref{threeonetwoa} and $\rho$ is a Heisenberg picture density
matrix representing the initial condition,
 then the decoherence functional is defined
by \cite{GH90a}
\begin{equation}
D\left(\alpha^\prime, \alpha\right) = Tr\,\bigl[C_{\alpha^\prime} \rho
C^\dagger_\alpha\bigr]\, .
\label{threeonefour}
\end{equation}

Sufficient conditions for probability sum rules can be defined in terms
of the decoherence functional.  For example, the condition that
generalizes the orthogonality of the branches discussed in Section II for
pure initial states is the {\it medium decoherence} condition that the
``off-diagonal'' elements of $D$ vanish, that is
\begin{equation}
D\left(\alpha^\prime, \alpha\right) \approx  0\quad , \quad \alpha^\prime \not=
\alpha \, .
\label{threeonefive}
\end{equation}
It is easy to see that \eqref{threeonefive} reduces to \eqref{foursix} 
when $\rho$ is
pure, $\rho = |\Psi\rangle\langle\Psi|$, and the $C$'s are chains of
projections.

The probabilities $p(\alpha)$ for the individual histories in a
decohering set are the diagonal elements of the decoherence functional
so that the condition for medium decoherence and the definition of
probabilities may be summarized in one compact fundamental formula:
\begin{equation}
D\left(\alpha^\prime, \alpha\right) \approx \delta_{\alpha^\prime\alpha} p
(\alpha)\, . 
\label{threeonesix}
\end{equation}
The decoherence condition \eqref{threeonefive} is easily seen to be a 
sufficient
condition for the most general probability sum rules.  Unions of
histories that are again chains of projections give coarser-grained
histories.
The corresponding probability sum rules are the requirements that the
probabilities of the coarser-grained histories are the sums of the
individual histories they contain.  More precisely let $\{c_\alpha\}$ be a
set of histories and $\{c_{\bar\alpha}\}$ {\it any} coarse graining of it.
We require
\begin{equation}
p\left(\bar\alpha\right) \approx \sum\limits_{\alpha\epsilon\bar\alpha} p
(\alpha)\, . 
\label{threeoneseven}
\end{equation}
This can be established directly from the condition of medium
decoherence. The chains for the coarser-grained set $\{c_{\bar\alpha}\}$ are
related to the chains for $\{c_\alpha\}$ by
\begin{equation}
C_{\bar\alpha} = \sum\limits_{\alpha\epsilon\bar\alpha} C_\alpha\, .
\label{threeoneeight}
\end{equation}
Evidently, as a consequence of \eqref{threeonesix},
\begin{equation}
p\left(\bar\alpha\right)  = Tr\,\bigl[C_{\bar\alpha} \rho
C^\dagger_{\bar\alpha}\bigr]
  = \sum\limits_{\alpha^\prime\epsilon\bar\alpha}
\sum\limits_{\alpha\epsilon\bar\alpha} Tr\,\bigl[C_{\alpha^\prime} \rho
C^\dagger_\alpha\bigr]
  \approx  \sum\limits_{\alpha\epsilon\bar\alpha} Tr\,\bigl[C_\alpha \rho
C^\dagger_\alpha\bigr] = \sum\nolimits_\alpha p (\alpha)\, .
\label{threeonenine}
\end{equation}
which establishes the sum rule.

Medium decoherence is not a necessary condition for the probability sum
rules.  The weaker necessary condition is the {\it weak
decoherence condition}.
\begin{equation}
Re\, D\left(\alpha^\prime, \alpha\right) \approx \delta_{\alpha^\prime\alpha}
p (\alpha)\, . 
\label{threeoneten}
\end{equation}
To see this, note that the simplest operation of coarse graining is to
combine just two histories according to the logical operation ``or'' as
represented in \eqref{threeoneonea}.  Write out \eqref{threeonenine} 
to see that
the probability that the system follows one or the other history is the
sum of the probabilities of the two histories if and only if the sum of
the interference terms represented by \eqref{threeoneten} vanishes. Applied
to all pairs of histories this argument yields the weak decoherence
condition.
However,
realistic mechanisms of decoherence such as those
illustrated in Section II.5 seem to imply medium decoherence (see also
Section III.3.2)  and for
concrete problems such as characterizing quasiclassical domains we shall
employ this stronger condition.

\subsubsection{Prediction, Retrodiction, and States}

We mentioned that  considering conditional probabilities
based on known information
is one strategy for identifying definite predictions with
probabilities near zero or one. We shall now consider the construction of
these conditional probabilities in more detail.  Suppose that we are
concerned with a decohering set of coarse-grained histories
that consist
of sequences of alternatives $\alpha_1, \cdots, \alpha_n$ at definite
moments of time $t_1, \cdots, t_n$ and whose individual histories are
therefore represented by class operators $C_\alpha$ which are
chains of the corresponding projections [and
not sums of such chains as in \eqref{threeonetwoa}].  The joint probabilities
of these histories,
$p(\alpha_n, \cdots, \alpha_1)$, are given by the
fundamental formula \eqref{threeonesix}.  Let us consider the various 
conditional
probabilities that can be constructed from them.

The probability for predicting a future sequence of alternatives
$\alpha_{k+1}, \cdots, \alpha_n$ given that alternatives $\alpha_1,
\cdots, \alpha_k$ have already happened up to time $t_k$ is
\begin{equation}
p\bigl(\alpha_n, \cdots, \alpha_{k+1}\big|\,\alpha_k, \cdots,
\alpha_1\big) = \frac{p\left(\alpha_n,\cdots,
\alpha_1\right)}{p\left(\alpha_k,\cdots, \alpha_1\right)}
\label{threeoneeleven}
\end{equation}
where $p(\alpha_k, \cdots, \alpha_1)$ can be calculated either directly
from the fundamental formula or as
\begin{equation}
p\left(\alpha_k, \cdots, \alpha_1\right) = \sum\limits_{\alpha_n,
\cdots, \alpha_{k+1}} p\,\left(\alpha_n, \cdots, \alpha_1\right)
\, . 
\label{threeonetwelve}
\end{equation}
These alternative computations are consistent because decoherence
implies the probability sum rule \eqref{threeonetwelve}.

If the known information at time $t_k$ just consists of alternative
values of present data then the probabilities for future prediction are
conditioned just on the values of this data, \viz
\begin{equation}
p\,\bigl(\alpha_n, \cdots, \alpha_{k+1}\big|\,\alpha_k\bigr) =
\frac{p\left(\alpha_n, \cdots, \alpha_k\right)}{p\left(\alpha_k\right)}
\, . \label{threeonethirteen}
\end{equation}
similarly the probability that alternatives $\alpha_1, \cdots,
\alpha_{k-1}$ happened in the past given present data $\alpha_k$ is
\begin{equation}
p\,\bigl(\alpha_{k-1}, \cdots, \alpha_1 \big | \alpha_k\bigr)
= \frac{p\,\left(\alpha_k, \cdots,
\alpha_1\right)}{p\left(\alpha_k\right)}\, . 
\label{threeonefourteen}
\end{equation}
It is through the evaluation of such conditional probabilities that
history is most honestly reconstructed in quantum mechanics.  We say
that particular alternatives $\alpha_1, \cdots, \alpha_k$ {\it happened}
in the past when the conditional probability \eqref{threeonefourteen} is near
unity for those alternatives given our present data.  The present data
$\alpha_k$ are then said to be good records of the past events
$\alpha_1, \cdots, \alpha_{k-1}$.

Future predictions can be obtained from an effective density matrix in
the present that summarizes what has happened.  If $\rho_{\rm eff}
(t_k)$ is defined by
\begin{equation}
\rho_{\rm eff}(t_k) = \frac{P^k_{\alpha_k}(t_k) \cdots
P^1_{\alpha_1}(t_1)\,\rho\, P^1_{\alpha_1}(t_1) \cdots P^k_{\alpha_k}
(t_k)}{Tr\left[P^k_{\alpha_k}(t_k) \cdots P^1_{\alpha_1}(t_1)\,\rho
\, P^1_{\alpha_1}(t_1) \cdots P^k_{\alpha_k}(t_k)\right]}
\label{threeonefifteen}
\end{equation}
then
\begin{equation}
p\bigl(\alpha_n, \cdots, \alpha_{k+1}\big|\alpha_k, \cdots,
\alpha_1\bigr) = Tr\,\left[P^n_{\alpha_n}(t_n) \cdots
P^{k+1}_{\alpha_{k+1}}(t_{k+1})\, \rho_{\rm eff}(t_k)\,
P^{k+1}_{\alpha_{k+1}}(t_{k+1}) \cdots P^n_{\alpha_n}(t_n)\right]\ .
\label{threeonesixteen}
\end{equation}
This effective density matrix represents the usual notion of
``state-of-the-system at the moment of time $t_k$''.

The effective
density matrix may be thought of as evolving in time in the following
way: Define it to be constant between the projections at $t_k$ and
$t_{k+1}$ in this Heisenberg picture.
Its Schr\"odinger
picture representative
\begin{equation}
e^{-iH(t-t_k)/\hbar}\, \rho_{\rm eff} (t_k) e^{iH(t-t_k)/\hbar}
\label{threeoneseventeen}
\end{equation}
then evolves unitarily between $t_k$ and $t_{k+1}$. At $t_{k+1}$, $\rho_{\rm
eff} (t)$, is ``reduced'' by the action of the projection
$P^{k+1}_{\alpha_{k+1}} (t_{k+1})$.
It then evolves unitarily to the time of the
next projection.  The action of the projections in this picture is the
notorious ``reduction of the wave packet''.  In this quantum mechanics
of a closed system it is not necessarily associated with a measurement
situation but is merely part of the description of
histories.\footnote{For further discussion see, \cite{Har91a}
(Appendix) and \cite{Har93b}.} If we consider alternatives that are sums
of chains of projections, or the spacetime generalizations of
Hamiltonian quantum mechanics to be discussed in subsequent sections, it
is not possible to summarize prediction by an effective density matrix
that evolves in time.

In contrast to probabilities for the future, there is no effective
density matrix representing present information from which probabilities
for the past can be derived.  As \eqref{threeonefourteen} shows, probabilities
for the past require {\it both} present records {\it and}
the initial condition of the system.  In this respect the quantum
mechanical notion of state at a moment of time is different from the
classical notion which is sufficient to specify {\it both} future and past.
This is an aspect of the arrow of time in quantum mechanics which we
shall discuss more fully in the Section IV.7.

\subsubsection{The Decoherence Functional in Path Integral Form}

Feynman's path integral provides a useful alternative representation of
unitary quantum dynamics for certain systems.  These are characterized
by a configuration space spanned by generalized coordinates $\{q^i\}$
and a Hilbert space of square-integrable functions on this configuration
space.  The path integral can also be used to represent the ``second law
of evolution'' --- that is the action of chains of projection operators
--- for alternatives that consist entirely of projections onto
alternative ranges $\{\Delta^k_\alpha(t_k\}$ of the $q$'s at a sequence
of times $t_1, \cdots, t_n$.
The key identity in establishing this representation is the following
\cite{Cav86, Sta86}:
\begin{equation}
\left\langle q_f T\big| P_{\Delta_n}(t_n) \cdots
P_{\Delta_1}(t_1)\big| q_00\right\rangle =
\int\nolimits_{\left[q_0\Delta_1 \cdots \Delta_n q_f\right]} \delta q\,
e^{iS\left[q(\tau)\right]/\hbar}
\label{threeoneeighteen}
\end{equation}
where we have omitted coordinate indices.
On the left is the matrix element of Heisenberg projections at times
$t_1 \cdots, t_n$ onto ranges of the $q's$ $\Delta_1, \cdots, \Delta_n$ taken
between
localized Heisenberg states at initial and final times $0$ and $T$.
On the right is a path integral over all paths that begin at $q_0$ at
time $0$, pass through the ranges $\Delta_1, \cdots, \Delta_n$ at times
$t_1, \cdots t_n$ respectively and end at $q_f$ at time $T$.  To see
how to prove \eqref{threeoneeighteen} consider just one interval
$\Delta_k$ at time
$t_k$.  The matrix element on the left of \eqref{threeoneeighteen}
 may be further expanded
as
\begin{equation}
\left\langle q_f T |P_{\Delta_k} (t_k) | q_0 0 \right\rangle
 = \int\nolimits_{\Delta_k}
dq_k
\left\langle q_f T | q_k t_k\right\rangle
\ \left\langle  q_k t_k | q_0 0 \right\rangle\, .
\label{threeonenineteen}
\end{equation}
Since the paths cross the surface of time $t_k$ at a single point $q_k$,
the sum
on the right of \eqref{threeoneeighteen} may be factored as shown in Figure 5,
\[
\int\nolimits_{\left[q_0\Delta_k q_f\right]} \delta q\ e^{iS [q
(\tau)]/\hbar} =
\int\nolimits_{\Delta_k} dq_k\left(\int\nolimits_{[q_k q_f]} \delta q
e^{iS [q (\tau)]/\hbar}\right)
\]
\begin{equation}
\times \left(\int\nolimits_{[q_0 q_k]} \delta q e^{iS
[q (\tau)]/\hbar}\right)\, .
\label{threeonetwenty}
\end{equation}
But, it is an elementary calculation to verify that
\begin{equation}
\left\langle q^{\prime\prime}t^{\prime\prime}|q^\prime t^\prime
\right\rangle =
\int_{\left[q^\prime
q^{\prime\prime}\right]} \delta q e^{iS [q(\tau)]/\hbar}
\label{threeonetwentyone}
\end{equation}
and that inverting the time order on the right is the same as complex
conjugation.  Thus, \eqref{threeonenineteen} is true and, by extension,
also the equality \eqref{threeoneeighteen}.

\begin{figure}[t]
\begin{center}
\includegraphics[width=6in]{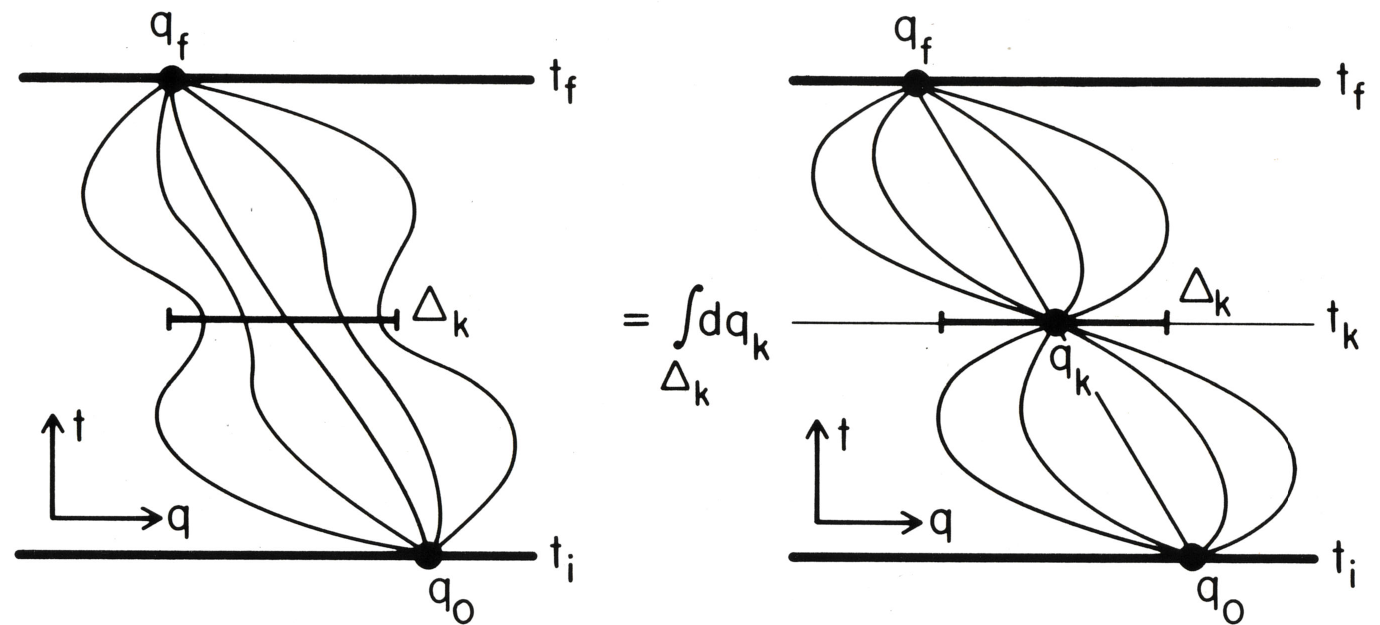}
\caption{Factoring a sum over paths
single-valued in time across a
surface of constant time.  Shown at left is the sum over paths defining
the
amplitude to start
from $q_0$ at time $t=0$, proceed through interval $\Delta_k$ at time
$t_k$, and
wind up at $q_f$ at time $T$.  If the histories are such that each
path
intersects each surface of constant time once and only once, then the
sum on
the left can be factored as indicated at right. The factored sum
consists of a
sum over paths before time $t_k$, a sum
over paths after time $t_k$, followed by a sum over the values of $q_k$
at time
$t_k$ inside the interval $\Delta_k$.  The possibility of this
factorization
is what allows the Hamiltonian form of quantum mechanics to be derived
from
 a sum-over-histories formulation.  The sums over paths before and after
$t_k$
define wave functions on that time-slice and the integration over $q_k$
defines
their inner product.  The notion of state at a moment of time and the
Hilbert
space of such states is thus recovered.\\
If the sum on the left were over paths that were multiple valued in
time,
the factorization on the right would not be possible.}
\end{center}
\end{figure}

Using this identity, the decoherence functional may be rewritten in path
integral form for coarse grainings defined by ranges of configuration
space.  Let $\alpha$ denote the history corresponding to the sequence of
ranges $\Delta^1_{\alpha_1}, \cdots, \Delta^n_{\alpha_n}$ at times $t_1,
\cdots, t_n$ and let $C_\alpha$ denote the corresponding chain of
projections.  The decoherence functional can be written
\[
D\bigl(\alpha^\prime, \alpha\bigr) = Tr\, \bigl(C_{\alpha^\prime} \rho
C^\dagger_\alpha\bigr) = \int \, dq^\prime_f \int\, dq_f \int\,
dq^\prime_0 \int \, dq_0
\]
\begin{equation}
\times\delta\bigl(q^\prime_f - q_f\bigr)\,
\bigl\langle q^\prime_f T \bigl| C_{\alpha^\prime} \bigr| q^\prime_0 0
\bigr\rangle
 \bigl\langle q^\prime_0 0 \bigl| \rho \bigr| q_0 0
\bigr\rangle\, \bigl\langle q_00\bigl | C^\dagger_\alpha \bigr| q_f
T \bigr\rangle \, , \label{threeonetwentytwo}
\end{equation}
where we have suppressed the indices on the $q$'s and written $dq$ for
the volume element in configuration space.  This can be rewritten using
\eqref{threeoneeighteen} as
\begin{equation}
D\left(\alpha^\prime, \alpha\right) = \int\nolimits_{\alpha^\prime}
\delta q^\prime \int\nolimits_\alpha \delta q\, \delta \bigl(q^\prime_f
- q_f\bigr)\, \exp\left\{i\left(S\left[q^\prime(\tau)\right] -
S\left[q(\tau)\right]\right)\right\}\, \rho \left(q^\prime_0,
q_0\right)\, . \label{threeonetwentythree}
\end{equation}
Here, the integrals are over paths $q^i(\tau)$ that begin at $q_0$ at
time $0$, pass through the regions $\alpha = (\alpha_1, \cdots,
\alpha_n)$ and end at $q_f$ at time $T$.  The integral over
$q^{{\prime}i}(\tau)$ is similar but restricted by the coarse graining
$\alpha^\prime$.  We have written $\rho(q^\prime_0, q_0)$ for the
configuration space matrix elements of the initial density matrix
$\rho$.  This expression allows us to identify the decoherence
functional for the completely fine-grained set of histories specified by
paths $\{q^i(t)\}$ on the interval $t=0$ to $t=T$ as
\begin{equation}
D\left[q^\prime(\tau), q(\tau)\right] = \delta\bigl(q^\prime_f -
q_f\bigr) \exp \left\{i\left(S\left[q^\prime(\tau)\right] -
S[q(\tau)]\right)/\hbar\right\}\, \rho \left(q^\prime_0, q_0\right)
\, . \label{threeonetwentyfour}
\end{equation}
Evidently the set of fine-grained histories defined by coordinates does not
decohere.  In Section
III.3
we will discuss  models in which suitable coarse
grainings of these histories do decohere.

\subsection{The Emsch Model}

To make the formalism we have introduced more concrete we shall
illustrate it with a few tractable models.  The first of these, the
Emsch model,  is not
very realistic but has the complementary virtue of being exactly
solvable.  It will chiefly serve to illustrate the notation in a
concrete case.

We consider the quantum theory of a particle moving in one-dimension
whose Hilbert space is ${\cal H}=L_2({\bf R})$. The simplifying
feature of the model is its Hamiltonian.  This we take to be linear in
the momentum
\begin{equation}
H=v p\label{threetwoone}
\end{equation}
where $v$ is a constant with the dimensions of velocity.

We consider coarse grainings that at times $t_1, \cdots, t_n$ divide
the real line up into exhaustive sets of intervals
$\{\Delta^k_{\alpha_k}\}, k=1, \cdots, n$.  The index $k$ allows
different sets of intervals to be used at different times.  In each set,
$\alpha_k$ is an integer that ranges over the possible intervals.

In the Schr\"odinger picture the alternative that the particle is in a
particular interval $\Delta^k_{\alpha_k}$ is represented by the
projection operator
\begin{equation}
P^k_{\alpha_k} = \int\nolimits_{\Delta^k_{\alpha_k}} dx|
x \rangle\langle x |\, .
\label{threetwotwo}
\end{equation}
The corresponding Heisenberg picture projections are, of course,
\begin{equation}
P^k_{\alpha_k} (t_k) = e^{iHt_k/\hbar} \, P^k_{\alpha_k} e^{-iHt_k/\hbar}
\label{threetwothree}
\end{equation}
where $H$ is given by \eqref{threetwoone}.  With the Hamiltonian
\eqref{threetwoone},  the action of the unitary
evolution operators in \eqref{threetwothree} are equivalent to spatial
translations by a distance $vt_k$. We therefore have
\begin{equation}
P^k_{\alpha_k} (t_k) = P^{(vt_k)k}_{\alpha_k} 
\label{threetwofour}
\end{equation}
where $P^{(vt_k)k}_{\alpha_k}$ denotes the Schr\"odinger picture
projection on the $\alpha_k$th interval in the set $k$ translated by a
distance $(vt_k)$. The $P^{(vt_k)k}_{\alpha_k}$ thus all commute.

Chains of projections corresponding to histories are
\begin{equation}
C_\alpha = P^{(vt_n)n}_{\alpha_n}\cdots P^{(vt_1)1}_{\alpha_1}
\label{threetwofive}
\end{equation}
and are thus themselves projections onto the interval $\alpha$ which is
the intersection of the intervals $\alpha_1,\cdots, \alpha_n$ in the
translated sets.  When $n$ is large and the coarse grainings are
reasonably fine, many of the $C$'s will vanish identically.  The
non-vanishing $C$'s are projections onto disjoint intervals in $x$.  As
a consequence we have, since $C^\dagger_\alpha=C_\alpha$,
\begin{equation}
C^\dagger_{\alpha^\prime} C_\alpha =
\delta_{\alpha^\prime\alpha} C_\alpha\, . 
\label{threetwosix}
\end{equation}
Decoherence, as defined by \eqref{threeonesix}, is thus {\it
exact} and automatic for these histories whatever the initial $\rho$.
The probabilities of individual histories \eqref{threeonesix} may be written
\begin{equation}
p(\alpha) = Tr\left(C_\alpha\rho\right)\, . 
\label{threetwoseven}
\end{equation}
Evidently all the probability sum rules are satisfied because of the
linearity of \eqref{threetwoseven} in $C_\alpha$.  The property of exact
decoherence independent of initial state, is, of course, neither
general nor realistic.  It is a special consequence of the Hamiltonian
\eqref{threetwoone}.

Parenthetically, we note that if the histories are refined by adding
further partitions of ${\bf R}$ at more and more times, the
non-vanishing $C_\alpha$ will generically project onto smaller and
smaller intervals of ${\bf R}$.  If the initial density matrix is pure,
$\rho=|\Psi\rangle\langle\Psi|$, the non-vanishing vectors
\begin{equation}
C_\alpha|\Psi\rangle \label{threetwoeight}
\end{equation}
tend to a dense, orthogonal set in ${\cal H}$.  This has been called a
full set of histories \cite{GH90b}.

Suppose the intervals defining the coarse graining
 are of equal length $\Delta$.  A point $x$ in
${\bf R}$ may be located by the number, $\alpha$, of its interval and
its relative coordinate $\xi$ within that interval
\begin{equation}
x = \Delta (\alpha + \xi)\ ,\ -1\leq\xi\leq 1\, . 
\label{threetwonine}
\end{equation}
Correspondingly the Hilbert space ${\cal H}$ may be factored into ${\cal
H}^{(\alpha)}\otimes {\cal H}^{(\xi)}$ where ${\cal H}^{(\xi)}$ is
$L_2(-1, 1)$ --- the space of square integrable functions on the
interval defined by the range of $\xi$ --- and ${\cal H}^{(\alpha)}$ is
the space of square summable functions  of the integers.
Thus coarse grainings by {\it equal} intervals may be described as
distinguishing which interval the particle is in while ignoring the
relative position within the interval.
A similar factorization can be exhibited when the intervals are of
unequal length, but the relevant variables are not simple linear
functions of the basic coordinates.

\subsection{Linear Oscillator Models}

\subsubsection{Specification}

A useful class of models, in which the decoherence of histories can be
explored analytically, are the linear oscillator models. These have been
 studied from the
point of view of histories by
Feynman and Vernon \cite{FV63}, Caldeira and Leggett \cite{CL83}, Unruh
and Zurek \cite{UZ89}, Dowker and Halliwell \cite{DH92},
Gell-Mann and Hartle \cite{GH93}, and many
others. The simplest model consists of a distinguished oscillator moving in one
dimension and interacting linearly with a large number of other
independent oscillators.  The models are studied with coarse grainings
that follow the coordinates of the distinguished oscillator and ignore
all the rest.  An initial condition is
assumed whose density matrix factors into an arbitrary density matrix
for the distinguished oscillator and a thermal density matrix at
temperature $T_B$ for the rest.  The model thus captures in the most
elementary way the idea of a system interacting with a bath of other
systems that can carry away phases and effect decoherence.  The model is
soluble because the linearity of the interactions, and the thermal
nature of the bath, mean that the trace in the decoherence functional
can be reduced to Gaussian functional integrals and evaluated explicitly.
We now show how to do this.

To define the model more precisely let $x$ denote the coordinate of the
distinguished oscillator and $Q_k$ the coordinates of the rest.  The
Hamiltonian of the distinguished oscillator is
\begin{equation}
H_{\rm free} (p, x) = \frac{1}{2M}\left(p^2+\omega^2x^2\right)
\label{threethreeone}
\end{equation}
and
\begin{equation}
H_0 = \sum\nolimits_k H_k = \frac{1}{2m}\ \sum\nolimits_k \left(P^2_k +
\omega^2_k Q^2_k\right) \label{threethreetwo}
\end{equation}
for the rest. The interaction is linear
\begin{equation}
H_{\rm int} \left(x, Q_k\right) = x\, \sum\nolimits_k C_k Q_k
\label{threethreethree}
\end{equation}
defining coupling constants $C_k$. The initial density matrix is assumed
to be of the form
\begin{equation}
\bigl\langle x^\prime, Q^\prime_k\bigl| \rho\bigr| x, Q_k\bigr\rangle
= \bar\rho \bigl(x^\prime, x\bigr)\rho_B\bigl(Q^\prime_k, Q_k\bigr)
\label{threethreefour}
\end{equation}
where $\rho_B(Q^\prime_k, Q_k)$ is a product of thermal density matrix
$\rho^\beta_k(Q^\prime_k, Q_k)$ for each oscillator in the bath all at
one temperature $T_B =1 /(k\beta)$.  Explicitly the $\rho^\beta_k$ have
the form
\[
\rho^\beta_k \left(Q_k^\prime, Q_k\right) = \bigl\langle
Q^\prime_k\bigl|e^{-\beta H_k}\bigr|Q_k\bigr\rangle\,/\, Tr
\,\bigl(e^{-\beta H_k}\bigr) =
 \left[\frac{m\omega_k}{\pi \hbar}\ \tanh
\ \left(\frac{\hbar\omega_k\beta}{2}\right)\right]^\half
\]
\begin{equation}
 \times \exp\biggl[-
\biggl\{\frac{m\omega_k}{2\hbar
\sinh\,\left(\hbar\beta\omega_k\right)}
\left[\left(Q^{\prime2}_k + Q^2_k\right)\,
\cosh\,\left(\hbar\beta\omega_k\right) - 2Q^\prime_k
Q_k\right]\biggr\}\biggr]\, .
\label{threethreefive}
\end{equation}
It is the quadratic form of the exponent in this expression, together with the
quadratic
actions that correspond to the Hamiltonians
\eqref{threethreeone} -- \eqref{threethreethree},
that make the model explicitly soluble.

\subsubsection{The Influence Phase and Decoherence}

We consider a special class of coarse grainings that follow the
coordinate $x(t)$ of the distinguished oscillator over a time interval
$[0,T]$ and ignore the coordinates $Q_k(t)$ of the rest.  As this
model has a configuration space description with coordinates $q^i=
\left(x,Q_k\right)$, the decoherence functional for these coarse
grainings is conveniently computed in its sum-over-histories form.  From
\eqref{threeonetwentythree} we have
\[
D\left[x^\prime(\tau), x(\tau)\right] = \delta\bigl(x^\prime_f -
x_f\bigr)
\]
\begin{equation}
\times \exp\biggl\{i\biggl(S_{\rm free} \left[x^\prime
(\tau)\right]
-S_{\rm free} [x(\tau)]
+ W\left[x^\prime(\tau)\, ,
x(\tau)\right]\biggr)/\hbar\biggr\} \bar\rho \left(x^\prime_0, x_0\right)
\label{threethreesix}
\end{equation}
where $W$ is defined by
\[
\exp  \bigl(iW\left[x^\prime (\tau), x(\tau)\right]\bigr) \equiv \int
\delta Q^\prime\int \delta Q\ \delta\bigl(Q^\prime_f - Q_f\bigr)
\]
\begin{equation}
\times\exp  \biggl\{i\biggl(S_0\left[Q^\prime(\tau)\right] + S_{\rm int}
\left[x^\prime(\tau), Q^\prime(\tau)\right] - S_0 [Q(\tau)]
 - S_{\rm int}[x(\tau), Q(\tau)]\biggr)/\hbar\biggr\}\,\rho_B
\left(Q^\prime_0, Q_0\right)\, . 
\label{threethreeseven}
\end{equation}
In these expressions, $S_{\rm free}$, $S_0$, and $S_{\rm int}$ are the
actions corresponding to the Hamiltonians $H_{\rm free}$, $H_0$, and
$H_{\rm int}$.  The functional $W[x^\prime(\tau), x(\tau)]$ is called
the Feynman-Vernon influence phase and summarizes for the behavior of
the distinguished oscillator all information about the rest.

The important point about the model is that, since the $Q$'s are not
restricted by the coarse grainings, the integrations defining the
influence phase in \eqref{threethreeseven} are over a complete range.  
Since the
actions are quadratic in the $Q$'s, and since $\rho_B$ is the
exponential of a quadratic form, all the integrations can be carried out
explicitly.  The resulting influence phase is necessarily a quadratic
functional of the $x^\prime(\tau)$ and $x(t)$.  It has the form
\[
W \left[x^\prime(\tau), x(\tau)\right] =
{1\over 2}\int^T_0 dt \int^t_0
 dt^\prime\left[x^\prime(t)-x(t)\right]^{\dagger}
\bigl\{k_R\left(t,t^\prime\right)\left[x^\prime\left(t^{\prime}\right)+
x\left(t^\prime\right)\right]
\]
\begin{equation}
+ ik_I\left(t,t^\prime\right)\left[x^\prime(t^\prime) -
x(t^\prime)\right]\bigr\}\, .
\label{threethreeeight}
\end{equation}
General arguments of symmetry and quantum mechanical causality are
enough to show that $W$ has this form \cite{FV63, FH65},
but in the present case it also
follows from explicit computation which shows the kernels to be
\cite{FV63, CL83}:
\begin{eqnarray}
k_R\left(t,t^\prime\right) & = & -\sum_k\ \frac{C^2_k}{m\omega_k}\ \sin
\left[\omega_k
\left(t-t^\prime\right)\right]\, ,  \label{threethreenine}\\
k_I \left(t,t^\prime\right)& = \sum_k\ \frac{C^2_k}{m\omega_k} \coth
\left(\half
\hbar\beta\omega_k\right)\cos\left[\omega_k\left(t-t^\prime\right)\right]
\, . \label{threethreeten}
\end{eqnarray}
The imaginary part of the influence phase effects decoherence.  To see
this define, $\xi(t) = x^\prime(t)-x(t)$, and write
\begin{equation}
Im\,W\left[x^\prime(\tau), x(\tau)\right] = \sum_k \frac{C^2_k}{4m\omega_k}
\ \coth\left(\half\,\hbar\beta\omega_k\right)\int\nolimits^T_0 dt
\int\nolimits^T_0 dt^\prime
\xi(t)\,\cos\left[\omega_k(t-t^\prime)\right]\,\xi(t^\prime)\ .
\label{threethreeeleven}
\end{equation}
Alternatively, defining
\begin{equation}
\tilde\xi(\omega) = \int\nolimits^T_0 dt\ e^{i\omega t}\xi(t)
\label{threethreetwelve}
\end{equation}
we have
\begin{equation}
Im\, W\left[x^\prime(\tau), x(\tau)\right] = \sum\nolimits_k
\ \frac{C^2_k}{4m\omega_k}\ \coth\left(\half\hbar\beta\omega_k\right)
\left| \tilde\xi \left(\omega_k\right) \right|^2\, ,
\label{threethreethirteen}
\end{equation}
showing that $Im\,W$ is strictly positive.  What either
\eqref{threethreeeleven}
or \eqref{threethreethirteen} show is that, as $\xi(t)$, the difference 
between the
fine-grained histories $x^\prime(t)$ and $x(t)$, becomes large, the
corresponding ``off-diagonal'' elements of the decoherence functional
are increasingly exponentially suppressed.  This is the source of
decoherence in further coarse grainings of $x$.

For sets of histories of the distinguished oscillator that are coarse
grained by exhaustive sets of intervals of $x$, $\{\Delta^k_{\alpha_k}
\}$, at times $\{t_k\}$, the decoherence functional is given by
\begin{equation}
D\left(\alpha^\prime, \alpha\right) = \int\nolimits_{\alpha^\prime}
\delta x^\prime \int_{\alpha} \delta x\ D\left[x^\prime(\tau),
x(\tau)\right] 
\label{threethreefourteen}
\end{equation}
where $\alpha$ is a chain of particular intervals $(\alpha_1, \cdots,
\alpha_n)$ and the integrals are over the paths on the time-range
$[0,T]$ that pass through those intervals. This set of alternatives will
decohere provided that the characteristic size of the intervals in the
sets $\{\Delta^k_{\alpha_k}\}$ and the spacing between these sets in
time are both large enough that sufficient $Im\,W$ is built up to
suppress all of the off-diagonal elements of $D(\alpha^\prime, \alpha)$.

A simple criterion for decoherence can be given in the important case of
a cutoff continuum of oscillators with density of states
$\rho_D(\omega)$ and couplings
\begin{equation}
\rho_D(\omega) C^2(\omega) = \left\{\begin{array}{lr}
                            4Mm\gamma\omega^2/\pi, & \omega <
                                               \Omega,\\
                            0\qquad\qquad, & \omega > \Omega,
				\end{array} \right.
\label{threethreefifteen}
\end{equation}
where $\gamma$ is an effective coupling strength.  In the Fokker-Planck
limit, $kT>>\hbar\Omega >> 0$, the imaginary part of the influence phase
becomes purely local in time, \viz
\begin{equation}
Im\, W\left[x^\prime(\tau), x(\tau)\right] = \frac{2M\gamma
kT_B}{\hbar}\ \int\nolimits^T_0 dt\,\xi^2(t)\, .
\label{threethreesixteen}
\end{equation}
Then, if the characteristic size of the intervals in the sets
$\{\Delta^k_{\alpha_k}\}$ is $d$, this set of histories will decohere
provided the sets are spaced in time by intervals longer than
\begin{equation}
t_{\rm decoherence} \sim
\frac{1}{\gamma}\,\left[\frac{\hbar}{\sqrt{2MkT_B}}\cdot
\left(\frac{1}{d}\right)\right]^2\, . 
\label{threetwentyseven}
\end{equation}
As stressed by Zurek \cite{Zur84}, for typical ``macroscopic'' parameters
this minimum time for decoherence is many orders of magnitude smaller
than characteristic dynamical times, for example $1/\gamma$.  For $M\sim
1$gm, $T_B \sim 300^\circ$K, $d\sim$cm the ratio is around $10^{40}$!
Decoherence in the realistic situations approximated by these models is very
effective.

\subsection{The Emergence of a Quasiclassical Domain}

As discussed in Section II.7 the quasiclassical domain of familiar
experience is a set of decohering, coarse-grained alternative histories
of the universe (or a class of roughly equivalent sets) that is
maximally refined consistent  with decoherence, is coarse-grained mostly
by values of a small class of quasiclassical variables at different
times, and exhibits a high degree of deterministic correlations among
these variables in time.

Providing a satisfactory criterion that would differentiate among all
possible decohering sets of coarse-grained histories of a closed system
by their degree of classicality is, at the time of writing, still an
unsolved problem. Such a criterion would enable us to derive (rather
than posit) the habitually decohering variables that characterize the
quasiclassical domain of everyday experience.  Such a criterion would
enable us to determine whether that quasiclassical domain is essentially
unique or but one of a number of essentially different possibilities
exhibited by the initial condition of the universe and its dynamics.

Whatever the exact nature of such a general criterion, or even whether
it exists, one feature of the description of quasiclassical behavior
cannot be stressed too strongly: Classical, deterministic behavior of a
quantum mechanical system is defined in terms of the probabilities of
its time histories.  The statement that the moon moves in an orbit that
obeys Newton's laws of motion is the quantum-mechanical statement that
successive determinations of the position of the moon are correlated in
time according to Newton's laws with a probability near unity.  More
precisely, a set of decohering, alternative, coarse-grained histories
defined by ranges of position of the moon's center of mass at a
succession of times exhibits classical behavior if the
probabilities are low for those histories where the positions are not
correlated in time by Newton's law.  The time dependence of expected
values is not enough; deterministic behavior in quantum mechanics is
defined through the probabilities of histories.

Even in the absence of a general measure, considerable insight into the
problem of classicality can be obtained by restricting attention to
special classes of coarse grainings and identifying
those that have high levels of classical correlations.  In such models
an assumption is being made as to the class of coarse grainings that
characterize the quasiclassical domain. Thus,
some parts of the general answer is being put in by hand.
Which of the class is
the most classical is being derived.  In this subsection we shall
examine one such class of  models. We shall introduce a powerful
technique for calculating the probabilities of decohering sets of
histories, namely a systematic expansion of the decoherence functional
in the {\it difference} between the two histories which are its
arguments.  This will enable us to derive the classical deterministic
laws that govern even highly non-linear systems in the class, including
the modifications that arise because of the mechanisms that produce
decoherence. We shall also be able to discuss quantitatively the
connections between decoherence noise, dissipation and the amount of
coarse graining necessary to achieve classical
predictability.\footnote{We follow
the discussion in \cite{GH93}.}

We consider model systems whose dynamics are describable by paths in a
configuration space spanned by (generalized) coordinates $\{q^i\}$ and
a Lagrangian that is the difference between a kinetic energy quadratic
in the velocities and a potential energy independent of velocities but
otherwise arbitrary.  We consider coarse grainings that distinguish a fixed
subset of coordinates, $\{x_a\}$, while ignoring the rest $\{Q_k\}$.
The initial density matrix of the closed system is assumed to factor
into a product of a density matrix of the distinguished variables and
another density matrix for the rest.  The linear oscillator models
discussed in the preceding subsection are special cases of this class of
models but the whole class is much more general because it is not
restricted to linear interactions.\footnote{For an extensive and
explicit discussion of the linear case from the point of view of the
decoherence functional see Dowker and Halliwell \cite{DH92}}
  Most non-relativistic systems of
interest fall into this class as far as dynamics are concerned.  What
is more special is the nature of the coarse graining and the factored
nature of the initial condition.  The anticipated repeated nature of
quasiclassical variables has been put in by hand by fixing a set of
coordinates distinguished by the coarse grainings for all time.  The
hydrodynamic variables that we expect to characterize at least one
realistic quasiclassical domain do not correspond to such a fixed
division of fundamental coordinates.  The same fixed division means that
the model coarse grainings do not  incorporate the branch
dependence expected to characterize realistic
 quasiclassical domains (see Section
III.1). Set off against these shortcomings, however, is the great advantage
of the model class of coarse grainings and initial conditions that  we can
relatively easily and explicitly exhibit which members of the class have
high classicality.

The first stages of an analysis of these models proceeds exactly as in
the linear oscillator models discussed in the preceding subsection.
The action can be written
\begin{equation}
S[q(\tau)] = S_{\rm free} [x(\tau)] + S_0 [Q(\tau)] + S_{\rm int}
[x(\tau), Q(\tau)] 
\label{threefourone}
\end{equation}
where $S_{\rm free}$ and $S_0$ are of kinetic minus potential energy form
and $S_{\rm int}$ is independent of velocities but otherwise arbitrary.
The variables $x$ now refer to a set of coordinates $x_a$
but we have suppressed the indices on them as we have on the $Q_k$.  For
the coarse grainings of interest that distinguish only the $x$'s the
unrestricted integrations over the $Q$'s can be carried out yielding a
decoherence functional $D[x^\prime(\tau), x(\tau)]$ of the same form as
\eqref{threethreesix} incorporating an influence phase defined by
\eqref{threethreeseven}.

Of course, the influence phase defined by \eqref{threethreeseven} does not have
the simple quadratic form \eqref{threethreeeight} appropriate to linear
interactions.  A useful operator expression for $W$ can be derived by
noting that the path integrals in the defining relation
\eqref{threethreeseven}
correspond to unitary evolution on the Hilbert space ${\cal H}^Q$ of
square-integrable functions of the $Q$'s generated by the Hamiltonian
corresponding to the action
\begin{equation}
S_Q[x(\tau), Q(\tau)] = S_0 [Q(\tau)] + S_{\rm int} [x(\tau), Q(\tau)]
\label{threefourtwo}
\end{equation}
which depends on the path $x(\tau)$ as an external parameter.  Since we
have assumed that the interaction is local in time, specifically of the
form
\begin{equation}
S_{\rm int} [x(\tau), Q(\tau)] = \int^T_0\ dt\ L_{\rm int} \bigl(x(t),
Q(t)\bigr)\, ,\label{threefourthree}
\end{equation}
the corresponding Hamiltonian
\begin{equation}
H_Q\bigl(x(t)\bigr) = H_0 + H_{\rm int} \bigl(x(t)\bigr)
\label{threefourfour}
\end{equation}
depends only on the instantaneous value of $x(t)$.  The operator
effecting unitary evolution generated by this Hamiltonian between times
$t^\prime$ and $t^{\prime\prime}$ is
\begin{equation}
U_{t^{\prime\prime}, t^\prime} [x(\tau)] = {\bf T}\  \exp
\Bigl[-\frac{i}{\hbar}\ \int^{t^{\prime\prime}}_{t^\prime}\ dt
\ H_Q\bigl(x(t)\bigr)\Bigr]\label{threefourfive}
\end{equation}
where ${\bf T}$ denotes the time-ordered product.  In terms of this,
eq.~\eqref{threethreeseven} becomes
\begin{equation}
\exp \Bigr(i\ W\left[x^\prime(\tau), x(\tau)\right]/\hbar\Bigr)=
Sp\Bigl\{U_{T,0} [x^\prime(\tau)]
 \rho_B U^\dagger_{T,0} \bigl[x(\tau)]
\Bigr\}
\label{threefoursix}
\end{equation}
where $Sp$ denotes the trace on ${\cal H}^Q$ and $\rho_B$ is the density
operator on ${\cal H}^Q$ whose matrix elements are $\rho_B(Q^\prime_0,
Q_0)$.

We will now assume that the influence phase is strongly peaked about
$x^\prime(\tau) \approx x(\tau)$ so as to produce the decoherence of
histories further coarse grained by suitable successions of regions of
$x$.  We will then analyze the circumstances in which the probabilities
of these decohering sets of histories predict classical, deterministic
correlations in time.

To make this program more precise it is convenient to introduce new
coordinates that measure the average of and the difference between
$x^\prime(t)$ and $x(t)$.  We define
\begin{subequations}
\label{threefourseven}
\begin{eqnarray}
\xi(t) & = & x^\prime(t) - x(t)\, , 
\label{threefourseven a}\\
X(t) & = & \half \left[x^\prime(t) + x(t)\right]\, . 
\label{threefourseven b}
\end{eqnarray}
\end{subequations}
We assume that $Im\,W$ increases with increasing $\xi(\tau)$, so that
$\exp(iW)$ is non-negligible only when $x^\prime(\tau) \approx x(\tau)$
for $0<\tau<T$.
This leads to decoherence of sets of histories further
coarse-grained by suitable regions of $x$'s.  Specifically, consider a set of
alternative coarse-grained histories specified at a sequence of times
$t_1, \cdots, t_n$ by exhaustive sets of exclusive regions of the $x$'s
which we denote by $\{\Delta^1_{\alpha_1}\},
\{\Delta^2_{\alpha_2}\}, \cdots, \{\Delta^n_{\alpha_n}\}$. The
decoherence functional for such a set is given by
\eqref{threethreefourteen}.
Evidently, (Fig.~6), if the characteristic sizes of these regions are
large compared to the width in $\xi(\tau)$ over which $\exp(iW)$ is
non-vanishing, the ``off-diagonal'' elements $D(\alpha^\prime, \alpha)$
will be very small.  That is decoherence.  The probabilities $p(\alpha)$
of the individual histories in this decohering set are the diagonal
elements $D(\alpha, \alpha)$ which, from \eqref{threethreefourteen} and
\eqref{threethreeseven}, are
\[
p(\alpha) = \int\nolimits_\alpha \delta X\delta\xi\ \delta
\left(\xi_f\right)\
\exp\biggl\{i\biggl(S_{\rm free}[X(\tau) +
\xi(\tau)/2]
\]
\begin{equation}
 - S_{\rm free} [X(\tau) - \xi(\tau)/2] + W [X(\tau),
\xi(\tau)]\biggr)/\hbar\biggr\}
 \bar\rho \left(X_0 + \xi_0/2, X_0 - \xi_0/2\right)\ .
\label{threefoureight}
\end{equation}

The assumed negligible values of $\exp(-Im\,W[X(\tau), \xi(\tau)])$ for values
of $\xi(\tau)$ much different from zero allows two further approximations
to the probabilities \eqref{threefoureight} which are useful in exhibiting
classical behavior.  First we may neglect the restrictions on the
$\xi(\tau)$ integration arising from the coarse graining with negligible
error (see Fig.~6).  Second, we may expand the exponent in
\eqref{threefoureight} in powers of $\xi(\tau)$ and get a good approximation 
to the
integral by neglecting higher than quadratic terms.  The result of these
two approximations is a Gaussian integral in $\xi(\tau)$ that can be
explicitly evaluated.

\begin{figure}[t]
\begin{center}
\includegraphics[width=4in]{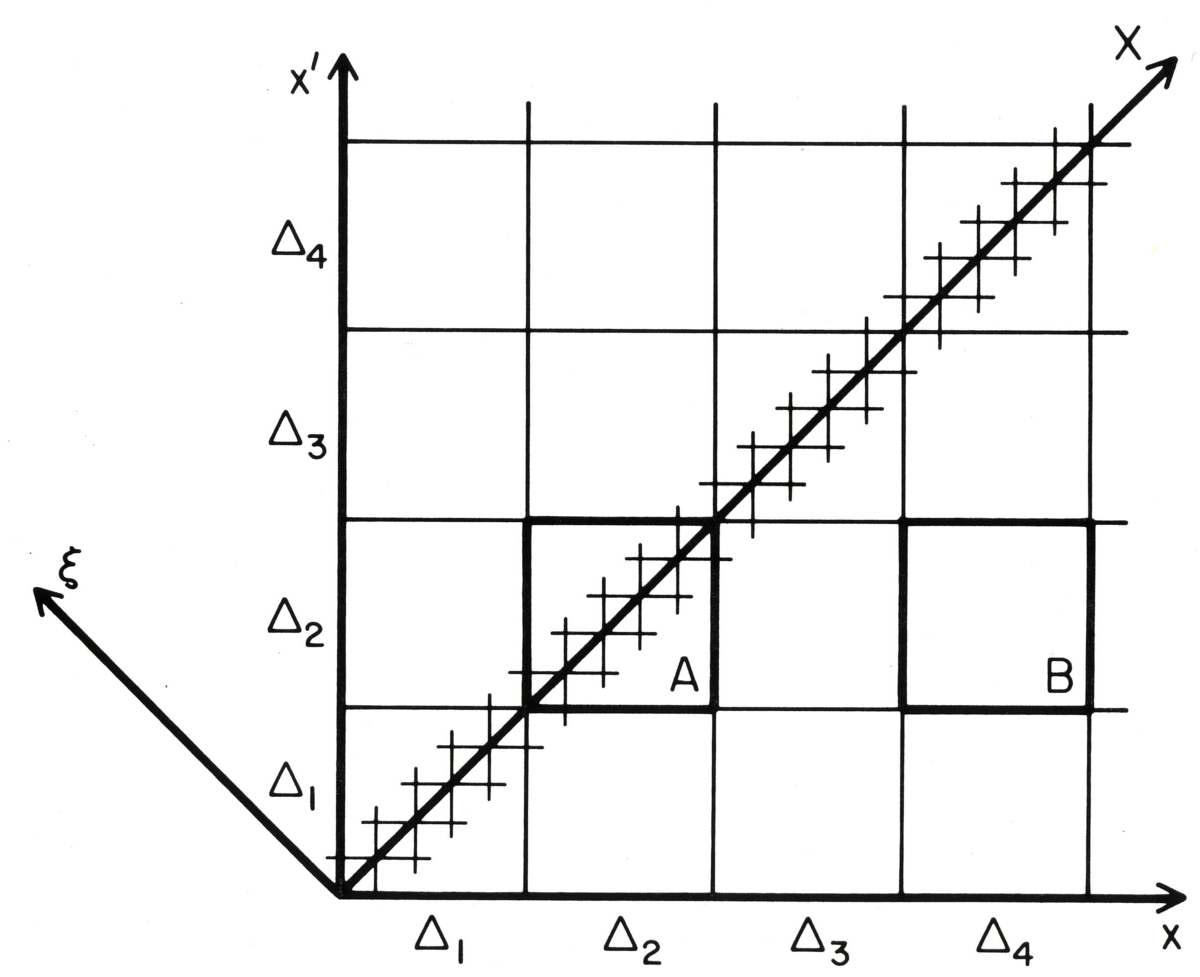}
\caption{The decoherence of histories coarse-grained by
intervals of a distinguished set of configuration space coordinates.
The decoherence functional for such sets of histories is defined by the
double path integral of \eqref{threefoureight} over paths $x^\prime(t)$ 
and $x(t)$
that
are restricted by the coarse graining. These path integrals may be
thought of
as the limits of multiple integrals over the values of
$x^\prime$ and $x$ on a series of discrete time slices of the interval
$[0,T]$.  A typical slice at a time when the range of integration is
constrained by the coarse graining is illustrated. Of course, only
 one of the
distinguished coordinates $x_a$ and its corresponding $x^{\prime}_a$
can be shown and we have assumed for illustrative purposes that the
regions defining the coarse-graining correspond to a set of intervals
$\Delta_\alpha, \alpha = 1, 2, 3, \cdots$ of this coordinate.
On each slice where there is
a restriction from the coarse graining, the integration over $x^\prime$
and $x$ will be restricted to a single box.  For the ``off-diagonal''
elements of the decoherence functional corresponding to distinct
histories, that box will be off the diagonal (\eg B) for {\it some}
slice.  For the diagonal elements, corresponding to the same histories,
the box will be on the diagonal (\eg A) for all slices.
If the imaginary part of the influence phase $W[x^\prime(\tau), x(\tau)]$
grows as a functional of the difference $\xi(\tau) = x^\prime (\tau) -
x(\tau)$,
as it does in the oscillator models [\cf \eqref{threethreethirteen}], then the
integrand of the
decoherence functional will be negligible except when $x^\prime(\tau)$ is
close to $x(\tau)$ a regime
illustrated by the shaded band about the diagonal in the
figure.  When the characteristic sizes of the intervals $\Delta_\alpha$
are large compared to the width of the band in which the integrand is
non-zero   the off-diagonal elements of the
decoherence functional will be negligible because integrals over those
slices where the histories are distinct is negligible (\eg over box B).
That is decoherence of the coarse-grained set of histories.  Further,
the evaluation of the diagonal elements of the decoherence
functional that give the probabilities of the individual histories in
decoherent set can be simplified.
If the integrations over $x^\prime$ and $x$ are
transformed to integrations over $\xi = x^\prime - x$ and $X=(x^\prime +
x)/2$ the restrictions on the range of the $\xi$-integration to one
diagonal box may be neglected with negligible error to the probability.}
\end{center}
\end{figure}

Expanding the free action terms in the exponent of \eqref{threefoureight} is
elementary.  Assuming the Lagrangian $L_{\rm free}$ consists of a
kinetic energy quadratic in the velocities $\dot x_a$ minus a potential
energy independent of the velocities, there is only a contribution from
the linear terms in
$\xi$ to quadratic order
\begin{equation}
S_{\rm free} [X(\tau) + \xi(\tau)/2] - S_{\rm free} [X(\tau) -
\xi(\tau)/2] = \xi^\dagger_0 P_0 (\dot X_0) +
\int\nolimits^T_0 dt\ \xi^\dagger(t)\ \left(\frac{\delta S_{\rm
free}}{\delta X(t)}\right) +\cdots\, . 
\label{threefournine}
\end{equation}
Here, $P_0(\dot X_0)$ is the canonical momentum, $\partial L_{\rm
free}/\partial \dot X$, evaluated at the endpoint $t=0$ and expressed in
terms of the velocities. $\delta S_{\rm free}/\delta X(t)$ is the usual
equation of motion.  We are using an obvious matrix notation in which
$y^\dagger z = \Sigma_a y_a z_a$ and we have used the fact that $\xi_f =
0$ in \eqref{threefoureight} to eliminate one surface term.

The general form of the expansion of $W[X(\tau), \xi (\tau)]$ in powers
of $\xi(\tau)$ is
\[
W[X(\tau), \xi(\tau)] = W[X(\tau), 0] + \int\nolimits^T_0 dt
\ \xi^\dagger(t)\, \left(\frac{\delta W}{\delta\xi(t)}\right)_{\xi(\tau)=0}
+ \half \int\nolimits^T_0 dt \int\nolimits^T_0 dt^\prime \xi^\dagger(t)
\]
\begin{equation}
\left(\frac{\delta^2W}{\delta\xi(t)
\delta\xi(t^\prime)}\right)_{\xi(\tau)=0} \xi(t^\prime) + \cdots\, .
\label{threefourten}
\end{equation}
The coefficients in the expansion can be computed from \eqref{threefoursix}.
First,
\begin{equation}
\exp\bigl(iW[X(\tau), 0]/\hbar\bigr) = Sp\left\{U_{T,0} [X(\tau)]
\rho_B U^\dagger_{T,0} [X(\tau)]\right\} = Sp\,\rho_B=1\, .
\label{threefoureleven}
\end{equation}
Thus, the leading term in \eqref{threefournine} vanishes,
\begin{equation}
W[X(\tau), 0]=0\, .\label{threefourtwelve}
\end{equation}

To evaluate the next terms we must consider the derivatives $\delta
U_{T,0} [(X(\tau) + \xi(\tau)/2]/\delta\xi(t)$. To do this introduce the
definition:
\begin{equation}
F\bigl(x(t)\bigr) \equiv -\frac{\partial H_Q(t)}{\partial x(t)} =
\frac{\partial L_{\rm int}\bigl(x(t), Q(t)\bigr)}{\partial x(t)}\, .
\label{threefourthirteen}
\end{equation}
The operator $F(x(t))$ is an operator in the Schr\"odinger picture in
which we have been working.  It is a function of $x$ because
$L_{\rm{int}}$ is a function of $x$ and becomes a function of $t$
because $x$ is a function $t$.  It represents the force on the
distinguished subsystem arising from the rest of the closed system.

Carrying out the indicated differentiations of $U$ gives
\begin{eqnarray}
\Bigl(\delta U_{T,0}\bigl[X(\tau) \pm
\xi(\tau)/2\bigr]/\delta\xi(t)\Bigr)_{\xi(\tau)=0}
& = & \pm (i/2\hbar) U_{T,t} \left[X(\tau)\right]\ F\left(X(t)\right)
\, U_{t,0}
\left[X(\tau)\right]\nonumber\\
&\equiv & \pm (i/2\hbar)\ F\bigl(t, X(\tau)\bigr]\, .
\label{threefourfourteen}
\end{eqnarray}
The operator $F(t, X(\tau)]$ is the representative of
the Schr\"odinger operator \eqref{threefourthirteen} in a picture something
like the Heisenberg picture. However, it is not the usual Heisenberg
picture because its connection to the Schr\"odinger picture involves
unitary evolution over future ranges of time to time $T$ as well as past
ones.  In fact, however, it can be shown \cite{GH93} that
probabilities are independent of $T$.  The operator $F(t,X(\tau)]$
  is a function of time
but also a functional of the path $X(\tau)$.  This dual dependence we
have indicated with round and square brackets.

It is then only a short calculation to find for the next coefficient in
\eqref{threefourten}:
\begin{equation}
\bigl(\delta W/\delta\xi(t)\bigr)_{\xi(\tau)=0} = \bigl\langle F\big(t,
X(\tau)\big]\bigr\rangle 
\label{threefourfifteen}
\end{equation}
where the expected value is defined by
\begin{equation}
\langle A\rangle = Sp \left(A\ \rho_B\right)\, .
\label{threefoursixteen}
\end{equation}
In a similar manner the next coefficient in \eqref{threefourten} may be
calculated.  One finds after slightly more trouble (for more details see
\cite{GH93}):
\begin{equation}
\bigl(\delta^2W/\delta\xi(t)\delta\xi(t^\prime)\bigr)_{\xi(\tau)=0}
=(i/2\hbar) \bigl\langle\bigl\{\Delta F\bigl(t,X(\tau)\bigr]\, ,\ \Delta
F\bigl(t^\prime, X(\tau)\bigr]\bigr\}\bigr\rangle 
\label{threefourseventeen}
\end{equation}
where $\{,\}$ denotes the anticommutator and $\Delta F$ is the
operator
\begin{equation}
\Delta F\bigl(t,X(t)\bigr] = F\bigl(t,X(\tau)\bigr] - \bigl\langle
F\bigl(t,X(\tau)\bigr]\bigr\rangle 
\label{threefoureighteen}
\end{equation}
representing fluctuations in the force $F$ about its mean.  We note that
$(\delta W/\delta\xi)_{\xi=0}$ is purely real and
$(\delta^2W/\delta\xi\delta\xi)_{\xi=0}$ is purely imaginary.

With these definitions the Gaussian integral that results from inserting
\eqref{threefourten}  into \eqref{threefoureight} can be carried out.
The result is
\[
p(\alpha)\cong\int_\alpha\ \delta X\ \bigl[ {\rm det}
\left(K_I/4\pi\right)\bigr]^{-\half}
\]
\begin{equation}
         \times \exp \Bigl[-\frac{1}{\hbar}\ \int^T_0\ dt
\ \int^T_0\ dt^\prime\ {\cal E}^\dagger(t,X(\tau)]\ K^{\rm inv}_I
(t,t^\prime; X(\tau)]\ {\cal E}(t^\prime,X(\tau)]\Bigr]
w(X_0, P_0)\, .
\label{threefournineteen}
\end{equation}
The ingredients of this expression are as follows: ${\cal E}$ is the
expression
\begin{equation}
{\cal E}(t,X(\tau)] = \frac{\delta S_{\rm free}}{\delta
X(t)} + \bigl\langle F\bigl(t, X(\tau)\bigr]\bigr\rangle\ .
\label{threefourtwenty}
\end{equation}
The kernel $K_I$ is
\begin{equation}
K_I(t, t^\prime; X(\tau)] = \hbar^{-1}\bigl\langle\bigl\{\Delta F\bigl(t,
X(\tau)\bigr], \Delta F(t^\prime, X(\tau)\bigr]\bigr\}\bigr\rangle\, ,
\label{threefourtwentyone}
\end{equation}
and $K^{\rm inv}_I$ is its inverse on the interval $[0,T]$.  The
function $w$ is the Wigner distribution associated with the initial
density matrix $\bar\rho$ defined by
\begin{equation}
w\left(X_0, P_0\right) = \int d\xi_0\ e^{i(\xi^\dagger_0 P_0)/\hbar} \bar\rho
\left(X_0 + \xi_0/2, X_0 - \xi_0/2\right)\, .
\label{threefourtwentytwo}
\end{equation}
For the explicit form of the measure see \cite{GH93}.

The expression \eqref{threefournineteen} for the probabilities $p(\alpha)$ 
has a
simple physical
interpretation.  The kernel $K_I(t, t^\prime)$ is positive because it is
the expected value of an anticommutator.  The probabilities of histories
are therefore peaked about ${\cal E}(t)=0$, that is, about histories
which satisfy the equation
\begin{equation}
{\cal E}(t) = \frac{\delta S_{\rm free}}{\delta X(t)} + \bigl\langle
F\bigl(t, X(\tau)\bigr]\bigr\rangle = 0\, .
\label{threefourtwentythree}
\end{equation}
This is the equation of motion of the free action modified by effective
forces arising from the interaction of the $x$'s with the rest of the
system.  In general, these forces will be non-local in time and
non-conservative representing such familiar phenomena as friction.  As
an exercise the reader can show that these forces are causal, that is
$\langle F(t, X(\tau)]\rangle$ depends on $X(\tau)$ only for $\tau < t$.
The initial positions and momenta are distributed according to the
Wigner distribution.  The Wigner distribution is not necessarily
positive, but the probabilities $p(\alpha)$ are positive by construction
apart from small errors that may have been introduced by the
approximations mentioned above \cite{Hal92}.

Eq.~\eqref{threefournineteen} therefore describes the probabilities of a set of
histories whose initial conditions are distributed but for which, given
an initial condition, the probabilities are peaked about histories
satisfying the equation of motion \eqref{threefourtwentythree}.  Of course,
eq.~\eqref{threefournineteen} also shows that there are probabilities for
deviations from the equation of motion governed by $K^{\rm inv}_I$.
These represent noise --- both classical and quantum --- arising from
the interactions of the distinguished system with the rest.  Indeed, in
this approximation, the probabilities $p(\alpha)$ are identical to those
of a classical system obeying a Langevin equation
\begin{equation}
{\cal E}(t,X(\tau)] + {\cal L}\bigl(t, X(\tau)\bigr] = 0
\label{threefourtwentyfour}
\end{equation}
with a Gaussian distributed stochastic classical force whose spectrum is
fixed by the correlation function
\begin{equation}
\left\langle{\cal L}(t, X(\tau)] {\cal L} (t^\prime,
X(\tau)]\right\rangle_{\rm classical} =
\hbar K_I \bigl(t, t^\prime ; X(\tau)\bigr]\,  .
\label{threefourtwentyfive}
\end{equation}

If the noise is small, or alternatively, if the parameters of the
actions are such that the ``width'' of the Gaussian distribution of
paths is small, then there will be vanishing probabilities for all
sufficiently coarse-grained histories $\alpha$ except those correlated
in time by the deterministic equation ${\cal E}(t,X(\tau)] = 0$. That is
classical behavior.

The linear oscillator models of the preceding subsection provide a
simple example of the above general analysis.  The coefficients $(\delta
W/\delta \xi)_{\xi=0}$ and $(\delta^2W/\delta\xi\delta\xi)_{\xi=0}$ may
be computed directly from \eqref{threethreeeight} and the subsequent 
expressions
for $k_R(t, t^\prime)$ and $k_I(t, t^\prime)$.  One finds for the
equation of motion
\begin{equation}
-M\ddot X - M\omega^2X + \int\nolimits^t_0 dt^\prime\, k_R \left(t,
t^\prime\right) X\left(t^\prime\right)=0
\label{threefourtwentysix}
\end{equation}
where $k_R(t, t^\prime)$ is given by \eqref{threethreenine}.  
The spectrum of the
noise is simply
\begin{equation}
K_I \bigl(t, t^\prime ; X(\tau)\bigr] = k_I \left(t, t^\prime\right)
\label{threefourtwentyseven}
\end{equation}
given explicitly by \eqref{threethreeten}. These expressions are even simpler 
in the high temperature Fokker-Planck limit defined by
\eqref{threethreesixteen}.
Then one finds for the equation of motion (away from $t=0$)
\begin{equation}
{\cal E}(t,X(\tau)] = -M\ddot X(t) - M\omega^2 X(t) - 2M\gamma \dot X(t) = 0
\label{threefourtwentyeight}
\end{equation}
explicitly exhibiting dissipation.  In the same limit the spectrum of
noise is
\begin{equation}
K_I \bigl(t, t^\prime ; X(\tau)\bigr] = \frac{8M\gamma k T_B}{\hbar}
\ \delta\left(t-t^\prime\right)\, . 
\label{threefourtwentynine}
\end{equation}
Thus, in this limit, the exponent in the probability formula
\eqref{threefoureight} can be written
\begin{equation}
-\frac{M}{8\gamma kT_B} \int\nolimits^T_0 dt\,\left[\ddot X + \omega^2 X
+ 2\gamma \dot X\right]^2\, . 
\label{threefourthirty}
\end{equation}
This expression exhibits explicitly the requirements necessary for
classical behavior.  Large values of $\gamma T_B$ lead to effective
decoherence as \eqref{threethreesixteen} shows.  However, large values of
$\gamma T_B$ also lead to significant noise \eqref{threefourtwentynine} and
therefore deviations from classical predictability in
\eqref{threefourthirty}).  To
obtain classical predictability, a large coefficient in front of
\eqref{threefourthirty} is needed, $M/\gamma T_B$ therefore
must also be large in the limit that
$\gamma T_B$ is becoming large.  This is a general and physically
reasonable result.  Stronger coupling to the ignored variables produces
more rapid dispersal of phases and more effective decoherence.  The same
stronger coupling produces greater noise.  A high level of inertia is
needed to resist this noise and achieve classical predictability.

What this class of models argues for generally is that the classical
behavior of a quantum system is an emergent property of its initial
condition described by certain decohering sets of alternative coarse-grained
histories.  Histories of suitable sets are, with high
probability, correlated in time by classical deterministic laws with
initial data probabilistically distributed according to the system's
initial condition. Coarse graining is required for decoherence and
coarse graining beyond that is required to provide the inertia to resist the
noise
that typical mechanisms of decoherence produce.  We may hope to exhibit
these conclusions in more general models than those discussed
here.\footnote{See \cite{GH93} for suggestions on how to do so.}
In particular in quantum cosmology we hope to exhibit the quasiclassical
domain, including the classical behavior of spacetime geometry, as an
emergent property of the initial condition of the universe.  We shall
discuss this further in Section IX but first we must develop a quantum
mechanics general enough to deal with spacetime.

\section{Generalized Quantum Mechanics}\footnotemark{Some of
the material
in this section has
been adapted from the author's lectures at the 1989 Jerusalem Winter
School on Quantum Cosmology and Baby Universes \cite{Har91a}
where the notion of a generalized quantum mechanics was originally
introduced.}
\setcounter{footnote}{0}

\subsection{Three Elements}

As described in the Introduction, these lectures are concerned with two
generalizations of the usual flat spacetime
quantum mechanics of measured subsystems
that are needed to apply quantum mechanics to cosmology.
The first was the generalization to the quantum mechanics of closed
systems in which ``measurement'' does not play a fundamental role.  That
generalization has been described in the preceding two sections. The
remainder of these lectures are concerned with the generalization needed
to deal with a quantum theory of gravity in which there is no fixed
background spacetime geometry and therefore no fixed notion of time.

We begin, in this section, by abstracting
some general principles that define a quantum mechanical theory from the
preceding discussion.
The resulting framework, called generalized quantum
mechanics,
 provides a general arena for discussing many different generalizations
of familiar Hamiltonian quantum mechanics. Among these will be the
particular generalization we shall develop for a quantum theory of
spacetime.

Roughly speaking, by a generalized quantum mechanics we
mean a quantum
theory of a closed system
 that admits a notion of fine- and coarse-grained histories, the
decoherence functionals for which are
connected by the principle of superposition and for which there is a
decoherence condition
 that determines when coarse-grained histories can be assigned
 probabilities obeying the
sum rules of probability calculus.  More precisely,
a generalized quantum theory is defined by the following
elements:

\begin{enumerate}

\item {\sl Fine-Grained Histories:} The fine-grained histories are the sets of
exhaustive, alternative histories of the
closed system
$\{f\}$ which are the most refined description of its dynamical
evolution to which one can
contemplate assigning probabilities.  Examples are the set of particle
paths in non-relativistic quantum mechanics, the set of four-dimensional
field configurations in field theory and the set of four-geometries in
general relativity as described and qualified in the rest of these
lectures.  For generality, however, we take $\{f\}$ to be {\it any} set
here and leave its connection with evolution in spacetime to the
specific examples.  As the example of non-relativistic quantum mechanics
illustrates, there may be many different sets of fine-grained
histories.

\item {\sl Allowed Coarse Grainings:}
A set of fine-grained histories may be partitioned into an
exhaustive set of exclusive classes $\{c_\alpha\}$. That is an operation
of {\it coarse graining}; each class is a coarse-grained history, and
the set of classes is a set of coarse-grained histories.  Further
partitions of a coarse-grained set are further operations of
coarse-graining and yield coarser-grained sets of alternative histories.
Conversely, the finer sets are {\it fine grainings} of the coarser ones.
The process of coarse graining terminates in the trivial case of a set
with only a single member --- the class $u$ of all fine-grained
histories --- which we assume to be a common coarse graining for
all fine-grained sets.

The sets of exclusive histories arrived at by operations of coarse
graining exhaust the alternatives of the closed system to which
generalized quantum mechanics potentially assigns probabilities.  The
set of all sets of histories is partial ordered by the operations of
coarse and fine graining because two given sets need not be either fine
or coarse grainings of each other.  For convenience we may regard the
fine-grained sets as coarse-grained sets with a trivial coarse graining.
The set of coarse-grained sets of histories is then a semi-lattice.

\item {\sl Decoherence Functional}: Interference between the members of a
coarse-grained set of histories is measured by the decoherence
functional.
The decoherence functional is a complex-valued functional,
$D (\alpha^\prime,
\alpha)$,  defined for each pair of histories in a coarse-grained set
$\{\alpha\}$. The decoherence functional for each set of alternative
coarse-grained histories must satisfy the following
conditions:
\begin{enumerate}
\item {\sl Hermiticity:}
\begin{subequations}
\label{fouroneone}
\begin{equation}
D(\alpha^\prime, \alpha) = D^*(\alpha, \alpha^\prime)\, ,
\label{fouroneone i}
\end{equation}
\item {\sl Positivity:}
\begin{equation}
D(\alpha^\prime, \alpha)\geq 0\, ,
\label{fouroneone ii}
\end{equation}
\item {\sl Normalization:}
\begin{equation}
\sum\limits_{\alpha^\prime, \alpha}D(\alpha^\prime, \alpha) =1
\, .\label{fouroneone iii}
\end{equation}
In addition, and most importantly, the decoherence functional for
different coarse-grained sets must be related by the principle of
superposition:
\item {\sl The principle of superposition:}
\begin{equation}
D(\bar\alpha^\prime, \bar\alpha) =
\sum\limits_{\alpha^\prime\epsilon\bar\alpha^\prime}
\sum\limits_{\alpha\epsilon\bar\alpha}
 D(\alpha^\prime, \alpha)\, .
\label{fouroneone iv}
\end{equation}
\end{subequations}

\end{enumerate}
\end{enumerate}

The superposition principle means that once the decoherence
functional is defined for any fine-grained set of histories, $\{f\}$,
the decoherence functional for any coarse-graining of it may be
determined by \eqref{fouroneone iv}.  If there is a unique most fine-grained
set, as in a sum-over-histories formulation of quantum mechanics, then
the specification of a $D(f^\prime, f)$ consistent with 
\eqref{fouroneone i} -- \eqref{fouroneone iii}
specifies all other decoherence functionals.  If there is more than one
most fine-grained set $\{f\}$, then the decoherence functional must be
specified consistently so that if a set of alternative histories is a
coarse-graining of two {\it different} fine-grained sets of the same
decoherence results from \eqref{fouroneone iv} applied to the different
fine-grained sets.

The specification of a generalized quantum mechanics is completed by
giving a {\it decoherence condition} that specifies which sets of alternative
coarse-grained histories are assigned probabilities in the theory.  Such
sets of histories are said to decohere.  The probabilities of the
individual histories in a decoherent set are the ``diagonal'' elements
of the decoherence functional
\begin{equation}
p(\alpha) = D(\alpha, \alpha)\, . 
\label{fouronetwo}
\end{equation}
These must satisfy the general rules of probability theory (\eg as in
\cite{Fel57}).  They must be real numbers between zero and one defined
on the sample space supplied by a set of alternative
coarse-grained histories.  The probabilities must be additive on
disjoint sets of the sample space which in the present instance means
\begin{equation}
p(\bar \alpha) = \sum\limits_{\alpha\epsilon \bar\alpha} p (\alpha)\, ,
\label{fouronethree}
\end{equation}
for {\it any} coarse-graining $\{c_{\bar\alpha}\}$ of $\{c_{\alpha}\}$, to the
approximation with which the probabilities are used. The probability of
the empty set, $\phi$, must be zero and the probability of the whole set,
$u$, must be one.

The simplest decoherence condition is the requirement that the
``off-diagonal'' elements of the decoherence functional be sufficiently
small
\begin{equation}
D\left(\alpha^\prime, \alpha \right) \approx 0\ ,\quad \alpha^\prime
\not= \alpha\, . 
\label{fouronefour}
\end{equation}
This was the sufficient condition for the probability sum rules used in
Sections II and III and the decoherence condition we shall assume in the
rest of this paper.  As a consequence, the conditions 
\eqref{fouroneone i}--\eqref{fouroneone iv} of
\eqref{fouroneone} the numbers {\it defined} by \eqref{fouronetwo} 
obey the rules
of probability theory  for sets of histories obeying the decoherence condition
\eqref{fouronefour}.
They are real and positive because of the
hermiticity and positivity conditions.  They sum to unity by the
normalization condition and they obey the sum rules \eqref{fouronethree}
because of the principle of superposition, \viz
\begin{equation}
p(\bar\alpha) = D(\bar\alpha, \bar\alpha) =
\sum\limits_{\alpha^\prime\epsilon\bar\alpha}
\,\sum\limits_{\alpha\epsilon\bar\alpha}\, D(\alpha^\prime, \alpha)
\approx \sum\limits_{\alpha\epsilon\bar\alpha}\, D(\alpha, \alpha)
= \sum\limits_{\alpha\epsilon\bar\alpha}\, p(\alpha)\, .
\label{fouronefoura}
\end{equation}

Decoherence conditions both stronger and weaker than \eqref{fouronefour} have
been investigated.  (See \cite{GH90b} and \cite{GH93} for
discussion.) If {\it arbitrary}   unions of coarse-grained histories into new,
mutually exclusive classes are allowed operations of coarse graining,
then it is not difficult to see that the {\it necessary} as well as
sufficient condition for the probability sum rules \eqref{fouronethree} to be
satisfied is [\cf \eqref{threeoneten}]
\begin{equation}
Re\, D\left(\alpha^\prime, \alpha\right) \approx 0 \ ,\quad \alpha^\prime
\not= \alpha\, . 
\label{fouronefive}
\end{equation}
Conditions \eqref{fouronefive} and \eqref{fouronefour} are called the 
{\it weak}
and {\it medium decoherence conditions}, respectively.  An even weaker
condition was used by Griffiths \cite{Gri84} and Omn\`es
\cite{Omnsum} in their original investigations, and conditions stronger
than medium decoherence have been investigated in the efforts to
precisely characterize quasiclassical domains \cite{GH93}.

A choice between weak, medium, or other forms of the decoherence
condition is not really needed for the rest of the discussion of
generalizations of quantum mechanics that are free from the problem of
time since we shall not carry out explicit calculations of the decoherence
of specific sets of histories.  All that is necessary is that the
condition be expressed in terms of the decoherence functional.  For
simplicity, the reader can keep in mind the medium decoherence condition
\eqref{fouronefour}.  Realistic mechanisms of decoherence, such as those
illustrated in Section III, lead to medium decoherence.

These three elements --- fine-grained histories, coarse-graining, and a
 decoherence
functional together with a decoherence condition --- capture the
essential features of quantum mechanical prediction.  In the following
we shall see that Hamiltonian quantum mechanics is one way of specifying
these elements but not the only way.  Alternative specifications lead to
generalizations of Hamiltonian quantum mechanics.  In the remainder of
this section we discuss some familiar formulations of quantum mechanics
from the generalized quantum mechanics point of view.

\subsection{Hamiltonian Quantum Mechanics as a Generalized
Quantum Mechanics}

First, we consider Hamiltonian quantum mechanics
as a generalized quantum mechanics.
 In Hamiltonian quantum mechanics  sets of histories are represented by
chains of projections onto exhaustive sets of orthogonal subspaces of
a Hilbert space.  The fine-grained histories, coarse graining, and
decoherence functional are specified as follows:

\begin{enumerate}

\item {\sl Fine-Grained Histories:} These correspond to the possible
sequences of sets
of projections onto a {\sl complete} set of states, one set at every
time.
  There are thus {\sl many}
different sets of fine-grained histories corresponding to the various
possible
complete sets of states at each and every time. The many possible
fine-grained starting points in Hamiltonian quantum
mechanics
are a reflection of the democracy of transformation theory.  No one
basis is
distinguished from any other.

\item {\sl Allowed Coarse Grainings:} For definiteness, we take the allowed 
sets of
coarse-grained histories of Hamiltonian quantum mechanics to consist of
sequences of independent alternatives at definite moments of time so
that every history can be represented as a chain of projections as in
\eqref{fourtwo}. A set of such histories is a coarse graining of a finer
set if each projection in the coarser grained set is a sum of
projections in
the finer grained set.  The projections constructed as sums define a
partition
of the histories in the finer grained set.

By way of example, consider the
quantum mechanics of a particle. A very fine-grained set of histories
can be specified by very small position intervals at a great many times
thus approximately specifying the particle's path in configuration
space.
 An example of
coarse
graining of these consists of projections onto an exhaustive set of ranges of
position
at, say, three different times defining a partition of the configuration
space
paths into those that pass through the various possible combinations of
ranges
at the different times.

Given the discussion in Section III.1.1 on the importance of branch
dependence, it may seem arbitrary to limit the coarse-grained sets of
histories of Hamiltonian quantum mechanics to be always represented by chains
of projections and not
{\it sums} of chains of projections. We do so to ensure, as discussed in
Section III.1.3, that evolution in Hamiltonian quantum mechanics can be
formulated in the familiar terms of a state vector that evolves
unitarily in between alternatives and is reduced at them. Incorporating
branch dependent histories represented by sums of chains of projections
we consider as a generalization of Hamiltonian quantum mechanics.

\begin{figure}[t]
\begin{center}
\includegraphics[width=6in]{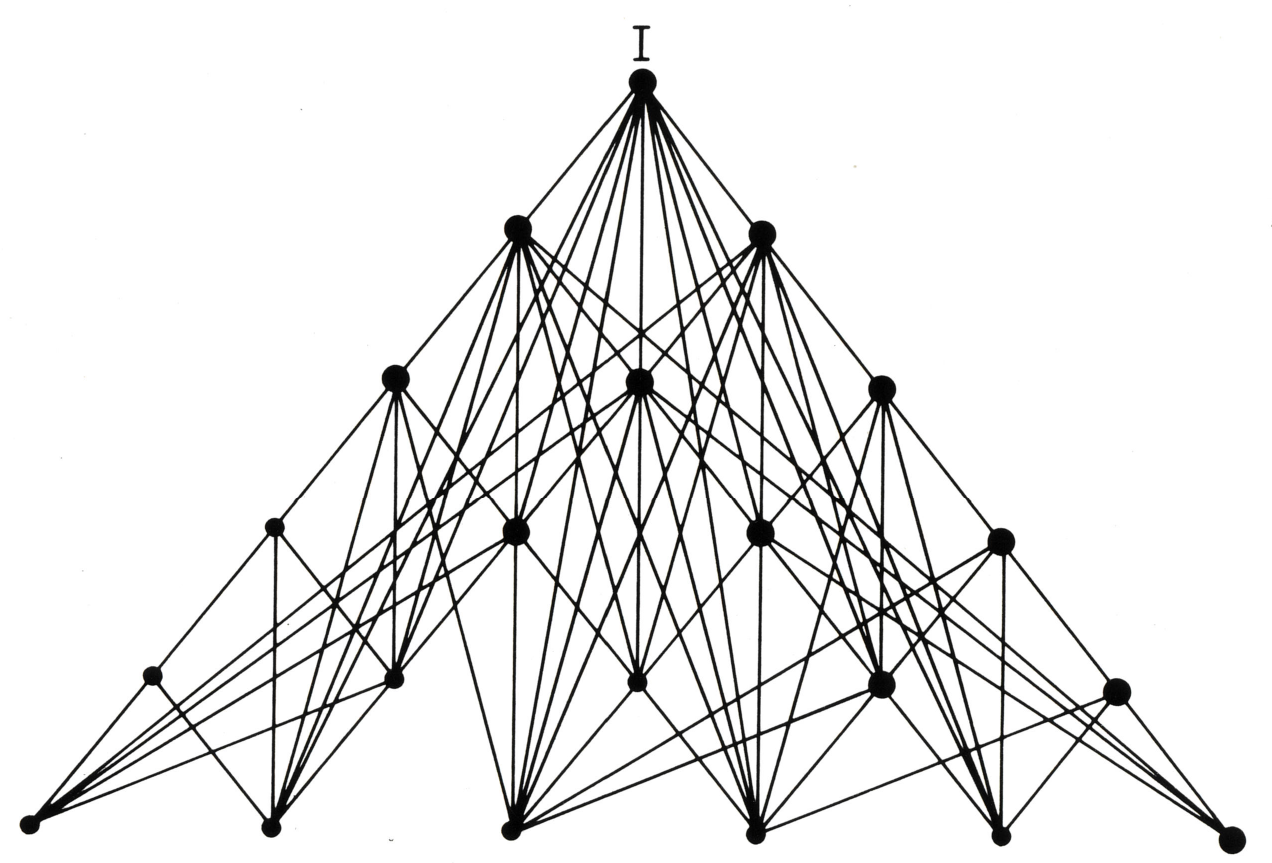}
\caption{The schematic structure of
the space of {\it sets} of possible histories in Hamiltonian quantum
mechanics.
Each dot in this diagram represents an
exhaustive {\it set} of alternative histories for the universe.  (This
is not a
picture of the branches defined by a given set!)~~Such sets
correspond in the Heisenberg picture to time
sequences ($P^1_{\alpha_1} (t_1),\ P^2_{\alpha_2} (t_2),\ \cdots
\ P^n_{\alpha_n} (t_n))$ of sets of projection operators, such that at
each time
$t_k$ the alternatives $\alpha_k$ are an orthogonal and exhaustive set
of possibilities for the universe.  At the bottom of the diagram are the
completely fine-grained sets of histories each arising from taking
projections
onto eigenstates of a {\it complete set} of observables for the universe
at {\it every time}.\\
The dots above the bottom row are coarse-grained sets of alternative
histories.
If two dots are connected by a path, the one above is a coarse graining
of the
one below ---  that is, the projections in the set above are {\it sums}
of those in
the set below.  A line, therefore, corresponds to an operation of coarse
graining.  At the very top is the degenerate case in which  complete
sums
are taken at every time, yielding  no projections at all other than the
unit operator!
The space of sets of alternative histories is thus partially ordered by
the operation of coarse graining.\\
The heavy dots denote the decoherent sets of alternative histories.
Coarse grainings of decoherent sets remain decoherent.}
\end{center}
\end{figure}

\item {\sl Decoherence Functional:} For Hamiltonian quantum mechanics
this is \eqref{threeonefour}.  In the present
notation
$\alpha$ stands for the history corresponding to a particular chain of
projections $C_\alpha$.  Thus,
\begin{equation}
D(\alpha^\prime, \alpha) = Tr\,\bigl[C_{\alpha^\prime} \rho
C^\dagger_\alpha\bigr] = Tr \bigl[P^n_{\alpha^\prime_n} (t_n) \cdots
P^1_{\alpha^\prime_1}(t_1)
\rho P^1_{\alpha_1}(t_1)\cdots
P^n_{\alpha_n}(t_n)\bigr]
\label{fourtwoone}
\end{equation}
which is easily seen to satisfy properties \eqref{fouroneone
i}--\eqref{fouroneone iv} above.

The structure of sets of alternative coarse-grained histories of
Hamiltonian
quantum mechanics is shown
schematically in Fig.~7.  The sets of coarse-grained histories form a
partially ordered set
defining a semi-lattice.  For any pair of sets of histories, the least
coarse
grained set of
which they are both fine grainings can be defined.  However, there is
not, in
general, a unique most fine-grained set of which two sets are a coarse
graining.
\end{enumerate}

\subsection{Sum-Over-Histories Quantum Mechanics for Theories
with a Time.}

The fine-grained histories, coarse graining, and decoherence functional
of a sum-over-histories quantum
mechanics
of a theory with a well defined physical time are specified as follows:

\begin{enumerate}

\item {\sl  Fine-Grained Histories:}  The fine-grained histories are the
possible
paths in a configuration space of generalized coordinates $\{q^i\}$
expressed as
{\sl single-valued} functions of the physical time.  Only one
configuration is
possible at each instant.  Sum-over-histories quantum mechanics,
therefore,
starts from a {\sl unique} fine-grained set of alternative histories of
the universe
in contrast to Hamiltonian quantum mechanics that starts from many.

\item {\sl Allowed Coarse Grainings:}  There are many ways of partitioning the
fine-grained paths into exhaustive and exclusive classes, $\{c_\alpha\}$.
However, the
existence of
a physical time allows an especially natural coarse graining because
paths
cross a constant time surface in the extended configuration space $(t,
q^i)$
once and only once.  Specifying an exhaustive set of regions
$\{\Delta_\alpha\}$ of the $q^i$ at one time, therefore, partitions the
paths
into the class of those that pass through $\Delta_1$ at that time, the
class of
those that pass through $\Delta_2$ at that time, etc.  More generally,
different exhaustive sets of regions $\{\Delta^k_{\alpha_k}\}$ at times
$\{t_k\}$,
$k=1, \cdots, n$ similarly define a partition of the fine-grained
histories
into exhaustive and exclusive classes.  More general partitions of the
configuration space paths corresponding to alternatives that are not at
definite moments of time will be described in Section V.

\item {\sl Decoherence Functional:}  The decoherence functional for
sum-over-histories quantum mechanics for theories with a well-defined
time is
\begin{equation}
D(\alpha^\prime, \alpha) = \int\nolimits_\alpha \delta q^\prime
\int\nolimits_\alpha
\delta
q \delta (q^{\prime }_f-q_f)\ \exp\biggl\{i\left(S[q^{\prime }(\tau)] -
S[q(\tau)]\right)/\hbar\biggr\}\rho(q^{\prime }_0, q_0)\, .
\label{fourthreeone}
\end{equation}
Here, we consider an interval of time from an initial instant $t=0$ to
some
final
time $t=T$.  The first integral is over paths $q(t)$ that begin at
$q_0$,
end at $q_f$, and lie in the class  $c_\alpha$.  The integral includes an
integration over $q_0$ and $q_f$.  The second integral over paths
$q^{\prime }(t)$ is similarly defined.  If $\rho(q^{\prime}, q)$ is
a
density matrix, then it is easy to verify that $D$ defined by 
\eqref{fourthreeone}
satisfies
conditions (i)--(iv) of \eqref{fouroneone}.  When the coarse graining is
defined by
sets of configuration space regions $\{\Delta^k_{\alpha_k}\}$ as discussed
above,
then \eqref{fourthreeone} coincides with the sum-over-histories
decoherence functional
previously introduced in \eqref{threeonetwentythree}.  However, more general
partitions are
possible.
\end{enumerate}

The structure of the collection of sets of coarse-grained histories in
sum-over-histories quantum
mechanics is illustrated in Fig. 8.  Because there is a {\sl unique}
fine
grained
set of histories, many fewer coarse grainings are possible in a
sum-over-histories formulation than in a Hamiltonian one, and the space
of sets
of coarse-grained histories is a lattice rather than a semi-lattice.

\subsection{Differences and Equivalences between Hamiltonian and
Sum-Over-Histories\\
 Quantum Mechanics for Theories with a Time}

From the perspective of generalized quantum theory, the
sum-over-histories
quantum mechanics of Section IV.3 is different from the Hamiltonian
quantum
mechanics of Section IV.2.  Even when the action of the former gives
rise to
the Hamiltonian of the latter, the two formulations differ in their
notions of
fine-grained histories, coarse graining and in the resulting space of
coarse-grained sets of histories as Figs 7 and 8 clearly show.  Yet,
as we demonstrated in Section III.1.4,
the sum-over-histories formulation and the Hamiltonian
formulation are
equivalent for those particular coarse grainings in which the histories
are
partitioned according to exhaustive sets of configuration space regions,
$\{\Delta^k_{\alpha_k}\}$, at various times $t_k$.  More precisely
 the sum-over-histories expression for the decoherence
functional, \eqref{fourthreeone}, is {\sl equal} to the Hamiltonian expression,
(IV.2.1),
when the latter is evaluated with projections onto the ranges of
coordinates
that occur in the former.  Crucial to this equivalence, however, is the
existence of a well-defined physical time in which the paths are
single-valued which permitted the factorization of the path integral in
\eqref{threeonetwenty} that led to the identity \eqref{threeoneeighteen} which
connected the two formulations.

\begin{figure}[t]
\begin{center}
\includegraphics[width=4in]{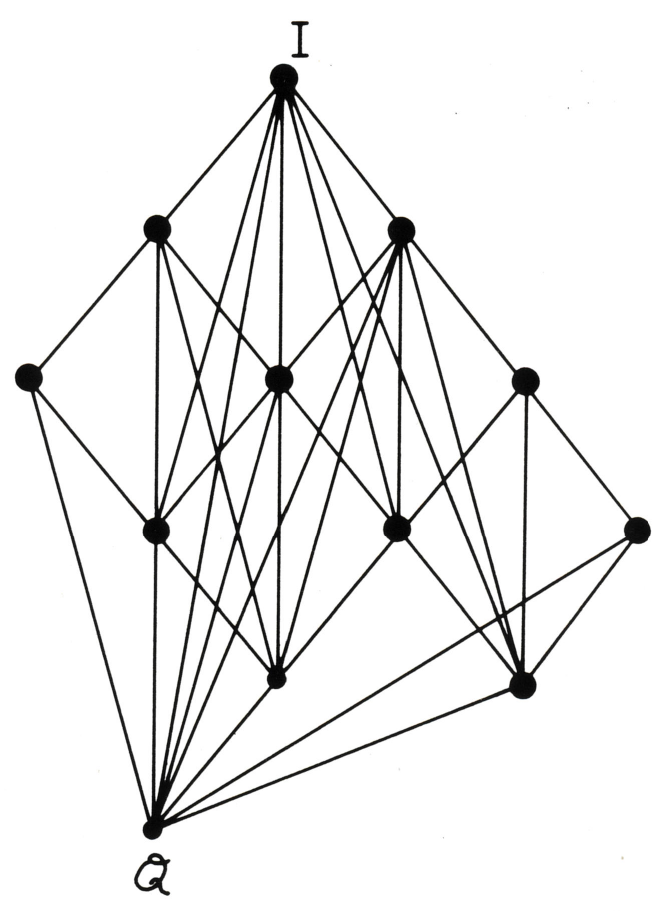}
\caption{The schematic structure
of the space of sets of histories in sum-over-histories quantum
mechanics.
The completely
fine-grained histories arise from a
single complete set of observables, say the set ${\cal Q}$ of field
variables
$q^i$ at each point in space and every time.}
\end{center}
\end{figure}

To see this more clearly let us sketch the derivation of a Hamiltonian
formulation from a sum-over-histories one --- the inverse of the
construction described in Section III.1.4.  For simplicity consider a
partition of the paths by an exhaustive set of configuration space
regions $\{\Delta_\alpha\}$ at a single time intermediate time, $t$.  We
could note that the decoherence functional \eqref{fourthreeone} could be
rewritten in the form \eqref{threeonetwenty} where the quantities $\langle
q_f T|C_\alpha|q_00\rangle$ are now defined by
\begin{equation}
\bigl\langle q_f T|C_\alpha | q_0 0\bigr\rangle = \int\nolimits_{[q_0
\alpha q_f]} \delta q\, e^{iS[q(\tau)]/\hbar}\, .
\label{fourfourone}
\end{equation}
{\it If} the paths $q(\tau)$ are single valued in time, the path-integral
may be factored using the identity \eqref{threeoneeighteen} (\cf Fig.~7) into
an integral over paths before $t$, an integral over paths after $t$,
and an integral over the region $\Delta_\alpha$ at $t$.  The integrals
over paths before $t$ may be taken to define a wave function on the
surface $t=t_k$, \viz
\begin{equation}
\psi_{(q_00)} (q,t) = \int_{[q_0\,q_k]} \delta q\, e^{iS[q(\tau)]/\hbar}
\, . \label{fourfourtwo}
\end{equation}
The integral over the paths after $t$ may be taken to define the complex
conjugate of a wave function $\psi_{( q_f T)}(q\, t)$.  The matrix
elements \eqref{fourfourone} are then given by
\begin{equation}
\bigl\langle q_f T\left| C_\alpha\right| q_0 0 \bigr\rangle =
\int\nolimits_{\Delta_\alpha} dq \psi^*_{(q_fT)} (q,t) \psi_{(q_0
0)} (q,t) \label{fourfourthree}
\end{equation}
thus defining an inner product on wave functions.  If we vary the time
$t$ it is an elementary consequence of the definition
\eqref{fourfourtwo} that the wave functions satisfy the Schr\"odinger
equation and the inner product is preserved \cite{FH65}.
In this way we would be led
to a Hamiltonian quantum mechanics of states on spacelike surfaces
evolving unitarily and by reduction of the wave packet.  Two things were
crucial to this derivation.  First, the existence of a set of surfaces
in the extended configuration space $(t, q^i)$ which the paths crossed
once and only once thus defining a notion of time.  Second, a coarse
graining that restricted the paths only on constant time surfaces.  In the
subsequent sections we shall discuss more general coarse grainings and
cases where there are no such surfaces and no associated time.  An
equivalent Hamiltonian formulation cannot then be expected.

Thus, despite their equivalence on certain coarse-grained sets of alternative
histories, Hamiltonian quantum mechanics and sum-over-histories quantum
mechanics are different because their underlying sets of fine-grained
histories
are different.\footnote{For more discussion see \cite{Har91b},
\cite{Har92}, and \cite{Gel89}.} Indeed, as we have presented them, the
fine-grained histories are defined in different spaces in the two cases
--- a space of paths in the sum-over-histories formulation and the space
of chains of projections on ${\cal H}$ in the Hamiltonian formulation.
  Are the more limited coarse grainings of
sum-over-histories
quantum mechanics
adequate for physics?  They are if all testable statements can be
reduced to statements about configuration space variables --- positions,
fields
of integer and half-integer spin, etc.  Certainly this would seem
sufficient
to describe the coarse graining associated with any classical domain.

In the following we shall see that the sum-over-histories formulation of
quantum mechanics provides an accessible route for investigating
generalizations of Hamiltonian quantum mechanics that covariantly
resolve the problem of time in quantum gravity.  The route is accessible
because the main objective of these generalizations will be to cast
quantum mechanics into fully four-dimensional form that does not require
a preferred time.  In sum-over-histories quantum mechanics the
fine-grained histories are spacetime paths and dynamics in summarized by
an action functional on these paths.  One is thus well along the way to
the desired objective and the conceptual clarity afforded by the
sum-over-histories formulation is of considerable help with the rest.
Because of this conceptual clarity, because a sum-over-histories
formulation may be general enough for all realistic applications of
quantum mechanics, and for reasons of simplicity and economy we shall
focus on sum-over-histories generalized quantum mechanics in what
follows.

This focus should not be interpreted to mean that we eschew operator
methods in quantum mechanics.  As we shall describe in Section V,
continuous operator products can be used to rigorously explore the
limits that define certain path integrals.  More importantly, such products can
be used to define generalizations of at least some of the
sum-over-histories frameworks that we explore which incorporate the
richer variety of coarse grainings of transformation theory. Operators
and path integrals are therefore not in conflict and often
complementary.  It is for clarity and simplicity that we focus on
sum-over-histories formulations in these lectures.

\subsection{Classical Physics and the Classical Limit of Quantum
Mechanics.}

Classical physics may be regarded as a trivial
generalized
quantum
mechanics.  The basic elements are:

\begin{enumerate}

\item {\sl Fine-grained histories:}  The fine-grained histories are paths
in phase space, $(p_i(t),\ q^i(t))$, parametrized by the physical time.

\item {\sl Allowed Coarse grainings:} The most familiar type of coarse 
graining is
specified by cells in phase space at discrete sequences of time.  The
paths are
partitioned into classes defined by which cells they pass through.

\item {\sl Decoherence Functional:}  From the perspective of quantum
theory, the distinctive features of classical physics are that the
fine-grained
histories are {\sl exactly} decoherent and exactly correlated in time
according
to classical dynamical laws.  A decoherence functional that captures
these
features may be constructed as follows:

Let $z^i=(p_i,\ q^i)$ serve as a compact notation for a point in phase
space.
 $z^i(t)$ is a phase space path.  Let $z^i_{cl}(t; z^i_0)$ denote the
path that is the classical evolution of the initial condition $z^i_0$ at
time
$t_0$. The path $z^i_{cl}(t) = (p^{cl}_i(t),\ q^i_{cl}(t))$ satisfies
the classical equations of motion:
\begin{equation}
\dot p^{cl}_i = -\frac{\partial H}{\partial q^i_{cl}}\quad ,\quad \dot
q^i_{cl} = \frac{\partial H}{\partial p^{cl}_i}\, ,
\label{fourfiveone}
\end{equation}
where $H$ is the classical Hamiltonian, with the initial
condition
$z^i(t_0; z^i_0) = z^i_0$.  Define a classical decoherence functional,
$D_{cl}$, on pairs of fine-grained histories as
\begin{equation}
D_{cl}  [z^{\prime i}(t),\ z^i(t)]\equiv \delta [z^{\prime i}(t) -
z^i(t)]
 \int d \mu(z^i_0) \delta  [z^i(t) - z^i_{cl}(t; z_0)]
f(z^i_0)\, .
\label{fourfivetwo}
\end{equation}
Here $\delta[\cdot]$ denotes a functional $\delta$-function on the space
of phase space paths, and $d\mu(z^i)$ is the usual Liouville measure,
$\Pi_i[dp_i\, dq^i/(2\pi\hbar)]$.  The function $f(z^i_0)$ is a real,
positive,
normalized distribution function on phase space which gives the initial
condition of the closed classical system.  The first $\delta$-function
in \eqref{fourfivetwo}
enforces the exact decoherence of classical histories; the second
guarantees correlation in time according to classical laws.

A coarse graining of the set of alternative fine-grained histories may
be defined by giving exhaustive partitions of phase space into regions
$\{R^k_{\alpha_k}\}$ at a sequence of times $t_k$, $k=1$, $\cdots, n$.
Here,
$\alpha$ labels the region and $k$ the partition.  The decoherence
functional for
the corresponding
 set of coarse-grained alternative  classical histories is
\begin{equation}
D_{cl} \left(\alpha^\prime, \alpha\right) =
\int\nolimits_{\alpha^\prime} \delta z^\prime \int\nolimits_\alpha
\delta
z D_{cl} [z^{\prime i}(t), z^i(t)]\, ,
\label{fourfivethree}
\end{equation}
where the integral is over pairs of phase space paths restricted by the
appropriate regions and the integrand is \eqref{fourfivetwo}.
It is then also easy to
see that
\eqref{fourfivethree} and \eqref{fourfivetwo}
satisfy the conditions \eqref{fouroneone i} -- \eqref{fouroneone iv}
 of Section IV.1 for
decoherence functionals.
 For all coarse
grainings one has
\begin{equation}
D_{cl}\left(\alpha^\prime, \alpha\right) =
\delta_{\alpha^\prime_1\alpha_1} \cdots \delta_{\alpha^\prime_n\alpha_n}
\ p_{cl} (\alpha_1, \cdots, \alpha_n)\, ,
\label{fourfivefour}
\end{equation}
where $p_{cl}(\alpha_1, \cdots, \alpha_n)$ is the classical probability
to find
the system in the sequence of phase space regions
$\alpha_1,\cdots,\alpha_n$ given
that it is
initially distributed according to $f(z^i_0)$.

It is not just as an academic exercise that we reformulate classical
mechanics as a trivial generalized quantum mechanics. This reformulation
enables us to give a more precise statement of the classical limit of
quantum mechanics.
In certain situations the decoherence functional of a quantum mechanics
may be
well approximated by a classical decoherence functional of the form
\eqref{fourfivethree}.
For example, in Hamiltonian quantum mechanics it may happen that for
some coarse
grained set of alternative histories $\{\alpha\}$
\begin{equation}
D\left(\alpha^\prime, \alpha\right) = Tr
[C_{\alpha^\prime} \rho\, C^\dagger_\alpha]
\cong D_{cl}
\left(\alpha^\prime,
\alpha\right)\, ,
\label{fourfivefive}
\end{equation}
for some corresponding coarse graining of phase space $\{R^k_{\alpha_k}\}$
and distribution function $f$.  One has then exhibited the classical limit
of quantum mechanics.

Some coarse graining is needed for a relation like \eqref{fourfivefive}
to hold because
otherwise the histories, $\{c_\alpha\}$,
 would not decohere.  Moreover, a relation like
\eqref{fourfivefive}
cannot be expected to hold for {\sl every} coarse graining.  Roughly, we
expect
that
the projections $\{P^k_{\alpha_k}\}$ must correspond to phase space regions,
for
example, by projecting onto sufficiently crude intervals of
configuration space
and momentum space or
onto coherent states corresponding to  regions of phase space. (See,
\eg \cite{Hep74}, \cite{Omn89}, and \cite{Cav94} for more on this.)  
Moreover, for a fixed coarse 
graining,  a relation like \eqref{fourfivefive} cannot hold for every initial
condition
$\rho$.
Only for particular coarse grainings and particular $\rho$ do we recover
the classical limit of a quantum mechanics in the sense of
\eqref{fourfivefive}
\end{enumerate}

\subsection{Generalizations of Hamiltonian Quantum Mechanics}

As the preceding example of classical physics illustrates, there are
many
examples of generalized quantum mechanics that do not coincide with
Hamiltonian quantum mechanics.  The requirements for a generalized
quantum
mechanics are weak.  Fine-grained histories, a notion of coarse
graining, and a
decoherence functional and decoherence condition are all that is needed.  There
are probably many
such
constructions.
It is
thus important to search for further physical principles with which to
winnow
these possibilities.  In this search there is also the scope to
investigate
whether the familiar Hamiltonian formulation of quantum mechanics might
not
itself be an approximation to some more general theoretical framework
appropriate only for certain coarse grainings and particular
initial conditions of the universe.
If $D$ were the decoherence functional of the
generalization then
\begin{equation}
D(\alpha^\prime, \alpha) \cong Tr\bigl[C_{\alpha^\prime}
\, \rho\,  C^\dagger_\alpha\bigr]
\label{foursixone}
\end{equation}
only for certain $\{c_{\alpha}\}$'s and corresponding $C$'s and for
 a limited class of $\rho$'s.  Thus, in cosmology it is possible to
investigate
which
features of Hamiltonian quantum mechanics are fundamental and which are
``excess baggage'' that only appear to be fundamental because of our
position
late in a particular universe  able to employ only limited coarse
grainings.\footnote{For more along these lines see
\cite{Har90b}.}  In the
next sections we shall argue that one such feature is the preferred time
of
Hamiltonian quantum mechanics.

\subsection{A Time-Neutral Formulation of Quantum Mechanics}

The Hamiltonian quantum mechanics based on the decoherence functional
\eqref{fourtwoone} is not time neutral.  The future is treated differently
from the past so that the theory incorporates a fundamental,
quantum-mechanical arrow of time. As a first serious example of a
generalized quantum mechanics we shall describe a time-neutral
generalization of quantum mechanics that does not single out an arrow of
time.

The quantum-mechanical arrow incorporated into the decoherence functional
\eqref{fourtwoone} does not arise because of the time ordering of the chains
of projection operators.  Field theory is invariant under $CPT$ and the
ordering can be reversed by a $CPT$ transformation of the projection
operators and density matrix.  To see this, let $\Theta$ denote the
antiunitary  $CPT$ transformation and, for simplicity, consider alternatives
$\{P^k_{\alpha_k} (t_k)\}$ such that their $CPT$ transforms, $\{\widetilde
P^k_{\alpha_k} (-t_k)\}$, are given by
\begin{equation}
\widetilde P^k_{\alpha_k}(-t_k) = \Theta^{-1} P^k_{\alpha_k} (t_k)
 \Theta\, .\label{foursevenone}
\end{equation}
Since the Hamiltonian is invariant under $CPT$ these $CPT$ transforms
continue to be related to each other at different times by
\eqref{threeoneonea}.
Under $\Theta$, a sequence of alternatives at times $t_1<t_2 < \cdots <
t_n$ that is represented by the chain
\begin{equation}
C_\alpha = P^n_{\alpha_n} (t_n) \cdots P^1_{\alpha_1}(t_1)
\label{fourseventwo}
\end{equation}
is transformed into a sequence of $CPT$ transformed alternatives with the {\it
reversed time
ordering} $-t_n < \cdots < -t_2 <-t_1$ represented by the chain
\begin{equation}
\widetilde C^{\rm rev}_\alpha\equiv \Theta^{-1}C_\alpha\Theta =
\widetilde P^n_{\alpha_n}(-t_n)\cdots \widetilde P^1_{\alpha_1}
(-t_1) 
\label{fourseventhree}
\end{equation}
and similarly for alternatives represented by sums of chains.
If the density matrix is also transformed
\begin{equation}
\tilde\rho = \Theta^{-1}\rho \Theta\, ,
\label{foursevenfour}
\end{equation}
then the decoherence functional is complex conjugated
\[
\widetilde D^{\rm rev} \left(\alpha^\prime, \alpha\right) = Tr\bigl[\widetilde
C^{\rm rev}_{\alpha^\prime} \tilde\rho\, \widetilde C^{{\rm
rev}\dagger}_\alpha\bigr] =
Tr[\Theta^{-1} C_{\alpha^\prime}\Theta\Theta^{-1}\rho\Theta\Theta^{-1}
C^\dagger_\alpha \Theta]
\]
\begin{equation}
= Tr \bigl[\Theta^{-1}C_{\alpha^\prime} \rho\, C^\dagger_\alpha\Theta\bigr] =
D^*\left(\alpha^\prime, \alpha\right)\, . 
\label{foursevenfive}
\end{equation}
In the last step the antiunitarity of $\Theta$ which implies
$(\psi,\Theta^{-1}\phi)=(\Theta\psi,\phi)^*$ has been used.
Decoherent sets of histories are thus transformed into decoherent sets of
histories, their
 probabilities are unchanged, but the time ordering has been
reversed.  Either time ordering may therefore be used in formulating
quantum mechanics.
It is {\it by  convention} that we use the ordering in which the
projection
with the earliest time is closest to the density matrix in
\eqref{fourtwoone},
that is, the ordering in which  the density matrix is in the past.

The difference between the future and the past in the
usual formulation of quantum mechanics
arises therefore,  not from the time-ordering of the projections representing
histories, but rather because the ends of the histories are treated
asymmetrically in
\eqref{fourtwoone}.  At one end of the chains of projections (conventionally
the past) there is a density matrix.  At the other end (conventionally
the future) there is the trace.  Whatever conventions are used for time
ordering there is thus an asymmetry between future and past exhibited by
\eqref{fourtwoone}. That asymmetry is the arrow of time in quantum
mechanics.

The observed universe exhibits general time asymmetries.
These include\footnote{For clear reviews and further discussion
see Davies \cite{Dav76}, Penrose \cite{Pen79}, and Zeh \cite{Zeh89}.}

\begin{itemize}

\item The thermodynamic arrow of time --- the fact that
approximately isolated systems are now almost all evolving towards
equilibrium
in the same direction of time.

\item The psychological arrow of time --- we remember the
past, we predict the future.

\item The arrow of time of retarded electromagnetic
radiation.

\item The arrow of time supplied by the $CP$ non-invariance
of the weak interactions and the $CPT$ invariance of field theory.

\item The arrow of time of the approximately uniform
expansion of the universe.

\item The arrow of time supplied by the growth of
inhomogeneity in the expanding universe.

\end{itemize}

All of the time asymmetries on this list
 could arise from time-symmetric dynamical
laws solved with time-asymmetric boundary conditions.  The thermodynamic
arrow of time, for example, is implied by an initial condition in
which the progenitors of today's approximately isolated systems were all
far from
equilibrium at an initial time.  The $CP$ arrow of time could arise as a
spontaneously broken symmetry of the Hamiltonian \cite{Lee88}.
The
approximate uniform expansion of the universe and the growth of
inhomogeneity
follow from an initial ``big bang'' of sufficient spatial homogeneity
and isotropy, given the attractive nature of gravity.
Characteristically such arrows of time
can be reversed temporarily, locally, in isolated subsystems,
although typically at an expense so great that the experiment can
be carried out only in our imaginations.  If we could, in the
classical example of Loschmidt \cite{Los76},
reverse the momenta of all particles and
fields of an isolated subsystem, it would ``run backwards'' with
thermodynamic and electromagnetic arrows of time reversed.

In contrast to the time asymmetries mentioned above, in the
quantum mechanics of closed systems a quantum
mechanical arrow of time would be fundamental and not
reversible.\footnote{The arrow of time in the approximate quantum
mechanics of measured
subsystems is sometimes assumed
 to be deducible from the the
thermodynamic arrow of time and the nature of a measuring
apparatus (see, \eg Bohm \cite{Boh51}). This is a problematical
association, (see the
remarks in \cite{GH93b}) and in any case not germane to the present
discussion of the quantum mechanics of closed systems in which
measurement does not play a fundamental role.}
  That
is not inconsistent with observation because, as we have just described, all of
the observed arrows of time could be explained by special properties of
the initial $\rho$ in the usual formulation of quantum mechanics.
All such arrows of time would
therefore coincide with the fundamental quantum mechanical arrow of
time.  However, as we shall now show, all the arrows of time, including
the quantum mechanical one, can be put
on the same footing in a time-neutral generalization of quantum
mechanics.

Nearly thirty years ago, Aharonov, Bergmann, and Lebovitz \cite{ABL64}
showed how to cast the quantum mechanics of measured subsystems into
time-neutral form by considering final conditions as well as initial
ones.\footnote{For examples of further interesting discussions
of the time-neutral formulation of the quantum mechanics of measured
subsystems see Aharonov and Vaidman \cite{AV91} and Unruh
\cite{Unr86}.}  The same type of framework for the quantum mechanics of
closed
systems has been discussed by Griffiths \cite{Gri84} and by Gell-Mann
and the author
\cite{Har91a} and \cite{GH93b}  as an
example of generalized quantum mechanics.
The fine-grained histories and coarse grainings of this generalized
quantum mechanics are the same as for usual Hamiltonian quantum
mechanics as described in Section IV.2.  Only the decoherence functional
differs by employing both initial and final density matrices.  It is
\begin{subequations}
\label{foursevensix}
\begin{equation}
D\left(\alpha^\prime, \alpha\right) = {\cal N}\, Tr\left[\rho_f
C_{\alpha^\prime} \rho_i C^\dagger_\alpha\right] 
\label{foursevensix a}
\end{equation}
where
\begin{equation}
{\cal N}^{-1} = Tr\left(\rho_f\rho_i\right)\, . 
\label{foursevensix b}
\end{equation}
\end{subequations}
Here, $\rho_i$ and $\rho_f$ are Hermitian, positive operators that we
may
conventionally call Heisenberg operators representing the
 initial and final conditions.
  They need not be normalized as density matrices
with
$Tr(\rho)=1$ because \eqref{foursevensix} is invariant under
changes of normalization. It is easy to verify that
\eqref{foursevensix} satisfies the four requirements \eqref{fouroneone}. 
There is a
similar generalization for sum-over-histories quantum mechanics found by
replacing $\delta(q^\prime_f-q_f)$ in \eqref{fourthreeone} 
by a final density matrix
in configuration space $\rho_f(q^\prime_f, q_f)$ and multiplying by the
same normalizing factor.

The decoherence functional \eqref{foursevensix} is time-neutral.
There is a density
matrix at both ends of each history.  Initial and final conditions may
be interchanged by making use of the cyclic property
of the trace.  Therefore, the quantum mechanics of closed systems based
on \eqref{foursevensix}
does not have a fundamental arrow of time.
Different quantum-mechanical theories of cosmology are specified by
different
choices for the initial and final conditions $\rho_i$ and $\rho_f$.  For
those cases with $\rho_f\propto I$, where $I$ is the unit matrix, this
formulation
reduces to the usual one
because then \eqref{foursevensix} coincides with
\eqref{fourtwoone}.

Lost in this generalization is a built-in notion of causality in quantum
mechanics.  Lost also, when  $\rho_f$ is not proportional to $I$, is any
notion
of a unitarily evolving
``state of the system at a moment of time''.
We cannot construct an
effective density matrix at one time analogous to \eqref{threeonefifteen}
from which {\it alone} probabilities for both
future and past can be calculated.
What is gained is a quantum mechanics without a fundamental
arrow of
time in which {\it all}
 time asymmetries could arise in particular cosmologies
because
of differences between $\rho_i$ and $\rho_f$ or at particular epochs
from their being near the beginning or the end.
That generalized quantum mechanics
embraces a richer variety of possible universes, allowing for the
possibility of violations of causality and advanced as well as
retarded effects.  These, therefore, become testable features of the
universe rather than axioms of the fundamental quantum framework.

From the perspective of this generalized quantum mechanics, the task of
quantum cosmology is to find a theory of {\it both}  the initial and
final conditions that is theoretically compelling and fits our existing
data as well as possible. A final condition of indifference $\rho_f = I$
and a special initial condition $\rho_i$ would seem to fit well and give
rise to the observed arrows of time including the quantum mechanical
one.  More general conditions can be considered.  In the following we
shall adopt this more general and symmetric approach to quantum
cosmology.

\section{The Spacetime Approachto Non-Relativistic Quantum\\
 Mechanics}
\setcounter{footnote}{0}

\subsection{A Generalized Sum-Over-Histories Quantum Mechanics
for Non-Relativistic\\
 Systems}

As mentioned in the Introduction, an objective of these lectures is to
generalize usual quantum mechanics to put it in fully spacetime form so
that it can provide a covariant quantum theory of spacetime. We shall
employ the strategy of first developing these ideas in a series of model
problems which illuminate various aspects of the general relativistic
case.

The most elementary model is non-relativistic particle quantum mechanics
which we consider in this section.  We discussed non-relativistic,
sum-over-histories quantum mechanics as a generalized quantum mechanics
in Section IV.3.  However, we did not exhibit the theory in fully
spacetime form. The sum-over-histories formulation did cast quantum
dynamics into spacetime form involving spacetime histories directly and
summarized by an action that is a functional of particle paths.
However, our discussion of the coarse grainings to which the theory
potentially assigns probabilities was limited to those defined by
alternative ranges of coordinates at definite moments of time. Were
these the most general alternatives for which a quantum theory could
predict probabilities it would inevitably involve a preferred notion of
time.  More general spacetime coarse grainings are easy to imagine.  For
instance, we may partition the paths by their behavior with respect to a
spacetime region $R$ with extent both in space and time (Figure 9).  The
particle's path may never cross $R$ or, alternatively, it may cross $R$
sometime, perhaps more than once.  These two possibilities
are an exhaustive set of
spacetime alternatives for the systems that are not ``at a moment of
time''. In this section, we shall consider such spacetime alternatives
and cast non-relativistic quantum mechanics into fully spacetime form.

\begin{figure}[t]
\begin{center}
\includegraphics[width=4in]{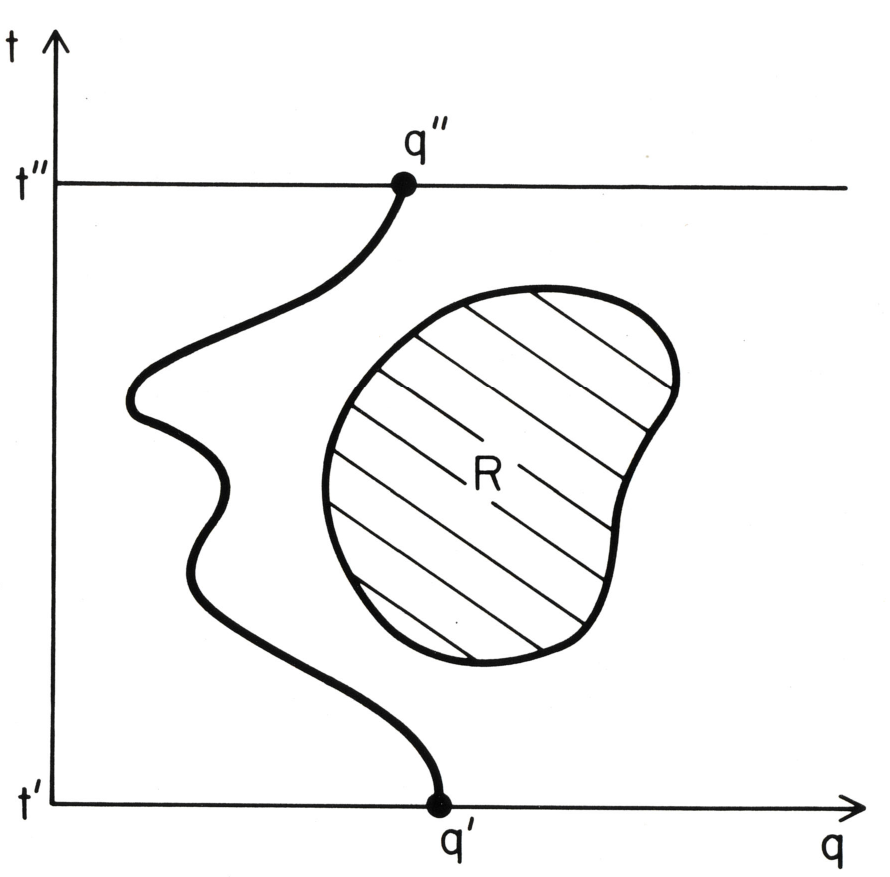}
\caption{\sl Coarse graining by the behavior of paths
with respect to a single spacetime region. The paths which pass between
$q^\prime$ at time $t^\prime$ and $q^{\prime\prime}$ at time
$t^{\prime\prime}$ may be partitioned into two classes. First, the class
of all paths which never cross the region $R$ one of which is
illustrated. Second, the class of paths which intersect $R$ sometime,
generally more than once. This partition defines a set of spacetime
alternatives for the particle which are not at a moment of time. }
\end{center}
\end{figure}

In his original paper on the sum-over-histories formulation of quantum
mechanics, Feynman \cite{Fey48} discussed alternatives defined by
spacetime regions such as we have described above.  In particular, he offered
a sum-over-histories prescription for the probability that ``if an ideal
measurement is performed to determine whether a particle has a path
lying in a region of spacetime\dots the result will be affirmative''.
However, that discussion, as well as more recent ones \cite{Men79},
\cite{Cav86}, \cite{Cav87},  \cite{Sch87}, \cite{Har88a}, \cite{Sor89},
and \cite{YT91a}, were incomplete because they did not specify precisely
what such an ideal measurement consisted of or what was to replace the
reduction of the state vector following its completion.
It is possible
to incorporate spacetime alternatives in a generalized non-relativistic
quantum mechanics in which ``measurement'' does not play a fundamental
role. We now specify more precisely the three elements ---
fine-grained histories, allowed coarse grainings and decoherence
functional --- for such a generalized sum-over-histories quantum
mechanics of the closed, non-relativistic models we shall
consider.\footnote{We shall follow the development in
\cite{Har91a} and \cite{Har91b}.  A very similar formulation was
arrived at independently by Yamada and Takagi in \cite{YT91b} and
\cite{YT92}.}

We consider systems described by an $\nu$-dimensional
configuration space ${\bf R}^\nu$.  The {\it fine-grained histories} are
paths in this configuration space parametrized by the physical time $t$
between times $t=0$ and $t=T$.
We denote the paths by $q(t)$ or by $(q^1(t), q^2(t), \cdots, q^\nu (t))$ 
when it is necessary to specify the individual coordinates.
A defining feature of a non-relativistic system is that its fine-grained
histories
are {\it single-valued} functions of the
physical time --- one and only one $q$ for each value of $t$.  It is a
characteristic feature of sum-over-histories formulations of quantum
mechanics that a unique most fine-grained set of histories is assumed.
In this case it is the set of paths in configuration space.

The {\it
allowed coarse-grainings} are any partition of the class $u$ of all paths
on the time interval $[0,T]$ into an exhaustive set of exclusive classes
$\{c_\alpha\}$.
\begin{equation}
\cup_\alpha c_\alpha = u\ , \qquad\qquad c_\alpha \cap c_\beta =
\phi\ ,\quad \alpha \not= \beta\, . 
\label{fiveoneone}
\end{equation}

The central element is the {\it decoherence functional} for the set of
alternative coarse-grained histories $\{c_\alpha\}$. Since we are
constructing a sum-over-histories formulation with a unique most
fine-grained set of histories (the particle paths) we could proceed by
simply writing down the decoherence functional for this fine-grained
set.  Decoherence functionals for coarse-grained sets are superpositions
of this [\cf (IV.1.1iv)].
However, for better analogy with later models we begin by
constructing the class operators that are the analogs for spacetime
coarse grainings of the chains of projections \eqref{fourtwo} that represent
sequences of alternatives at definite moments of time.

The class operator $C_\alpha$ corresponding to a coarse-grained history
$c_\alpha$
is defined by giving its matrix elements:
\begin{equation}
\left\langle q^{\prime\prime}\left|C_\alpha\right| q^\prime\right\rangle =
\int_{[q^\prime\alpha
q^{\prime\prime}]}
\delta q\, e^{iS[q(\tau)]/\hbar}
\label{fiveonetwo}
\end{equation}
where the sum is over all paths in the class $c_\alpha$ that start at
$q^\prime$ at time $t=0$ and end at $q^{\prime\prime}$ and time $T$.  The class
operators
incorporate the dynamics specified by the action $S[q(\tau)]$.  We assume this
is of standard non-relativistic form:
\begin{equation}
S[q(\tau)] = \int^T_0  dt\ \left[{\cal T}
 (\dot q) - V(q)\right]\, ,
\label{fiveonethreea}
\end{equation}
where ${\cal T}$ is the kinetic energy quadratic form
\begin{equation}
{\cal T}(V) = \half\sum\nolimits^\nu_{i=1} M_i (V^i)^2\ .
\label{fiveonethree}
\end{equation}

We shall return shortly to the mathematical definition of path
integrals like \eqref{fiveonetwo} including the specification of the
``measure''.  For the moment, we note a few consequences of the
definition of class operators.  If $\{c_{\bar\alpha}\}$ is any coarse
graining of the set $\{c_\alpha\}$ so that
\begin{equation}
c_{\bar \alpha} = \cup_{\alpha\epsilon\bar\alpha}\ c_\alpha \quad
, \quad c_{\bar\alpha} \cap c_{\bar\beta} = 0,\, \bar\alpha\not=
\bar\beta\, , 
\label{fiveonefour}
\end{equation}
then it follows immediately from the linearity of the integral in
\eqref{fiveonetwo} that
\begin{equation}
C_{\bar\alpha} = \sum\limits_{\alpha\epsilon\bar\alpha} C_\alpha\, .
\label{fiveonefive}
\end{equation}
If we completely coarse grain, then we have
\begin{equation}
\left\langle q''|C_u  |q'\right\rangle=
 \sum\nolimits_\alpha \left\langle q^{\prime\prime}
\left| C_\alpha \right| q^\prime
\right\rangle = \int_{[q^\prime, q^{\prime\prime}]}
\delta q\, e^{iS[q(\tau)]/\hbar}
\label{fiveonesix}
\end{equation}
where the integral is over the class of
 {\it all} the paths $q(t)$ between $q^\prime$ at
$t=0$ and $q^{\prime\prime}$ at $t=T$. This is the
propagator between $t=0$ and $t=T$ [\cf \eqref{threeonetwentyone}].
Thus, we have
\begin{equation}
\sum\nolimits_\alpha C_\alpha = e^{-iHT/\hbar}\, . 
\label{fiveoneseven}
\end{equation}

The result \eqref{fiveoneseven} points to a difference in normalization
between the class operators defined by \eqref{fiveonetwo} and the chains of
Heisenberg picture projections, \eg \eqref{fourtwo}, used in the
preceding sections. The latter add to unity when summed over all
alternatives while the $C_\alpha$'s of this section add to the unitary
evolution operator over the time interval $T$.  In dealing with path
integrals, the Schr\"odinger picture is more natural than the Heisenberg
one, and as a result the normalization \eqref{fiveoneseven} is more
convenient.  The normalization, however, is only a convention and we
could restore the Heisenberg picture normalization by multiplying all
$C$'s by $\exp(iHT)$.

To construct the decoherence functional $D(\alpha^\prime, \alpha)$ we
must specify not only the class operators but also initial and final
conditions.  In this non-relativistic example, an initial condition is
specified by giving a family of orthonormal wave functions
$\{\psi_j(q)\}$ in the Hilbert space ${\cal H}$ of square integrable
functions on ${\bf R}^\nu$ together with their probabilities
$\{p^\prime_j\}$.  Equivalently and more compactly, we can summarize the
initial condition by the density matrix
\begin{equation}
\rho_i\left(q^\prime_0, q_0\right) = \sum_j
\psi_j\left(q^\prime_0\right) p_j\psi^*_j \left(q_0\right)\, .
\label{fiveoneeight}
\end{equation}
A final condition is similarly specified by a family of orthonormal wave
functions $\{\phi_i(q)\}$ and their probabilities
$\{p^{\prime\prime}_i\}$ or equivalently by a final density matrix
$\rho_f$.

Initial and final conditions are adjoined to the class operators by the
usual inner product in ${\cal H}$
\begin{eqnarray}
\left\langle\phi_i \left| C_\alpha \right| \psi_j \right\rangle & = & \int
dq^{\prime\prime} \int dq^\prime \phi^*_i \left(q^{\prime\prime}\right)
\left\langle
q^{\prime\prime} \left| C_\alpha \right| q^\prime \right\rangle \psi_j
\left(q^\prime\right)\nonumber\\
& = & \int_\alpha \delta q\, \phi^*_i \left(q^{\prime\prime}\right)
\, e^{iS[q(\tau)]/\hbar}\,\psi_j\left(q^\prime\right) 
\label{fiveonenine}
\end{eqnarray}
where in the second line of \eqref{fiveonenine} we understand the path
integral to include an integration over the endpoints $q^\prime$ and
$q^{\prime\prime}$.

We now define the decoherence functional $D(\alpha^\prime, \alpha)$
by:
\begin{subequations}
\label{fiveoneten}
\begin{eqnarray}
D\left(\alpha^\prime, \alpha\right) & = & {\cal N} \sum_{ij}
p^{\prime\prime}_i \left\langle \phi_i \left| C_{\alpha^\prime} \right|
\psi_j \right\rangle p^{\prime}_j \left\langle\psi_j\left|
C_\alpha \right| \phi_i \right\rangle 
\label{fiveoneten a}\\
& = & {\cal N}\, Tr\left[\rho_f C_{\alpha^\prime} \rho_i C^\dagger_\alpha
\right] 
\label{fiveoneten b}
\end{eqnarray}
where
\begin{equation}
{\cal N}^{-1} = Tr\left[\rho_f e^{-iHT} \rho_i e^{iHT}\right]\, .
\label{fiveoneten c}
\end{equation}
\end{subequations}
Equation \eqref{fiveoneten} is the same as \eqref{foursevensix} but using
Schr\"odinger picture representatives of the initial and final
conditions rather than Heisenberg ones.

It is straightforward to verify that the decoherence functional defined
by \eqref{fiveoneten} satisfies the four requirements \eqref{fouroneone} of a
decoherence functional of a generalized quantum mechanics.  It is
Hermitian because $\rho_i$ and $\rho_f$ are Hermitian, and positive
because they are positive. It is normalized because of \eqref{fiveoneseven}.
It obeys the superposition principle because of \eqref{fiveonefive}.

The generalized quantum mechanics we have just constructed appears to
depend explicitly on the time interval $T$.  Paths were considered on
the time interval $[0,T]$ and the class operators and decoherence
functional depend explicitly on its length.  However, any partition of
the paths on the interval $[0,T]$ is also trivially a partition of the
paths on a longer interval $[0,\widetilde T]$, $\widetilde T\geq T$. The
class operators are related by
\begin{equation}
\widetilde C_\alpha = e^{-iH(\widetilde T-T)/\hbar}\, C_\alpha\, .
\label{fiveoneeleven}
\end{equation}
If the final density matrix at $\widetilde T$ is related to that at $T$
by Schr\"odinger evolution
\begin{equation}
\tilde \rho_f = e^{-iH(\widetilde T-T)/\hbar}\,\rho_f e^{iH(\widetilde
T-T)/\hbar}\, ,
\label{fiveonetwelve}
\end{equation}
then the decoherence functional is independent of $\widetilde T$.  A
similar argument shows independence of the time of the initial condition,
provided $\rho_i$ evolves according to the Schr\"odinger equation.

We have given the decoherence functional in its general time-neutral
form with both initial and final conditions. As discussed in Section
IV.7, a final condition of indifference with respect to final state,
$\rho_f=I$, is likely to be an accurate representation of the final
condition of our universe.  In that case the decoherence functional
takes the more familiar form
\begin{equation}
D\left(\alpha^\prime, \alpha\right) = Tr\bigl[C_{\alpha^\prime} \rho
C^\dagger_\alpha\bigr] 
\label{fiveonethirteen}
\end{equation}
where $\rho_i =\rho$ is a normalized density matrix representing the
initial condition.

The generalized sum-over-histories, non-relativistic quantum mechanics we
have just constructed is in fully spacetime form.  Dynamics are expressed
as a sum-over-fine-grained-spacetime histories involving an action
functional of these histories and coarse-grained alternatives are
defined by spacetime partitions of these histories.  The class of
alternatives considered by this generalized quantum mechanics is thus
greatly enlarged beyond the usual alternatives at definite moments of
time.  It is this extension of the ``observables'' that will be
important in constructing quantum mechanics for theories where there is
no well defined notion of time.  In the following we shall illustrate
some of these more general spacetime alternatives explicitly in
non-relativistic quantum mechanics.  First, however, we consider how to
define the path integrals involved.

\subsection{Evaluating Path Integrals}

\subsubsection{Product Formulae}

We are interested in the path integrals that define the class operators
of the form
\begin{equation}
\left\langle \phi_i\left| C_\alpha \right| \psi_j \right\rangle =
\int_\alpha \delta q\, \phi^*_i \left(q^{\prime\prime}\right) e^{iS[q(\tau)]}
\psi_j
\left(q^\prime\right)\, , 
\label{fivetwoone}
\end{equation}
units having been chosen for this and subsequent sections so that $\hbar
= 1$.
How are they defined and how do we compute them?

General arguments \cite{Cam60} show that it is not possible to introduce a
complex measure on the space of paths to define the Feynman integral.
However, path integrals may be defined and computed by other
means \cite{DMN79}. Here, we take the point of view, introduced by
Feynman \cite{Fey48}, that expressions like \eqref{fivetwoone} are to be
{\it defined}
by the
limits of their values on polygonal (skeletonized) paths on a
time slicing of the interval $[0, T]$.  Suppose that this  interval is
divided
into $N$ sub-intervals of equal length $\epsilon = T/N$ with boundaries
at $t_0 = 0, t_1, t_2, \cdots, t_N = T$.  A polygonal path is specified
by giving the values $(q_0, \cdots, q_N)$ of $q(t)$ on the $N+1$ time
slices
including the value $q_0 (\equiv q^\prime)$ at the initial time $t=0$ and the
value $q_N$
$(\equiv q^{\prime\prime})$ at the final time $t_N=T$.  The polygonal paths
consist
of
straight line segments joining the points $(q_0, \cdots, q_N)$ at the
times defining the subdivision.  The non-relativistic action 
\eqref{fiveonethreea} is
straightforwardly evaluated on polygonal paths when the spacing
$\epsilon$ is small.
\begin{equation}
S(q_N, \cdots, q_0) \approx \sum\limits^{N-1}_{k=0} \epsilon
\Biggl[{\cal T}\biggl(\frac{q_{k+1}-q_k}{\epsilon}\biggr) -
V(q_k)\Biggr]\, .
\label{fivetwotwo}
\end{equation}

Any partition of continuous paths will also partition the polygonal
paths.  Let $e_\alpha (q_N, \cdots, q_0)$ be the function which is unity
on all polygonal paths in the class $c_\alpha$ and zero otherwise.
Then, with these preliminaries, we define an expression like
\eqref{fivetwoone}
 as
the following limit
\[
\left\langle \phi_i \left| C_\alpha \right| \psi_j\right\rangle
= \lim\limits_{N\to\infty} \int dq_N \int dq_{N-1}
\cdots \int dq_0\ \mu(N)
\]
\begin{equation}
\times \phi^*_i (q_N) e_\alpha (q_N,\cdots,q_0) e^{iS(q_N,\cdots,q_0)}
\psi_j (q_0)\, .
\label{fivetwothree}
\end{equation}
where $\mu(N)$ is an $N$-dependent constant ``measure'' factor and the
integrals are all over ${\bf R}^\nu$.

The definition \eqref{fivetwothree} is not, by itself,
a computationally effective way
of evaluating Feynman integrals.  Operator methods provide a more
efficient tool.
As was first recognized by Nelson \cite{Nel64}, operator product formulae
provide both a way of demonstrating the existence of limits like
\eqref{fivetwothree}
and of evaluating the class operators $C_\alpha$ to which they
correspond.\footnote{For further discussion of the definition of path
integrals see Simon \cite{Sim79} and DeWitt, Maheshwari and Nelson
\cite{DMN79}.}  As the most familiar example, consider the propagator which
is the path integral \eqref{fivetwoone}
evaluated over the class, $u$, of all paths on
the time interval $[0,T]$. Then $e_\alpha = 1$.  Divide the total
Hamiltonian $H$ following from the action \eqref{fiveonethreea} into a free
part $H_0$
corresponding to the kinetic energy ${\cal T}$ and the potential $V$:
\begin{equation}
H = \sum\nolimits^\nu_{i=1} \frac{p^2_i}{2M_i} + V(q^k) \equiv H_0 +V
\, .\label{fivetwofour}
\end{equation}
The propagator for the free part of the Hamiltonian is an elementary
calculation,
\begin{equation}
\langle q^{\prime\prime}|e^{-iH_0t}|q^\prime\rangle
 = F(t) \exp \Biggl[it{\cal T}
\biggl(\frac{q^{\prime\prime}-q^\prime}{t}\biggr)\Biggr]\, ,
\label{fivetwofive}
\end{equation}
where
\begin{equation}
F(t) = \prod\limits^\nu_{i=1} (M_i/2\pi it)^\half\, .
\label{fivetwosix}
\end{equation}
It follows that, if the constant $\mu$ in \eqref{fivetwothree} happens to be
$[F(\epsilon)]^N$, then we can write
\begin{equation}
 \left\langle \phi_i \left| C_\alpha \right| \psi_j\right\rangle
 = \lim\limits_{N\to\infty} \langle\phi_i|\Bigl(e^{-iH_0(T/N)}
\ e^{-iV(T/N)}\Bigr)^N|\psi_j\rangle\, .
\label{fivetwoseven}
\end{equation}
This fixes the ``measure'' in the path integral. If $H_0$ and $V$ are
densely defined, self-adjoint and bounded from
below, the Trotter product formula \cite{Tro59} states
\begin{equation}
\lim\limits_{N\to\infty}\Bigl(e^{-iH_0(T/N)}\ e^{-iV(T/N)}\Bigr)^N
= e^{-i(H_0 + V)T}\, .
\label{fivetwoeight}
\end{equation}
Thus, the limit in \eqref{fivetwothree} exists, and the path integral 
$\left\langle
\phi_i \left| C_\alpha \right| \psi_j\right\rangle$ is
evaluated as
\begin{equation}
  \left\langle \phi_i \left| C_\alpha \right| \psi_j\right\rangle
 = \langle\phi_i|e^{-iHT}|\psi_j\rangle\, .
\label{fivetwonine}
\end{equation}
The relation \eqref{fivetwonine} is hardly a surprise.  It is the path integral
expression for the propagator originally derived by Feynman
\cite{Fey48}.

From this perspective formulating quantum mechanics in terms of path
integrals does not eliminate the need for Hilbert space.  Indeed, a
Hilbert space is central to the product formulae approach to defining
path integrals.  In
non-relativistic quantum mechanics, the Hilbert space used to define
path integrals coincides with the Hilbert space of states on a constant
time surface.  In general, however, we shall see that there is no such
connection, not least because it is not always possible to define
states on a spacelike surface.

\subsubsection{Phase-Space Path Integrals}

In \eqref{fivetwoseven} we used a product formula to fix the ``measure''
factors \eqref{fivetwosix} in the path integral.  Evaluated in a different
way, the product formula can be interpreted as an integral over phase
space paths which gives a more convenient and physically suggestive way
of summarizing these factors.  As such phase-space path integrals are
a natural way of fixing the measure in the generalizations of quantum
mechanics we shall consider later we briefly pause to consider such
integrals here.

We recover the Lagrangian path integral for $C_u$ [Eq.~\eqref{fivetwothree}
with $e_u=1$] if the resolution of the identity
\begin{equation}
I = \int dq |q\rangle\,\langle q| 
\label{fivetwoten}
\end{equation}
is inserted between each factor of the product in \eqref{fivetwoseven} and
\eqref{fivetwofive} is used to evaluate the matrix elements involving $H_0$.
The same procedure used with the resolution
\begin{eqnarray}
I & = & \frac{1}{2\pi}\int dp\,dq | q\rangle\,\langle q | p\rangle\,\langle p|
\nonumber\\
 & = & \frac{1}{2\pi} \int dp\,dq\, e^{ip\cdot q} |q\rangle \, \langle p|
\label{fivetwoeleven}
\end{eqnarray}
yields the phase-space path integral.  The matrix elements are immediate
since $H_0$ is diagonal in $p_k$ and $V$ is diagonal in $q^k$.  The result
for the configuration space matrix elements of $C_u$ is the limit.
\[
\langle q^{\prime\prime} |C_u| q^\prime\rangle = \langle q^{\prime\prime} |
e^{-iHT} | q^\prime\rangle = \lim_{N\to\infty} \int
\frac{dp_N}{2\pi} \prod^{N-1}_{k=1} \left(\frac{dp_k dq_k}{2\pi}\right)
\]
\begin{equation}
\times \exp\Biggl[i\sum^N_{J=1} \epsilon\left\{p_J\cdot
\left(\frac{q_J-q_{J-1}}{\epsilon}\right) - \Bigl[H_0\left(p_J\right) +
V\left(q_{J-1}\right) \Bigr]\right\}\Biggr]\, .
\label{fivetwotwelve}
\end{equation}
Here, of course, $p\cdot q = p_iq^i\ ,\ H_0(p)$ is the {\it function}
defined by \eqref{fivetwofour}, and $dp$ and $dq$ represent the usual volume
elements on the $\nu$-dimensional momentum and configuration spaces
respectively.  The limit \eqref{fivetwotwelve} defines the phase-space path
integral
\begin{equation}
\left\langle q^{\prime\prime}\left| C_u\right| q^\prime\right\rangle = \int
\delta p
\, \delta q\,\exp\left[i \int^T_0 dt\, \Bigl[p\cdot (dq/dt) -
H(p,q)\Bigr]\right]\, . 
\label{fivetwothirteen}
\end{equation}

The phase-space path integral \eqref{fivetwothirteen} has been discussed by
many authors, \eg \cite{Fey51} and \cite{Gar66}. The construction can
be extended to define the $C_\alpha$'s for finer configuration space
coarse grainings simply by restricting the $q$-integral to the class of
paths $\{c_\alpha\}$.  It can also be extended to incorporate momentum
coarse-grainings (see, \eg \cite{GH93}).  The interpretation of
\eqref{fivetwothirteen} as an integral is not as straightforward as in the
Lagrangian case.\footnote{See Schulman \cite{Schul81} for a convenient
discussion.} Among other things, the momentum space paths are
discontinuous.

For our purposes, the utility of the phase-space path integral is that
it provides a physically transparent way of summarizing the ``measure''
in the path integral and a way of computing that measure in more general
cases.  The measure in \eqref{fivetwotwelve} is the canonical, Liouville
measure in phase-space ``$dpdq/(2\pi\hbar)$''. Since the momentum space
integrals like \eqref{fivetwotwelve} are unconstrained even when the
configuration space ones are restricted by a coarse graining, the
Gaussian integrals over the $p^i_J$ can be carried out explicitly. The
result is the Lagrangian path integral \eqref{fivetwothree} over
configuration space paths including the correct ``measure'' factor
$[F(T/N)]^N$ with $F$ given by \eqref{fivetwosix}.

\subsection{Examples of Coarse Grainings}

\subsubsection{Alternatives at Definite Moments of Time}

The most familiar type of coarse graining is by regions of configuration
space at successive moments of time (see Figure 10) described briefly in
Section IV.3.  Suppose, for
example,
we consider
 sets of exhaustive non-overlapping regions of ${\bf R}^\nu$,
$\{\Delta^1_{\alpha_1} \}$, $\{\Delta^2_{\alpha_2} \}$, $ \cdots$,
$\{\Delta^n_{\alpha_n} \}$ at a
discrete series of times $t_1, \cdots, t_n$.  At each time
$t_k$
\begin{equation}
\cup_{\alpha_k} \Delta^k_{\alpha_k}  = {\bf R}^\nu\ , \qquad \qquad
\Delta^k_{\alpha_k}  \cap \Delta^k_{\beta_k}  = \phi\ , \quad
\alpha_k \not= \beta_k\, .\label{fivethreeone}
\end{equation}

Since the paths are single valued in time, they pass through one and
only
one region at each of the instants $t_k$.  The class of all paths may be
partitioned into all possible ways they cross these
regions.  Coarse grained histories are thus labeled by the
particular sequence of regions $\Delta^1_{\alpha_1}, \cdots,
\Delta^n_{\alpha_n}$ at
times $t_1, \cdots, t_n$.  We write them as $c_{\alpha_n \cdots
\alpha_1}$.
The individual coarse-grained history $c_{\alpha_n \cdots \alpha_1}$
corresponds to the particle being localized in region
$\Delta^1_{\alpha_1}$ at
time $t_1$, $\Delta^2_{\alpha_2}$ at time $t_2$ and so forth.

The class operators $C_\alpha$ for coarse grainings defined by
alternative spatial regions at definite moments of time are readily
evaluated by the techniques of the last subsection.  The integrals in
\eqref{fiveonetwo} are restricted to the ranges $\Delta^1_{\alpha_1} ,
\cdots, \Delta^n_{\alpha_n} $ on the slices $t_1,\cdots, t_n$.  The
corresponding product formula analogous to \eqref{fivetwoseven} will consist of
unitarity evolution in between these times interrupted by projections on
these ranges at them.  Thus if $\alpha=(\alpha_n, \cdots, \alpha_1)$
denotes the
coarse-grained history in which the paths pass through regions
$\Delta^1_{\alpha_1}, \cdots, \Delta^n_{\alpha_n}$ at times
$0\leq t_1\leq t_2\leq \cdots \leq t_n \leq T$,  then
\begin{equation}
C_\alpha = e^{-iH(T-t_n)} P^n_{\alpha_n}
e^{-iH(t_n-t_{n-1})} P^{n-1}_{\alpha_{n-1}}\cdots P^1_{\alpha_1}
e^{-iHt_1}
\label{fivethreetwo}
\end{equation}
where $P^k_{\alpha_k}$ is the projection on the configuration
space region $\Delta^k_{\alpha_k} $ at time $t_k$.  The expression is
more compact with Heisenberg picture operators
\begin{equation}
C_\alpha = e^{-iHT} P^n_{\alpha_n} (t_n) \cdots
P^1_{\alpha_1} (t_1)\, .
\label{fivethreethree}
\end{equation}
This is enough to show that the $C_\alpha$ in general will neither be
unitary nor Hermitian.  Neither is it true that $C_\alpha C_\beta= 0$
for distinct histories.
The relations \eqref{fivethreeone} expressing the conditions that the regions of
configuration space are exhaustive and exclusive at each time translate
into
\begin{equation}
\sum\nolimits_{\alpha_k} P^k_{\alpha_k} (t_k) = 1, \quad P^k_{\alpha_k}
(t_k) P^k_{\alpha^\prime_k} (t_k) = \delta_{\alpha_k\alpha^\prime_k}
P^k_{\alpha_k}(t_k)\, .
\label{fivethreefour}
\end{equation}
These are enough to show explicitly that \eqref{fivetwofour}
is satisfied and further
that
\begin{equation}
\sum\nolimits_\alpha
C^\dagger_\alpha
C_\alpha =  I
\label{fivethreefive}
\end{equation}
for this particular class of coarse grainings. Thus, we recover from the
sum-over-histories formulation the usual Hamiltonian expressions for the
class operators of this kind of coarse graining [\cf \eqref{fourtwo} with
appropriate change in normalization].

\begin{figure}[t]
\begin{center}
\includegraphics[width=5in]{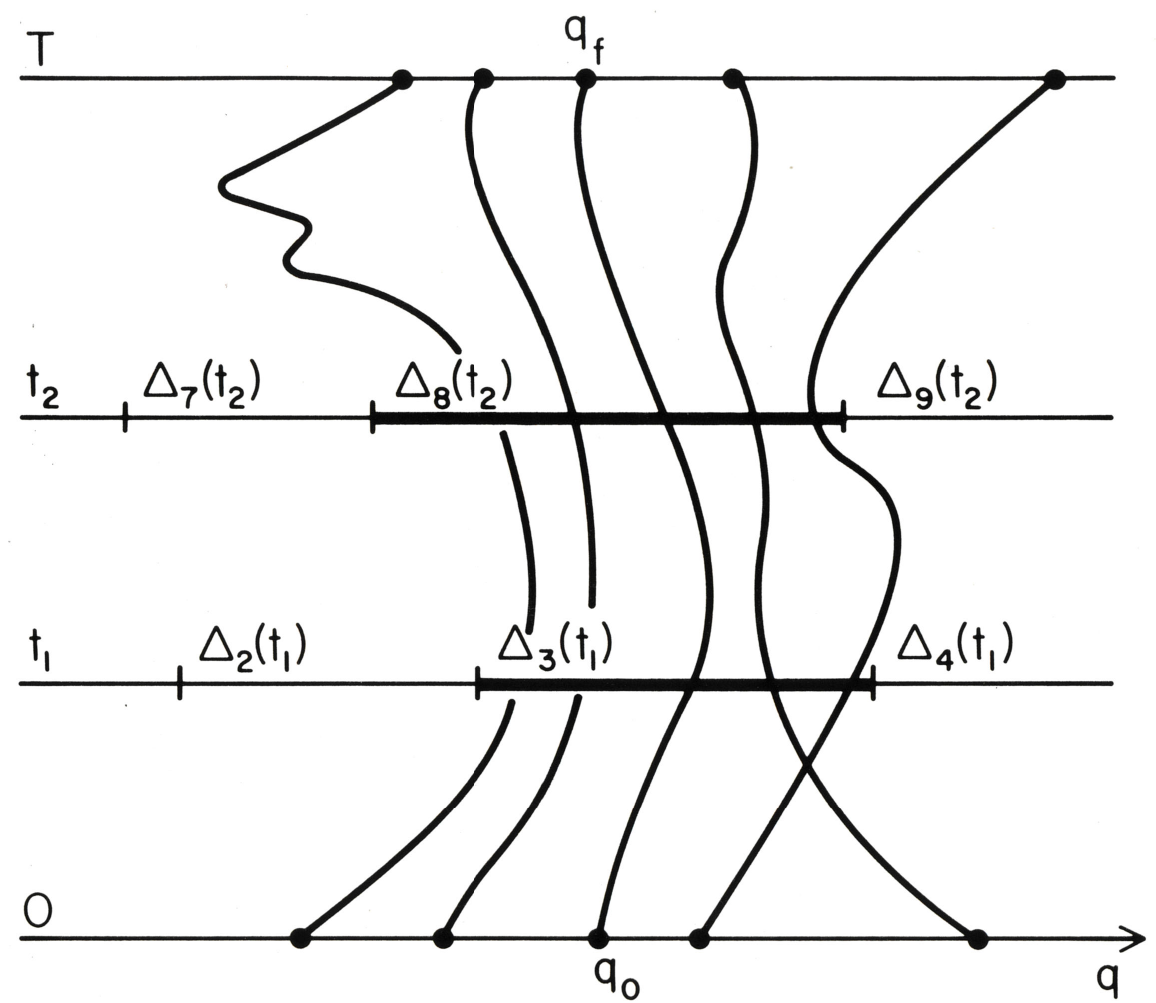}
\caption{\sl Coarse graining by regions of
configuration space at
successive moments of time.  The figure shows a spacetime that is a
product of a one-dimensional configuration space $(q)$ and the time
interval $[0,T]$.  At times $t_1$ and $t_2$ the configuration space is
divided into exhaustive sets of non-overlapping intervals:
$\{\Delta^1_{\alpha_1}\}$ at time $t_1$, and
$\{\Delta^2_{\alpha_2}\}$ at time $t_2$.  Some of these intervals are
illustrated. (The superscripts have been omitted from the $\Delta$'s for
compactness.)  The fine-grained histories are the paths which pass
between $t=0$ and $t=T$.  Because the paths are assumed to be
single-valued in time, the set of fine-grained histories may be
partitioned according to which intervals they pass through at times
$t_1$ and $t_2$.  The figure illustrates a few representative paths in
the class $c_{83}$ which pass through region $\Delta^1_3$ at time
$t_1$ and region $\Delta^2_8$ at time $t_2$.}
\end{center}
\end{figure}

\subsubsection{Alternatives Defined by a Spacetime Region}

Coarse grainings by spatial regions at definite moments of time are only
a very special case of the coarse grainings that are possible in
sum-over-histories quantum mechanics.  As an example of a more general
coarse grainings we consider partitions of the paths according to their
behavior with respect to a spacetime region $R$ (Figure 9).

Given a spacetime region $R$, the paths between $t=0$ and $t=T$ may be
partitioned into two exclusive classes: (1)~~The class $\bar r$
of all paths that never intersect $R$, and (2)~~the class $r$ of
paths that intersect $R$ at least once.  To evaluate the corresponding
class operators we begin with the path integral over the class $c_{\bar
r}$
\begin{equation}
\left\langle \phi_i \left| C_{\bar r}\right| \psi_j\right\rangle
= \int\nolimits_{\bar r} \delta q\ \phi^*_i
(q^{\prime\prime})\,\exp\bigl(iS[q(\tau)])\psi_j (q^\prime)\, .
\label{fivethreesix}
\end{equation}

The Feynman integral is the limit of the integral over polygonal paths
in $\bar r$ as in \eqref{fivetwothree}. Each constant-time cross section of
$R$ is a region of configuration space $\Delta(t)$. The paths of $\bar
r$ lie entirely in the complements of these regions, $\bar\Delta (t)$.
By introducing projection operators on the regions $\bar\Delta$ at the
various
times, the integrals over polygonal paths defining \eqref{fivethreesix}
may be expressed as matrix elements of operators.  Let
$P_{\bar\Delta(t)}$ denote the projection onto the complement of
$\Delta(t)$.  $P_{\bar\Delta(t)}$ is time dependent, not because it is a
Heisenberg picture operator, but because the region $\bar\Delta(t)$ is
time dependent.  Clearly
\begin{equation}
<q^{\prime\prime} | P_{\bar\Delta} | q^\prime > = \delta
(q^{\prime\prime} - q^\prime) e_{\bar\Delta} (q^\prime)\ .
\label{fivethreeseven}
\end{equation}
Using this and the free propagator, \eqref{fivetwofive}, the path integral
over
the class $\bar r$ can be written as the limit
\begin{equation}
\left\langle\phi_i \left| C_{\bar r} \right| \psi_j \right\rangle
= \lim\limits_{N\to\infty} <\phi_i |
{\bf T}\prod\limits^{N-1}_{k=0} \Biggl(e^{-iH_0(T/N)}
e^{-iV(T/N)} P_{\bar\Delta(kT/N)} \Biggr)|\psi_j>
\label{fivethreeeight}
\end{equation}
where the product is time ordered --- written with the earliest
$P_{\bar\Delta(t)}$'s to the right.
The projection, $P_{\bar\Delta(t)}$ can be written in the form
\begin{equation}
P_{\bar\Delta(t)} = e^{-E_R(t)\epsilon}
\label{fivethreenine}
\end{equation}
where $\epsilon$ is an arbitrary positive number and $E_R$ is the
{\it excluding potential} for the spacetime region $R$, that is
\begin{equation}
E_R(q,t)=\left\{ \begin{array}{lr}
                0 & (q,t)\ \notin \  R\, ,\nonumber\\
                +\infty & (q,t)\ \in \ R\, .
                \end{array} \right.
\label{fivethreeten}
\end{equation}
Choosing $\epsilon=T/N$ we may then write \eqref{fivethreeeight} as
\begin{equation}
\left\langle\phi_i \left| C_{\bar r} \right| \psi_j \right\rangle
= <\phi_i|\lim\limits_{N\to\infty}
{\bf T}\prod\limits^{N-1}_{k=0} \Biggl(e^{-iH_0(T/N)}
 e^{-i(V-iE_R(kT/N))(T/N)}\Biggr) |\psi_j>\, .
\label{fivethreeeleven}
\end{equation}
Again, the operators in \eqref{fivethreeeleven}
 are time ordered with the earliest on the
right.

As a generalization of the Trotter product formula \eqref{fivetwoeight}
 we expect
\[
\lim\limits_{N\to\infty}{\bf T} \prod\limits^{N-1}_{k=0}\left(e^{-iH_0(T/N)}
e^{-i(V-iE_R(kT/N))(T/N)}\right)
\]
\begin{equation}
= {\bf T} \exp \biggl\{-i\int\nolimits^T_0 dt\bigl[H_0
+V-iE_R(t)\bigr]\biggr\}
\label{fivethreetwelve}
\end{equation}
where ${\bf T}$ denotes the time ordered product.\footnote{The
author knows of no rigorous demonstration of a product formula general
enough to prove
\eqref{fivethreetwelve} at the time of writing.  The mathematical issues
concern the time dependence of $E_R(t)$ and the fact that it is not
self-adjoint because its domain is not dense in the Hilbert space.}
  That is,
the right hand side of \eqref{fivethreetwelve} may be interpreted as 
$U_R(T)$ where $U_R(t)$ is the solution of
\begin{equation}
i\frac{dU_R(t)}{dt} = \left[H_0+V-iE_R(t)\right]U_R(t)\, ,
\label{fivethreethirteen}
\end{equation}
with the boundary condition
\begin{equation}
U_R(0)=I\, .
\label{fivethreefourteen}
\end{equation}
Physically \eqref{fivethreethirteen} represents Schr\"odinger
evolution in the presence of a completely absorbing potential on the
spacetime region $R$.  Paths that once cross into the region $R$ do not
contribute to the final value of $U$.

Equation \eqref{fivethreetwelve} allows us to identify the class operators 
for the
coarse graining based on a single spacetime region, $R$.  There are two
coarse-grained histories in the set:  $r$, the class of fine-grained
histories which cross $R$ at least once and $\bar r$, class of the
fine-grained histories which never cross $R$.  For $\bar r$ we have
\begin{equation}
C_{\bar r} = U_R(T) = {\bf T} \exp\biggl\{-i
\int\nolimits^T_0 dt \left[H_0+V-iE_R(t)\right]\biggr\}\, .
\label{fivethreefifteen}
\end{equation}
The operator $C_r$ then follows from the fact that the set of paths
$r$ which cross $R$ at least once is the {\it difference} between
the set of all paths $u$ and the set which $\bar r$ which never
cross $R$:
\begin{equation}
r= u-\bar r
\label{fivethreesixteen}
\end{equation}
where, as usual, $a-b \equiv a\cap\bar b$.
The corresponding relation for the class operators is
\begin{equation}
C_r = e^{-iHT} - U_R(T)\, ,
\label{fivethreeseventeen}
\end{equation}
which is the same as \eqref{fiveoneseven}.

\subsubsection{A Simple Example of a Decoherent Spacetime Coarse Graining}

Consider a free particle in one
dimension and let the region $R$ be the whole region $x<0, \; 0<t<T$. Then
$C_{\bar r}$ is just the evolution operator in the presence of an
infinite potential wall at $q=0$,   that is
\[
<q^{\prime\prime} | C_{\bar r} | q^\prime > = \theta (q^{\prime\prime})
\theta (q^\prime) \biggl(\frac{M}{2\pi iT}\biggr)^\half
\]
\begin{equation}
\times\Biggl\{\exp\left[i\frac{M}{2T}
(q^{\prime\prime}-q^\prime)^2\right] - \exp
\left[i\frac{M}{2T} (q^{\prime\prime} + q^\prime)^2\right]\Biggr\}\, .
\label{fivethreeeighteen}
\end{equation}
From \eqref{fivethreeseventeen} the position matrix elements of $C_r$ are the
free propagator minus \eqref{fivethreeeighteen} or
\[
<q^{\prime\prime} | C_{ r} | q^\prime > = \Bigl[\theta
(q^{\prime\prime})
\ \theta (-q^\prime) + \theta (-q^{\prime\prime}) \theta
(q^\prime)\Bigr]
\left(\frac{M}{2\pi iT}\right)^\half
 \exp \Bigl[i\frac{M}{2T} (q^{\prime\prime}-q^\prime)^2\Bigr]
\]
\begin{equation}
+ \bigl[\theta (q^{\prime\prime}) \theta (q^\prime) + \theta
(-q^{\prime\prime}) \theta (-q^\prime)\bigr]
 \left( \frac{M}{2\pi iT}\right)^\half
 \exp \left[i\frac{M}{2T}
(q^{\prime\prime} +q^\prime)^2\right]\, .
\label{fivethreenineteen}
\end{equation}

Special choices of the initial condition can give examples in which the
alternatives $r$ and $\bar r$ are decoherent.  Such examples have been
investigated especially by Yamada and Takagi \cite{YT91b}.  A simple
case is obtained by considering a pure initial state with a wave
function $\psi(x)$. Write this as
\begin{equation}
\psi(x) = \alpha\,\phi_+(x) + \beta\,\phi_-(x)\ ,
\qquad\ \left|\alpha\right|^2 + \left|\beta\right|^2 = 1\, ,
\label{fivethreetwenty}
\end{equation}
where $\phi_+(x)$ and $\phi_-(x)$ are normalized wave functions
having support on $x>0$ and $x<0$ respectively.  The branch wave
functions corresponding to the alternatives $r$ and $\bar r$ may be
expressed in terms of the free unitary evolution operator for the time
interval $T$ which we denote by $U$.  Thus, for example, the branch,
$\psi_{\bar r}$, representing the alternative that the particle never crosses
into $x<0$ in the time interval $T$ is
\begin{equation}
\psi_{\bar r} (x) = \alpha P_+ \bigl[U\phi_+ (x) - U\phi_+ (-x)\bigr]
\label{fivethreetwentyone}
\end{equation}
where $P_+$ is the projection onto $x>0$. Eq.~\eqref{fivethreetwentyone} is 
just
the usual ``method of images'' solution of the Schr\"odinger equation is
the presence of an infinite barrier at $x = 0$ and is another way of
writing \eqref{fivethreeeighteen}.

The other branch is
\begin{equation}
\psi_r(x) = U\psi(x) - \psi_{\bar r} (x)\, . 
\label{fivethreetwentytwo}
\end{equation}
The condition for decoherence is
\begin{equation}
\left(\psi_r, \psi_{\bar r}\right) = 0\, . 
\label{fivethreetwentythree}
\end{equation}
Evidently this is a linear relation and $\alpha$ and $\beta$ of the form
\begin{equation}
\alpha c_+ + \beta c_- = 0 
\label{fivethreetwentyfour}
\end{equation}
where $c_\pm$ are coefficients completely determined by $\phi_\pm$ and
$U$.  Eq.~\eqref{fivethreetwentyfour} and the normalization condition
\eqref{fivethreetwenty} fix $\alpha$ and $\beta$.

The probabilities for the decoherent set of alternatives may also be
expressed directly in terms of $c_\pm$.  We have
\begin{subequations}
\label{fivethreetwentyfive}
\begin{eqnarray}
p_r = \left(\psi_r, \psi_r\right) & = & c^2_+/\bigl(c^2_+ + c^2_-\bigr)
\, , \label{fivethreetwentyfive a}\\
p_{\bar r} = \left(\psi_{\bar r}, \psi_{\bar r}\right)
& = & c^2_-/ \bigl(c^2_+ +
c^2_-\bigr)\, . \label{fivethreetwentyfive b}
\end{eqnarray}
\end{subequations}
It is not difficult to be convinced that, by different choices of
$\phi_\pm$, examples of the whole range of possible probabilities may be
obtained.  An especially simple example is to take \cite{YT91b}
\begin{equation}
\phi_+(x) = -\phi_{-}(-x) 
\label{fivefortyeightaa}
\end{equation}
and $\alpha=\beta=1/\sqrt2$.  Then, for any $\phi_-(x)$ decoherence is exact
and $p_r=p_{\bar r}=1/2$ --- both results which alternatively follow
from symmetry considerations.

These examples show that decoherence of spacetime coarse grainings can
be achieved in special examples and that these
alternatives can have non-trivial probabilities.

\subsection{Coarse Grainings by Functionals of the Paths}

\subsubsection{General Coarse Grainings}

The most general notion of coarse graining is given by partitions of the
paths by ranges of values of functionals of the paths.  Several
functionals are possible but for simplicity we shall just consider one.
Denote it by $F[q(\tau)]$ and consider an exhaustive set of intervals
$\{\Delta_\alpha\}$ of the real line.  The class $c_\alpha$ consists of
those paths for which $F[q(\tau)]$ lies in the interval $\Delta_\alpha$
\begin{equation}
c_\alpha = \bigl\{q(t) \big | F[q(\tau)] \epsilon \Delta_\alpha
\bigr\}\, .
\label{fivefourone}
\end{equation}
This is the most general notion of coarse graining because, given any
partition of the paths into classes $\{c_\alpha\}$, we could always take
$F$ to be the function that is $\alpha$ if the path is in class
$c_\alpha$
 and take $\{\Delta_\alpha\}$ to be unit intervals surrounding
the integers.

The class operators $C_\alpha$ corresponding to the classes $c_\alpha$
are defined, as always, by
\begin{equation}
\left\langle\phi_i\left | C_\alpha\right|\psi_j\right\rangle =
\int\nolimits_{\alpha} \delta q\,\phi^*_i (q^{\prime\prime})
\exp\bigl(iS[q(\tau)]\bigr)\psi_j (q^\prime)\, .
\label{fivefourtwo}
\end{equation}
They can be evaluated by introducing the characteristic functions for
the intervals $\Delta_\alpha$ on the real line:
\begin{equation}
e_\alpha (x) = \left\{\begin{array}{rl}
                1 & x\in\Delta_\alpha\, ,\nonumber\\
                0 & x\notin\Delta_\alpha\, ,
                \end{array} \right.
\label{fivefourthree}
\end{equation}
and their Fourier transforms $\tilde e_\alpha(\mu)$
\begin{equation}
e_\alpha (x) = \int^{+\infty}_{-\infty} d\mu\, e^{i\mu x}
\tilde e_\alpha(\mu)\, . 
\label{fivefourfour}
\end{equation}
Then, clearly
\begin{equation}
\left\langle\phi_i\left|C_\alpha\right| \psi_j\right\rangle =
\int^{+\infty}_{-\infty}
 d\mu\, \tilde e_\alpha(\mu)
\int\nolimits_u \delta q\,\phi^*_i (q^{\prime\prime})
\, \exp\biggl\{i\Bigl(S[q(\tau)] +
\mu F[q(\tau)]\Bigr)\biggr\}\psi_j(q^\prime)\, . 
\label{fivefourfive}
\end{equation}
When $F[q(\tau)]$ is a {\it local} functional, that is of the form
\begin{equation}
F[q(\tau)] = \int\nolimits^T_0 dt f\left(\dot q(t), q(t), t\right)
\, ,\label{fivefoursix}
\end{equation}
then there is an effective Hamiltonian $H_F(t,\mu)$
associated with the effective action $S+\mu F$.  Quantum
mechanically it may be difficult to determine the operator ordering of
\eqref{fivefoursix}  that reproduces the path integral \eqref{fivefourfive}
if one exists at all.  However, when this can be done the class
operators may be expressed formally as
\begin{equation}
C_\alpha = \int^{+\infty}_{-\infty} d\mu\, \tilde e_\alpha(\mu) {\bf
T}
\exp\Bigl[-i\int\nolimits^T_0 H_F (t,\mu) dt \Bigr]\, . 
\label{fivefourseven}
\end{equation}
Equations \eqref{fivefourfive} and \eqref{fivefourseven} are powerful tools for
the evaluation of the class operators of the most general
sum-over-histories spacetime coarse graining.

It should be stressed that the partitions by values of a functional such
as \eqref{fivefoursix} that we have defined here are not the same as
partitions by
the eigenvalues of the Heisenberg operator corresponding to
\eqref{fivefoursix}. The class operators for the latter are {\it projections}
onto ranges of the eigenvalues while the class operators
\eqref{fivefourfive}
are not projections in general. The two kinds of class operators
represent distinct quantum mechanical alternatives that coincide
classically  -- a familiar enough situation. In this sum-over-histories
approach to quantum mechanics we shall only consider the path integral
partitions of the type we have described.

\subsubsection{Coarse Grainings Defining Momentum}

We have introduced a large class of spacetime alternatives in the
sum-over-histories generalized quantum mechanics of a non-relativistic
system.  However, we have not mentioned some of the most familiar
alternatives of ordinary quantum mechanics, for example, alternative
values of momentum at a moment of time.  The reason
momentum has not been considered
 is that there is no obvious meaning to a partition
of non-differentiable, polygonal, paths by values of $M_i\dot q^i (t)$
at a moment of time.  Using the techniques of this section, we can,
however, consider partitions by the values of the {\it averages} of such
derivatives over a time and interval and define momentum with suitable
limits of these
coarse grainings (\cf \cite{FH65}).

For simplicity, restrict attention to the case of a free particle moving
in one dimension.  We consider  a coarse graining by values of the
momentum at time $t$ averaged over a time interval $s$, that is,
by values of the functional
\begin{equation}
F_s[q(\tau)] = \frac{1}{s} \int\limits^{t+s/2}_{t-s/2} dt' M\dot
q(t') = M\left(\frac{q(t+s/2) - q(t-s/2)}{s}\right)\, .
\label{fivefoureight}
\end{equation}
The class operator corresponding to the coarse-grained history in which
the value of this averaged momentum lies in a range $\tilde \Delta$ is
\begin{equation}
\bigl\langle \phi_i\bigl|C_{\tilde\Delta}\bigr| \psi_j\bigr\rangle =
\int\delta q \phi^*_i (q^{\prime\prime}) e_{\tilde\Delta}\left\{F_s
[q(\tau)]\right\}
\exp\bigl(iS[q(\tau)]\bigr) \psi_j (q^\prime)
\label{fivefournine}
\end{equation}
where $e_{\tilde\Delta}(x)$ is the characteristic function for the
interval $\tilde\Delta$ [\cf \eqref{fivefourthree}].  If we write
\begin{equation}
e_{\tilde\Delta}(x) = \int_{\tilde\Delta} dp\ \delta (x-p)
\label{fivefourten}
\end{equation}
then the path integral in \eqref{fivefournine} is over all paths between
$t=0$ and $t=T$ for which the difference in $q$'s in \eqref{fivefoureight}
is fixed by $p$.  Since the unrestricted path integration between two
times generates unitary evolution [\cf \eqref{threeonetwentyone}],
this may be written
\[
\bigl\langle \phi_i\bigl| C_{\tilde\Delta}\bigr| \psi_j\bigr\rangle =
\int\nolimits_{\tilde\Delta} dp \int\nolimits^{+\infty}_{-\infty} dq^\prime
\int\nolimits^{+\infty}_{-\infty} d q^{\prime\prime} \delta
\left[M(q^{\prime\prime} - q^\prime)/s-p\right]
\]
\begin{equation}
\times \left\langle\phi_i \bigl| q^{\prime\prime}, t+s/2\right\rangle
\left\langle q^{\prime\prime}, t+s/2\bigl| q^\prime, t-s/2\right\rangle
\left\langle q^\prime, t-s/2\bigl| \psi_j\right\rangle\, .
\label{fivefoureleven}
\end{equation}
Carry out the integration over $q^{\prime\prime}$ using the
$\delta$-function, insert the form of the free propagator from 
\eqref{fivetwofive},
insert complete sets of momentum eigenstates immediately before the
final state and after the initial one, and carry out the remaining
$q^\prime$ integration to find
\begin{equation}
\bigl\langle\phi_i \bigl| C_{\tilde\Delta} \bigr| \psi_j\bigr\rangle =
s\int\nolimits_{\tilde\Delta} dp \int\nolimits^{+\infty}_{-\infty}
\frac{dk}{2\pi}
\left(\frac{M}{2\pi i  s}\right)^\half \exp
\left[\frac{is}{2M}\ (p-k)^2\right]\ \tilde\phi^*_i (k)
\tilde\psi_j (k) 
\label{fivefourtwelve}
\end{equation}
where $\tilde\phi_i(k)$ and $\tilde\psi_j(k)$ are the momentum space
representatives of the final and initial wave functions respectively.

We examine \eqref{fivefourtwelve} in the limits of short and long averaging
times $s$.  As $s\rightarrow 0$, it is evident that
\begin{equation}
\bigl\langle \phi_i \bigl|C_{\tilde\Delta} \bigr| \psi_j
\bigr\rangle\sim s^\half\, ,\ s\rightarrow 0 
\label{fivefourthirteen}
\end{equation}
so that the class operator becomes vacuous! This is another statement of
the non-differentiability of the paths.  The amplitude to find any
finite value for $M\dot q(t)$ at a moment of time is zero.

In the limit $s\rightarrow \infty$, the integral in \eqref{fivefourtwelve} can
be evaluated by the method of stationary phase yielding
\begin{equation}
\bigl\langle\phi_i \bigl| C_{\tilde\Delta} \bigr| \psi_j\bigr\rangle =
\int\nolimits_{\tilde\Delta} \frac{dp}{2\pi} \tilde\phi^*_i (p) \psi_j (p) =
\int\nolimits_{\tilde\Delta} \frac{dp}{2\pi} \bigl\langle \phi_i\big| p,
t\bigr\rangle \bigl\langle p, t\big| \psi_j\bigr\rangle\, .
\label{fivefourfourteen}
\end{equation}
In the limit of large averaging times, therefore, partition by average
values of $M\dot q$ reproduces the usual momentum alternatives of ordinary
quantum mechanics.  That such a limit is necessary to precisely define
momentum is easily understood from the uncertainty principle.  Coarse
graining by time averages of the velocity corresponds to determining
momentum
by time of flight.  Classically, the error in this procedure is $\Delta p
\sim M\Delta q/s$ where $\Delta q$ is the error in determining $q$.
However, quantum mechanically there is also the uncertainty $\Delta p
\sim\hbar/\Delta q$ (with $\hbar$ equaling one in the units of this
section).  For a precise determination of momentum both of
these uncertainties must go to zero.  This cannot be achieved if $s$
becomes small.  A precise determination of momentum is possible in the
limit of large $s$ provided $\Delta q$ goes to infinity in such a way
that $\Delta q/s$ goes to zero.

In more realistic situations, if we consider coarse grainings by values
of $M\dot q$ averaged over a time interval $s$ that is short compared to
the dynamical time scale, eq.~\eqref{fivefourtwelve} shows that we may
approximately replace them by usual partitions of momentum
\eqref{fivefourfourteen} making an error in the momentum of order 
$\Delta p \sim
(M\hbar/s)^\half$.  If $s$ can simultaneously be chosen short compared to the
dynamical time scale and long enough so that $\Delta p$ is small then we
have an accurate determination of momentum that can be approximately
represented by momentum projection operators.  It is in this limiting
sense that we recover the usual notion of momentum from sum-over-histories
quantum mechanics.

\subsection{The Relation Between the Hamiltonian and Generalized
Sum-Over-Histories\\
 Formulations of Non-Relativistic  Quantum Mechanics}

To what extent does the sum-over-histories formulation of
non-relativistic quantum mechanics developed in this section coincide or
differ from the more familiar Hamiltonian quantum mechanics
of states.  In making this comparison, I shall take a strict view of
what these formulations mean.  As described in Section IV, by Hamiltonian
quantum mechanics we mean a quantum mechanics
 of states and alternatives defined at moments in
time.  States evolve unitarily in between alternatives and by reduction
of the wave packet at them.
The sum-over-histories formulation is a
spacetime formulation in which alternatives are defined by partitions of
spacetime histories with associated
amplitudes computed directly in terms
of path integrals.  Path
integrals {\it vs} operators in Hilbert space is not the issue in the
comparison of the two.  As we have seen, path
integrals define operators and {\it vice versa}.  Rather the issues are: (1)
whether the alternatives to which the two formulations assign probabilities
are the same
and ~(2) whether the notion of state at a moment of time and its two
forms of evolution can be recovered from sum-over-histories quantum
mechanics.

As we discussed in Section IV.4, the two formulations coincide for
coarse grainings by regions of space
at definite moments of time.  That is evident from \eqref{fivethreethree}
 which shows
that the class operators for the individual coarse-grained histories
calculated from the sum-over-histories formulation coincide with those
of the Hamiltonian formulation [eq.~\eqref{fourtwo}] up to an overall
factor of $\exp (-iHT/\hbar)$ whose presence does not affect the value
of the decoherence functional [\cf \eqref{fiveonethirteen}].  Beyond this, 
however,
the
two formulations differ, not because they predict different answers for
the same alternatives, but because they deal with different
alternatives.

Any exhaustive set of orthogonal projection operators describes a set of
alternatives at one moment of time of the Hamiltonian
formulation of quantum mechanics.  Alternative values of $q$, of $p$, of
$q^3p+pq^3$ are just a small number of the many examples. By contrast,
at one moment of time,
the sum-over-histories formulation deals directly only with position
alternatives. Alternative values of $p$,
$q^3p+pq^3$, etc.~must first be expressed in spacetime form and then only
approximately or by limiting procedures as we discussed for momentum.
In this sense, because it employs a spacetime description, the
sum-over-histories deals with a less general set of alternatives at one
moment of time than does the Hamiltonian formulation.

The situation is reversed for spacetime coarse grainings that are not at
a moment of time such as the coarse graining by a spacetime region
discussed in Section V.3.2. These are directly accessible in the
sum-over-histories formulation but non-existent in the
Hamiltonian one. There is no single chain of projections,
for example, that can represent the class operator for the alternative
that the path crosses the region $R$ at least once in the example of
Section V.3.2.  In its access to spacetime coarse grainings, the
sum-over-histories formulation is more general than the
Hamiltonian formulation.  This generality will be important
for the quantum mechanics of spacetime geometry
where there is no covariant notion of
alternatives at one moment of time.

The quantum mechanics of the more general spacetime alternatives of
sum-over-histories quantum mechanics cannot be formulated in terms of
states on a spacelike surface and the two forms of evolution. If the
class operators cannot be represented as chains of projections, we
cannot construct the state of the system at a moment of time as we did
in Section III.1.3.  In the coarse graining by a spacetime region
discussed in Section V.3.2., as \eqref{fivethreefifteen}  shows,
there is neither unitary
evolution nor reduction of the wave packet in the time interval over
which the region extends.  The sum-over-histories formulation does not
permit the notion of state on a spacelike surface so central to the
Hamiltonian version of quantum mechanics.

For non-relativistic systems, the two formulations of quantum mechanics
may be unified in a common generalization.  The sets of fine-grained
histories are defined by exhaustive sets of one-dimensional projection
operators (\ie projections onto complete sets of states) at each and
every time.  Partitions of these fine-grained histories into an
exhaustive set of exclusive classes define coarse grainings as before.
Class operators would therefore be defined formally by
\begin{equation}
C_\alpha = \sum_{\alpha(t)\epsilon c_\alpha} \left(\prod_t\,
P^{k(t)}_{\alpha(t)} (t)\right)\, . 
\label{fivefiveone}
\end{equation}
What mathematical sense can be made out of such formal expressions, if
any, is an interesting question,
as is the identification of
invariant classes in theories with symmetries.

In the absence of a completed unification we shall develop the
sum-over-histories formulation in the rest of these lectures.  Its
spacetime coarse grainings offer the hope of there being realistic alternatives
in quantum gravity that are not ``at one moment of time''.

\section{Abelian Gauge Theories}
\setcounter{footnote}{0}

\subsection{Gauge and Reparametrization Invariance}

Einstein's general relativity is a dynamical theory of spacetime
geometry.  Geometry is described by a spacetime metric, but many
different metrics correspond to the same geometry.  Metrics
corresponding to the same geometry are connected by diffeomorphisms.
Diffeomorphisms are therefore a symmetry of any theory of spacetime
geometry; and physical predictions must be diffeomorphism invariant.  A
generalized quantum mechanics of spacetime geometry
whose fine-grained histories include a spacetime
metric will therefore assign probabilities to {\it diffeomorphism
invariant} {\it partitions of four-dimensional} {\it metrics and matter
field configurations}.

Since we lack a complete quantum theory of gravity, it is instructive
to discuss the quantum mechanics of model theories that exhibit similar
symmetries.  In this connection, it is useful to note that, when space
and time are separated,
the diffeomorphisms of spacetime theories contain two
familiar types of symmetries --- gauge symmetries corresponding to spatial
coordinate transformations  and reparametrizations
of the time.

\begin{figure}[t]
\begin{center}
\includegraphics[width=5.5in]{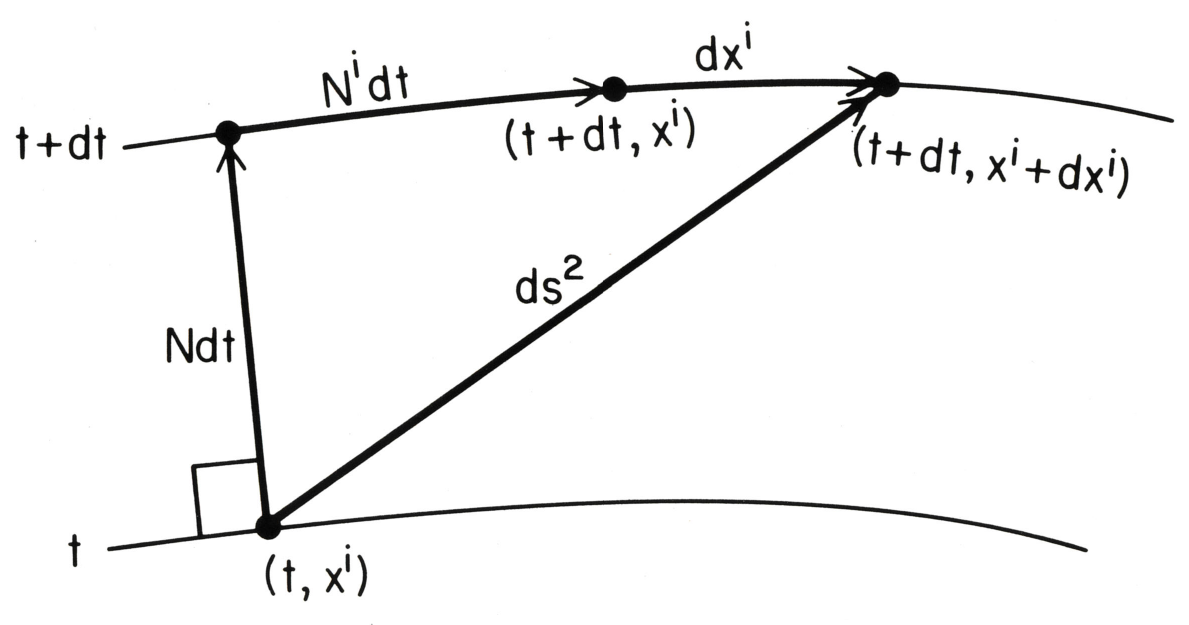}
\caption{\sl The $3+1$ decomposition of a spacetime
metric. The figure shows two nearby members of a family of spacelike
surfaces that foliate spacetime. The surfaces are labeled by a
continuous coordinate $t$; points in the surfaces are labeled by three
coordinates, $x^i$. The $3+1$ decomposition of a spacetime metric
with respect to these coordinates is achieved as follows: Connect the
two surfaces by a perpendicular line passing through the point $(t,x^k)$.
The lapse function $N(t,x^k)$ is defined so that the perpendicular
distance between the two surfaces separated by a coordinate interval
$dt$ is $Ndt$. The shift vector $N^i(t,x^k)$ is defined so that $N^{i}dt$
is the displacement between the intersection of the perpendicular with
the surface $t+dt$ and the point in that surface with the same spatial
coordinate $x^i$ as the point from which the construction started. The
distance between the points $(t,x^i)$ and $(t+dt,x^i+dx^i)$ is then
given by \eqref{sixoneone}}
\end{center}
\end{figure}

To make this distinction concrete recall the familiar $3+1$
decomposition of a four-dimensional metric defined by a foliating family of
spacelike
surfaces labeled by a coordinate $t$.  This is illustrated in Figure
11.
We write
\begin{equation}
ds^2 = -N^2\,dt^2 + h_{ij}\bigl(dx^i + N^i\, dt\bigr)
\bigl(dx^j + N^j\, dt\bigr)
\label{sixoneone}
\end{equation}
where the lapse $N$, shift vector $N^i$, and spatial metric, $h_{ij}$,
are all functions of $x^i$ and $t$.  There is a correspondence between
diffeomorphisms (maps of the manifold into itself) and coordinate
transformations
\begin{equation}
x^\alpha \longrightarrow \bar x^\alpha = \bar x^\alpha (x^\beta)
\, .\label{sixonetwo}
\end{equation}
In a $3+1$ decomposition, the coordinate transformations
\eqref{sixonetwo}
contain two special  cases of interest.  First, there are
reparametrizations of the time
\begin{equation}
t\longrightarrow \bar t = \bar t(t)\, .
\label{sixonethree}
\end{equation}
Second, there are spatial coordinate transformations
\begin{equation}
x^i\longrightarrow \bar x^i = \bar x^i (x^j)\, .
\label{sixonefour}
\end{equation}
Under an infinitesimal coordinate transformation of type
\eqref{sixonefour}, where
$\bar x^i = x^i + \xi^i (x^j)$, the three-metric transforms as
\begin{equation}
h_{ij} (x^k, t) \longrightarrow \bar h_{ij} (x^k, t) = h_{ij} (x^k, t) +
2D_{(i}\xi_{j)} (x^k,t)
\label{sixonefive}
\end{equation}
where $D_i$ is the spatial derivative.  Because of the similarity of
\eqref{sixonefive} with the symmetry transformations of gauge field theories,
spatial diffeomorphisms are often called (spatial) gauge
transformations.

Reparametrizations of the time and gauge transformations are combined in
the invariance group of dynamical theories of spacetime geometry.
However, from the point of view of the problem of time, these two types
of transformation have a considerably different status.  It is,
therefore, convenient to consider models in which they are exhibited
separately.  We shall consider the simplest two model theories: free
electromagnetism as an example with gauge symmetry and the free
relativistic particle as an example that is reparametrization invariant.
We begin with electromagnetism.

\subsection{Coarse Grainings of the Electromagnetic Field}

The {\it fine-grained histories} of the free electromagnetic field we
take to be specified by the various four-dimensional configurations of
the potential $A^\mu(x)$.  The allowed coarse grainings are partitions
of the potentials into exhaustive sets of exclusive {\it
gauge-invariant} classes, that is, classes invariant under gauge
transformations
\begin{equation}
A_\mu(x) \longrightarrow A_\mu(x) + \nabla_\mu\Lambda(x)
\label{sixtwoone}
\end{equation}
for arbitrary functions $\Lambda(x)$.  We denote sets of such classes by
$\{c_\alpha\}$, $\alpha = 1,2, \cdots$ and the entire class by $u=\cup_\alpha
c_\alpha$. A unique potential representing a class may be singled out by
imposing a gauge condition
\begin{equation}
\Phi(A)=0\, .\label{sixtwotwo}
\end{equation}
For example, the temporal gauge in which
\begin{equation}
A_0(x)=0\label{sixtwothree}
\end{equation}
is often convenient.  This condition does not fix the gauge entirely because
transformations of the form \eqref{sixtwoone} with $\Lambda$ independent of time
preserve the condition \eqref{sixtwothree}.  To fix the gauge completely a
further condition, say
\begin{equation}
\bigl(\vec\nabla\cdot\vec A\bigr)_\sigma = 0 
\label{sixtwofour}
\end{equation}
could be imposed on one spacelike surface $\sigma$. Both
\eqref{sixtwothree}
and \eqref{sixtwofour} are included in $\Phi(A)=0$.

We can now proceed with the definition of the decoherence functional for a set
of coarse-grained histories $\{c_\alpha\}$ that are a gauge-invariant
partition of the
potentials $A_\mu(x)$ defined on the region of spacetime between two
non-intersecting spacelike surfaces $\sigma^\prime$ and
$\sigma^{\prime\prime}$.  These spacelike surfaces do not
have to be planes, but for simplicity, let us consider only the case
where the initial surface $\sigma^\prime$ is the plane $t=0$ and the final
surface $\sigma^{\prime\prime}$ is
the plane $t=T$ in some Lorentz frame.  In that same frame there is a $3+1$
decomposition
of $A_\mu (x)$ into the temporal component $A_0 (x)$ and the transverse and
longitudinal components of the vector potential
\begin{subequations}
\label{sixtwofive}
\begin{equation}
\vec A(x) = \vec A^T(x) + \vec A^L(x)\, , 
\label{sixtwofive a}
\end{equation}
where
\begin{equation}
  \vec A^L(x) \cdot \vec A^T(x)
=0\, ,\ \vec\nabla\cdot \vec A^T(x) = 0\, . 
\label{sixtwofive b}
\end{equation}
\end{subequations}

The Hilbert space of states of the free electromagnetic field is
the space of square integrable functionals of transverse
vector potentials, ${\cal H}^T$.  This is defined by the inner product
\begin{equation}
(\psi,\chi) = \int\delta\vec A^T \psi^*\bigl[\vec A^T\bigr] \chi
\bigl[\vec A^T\bigr]\, . 
\label{sixtwosix}
\end{equation}
Here, $\psi$ and $\chi$ are functionals of $\vec A^T({\bf x})$
where ${\bf x}$ denotes the three spatial coordinates. The
measure is defined by
\begin{equation}
\delta \vec A^T = \prod\limits_{\bf k}\, \left(dA^1\bigl({\bf k}\bigr)
\, d\, A^2\bigl({\bf k}\bigr)\right) 
\label{sixtwoseven}
\end{equation}
where $A^1$ and $A^2$ are the two transverse components (polarizations)
of a decomposition of $\vec A^T(\bf x)$ into modes
\begin{equation}
\vec A^T\left(\bf x\right) = \int\, \frac{d^3k}{(2\pi)^{3}}\,
e^{i\bf k\cdot{\bf x}}\vec A^T\bigl(\bf k\bigr)\, . 
\label{sixtwoeight}
\end{equation}

Class operators $C_\alpha$ on ${\cal H}^T$ that correspond to the
individual classes $c_\alpha$ in a gauge-invariant partition are defined by
constructing their matrix elements
\begin{equation}
\bigl\langle\vec A^{T\prime\prime}\bigl|C_\alpha\bigr|\vec
A^{T\prime}\bigr\rangle = \bigl\langle\vec A^{T{\prime\prime}}\bigl|
\, \int_\alpha \delta A\Delta_\Phi [A]\,\delta[\Phi(A)]\, \exp\,
\left(iS[A]\right)\, \bigr| \vec A^{T\prime}\bigr\rangle\, .
\label{sixtwonine}
\end{equation}
The meaning of the right hand side is as follows: The functional
integral is over all potentials $A^\mu(x)$ that lie in the class
$c_\alpha$ and that match the transverse component of the vector
potential $\vec A^{T\prime}$ on the initial surface at $t=0$ and
similarly match $\vec A^{T{\prime\prime}}$ on the final surface at
$t=T$.  That which is not fixed is integrated over, so the integral
includes integrations over $A^0$ and $\vec A^L$ on the initial and
final surface.  The gauge-fixing $\delta$-function and its associated
Faddeev-Popov determinant, $\Delta_\Phi$,
 ensure that only one representative potential
in the gauge invariant class $c_\alpha$ contributes to the functional
integral. The action for the free
electromagnetic field is
\begin{equation}
S[A] = -\frac{1}{4}\,\int_M d^4 x F_{\alpha\beta} F^{\alpha\beta}
\label{sixtwoten}
\end{equation}
where $F_{\alpha\beta} = \nabla_\alpha A_\beta - \nabla_\beta A_\alpha$
and $M$ is the spacetime region between $t=0$ and $t=T$.  The measure in
a time slicing implementation of the functional integral is analogous to
that of Section V.2, namely
\begin{equation}
\delta A = \prod\limits_{t, {\bf k}}\left[dA^1(t,{\bf k})
\, dA^2(t,{\bf k})\, dA^L(t,{\bf k})\, dA^0 (t,{\bf k})\right]
\label{sixtwoeleven}
\end{equation}
in what, it is hoped, is an obvious notation. Since $\vec A^T$ is a
gauge invariant quantity and since the class $c_\alpha$ is gauge
invariant, it is a standard result \cite{Fad69} that the integral in
\eqref{sixtwonine} is independent of the gauge fixing condition
$\Phi$. Matrix elements of the $C_\alpha$ between arbitrary initial and
final states in ${\cal H}^T$ represented by wave functions $\phi_i [\vec
A^T]$ and $\psi_j [\vec A^T]$ may be constructed from \eqref{sixtwonine}
using the inner product \eqref{sixtwosix}, \viz:
\begin{equation}
\left\langle\phi_i|C_\alpha|\psi_j\right\rangle =
\int \delta \vec A^{T\prime\prime} \int \delta \vec A^{T\prime}
\phi^*_i [\vec
A^{T\prime\prime}]\ \langle\vec A^{T\prime\prime}| C_\alpha |
\vec A^{T\prime}
\rangle\, \psi_j\, [\vec A^{T\prime}]\, .
\label{sixtwoelevena}
\end{equation}

Were we to define the functional integral in \eqref{sixtwonine} by means of
a product formula for each mode we would be dealing with a larger
Hilbert space than ${\cal H}^T$.  This is most clearly illustrated in
the temporal gauge defined by \eqref{sixtwothree} and \eqref{sixtwofour}.
Choosing the initial surface at $t=0$ to be the
surface $\sigma$ on which the surface  gauge condition \eqref{sixtwofour} is
enforced, we have
\begin{equation}
\langle \vec A^{T^{\prime\prime}}\left\vert  C_\alpha\right\vert
\vec A^{T^\prime}\rangle = \int\,\delta \vec
A^{L^{\prime\prime}}\langle\vec A^{T^{\prime\prime}}\ ,\vec
A^{L^{\prime\prime}}\Vert\,\int_\alpha \delta\vec A\exp\,
(iS[\vec A])\Vert\,\vec A^{T^\prime}\, ,0 \rangle\, .
\label{sixtwotwelve}
\end{equation}
Here, we have used a double bar to denote states in the Hilbert space
${\cal H}^{(T,L)}$ of square integrable functionals of vector potentials
--- both transverse and longitudinal components.  The functional integral in
\eqref{sixtwotwelve} is over such vector potentials.  There is an integral
over the final value of $\vec A^{L^{\prime\prime}}$
but the initial value has been
set to zero by the surface gauge condition \eqref{sixtwofour}. The auxiliary
Hilbert spaces ${\cal H}^L$ and ${\cal H}^{(T,L)}$ will be useful in
evaluating functional integrals in what follows and in making contact
with Dirac Quantization in Section VI.5.

Having identified the class operators $C_\alpha$, the construction of the
decoherence functional $D(\alpha^\prime, \alpha)$ for the coarse-grained
set of histories $\{c_\alpha\}$ follows that for particle quantum
mechanics [\cf \eqref{fiveoneten}]:
\begin{subequations}
\label{sixtwothirteen}
\begin{equation}
D\left(\alpha^\prime, \alpha\right) = {\cal N} Tr\bigl[\rho_fC_{\alpha^\prime}
\, \rho_i \, C^\dagger_\alpha\bigr]
\label{sixtwothirteen a}
\end{equation}
where ${\cal N}$ is
\begin{equation}
{\cal N}^{-1} = Tr\left[\rho_f e^{-iH^TT} \rho_i e^{iH^TT}\right]\, .
\label{sixtwothirteen b}
\end{equation}
\end{subequations}
Here, $\rho_i$ and $\rho_f$ are
 density matrices describing the initial and final conditions of
the electromagnetic system.  These $\rho$'s, the $C_\alpha$, and the trace
are all defined on the Hilbert space, ${\cal H}^T$.
If we assume a final condition of indifference with respect to final
state we recover the standard
\begin{equation}
D\left(\alpha^\prime, \alpha\right) = Tr[C_{\alpha'}\rho_i
C^\dagger_\alpha]\, . 
\label{sixtwothirteena}
\end{equation}
Thus, the two gauge-{\it invariant} parts of the vector potential, $\vec
A^T({\bf x})$ and $\vec A^L({\bf x})$, are treated differently in the
construction of the decoherence functional.  Amplitudes, \eg
\eqref{sixtwotwelve} are summed over $\vec A^L({\bf x})$ on the final
surface; squares of amplitudes are summed over $\vec A^L({\bf x})$ in
\eqref{sixtwothirteen}.  For suitable coarse grainings this coincides with
usual Hamiltonian quantum mechanics as we shall shortly see.

It is not difficult to check that the decoherence functional
\eqref{sixtwothirteen} satisfies the requirements \eqref{fouroneone i}
-- \eqref{fouroneone iv} of Section IV.
Hermiticity and positivity are immediate from the general structure of
\eqref{sixtwothirteen} and the positivity of the $\rho$'s.  The superposition
principle is satisfied because of the linearity of the
sum-over-histories in \eqref{sixtwonine}.  It only remains to check the
normalization, and this involves the sum in \eqref{sixtwotwelve} over all
vector potentials $\vec A(x)$.  This factors into separate sums over
$\vec A^T$ and $\vec A^L$ involving the temporal gauge actions
\begin{subequations}
\label{sixtwofourteen}
\begin{eqnarray}
S\bigl[\vec A^T\bigr] & = & \half\,\int\, d^4 x\,\biggl[
\Bigl({\buildrel\textstyle{\buildrel\textstyle.\over\rightarrow}\over
{A^T}}\Bigr)^2 - \left(\vec\nabla \times \vec
A^T\right)^2\biggr]\, , \label{sixtwofourteen a}\\
S\bigl[\vec A^L\bigr] & = & \half\,\int\, d^4 x \Bigl(
{\buildrel\textstyle{\buildrel\textstyle.\over\rightarrow}\over {A^L}}
\Bigr)^2\, . \label{sixtwofourteen b}
\end{eqnarray}
\end{subequations}
Then, from the usual connection to Hamiltonian quantum mechanics
\begin{equation}
\bigl\langle\vec A^{T^{\prime\prime}}\left| C_u\right |\vec A^{T^\prime}
\bigr\rangle = \bigl\langle\vec A^{T^{\prime\prime}}\bigl|
e^{-iH^TT}\bigr| \vec A^{T^\prime}\bigr\rangle\, \int\, \delta\vec
A^{L^{\prime\prime}}\bigl\langle\vec A^{L^{\prime\prime}}
\bigl| e^{-iH^LT}\bigr|\vec A^{L^\prime} = 0 \bigr\rangle\, .
\label{sixtwofifteen}
\end{equation}
In this expression
where we have used a single bar to denote the inner product in either
${\cal H}^T$ or ${\cal H}^L$ and
 $H^T$ and $H^L$ are the Hamiltonians corresponding to the actions
\eqref{sixtwofourteen}, specifically:
\begin{subequations}
\label{sixtwosixteen}
\begin{eqnarray}
H^T & = & \half\,\int\, d^3 x\,\bigl[ (\vec\pi^T)^2
    + (\vec\nabla \times \vec
A^T)^2\bigr]\, , \label{sixtwosixteen a}\\
H^L & = & \half\,\int\, d^3 x\,(\vec\pi^L)^2
\, . \label{sixtwosixteen b}
\end{eqnarray}
\end{subequations}
The last factor in \eqref{sixtwofifteen},
 including the integral, can be written
\[
\left\{\int\,\delta \vec A^{L^{\prime\prime}} \exp\,\left[-i\, \int
\, d^3 x\bigl(\vec\pi^{L^{\prime\prime}}\cdot \vec
A^{L^{\prime\prime}}\bigr)\right] \bigl\langle\vec
A^{L^{\prime\prime}}\bigl|\, e^{-iH^LT}\bigr| \vec A^{L^\prime} =
0 \bigr\rangle \right\}_{\vec\pi_L = 0}
\]
\begin{equation}
= \bigl\langle\vec\pi^{L^{\prime\prime}} = 0\bigl|\, e^{-iH^LT}\bigr|
\vec A^{L^\prime} = 0\bigr\rangle\, \big/\,
\bigl\langle\vec\pi^{L^{\prime\prime}} = 0 \big | \vec A^{L^\prime} =
0\bigr\rangle\, . 
\label{sixtwoseventeen}
\end{equation}
But, since $H^L$ conserves
$\vec\pi^L$, the factor \eqref{sixtwofifteen} is just unity.  Thus,
\begin{equation}
C_u = e^{-iH^TT}\, , 
\label{sixtwoeighteen}
\end{equation}
and the normalization of the decoherence functional \eqref{sixtwothirteen}
follows immediately.

\subsection{Specific Examples}

Specific types of coarse grainings are of interest.  First, consider
partitions by ranges of values of $\vec A^T({\bf x})$ on a surface of
constant time $t$ between $0$ and $T$.  These are the usual
gauge-invariant, configuration-space observables of electromagnetism.
From the explicit forms \eqref{sixtwotwelve} and \eqref{sixtwofifteen} and a
repetition of the discussion in Section III.1.4, it follows that
\begin{equation}
C_\alpha = e^{-iH^T(T-t)} P_\alpha e^{-iH^Tt}
\label{sixthreeone}
\end{equation}
where the $P_\alpha$ are projections in ${\cal H}^T$ onto the ranges of
$\vec A^T({\bf x})$.  Similarly, for coarse grainings defined by sequences
of sets of alternative ranges of $\vec A^T({\bf x})$, at times $t_1,\cdots,
t_n$, one has
\begin{equation}
C_\alpha = e^{-iH^T(T-t_n)} P^n_{\alpha_n} e^{-iH^T(t_n-t_{n-1})}
P^{n-1}_{\alpha_{n-1}} \cdots P^1_{\alpha_1} e^{-iH^Tt_1}\ .
\label{sixthreetwo}
\end{equation}
In these expressions one recovers the familiar Hamiltonian
quantum mechanics of the ``true degrees of freedom'' of the
electromagnetic field.  These true degrees of freedom are the transverse
components of the vector potential.  As in Section III.1, the quantum theory
can be formulated in terms of states represented by wave functionals
$\psi [\vec A^T, t)$ that evolve unitarily in between projections
defining specific alternatives. When restricted to coarse grainings of
the true physical degrees of freedom on spacelike surfaces this
sum-over-histories quantum mechanics coincides with the usual
Hamiltonian quantum mechanics of the free electromagnetic field.

However, more general kinds of coarse-graining that are defined by
alternatives not at one moment of time are also possible. For example,
one can partition the potentials $A^\mu(x)$ by ranges of values
of particular field components averaged over a spacetime region that were
considered by Bohr and Rosenfeld \cite{BR33} in their discussion of the
measurability of the electromagnetic field. These
are partitions by the values of functionals of the potential of the form
\begin{equation}
F[A] = \frac{1}{V(R)}\, \int_R d^4 x\, F_{\mu\nu} (x)\, . 
\label{sixthreethree}
\end{equation}
where $V(R)$ is the volume of spacetime region $R$.
(We tolerate, just briefly, the use of $F$ for both field and
functional.) Partitions by values of averages of the magnetic field, say
\begin{equation}
F_{B_z}[A] = \frac{1}{V(R)}\, \int_R d^4 x\, B_z(x) = \frac{1}{V(R)}\, \int_R
d^4 x\left(\vec\nabla \times\vec A^T\right)_z 
\label{sixthreefour}
\end{equation}
are describable entirely in terms of the ``true physical degrees of
freedom'' of the electromagnetic field.  Their class operators may be
computed on ${\cal H}^T$ by the techniques described in Section V.
Indeed, since the free electromagnetic field is equivalent to an
assembly of oscillators we expect these class operators to be computable
explicitly.\footnote{We should mention again, as we did in Section
V.4.1, that the class operators for partitions by values of field
averages extended over time that are considered here are not the same as
the projections on the corresponding ranges of the average values of the
Heisenberg fields.  In general, they are not projections at all.}  As in
the case of spacetime alternatives for the non-relativistic particle, we
do not recover an alternative formulation of the generalized quantum
mechanics of these alternatives in terms of evolving states on a
spacelike surface reduced by the action of projections.

We cannot coarse grain by values of $\vec A^L$ at a moment of time
because $\vec A^L$ is not gauge-invariant.  We can, however coarse
grain by ranges of values of the electric field that involve $\vec A^L$,
for example, the following field average:
\begin{equation}
F_{E_z}[A] = \frac{1}{V(R)}\, \int_R d^4 x\, E_z(x) = \frac{1}{V(R)}\, \int_R
d^4 x \bigl(\vec\nabla A_0 - {\buildrel\textstyle{\buildrel\textstyle.
\over\rightarrow}\over A} \bigr)_z\, . 
\label{sixthreefive}
\end{equation}
Such coarse grainings are gauge-invariant, calculable by the techniques
in Section V, but not directly expressible in terms of the
``true physical degrees of freedom'' alone.  In the limit as the temporal size
of $R$ goes to zero such coarse grainings will be vacuous as were the
coarse grainings by $\dot q$ in Section V.  In the limit as the
temporal size of $R$ becomes large, however, such coarse grainings
coincide with coarse grainings by canonical field momenta as we shall
see next.

\subsection{Constraints}

Classically, the gauge invariance of electromagnetism implies a
constraint between its canonical coordinates $\vec A({\bf x})$ and the
corresponding canonical momenta $\vec \pi ({\bf x})$. The canonical
momenta are found from the Lagrangian density of the action
\eqref{sixtwoten}:
\begin{equation}
\vec\pi(x) = \frac{\partial{\cal L}}{\partial
{\buildrel\textstyle{\buildrel\textstyle.\over\rightarrow}\over A} (x)} = -\vec
E(x) = 
{\buildrel\textstyle{\buildrel\textstyle.\over\rightarrow}\over A}
(x) - \vec\nabla A_0(x)\, . 
\label{sixfourone}
\end{equation}
The constraint is the field equation
\begin{equation}
\vec\nabla \cdot \vec E({\bf x}) = 0\, , 
\label{sixfourtwo}
\end{equation}
or, what is the same thing,
\begin{equation}
\vec\pi^L({\bf x}) = 0\, .
\label{sixfourthree}
\end{equation}
Physical states are annihilated by  operator forms of the classical
constraints in the Dirac approach to quantization.  To what extent are
the constraints maintained in the present sum-over-histories quantization
of electrodynamics?

Whether a relation like \eqref{sixfourtwo} is satisfied in quantum theory
is not a question of definition, but a matter of probability. The
divergence of the electric field is a measurable quantity and a theory
that does not assign probabilities to its possible values is
incomplete. This
theory assigns probabilities to alternative values of $\vec\nabla\cdot \vec
E$ if they decohere.  The constraints can be said to be satisfied if the
probability vanishes for every value of $\vec\nabla \cdot \vec E$ except
zero.  We shall now compute the probabilities for various values of
the longitudinal component of the field momentum, $\vec \pi^L({\bf x})$

Momentum is accessible in a sum-over-histories formulation of quantum
field theory in essentially the same way that
we discussed for a sum-over-histories formulation of quantum particle
mechanics in Section V.4 .
Coarse grainings by average values of time derivatives of fields
become partitions by field momentum when the time over which the average is
taken becomes large.  We, therefore, consider partitions by values of
the gauge invariant functional:
\begin{equation}
\vec F_{\bf x}[A]=\frac{1}{\Delta t}\, \int^{t_2}_{t_1}dt
\, \vec\pi^L({\bf x},t)=\frac{1}{\Delta
t}\, \int^{t_2}_{t_1} dt\,\Bigl[
{\buildrel\textstyle{\buildrel\textstyle.\over\rightarrow}\over {A^L}}
\left({\bf x},
t\right) - \vec\nabla A_0 \left({\bf x}, t\right)\Bigr]\, .
\label{sixfourfour}
\end{equation}
where $0<t_1<t_2<T$ and $\Delta t= t_2-t_1$.
In the limit that $\Delta t$ becomes large, this becomes a partition by
$\vec\pi^L({\bf x})$. This is especially transparent in the temporal
gauge where the analogy with particle momenta is immediate.
As we shall show in more detail below, if we follow the analysis
of Section V.4, in the limit of large $\Delta t$
the sum over $A^\mu (x)$ in the class
with a particular range of values $\widetilde\Delta$ of the average
\eqref{sixfourfour} can be replaced by a projection, $P_{\tilde\Delta}$,
onto the  range of eigenvalues of the operator:
\begin{equation}
\vec\pi^L({\bf x}) = -i\delta/\delta \vec A^L({\bf x})\, . 
\label{sixfourfive}
\end{equation}
Further, the class operators for alternative
ranges of $\vec\pi^L(\bf x)$ will be shown to vanish except for ranges
which include $\vec \pi^L({\bf x})=0$, essentially as a consequence of
the gauge invariance of the construction of the decoherence functional.
A vanishing probability is thus predicted for every value of $\vec
\pi^L({\bf x})$ except zero, and it is in this sense that the constraint
is satisfied.

For simplicity let us consider a partition
by the time average of just a single mode of the scalar $\pi^L$, specifically
by the functional $F_{\bf k}[A]$ which in the temporal gauge is [\cf
\eqref{fivefoureight}]:
\begin{equation}
F_{\bf k}[A] = \frac{1}{\Delta t}\left[A^L({\bf k},t_2)-A^L({\bf
k},t_1)\right]\, . 
\label{sixfoursevena}
\end{equation}
The matrix elements of the class operator corresponding to $F_{\bf
k}[A]$ lying in the range $\tilde\Delta$ are
\begin{equation}
\bigl\langle \vec A^{T\prime\prime} |C_{\tilde\Delta}|\vec A^{T\prime}
\bigr\rangle
= \langle\vec A^{T\prime\prime}|\int\nolimits \delta A
\Delta_\Phi [A] \delta[\Phi(A)] e_{\tilde\Delta}(F_{\bf k}[A])
\exp(iS[A])|\vec A^{T\prime}\rangle 
\label{sixfoursevenb}
\end{equation}
where $e_{\tilde\Delta}(x)$ is the characteristic function for the
interval $\tilde\Delta$. The integral over the transverse parts of
$\vec A$ is unrestricted by the coarse graining as are the integrals
over longitudinal modes except those with wave-vector ${\bf k}$. The
class operator matrix elements may therefore be written
\begin{equation}
\langle \vec A^{T\prime\prime} |C_{\tilde\Delta}|\vec A^{T\prime}\rangle
=\langle \vec A^{T\prime\prime} |e^{-iH^T T}|\vec A^{T\prime}\rangle
{\cal C}_{\tilde\Delta} 
\label{sixfoursevenc}
\end{equation}
where
\begin{equation}
{\cal C}_{\tilde\Delta} =\int_{\tilde\Delta} df\, {\cal C}_f
\label{sixfoursevend}
\end{equation}
and ${\cal C}_f$ is the functional
integral over the mode $A^L({\bf k} ,t)$
restricted to those histories where $F_{\bf{ k}}[A]$ has the value $f$.
Specifically,
\[
{\cal C}_f = \int\nolimits dA^{L\prime\prime}\int\nolimits dA^L_2
\int\nolimits dA^L_1 \, \delta\left[(A^L_2 - A^L_1)/{\Delta t} -f \right]
\]
\begin{equation}
\langle A^{L\prime\prime}, T | A^L_2, t_2 \rangle
\, \langle A^L_2, t_2|A^L_1,
t_1\rangle\,\langle A^L_1,t_1|A^{L\prime}=0,0\rangle\, .
\label{sixfourten}
\end{equation}
In this expression $\langle A^{L\prime\prime}, t''|A^{L\prime},t'
\rangle$ is the propagator of the longitudinal part of the vector
potential constructed with the Hamiltonian $H^L$. We have suppressed all the
labels ${\bf k}$ that refer
to the particular mode summed over. We have used the surface gauge
condition \eqref{sixtwofour} to fix the initial integration. The final
integration turns the final propagator in the series of three into
$\langle \pi^{L\prime\prime}=0, T| A^L_2,t_2\rangle$ which is unity.
With the Hamiltonian \eqref{sixtwosixteen b} the remaining propagator in
\eqref{sixfourten}
is just that of a free particle.
Using the $\delta$-function to carry out the integral over $A^L_2$ and
making use of the invariance of this  propagator under translations both
in time and $A^L$, we
find that
\begin{equation}
{\cal C}_f = \Delta t \langle f\Delta t, \Delta t |0,0\rangle\, ,
\label{sixfoureleven}
\end{equation}
and more explicitly
\begin{equation}
{\cal C}_{\tilde\Delta} = \left(\frac{\Delta t}{2\pi i}\right)^\half
\int_{\tilde\Delta} \, df e^{i(\Delta t f^2)/2}\, .  
\label{sixfourtwelve}
\end{equation}

In the limit $\Delta t \to \infty$, ${\cal C}_{\widetilde\Delta}$ vanishes
unless $\widetilde\Delta$ contains $f=0$. In this limit the partition defines
momentum $\pi^L({\bf k})$ and this result is a more detailed
demonstration that the class operators vanish except when $\pi^L({\bf
k})= 0$. If we consider a partition of the real line by intervals
$\{\Delta_\alpha\}, \alpha=1,2,\cdots$, then the decoherence functional
is
\begin{equation}
D(\alpha',\alpha)={\cal C}_{\tilde\Delta_{\alpha'}}{\cal
C}_{\tilde\Delta_{\alpha}}.
\label{sixfourthirteen}
\end{equation}
Since the ${\cal C}$ 's are non-zero only for a {\it single} $\alpha$,
this coarse-grained set of alternatives decoheres and the probability is
zero for any value of $\alpha$ except that corresponding to the
interval containing $\pi^L({\bf k})=0$. If $\hbar$ is restored by
replacing $t$ with $t/\hbar$, then the same result is obtained for any $\Delta
t$ in the formal ``classical limit'' $\hbar \to 0$.  It is in these
precise probabilistic senses that the constraint is satisfied in this
generalized quantum mechanics of the electromagnetic field.
Thus, restricting the initial and final conditions to depend only on
the ``true physical degrees of freedom'', $\vec{A}^T$, means that
$\vec\pi^L=0$ with probability one at all other times. This is a
familiar result in the more usual quantum mechanics of states on
spacelike surfaces as we shall discuss below.

If $\hbar$ and $\Delta t$ are finite then the ${\cal C}$'s will be
non-zero for several different values of $\alpha$. Such alternatives
cannot decohere.  The decoherence functional \eqref{sixfourthirteen} factors
and the off-diagonal elements cannot vanish without the diagonal ones
vanishing also.  Probabilities are therefore not assigned to such
alternatives in the theory of the {\it free} electromagnetic field.

\subsection{ADM and Dirac Quantization}

The relation of the present generalized quantum mechanics of the
electromagnetic field with Arnowitt-Deser-Misner (ADM)
 and Dirac quantization is of interest.
In ADM quantization\footnote{We
use here the terminology of quantum gravity of ``ADM quantization''
for the quantization method in which the constraints are solved {\it
classically} for the ``true physical degrees of freedom'' which are
then quantized (\eg as in Arnowitt, Deser, and Misner \cite{ADM62}). That
method, of course, has
a much older history in the case of electromagnetism.}
 the constraints are solved classically.
Thus, $\vec\nabla\cdot\vec E = 0$ once and for all. Now, certainly
$\vec\nabla\cdot\vec E$ is a measurable quantity although in the ADM
approach it does not
correspond to an operator in Hilbert space.  However, as we
mentioned earlier, any quantum theory of the electromagnetic field must
predict a probability for $\vec\nabla \cdot \vec E$ since we can observe it
even on ``macroscopic'' scales.  The quantum theory would be incomplete
if it did not offer such a prediction. Presumably, ADM theory
predicts that a measurement of $\vec\nabla\cdot\vec E$
at a moment of time would yield its
classical value zero with probability one.\footnote{The
author is expressing some caution because he has received several
different authoritative versions of whether and what ADM theory predicts for
such quantities!}  One also presumes that ADM
quantization would predict zero probability for all but zero values of the
time average of $\vec\nabla\cdot\vec E$
represented by \eqref{sixfourfour}.
If so, then it differs in its
predictions from the present discussion
where the class operators given by \eqref{sixfoursevenb} and 
\eqref{sixfourtwelve} do {\it not} vanish for
values of these time averages of $\vec\nabla\cdot\vec E$ other than
zero.  The question of agreement is perhaps moot in the case of free
electromagnetism because alternatives defined by sets of ranges of
averages of $\vec\nabla \cdot \vec E$ over finite times do not decohere
and probabilities are not, therefore, predicted for them.  However, in
the presence of charges such alternatives might decohere and then the
predictions of the generalized quantum mechanics would differ from ADM
theory naively interpreted.  In assessing these contrasts, it should be
kept in mind that they concern predictions gauge-invariant quantities
which, although observable, are not constructed from the ``true physical
degrees of freedom''.  Further, the differences arise for
alternatives extended over time that are not usually
considered in quantum mechanics. A generalization of quantum  which includes
such quantities
is perhaps going beyond the domain of questions that ADM theory was
intended to answer.

Dirac quantization is another familiar approach to the quantum mechanics
of constrained Hamiltonian systems such as the free electromagnetic
field.\footnote{There are many reviews of Dirac quantization. Some
classics are \cite{Dir64, Kuc74, HRT76, Ash91}.
A lucid introduction is provided by the lectures of Ashtekar in this
volume.} Dirac quantization employs an extended linear space, ${\cal
L}^{(T,L)}$,  of
functionals of the vector potential, $\vec A ({\bf x})$.
 Observables commute with operator representations of
the constraints and physical states are represented by functionals that
are annihilated by them.  The linear space ${\cal L}^{(T,L)}$ cannot be the
Hilbert space ${\cal H}^{(T,L)}$ because solutions of the constraint
$\pi^L \psi=0$ are functionals of $\vec A^T$ alone and are therefore
not square integrable. For the electromagnetic field Dirac and ADM
quantization are fully equivalent as usually interpreted \cite{Kuc86}.
If that is true, Dirac quantization would share with ADM the differences
with the present approach for the predictions values of gauge-invariant
quantities that are not ``time degrees of freedom'' when extended over
time.  Despite this difference we can still ask whether we can construct
anything like the operators and states of Dirac quantization in the
present approach.  The following are possible:

Class operators corresponding to a set of coarse-grained histories
$\{c_\alpha\}$ may be introduced on ${\cal H}^{(T,L)}$ by
specifying their matrix elements by
\begin{equation}
\bigl\langle \vec A^{\prime\prime}\Vert\, C_\alpha \Vert
\vec A^\prime\bigr\rangle = \bigl\langle \vec A^{\prime\prime}
\Vert\, \int_\alpha \delta A\, \Delta_{\tilde\Phi}[A] \, \delta
\,\bigl[\tilde\Phi(A)\bigr]\, \exp\, \left(iS[A]\right)\Vert \vec
A^\prime\bigr\rangle\ .
\label{sixfiveone}
\end{equation}
The functional integral is over the potentials $A^\mu(x)$ that lie in
the class $c_\alpha$ and match the prescribed vector potentials on the
initial and final surfaces.  $\tilde\Phi$ is a gauge fixing condition
that does not include a surface gauge fixing condition as in 
\eqref{sixtwofour}
since the corresponding gauge freedom is already fixed by the
specification of the vector potentials on the initial and final
surfaces. Indeed, $\tilde\Phi$ must be such as to not restrict $\vec
A({\bf x})$ on the initial and final surfaces at all.

The operators so defined are independent of $\tilde\Phi$ in the class of
$\tilde\Phi$ generated from a given one by gauge transformations that
preserve the initial and final vector potentials. That is, they depend
only on the class of gauge fixing conditions of the form
$\tilde\Phi^\Lambda(A)=\tilde\Phi(A+\nabla\Lambda)$ for some fixed
$\tilde\Phi$ as $\Lambda$ ranges over such gauge transformations.

The class operators on ${\cal H}^{(T,L)}$ exhibited in \eqref{sixtwotwelve}
could be used as the starting
point for the construction of the decoherence functional
\eqref{sixtwothirteen}.  However, to incorporate initial and final conditions
represented by wave functions $\phi_i[\vec A^T]$ and $\psi_j[\vec A^T]$
that are solutions of the constraints we cannot use the inner product on
${\cal H}^{(T,L)}$ because such wave functions do not lie in that space.
Rather we must attach initial and final wave functions as in
\eqref{sixtwoelevena}
\begin{equation}
\left\langle\phi_i | C_\alpha | \psi_j \right\rangle = \int d\vec
A^{T\prime\prime} \int d\vec A^{T\prime} \int d\vec A^{L\prime\prime}
\phi^*_i [\vec
A^{T\prime\prime}]\,\langle\vec A^{T\prime},
\vec A^{L\prime\prime} \Vert C_\alpha
\Vert \vec A^{T\prime}, 0\rangle\,\psi_j [\vec A^{T\prime}]\, .
\label{sixfivetwo}
\end{equation}
essentially making use of the inner product on ${\cal H}^T$.  The
decoherence functional could then be constructed as in \eqref{fiveoneten}
and is equivalent to \eqref{sixtwothirteen}.  Such
constructions involving separate linear spaces for functional integrals
and initial and final conditions will be essential in defining the generalized
quantum
mechanics of reparametrization invariant systems.

Although the generalized quantum mechanics under discussion does usually
permit a notion of state on a spacelike surface, the above construction
suggests a way of associating a branch wave functional on ${\cal L}^{(T,L)}$
with
each branch of an initial pure state
$|\psi\rangle$
corresponding to a coarse-grained history $c_\alpha$.  Define the
extended wave functional by
\begin{equation}
\Psi_\alpha\bigl[\vec A^{\prime\prime}\bigr] = \int\, \delta \vec
A^\prime \bigl\langle\vec A^{\prime\prime}\left\Vert\,
C_\alpha\right\Vert \vec A^\prime\bigr\rangle \psi
\bigl[A^{T\prime}\bigr]\, .
\label{sixfivethree}
\end{equation}
This branch wave functional is independent of the gauge fixing condition
in the class generated from a given one by gauge transformations that
leave $\vec A''({\bf x})$ unchanged.

The branch wave functions $\Psi_\alpha[\vec A]$ may be thought of as
``states of the system'' on the final spacelike surface $t=T$. Indeed,
if we limit attention to coarse grainings that restrict the values of
$\vec A$ only on a family of spacelike surfaces labeled by $t$, then is
it is possible to define states on these surfaces represented by
wave functions $\Psi_\beta[\vec A, t]$ by following the construction
described in Section IV.4. These states would have the form of
\eqref{sixfivethree} but with the functional integrals defining $\langle
\vec A''\Vert C_\beta \Vert \vec A' \rangle$ limited to times less than $t$
and restricted only by the coarse graining there.

We are now in a position to ask whether the extended class operators
defined by \eqref{sixfiveone} commute with the constraint and whether the
wave functional of the individual branches are annihilated by it.
The simplest example of a gauge condition that does not restrict the
vector potentials on either the initial or final surfaces is the temporal
gauge. In this gauge, the question of commutation is easily analyzed
directly.  Shift the
variable of integration in \eqref{sixfiveone} by the gauge transformation
\begin{equation}
\vec A\left(t, {\bf x}\right) \longrightarrow \vec A \left(t, {\bf x}
\right) + \vec \nabla \epsilon \left({\bf x}\right) 
\label{sixfivefour}
\end{equation}
where
$\epsilon({\bf x})$ is independent of time.  Of course,
the integral is
not changed by this shift in integration variable.  But also, because
this shift
is a gauge transformation, the action and measure are left unchanged.
Because it is a {\it time-independent} gauge transformation the temporal gauge
is preserved.  Thus, \eqref{sixfiveone} is unchanged when the initial
and final $\vec A(\bf x)$ are shifted as in \eqref{sixfivefour}
 by the same amount.  Since $\vec\pi^L(\bf x)$ is the
operator that effects such a shift [{\sl cf.}~\eqref{sixfourfive}], this is
equivalent to
\begin{equation}
\bigl[\vec{\pi}^L\left({\bf x}\right), C_\alpha\bigr] = 0
\label{sixfivefive}
\end{equation}
so constraints commute with the extended class operators.  It is then an
immediate consequence of \eqref{sixfivefive} and $\vec\pi^L({\bf x})
\psi (\vec{A}^T) = 0$ that
\begin{equation}
\vec{\pi}^L({\bf x}) \Psi_\alpha \bigl[\vec A\bigr] = 0\, .
\label{sixfivesix}
\end{equation}
The wave functionals representing the branches of gauge invariant coarse
graining thus satisfy the Dirac constraint condition. Restricting the
initial and final conditions to wave functions that depend only on the
``true physical degrees of freedom'' means that wave functions
representing states at intermediate times also depend only on these.

We have derived these results in the temporal gauge. However, both
\eqref{sixfivefive} and \eqref{sixfivesix} are more general because the
functional integrals defining $C_\alpha$ and $\Psi_\alpha$ are
independent of the gauge fixing condition in the classes discussed
above.

There are thus two distinct ways in which the constraints can be said
to be satisfied in the generalized quantum mechanics of electromagnetism
under discussion. First $\vec{\pi}^L({\bf x})$
is a gauge-invariant quantity which can be given meaning in
a sum-over-histories formulation of quantum mechanics as average values
of field ``velocities'' over very long times.  The theory
predicts probabilities for alternative values of $\vec{\pi}^L({\bf x})$
when these alternatives decohere. The probability is zero for values
other than $\vec{\pi}^L({\bf x}) = 0$. Second, when class operators and branch
wave functionals
are defined on the configuration space of vector potentials as
described, then the class operators commute with the constraints and
the branch wave functions are annihilated by them. In these senses the
generalized quantum mechanics of electromagnetism  makes contact with
the ideas of Dirac quantization. When restricted to gauge invariant
partitions by potential or momenta at definite moments of time, the
predictions of the generalized quantum mechanics described here coincide
with those of the Dirac procedure. In considering gauge invariant
alternatives which are extended over time, however, it goes beyond either
Dirac or ADM quantization in their usual senses.

\section{Models with a Single Reparametrization Invariance}
\setcounter{footnote}{0}

\subsection{Reparametrization Invariance in General}

Generalized quantum mechanical theories are specified by their fine-grained
histories, their allowable coarse grainings, and their decoherence
functional.  In this section we shall construct examples of such
theories
for a class of models
whose unique set of
{\it fine-grained histories} are curves in a configuration
space ${\cal C}$ spanned by coordinates $Q^i, \;
 i=1, \cdots, \nu$.  The $Q^i$
include the variables describing the physical time, if there is
one. The most familiar example is the relativistic particle whose
fine-grained histories are curves in spacetime.

Curves may be described parametrically by giving the coordinates
as functions of a parameter $\lambda$, \viz
$Q^i(\lambda)$.  We shall
frequently suppress the coordinate labels and write $Q$ for a point in
the configuration space and $Q(\lambda)$ for a curve.  The
{\it curves}
are the fine-grained histories, not the functions $Q^i(\lambda)$ that
describe how the paths are parametrized.  For this reason these theories
are {\it reparametrization} {\it invariant}. The action summarizing dynamics
and the partitions defining allowed
coarse grainings may both be conveniently described in terms of the functions
$Q^i(\lambda)$, but they both must be invariant under reparametrizations:
\begin{equation}
\lambda \rightarrow \bar\lambda = f(\lambda)\, .
\label{sevenoneone}
\end{equation}

The most natural choice for the set of fine-grained histories is often
the set of {\it all} curves in ${\cal C}$ including those which cross and
recross the
surfaces of constant time if there is one.
However,
different theories can be obtained by restricting
the set of fine-grained histories, for example, to curves that intersect
hypersurfaces of a preferred time coordinate once and only once.  We
shall
illustrate the effects of such choices in the models below.

As we shall see below, reparametrization invariance implies a constraint
between the coordinates and their canonical momenta.  The quantum
mechanics of such a constrained theory is often most conveniently
formulated on an extended configuration space ${\cal C}_{\rm ext}$ of
coordinates $Q^i$ and a multiplier enforcing the constraint.  The free
relativistic particle provides the simplest example.  The configuration
space ${\cal C}$ is Minkowski spacetime and the fine-grained histories are
curves $x^\alpha(\lambda)$ in this spacetime.
A classical action for the
relativistic particle is the spacetime interval along its curve
\begin{equation}
S[x^\alpha] = m\int d\tau \equiv m\int^1_0 d\lambda
\left[-\eta_{\alpha\beta}\left(\frac{dx^\alpha}{d\lambda}\right)
\left(\frac{ dx^\beta}{d\lambda}\right)\right]^\half
\label{sevenonetwo}
\end{equation}
where $m$ is the particle's rest mass, $\eta_{\alpha\beta}$ is the
Minkowski metric and we have arbitrarily chosen $0$ and $1$ as the
values of the parameter labeling the ends of the curve. (In previous
sections we have used $\tau$ for a dummy argument variable. In this
section it means proper time.)   The action
\eqref{sevenonetwo} is manifestly reparametrization invariant and its extrema
satisfy the correct relativistic equations of motion.  However, it is
not the only classical action with these properties. Different actions
with the same extrema are equivalent classically, but
in quantum
mechanics it is not just the extrema of the action which are important.
The {\it value} of the action on non-extremal curves also contributes to
amplitudes through path-integrals of $\exp(iS)$.  Different forms of the
action will therefore generally lead to different sum-over-histories
quantum theories
assuming that the relevant sums over $\exp(iS)$ can be defined
at all.

The action \eqref{sevenonetwo} cannot easily
be used to formulate a sum-over-histories
quantum mechanics of the relativistic particle because it is not
quadratic in the velocities.  An action which does the job can be
formulated on the extended configuration space ${\cal C}_{\rm ext}$ of paths
$x^\alpha(\lambda)$ and multiplier $N(\lambda)$.
It is
\begin{equation}
S\left[x^\alpha, N\right] = \frac{m}{2} \int^1_0 d\lambda
N(\lambda)\left[\left(\frac{\dot x(\lambda)}{N(\lambda)}\right)^2 - 1\right]
\label{sevenonethree}
\end{equation}
where a dot denotes a derivative with respect to $\lambda$ and $(\dot
x)^2 = \eta_{\alpha\beta} \dot x^\alpha\dot x^\beta$.  The action
\eqref{sevenonethree} yields the correct equations of
motion when extremized with respect to $x^\alpha(\lambda)$ and
$N(\lambda)$ and it is invariant under the reparametrization
transformations
\begin{subequations}
\label{sevenonefour}
\begin{eqnarray}
x^\alpha(\lambda)\rightarrow \bar x^\alpha(\lambda) & = & x^\alpha
\bigl(f(\lambda)\bigr)\, , \label{sevenonefour a}\\
N(\lambda) \rightarrow \bar N(\lambda) & = & N\bigl(f(\lambda)\bigr)\dot
f(\lambda), \label{sevenonefour b}
\end{eqnarray}
\end{subequations}
provided $f(0)=0$ and $f(1)=1$ so the values of $x^\alpha$ and $\lambda$ at
the ends of the history are unchanged.  As we
shall show in detail in Section VII.4 and VII.5,
the action \eqref{sevenonethree} leads to
correct and manageable quantum theories of the relativistic particle.
Thus generally we take for the fine-grained histories of a
reparametrization-invariant theory curves $(Q(\lambda), N(\lambda))$ in
${\cal C}_{\rm ext}$.

The second element of a generalized quantum mechanics is the class of
{\it allowed coarse grainings}. For a reparametrization-invariant
theory, the general notion of a coarse graining
 is a partition of the fine-grained histories --- curves
in ${\cal C}_{\rm ext}$ --- into exclusive
reparametrization invariant classes $\{c_\alpha\}$.
More specifically, each
class must be invariant under the reparametrization transformation
\begin{subequations}
\label{sevenonefive}
\begin{eqnarray}
Q^i(\lambda) \rightarrow \bar Q^i(\lambda) & = & Q^i\bigl(f(\lambda)\bigr)
\, , \label{sevenonefive a}\\
N(\lambda) \rightarrow \bar N(\lambda) & = & N\bigl(f(\lambda)\bigr)\dot
f(\lambda)\, , \label{sevenonefive b}
\end{eqnarray}
\end{subequations}
for $f(\lambda)$ that leave the parameters of the endpoints of the curve
unchanged.
Examples of reparametrization invariant coarse grainings are readily
exhibited: Given a spacetime region $R$, the paths may partitioned into
the class of paths that never cross $R$ and the class of paths that
cross $R$ at least once.  Given a hypersurface in configuration space
the paths may be partitioned by the value of $Q$ at which they first
cross the hypersurface starting from one end.

Further examples can be constructed by introducing the arc-length along
a curve.
The multiplier $N(\lambda)$ allows reparametrization invariant
arc-length
\begin{equation}
\tau\bigl(\lambda^{\prime\prime}, \lambda^\prime, N(\lambda)\bigr] =
\int^{\lambda^{\prime\prime}}_{\lambda^\prime} N(\lambda) d\lambda 
\label{sevenonesix}
\end{equation}
to be defined between any two points
along a curve that are defined in a reparametrization invariant manner.
For instance, we might consider the arc-length $\tau$ of paths that
connect two points $Q'$ and $Q''$.
The paths may then be partitioned using this
additional invariant structure.  For example, the paths starting from
point $Q^\prime$ could be
partitioned by the positions $Q$ they have arrived at after a given length
$\tau$.

The most general notion
of coarse graining is a partition by ranges of values of
reparametrization invariant functionals of
the paths and multiplier $F[Q
(\lambda), N(\lambda)]$.   All of the above examples can be
characterized in this way.

A {\it decoherence functional} completes the specification of a generalized
quantum mechanics.  For a given coarse graining consisting of classes
$\{c_\alpha\}$ this will be constructed from path-integrals over the
classes of the form
\begin{equation}
\left\langle Q^{\prime\prime} \left\Vert C_\alpha\right\Vert Q^\prime
\right\rangle \equiv \sum\limits_{{\rm path}\ \in [Q'c_\alpha Q'']}
\exp\Bigl(iS[{\rm path}]\Bigr) 
\label{sevenoneseven}
\end{equation}
where the sum is over all paths in ${\cal C}_{\rm ext}$
 that begin at $Q^\prime$, end at
$Q^{\prime\prime}$, and are in the class $c_\alpha$.  To make this
precise we need to specify the action, measure, and the product formula
with which the sums in \eqref{sevenoneseven} are defined.  There is a canonical
way of doing this which is somewhat lengthy to describe so we shall take
it up separately in Section VII.2 below.  For the moment, we simply note
that, in cases where the path integral is defined by a product formula,
the most natural  Hilbert space involved
is ${\cal H}^Q$ --- the space of square-integrable
functions on the configuration space ${\cal C}$ spanned by the $Q^i$.  The
matrix elements \eqref{sevenoneseven} then define a class operator
$C_\alpha$ on ${\cal H}^Q$.  We have used a double bar to denote the
inner product on ${\cal H}^Q$.

Following the example of non-relativistic quantum mechanics discussed in
Section V, the next step in the construction of the decoherence
functional is to adjoin initial and final conditions represented respectively
by wave
functions $\{\psi_j(Q)\}$ and $\{\phi_i(Q)\}$ and their associated
probabilities.  In non-relativistic
quantum mechanics we did this using the same inner product that was used
to define the path-integrals.  However, it will prove to be important
for reparametrization invariant theories to allow a more general
construction.  We define
\begin{equation}
\left\langle \phi_i\left | C_\alpha\right| \psi_j\right\rangle = \phi_i
(Q^{\prime\prime}) \circ \left\langle Q^{\prime\prime}\left\Vert C_\alpha
\right\Vert Q^\prime\right\rangle \circ \psi_j (Q^\prime)
\label{sevenoneeight}
\end{equation}
where the $\circ$ denotes a Hermitian, but {\it not necessarily a
positive definite}, inner product.
For example, the Klein-Gordon inner product will
be useful in the case of the relativistic particle. We should stress
that the use of the notation $\langle \phi_i|C_\alpha|\psi_j\rangle$
does not mean that
we have defined a Hilbert space of states $|\psi_j\rangle$.
We take \eqref{sevenoneeight} to be the {\it definition} of $\langle
\phi_i|C_\alpha|\psi_j\rangle$.

The construction \eqref{sevenoneeight} may seem more familiar if we recall
its analogs in the cases of non-relativistic quantum mechanics and gauge
theories studied in Sections V and VI.   In non-relativistic
quantum mechanics the configuration space ${\cal C}$
was ${\bf R}^\nu$ and $\circ$ was the usual inner product on the space
of square-integrable functions on ${\bf R}^\nu$. In the
case of gauge theories we can take the configuration space ${\cal C}$ to be the
space of vector potentials $\vec A({\bf x})$, ($A^0$ is then a
multiplier).  To define the class operator matrix elements on ${\cal
H}^T$ in \eqref{sixtwoelevena} we used the analog of \eqref{sevenoneeight} with
$\circ$ being the inner product on ${\cal H}^{T}$.
Eq.~\eqref{sevenoneeight}
represents an even more general construction because of the weaker
conditions on $\circ$.

A decoherence functional may now be defined as follows: Specify a set
of initial wave functions $\{\psi_j (Q)\}$ together with probabilities
$\{p^\prime_j\}$.  Similarly, specify a set of final wave functions
$\{\phi_i(Q)\}$ together
with probabilities $\{p^{\prime\prime}_i\}$. Construct
\begin{equation}
D(\alpha^\prime, \alpha) = {\cal N}
\sum\nolimits_{ij} p^{\prime\prime}_i \left\langle
\phi_{i}\left |C_{\alpha^\prime}
\right|\psi_{j}\right\rangle\left\langle\phi_i\left|C_\alpha
\right|\psi_j\right\rangle^* p^\prime_j\, .
\label{sevenonenine}
\end{equation}
With an appropriate choice for ${\cal N}$, this construction satisfies the
requirements (i)--(iv) of Section IV.1
 for a decoherence functional.  It is manifestly
Hermitian with positive diagonal elements.  The linearity of the sum
over paths \eqref{sevenoneseven} ensures consistency with the principle of
superposition.  Normalization fixes ${\cal N}$ as
\begin{equation}
{\cal N}^{-1} = \sum\nolimits_{ij}p^{\prime\prime}_i \left|\left\langle
 \phi_i \left |
C_u\right| \psi_j\right\rangle \right|^2 p^\prime_j 
\label{sevenoneten}
\end{equation}
where the sum over {\it all} paths in \eqref{sevenoneseven} defines $C_u$.

The specification of a generalized quantum mechanics is now essentially
complete.  The fine-grained histories are parametrized paths in the
configuration space ${\cal C}_{\rm ext}$, the coarse-grained histories
are
reparametrization invariant partitions of these, and the decoherence
functional is \eqref{sevenonenine}.  There are still further choices to define
the theory --- the precise set of curves in ${\cal C}_{\rm ext}$ that are the
fine-grained histories,
the inner product $\circ$, the sets of initial and final wave
functions together with their probabilities, and the exact construction
of the path-integrals defining the class operators.  The general
framework is thus a loose one and many different theories are possible.
There is room for further principles to restrict these choices.
For the moment, in a course of lectures devoted to ways in which
Hamiltonian quantum mechanics might be generalized, it is perhaps
appropriate to illustrate the choices in explicit models rather than
search for further principles. We begin with a concrete prescription for
carrying out the path-integrals defining the class operators.

\subsection{Constraints and Path Integrals}

In a Hamiltonian formulation of dynamics,
reparametrization invariance implies a constraint between
the canonical coordinates $Q^i$ and their conjugate momenta $P_i$.  To
see this quickly\footnote{For more details see \cite{Dir64},
\cite{Kuc74}, and \cite{HRT76}.},
 suppose that the dynamics is summarized by a Lagrangian action
of the form
\begin{equation}
S\bigl[Q^i, N\bigr] = \int^1_0 d\lambda L \bigl[\dot Q^i
(\lambda), Q^i(\lambda), N(\lambda) \bigr] 
\label{seventwoone}
\end{equation}
that is invariant under the reparametrization
transformations \eqref{sevenonefive}.  Invariance under the infinitesimal
version of these transformations, with $f(\lambda) = 1+\xi(\lambda)$,
and $\xi(0) = \xi(1) = 0$, implies the following
 relation among the
equations of motion
\begin{equation}
\left[-\frac{d}{d\lambda}\left(\frac{\partial L}{\partial\dot Q^i}\right) +
\frac{\partial L}{\partial Q^i}\right] \dot Q^i + N\frac{\partial
L}{\partial N} = 0 
\label{seventwotwo}
\end{equation}
where we employ the summation convention.  This is an {\it identity}
which must be satisfied for arbitrary choice of the functions
$Q^i(\lambda)$ and $N(\lambda)$.  It can therefore  only be satisfied if the
coefficients of the various derivatives $\dot Q^i$, $\ddot Q^i$,
etc.~vanish separately.  In particular, the vanishing of the
coefficient of the second derivatives implies
\begin{equation}
\left(\frac{\partial^2L}{\partial\dot Q^j\partial \dot Q^i}\right) \dot
Q^i = 0\, . \label{seventwothree}
\end{equation}
This is the characteristic signature of a constrained Hamiltonian
theory.  Expressed in terms of the momenta
\begin{equation}
P_i = \frac{\partial L}{\partial\dot Q^i}\, ,
\label{seventwofour}
\end{equation}
\eqref{seventwothree} means there are linear relations of the form
\begin{equation}
\left(\partial P_j/\partial Q_i\right) \dot Q^i = 0\, . 
\label{seventwofive}
\end{equation}
The defining relations \eqref{seventwofour} thus cannot be inverted to find
the $\dot Q^i$ in terms of the $P_i$ because the $P_i$ are not
independent.  There must be a relation among them of the form
\begin{equation}
H\bigl(P_i, Q^i\bigr) = 0 
\label{seventwosix}
\end{equation}
and that is the constraint.  In the following we shall recover its
explicit from in particular examples.

The relations \eqref{seventwofour} and \eqref{seventwosix} {\it together} are
invertible to find the velocities in terms of the momenta and, with these
relations, the action may be re\"expressed in canonical form as the
integral of $[P_i\dot Q^i - ({\rm a\ function\ of}\ P_i {\rm ,}Q^i {\rm ,
and}\ N$)].
This
canonical action also must be invariant under reparametrization
transformations \eqref{sevenonefive} with the momenta transforming as
\begin{equation}
P_i \rightarrow \bar P_i = P_i \bigl(f(\lambda)\bigr)\, .
\label{seventwoseven}
\end{equation}
It can therefore only have the general form
\begin{equation}
S\bigl[P_i, Q^i, N\bigr] = \int^1_0 d\lambda\bigl[P_i\dot Q^i -
NH (P_i, Q^i)\bigr]\, . 
\label{seventwoeight}
\end{equation}
Reparametrization invariance forbids a term that is a function of $P_i$
and $Q^i$ but not proportional to $N$.  Variation of \eqref{seventwoeight}
with respect to $P_i, Q_i$ and $N$ yield the canonical equations of
motion and a constraint.  Since \eqref{seventwosix} is ambiguous up to a
multiplicative factor, we may take its form to coincide with the $H$ in
\eqref{seventwoeight} as we have anticipated in the notation.  The
Hamiltonian entering the canonical action \eqref{seventwoeight}
vanishes when the constraint is satisfied --- a general
feature of reparametrization invariant theories when the coordinates
and momenta transform as scalars under reparametrizations.

The canonical action \eqref{seventwoeight} is invariant under canonical
transformations of the $P$'s and $Q$'s generated by the constraint
under the Poisson bracket operation $\{,\}$,
 provided the multiplier is transformed suitably. Specifically the
canonical action is invariant under
\begin{subequations}
\label{seventwoeighta}
\begin{eqnarray}
\delta Q^i & = & \epsilon (\lambda) \bigl\{Q^i, H\bigr\}\, ,
\label{seventwoeighta a}\\
\delta P_i & = & \epsilon(\lambda)\left\{P_i, H\right\}\, ,
\label{seventwoeighta b}\\
\delta N & = & \dot\epsilon(\lambda)\, ,
\label{seventwoeighta c}
\end{eqnarray}
\end{subequations}
for arbitrary, infinitesimal $\epsilon(\lambda)$, vanishing at the
endpoints.  The transformations (VII.2.9ac) have the same form as
infinitesimal
reparametrization transformations \eqref{sevenonefive} with
$f(\lambda) = 1 + \dot\epsilon(\lambda)/N(\lambda)$.  In fact, the
transformations \eqref{seventwoeighta} are a larger group of symmetries than
reparametrizations because, for example, the requirement that
$f(\lambda)$ be single-valued, which is necessary for a reparametrization
need not be enforced to ensure the invariance of the canonical action
under \eqref{sevenonenine}.

The action \eqref{seventwoeight} is the basis for a canonical construction of
the path-integrals \eqref{sevenoneseven} defining the class operators,
$\{C_\alpha\}$, of a
reparametrization invariant coarse graining. We write the schematic
\eqref{sevenoneseven} out explicitly as
\begin{equation}
\left\langle Q^{\prime\prime}\left\Vert C_\alpha\right\Vert
Q^\prime\right\rangle = \int_{\alpha} \delta P\delta Q\delta N
 \Delta_\Phi [Q, N] \delta \bigl[\Phi [Q, N]\bigr]\exp\bigl(iS[P,Q,N]\bigr)\ .
\label{seventwonine}
\end{equation}
The action in this formula is \eqref{seventwoeight}.
The condition $\Phi[Q, N]=0$ fixes the symmetry \eqref{seventwoeighta}.
Here, for simplicity,  we assume it is independent of the momenta.
The quantity $\Delta_\Phi$ is the associated Faddeev-Popov determinant.
The measure is the Liouville measure on the extended phase space of
$P_i$ and $Q^i$. This is explicitly invariant under the canonical
transformation \eqref{seventwoeighta} and therefore reparametrization
invariant.
This  path-integral can be implemented, analogously to
the discussion in Section
V.2, as the limit of integrals over polygonal paths
defined on a slicing of the parameter range into $J$ equally spaced
intervals $\lambda_0=0,\lambda_2,\cdots,\lambda_J=1$ of
parameter length $\epsilon$.
The explicit form
of the measure is then
\begin{equation}
dN_{J}\left(\prod\limits^\nu_{i=1} \frac{dP_{iJ}}{2\pi}\right)
\left(\prod\limits^{J-1}_{K=1}
dN_K \prod\limits^\nu_{i=1}\frac{dP_{iK}dQ_K^i}{2\pi}\right)\, .
\label{seventwoten}
\end{equation}

The ranges of integration must be reparametrization invariant.  The
momenta are integrated from $-\infty$ to $+\infty$.  The coordinate
and multiplier integrations are
restricted by the reparametrization invariant class
$c_\alpha$.  If unrestricted by the coarse graining, several reparametrization
invariant ranges are available
for the multiplier $N$.  We could, for example, integrate from $-\infty$
to $+\infty$ on each slice or from $0$ to $+\infty$. Both are
reparametrization invariant [\cf \eqref{sevenonefive}].
Different ranges will in general yield different theories and we shall
explore several in the models discussed below.  With these choices for
action, measure, and range of integration the path-integrals defining the
class operators have been fixed.

In the models we shall consider, the canonical action will depend at
most quadratically on the momenta. Provided $N$ is positive, the
momenta may be integrated out of \eqref{seventwonine}
 to yield an integral for the
class operators over paths,
$(Q^i(\lambda)$, $N(\lambda))$,  in the extended configuration space, ${\cal
C}_{\rm ext}$.
When, as in the case of the relativistic
particle, the action is purely quadratic, this will be a Lagrangian path
integral of the form
\begin{equation}
\left\langle Q^{\prime\prime} \left\Vert C_\alpha \right\Vert
Q^\prime\right\rangle = \int\nolimits_{\alpha} \delta Q\delta N\
\Delta_\Phi [Q, N] \delta \bigl[\Phi [Q, N]\bigr] \exp \bigl(iS[Q,N]\bigr)
\label{seventwoeleven}
\end{equation}
where it is easily verified that the action is \eqref{seventwoone}. The
measure for $\delta Q$ that results from the integration over the $P$'s
now contains fixed factors of $\pi$, the separation $\epsilon$ between
slices will, in general, depend on the
multiplier.\footnote{ See, \eg \cite{HK86} for an explicit
construction in the case of the relativistic
particle.}

The construction of
the path-integrals spelled out in this subsection may not be the most
general consistent with the principles of generalized quantum mechanics
and reparametrization invariance.  However, it is an explicit
construction that will yield familiar results in the simple models to
which we now turn.

\subsection{Parametrized Non-Relativistic Quantum Mechanics}

The simplest reparametrization invariant model is parametrized
non-relativistic quantum mechanics \cite{Dir64, Kuc74}.
 To construct it we begin with the
action summarizing the dynamics of a non-relativistic particle, taken to
move
in only one dimension for simplicity,
\begin{equation}
S[X(T)] = \int^{T''}_{T'} dT\, \ell \left(\frac{dX}{dT}, X\right) \ .
\label{seventhreeone}
\end{equation}
We shall assume that the Lagrangian $\ell$ is of standard
quadratic kinetic energy minus potential energy form so that the
associated Hamiltonian can be written
\begin{equation}
h\left(P_X, X\right) = \frac{P^2_X}{2M} + V(X)\, . 
\label{seventhreetwo}
\end{equation}
The Newtonian time, $T$, may be elevated to the status of a dynamical
variable by introducing an arbitrary parameter $\lambda$ and writing the
action in parametrized form
\begin{equation}
S\bigl[X(\lambda), T(\lambda)\bigr] = \int\nolimits^1_0 d\lambda \dot
T\ell \bigl(\dot X/\dot T, X\bigr)\, . 
\label{seventhreethree}
\end{equation}
Here a dot denotes a derivative with respect to $\lambda$.  Since the
parameter $\lambda$ was arbitrary, the action is manifestly
reparametrization invariant.  It is thus an example of the kind
discussed in Section VII.1 with $Q^1 = X, Q^2 = T$ and
\begin{equation}
L(\dot Q, Q) = \dot T\ell (\dot X/\dot T, X)\, .
\label{seventhreefour}
\end{equation}
There is no multiplier.  The constraint implied by reparametrization
invariance is easily verified by direct computation to be
\begin{equation}
P_T + h\left(P_X, X\right) = 0\, . 
\label{seventhreefive}
\end{equation}

We now construct a generalized quantum mechanics for this model
according to the general schema of Section VII.1, specifying the fine-grained
histories, allowed coarse grainings, and decoherence functional.  We consider
two different
theories using, as starting
points, two different sets of {\it fine-grained histories}.  The first set is
the usual set of paths for which $X$ is a single-valued function of $T$.
Such paths are said to ``move forward in $T$''.  The second is the set
of arbitrary paths in the $(X,T)$ configuration space moving both
forward and backward in $T$.  These define ostensibly different theories
although we shall show that, in fact, they are both equivalent
to familiar non-relativistic quantum mechanics for certain classes of
coarse
grainings.

If the fine-grained histories are restricted to be single-valued in $T$,
the {\it allowed coarse grainings} are the familiar ones of the
non-relativistic theory discussed in Section V. However, if arbitrary
paths in the $(X,T)$ configuration space are allowed as fine-grained
histories, then these coarse grainings must be reconsidered because a
rule that partitions a subset does not necessarily partition a set which
contains it. For example, it is not possible to partition all paths by the
regions of $X$ through which they cross a sequence of constant-$T$
surfaces because the paths may cross each surface more than once.
Coarse grainings of the class of arbitrary paths will, of course, also
coarse-grain the subset of those that are single valued in $T$. For
example, given a
sequence of constant-$T$ surfaces divided into exclusive intervals in
$X$, the class of arbitrary paths could be partitioned by whether they
cross each of these regions at least once or not at all.  This is also a
partition of single-valued paths although those classes involving
multiple crossings of the same surface are vacuous.  In the following
when we speak of a coarse graining we mean a partition of the class of
arbitrary paths.

We begin the construction of the decoherence functional for these models
 by examining the path-integral \eqref{seventwonine} defining operators
corresponding to a partition $\{c_\alpha\}$ of the
fine-grained histories.
A convenient condition that fixes the
symmetry of \eqref{seventwoeighta} of
either set of fine-grained histories is\footnote{For more on the
requirements for suitable conditions that fix \eqref{seventwoeighta} see
Teitelboim \cite{Tei83aa} and Henneaux, Teitelboim, and Vegara
\cite{HTV92}.}
\begin{equation}
\Phi = \dot N = 0 \label{seventhreesix}
\end{equation}
so that $N$ is a constant.  The associated Faddeev-Popov determinant
is constant.

The explicit form of the canonical action in \eqref{seventwoeight} is
\begin{equation}
S\left[P_T, P_X, T, X\right] = \int\nolimits^1_0 d\lambda \left[P_T \dot
T + P_X \dot X- N\bigl(P_T + h(P_X, X)\bigr)\right]\, .
\label{seventhreeseven}
\end{equation}
Since the constraint is linear in $P_T$, the exponent in
\eqref{seventwonine}
is also linear, and the integration over $P_T$ produces a $\delta$-function.
The integral over $P_X$ can also be carried out explicitly to yield the
following expression for \eqref{seventwonine} in the gauge 
\eqref{seventhreesix}:
\begin{equation}
\left\langle X^{\prime\prime}, T^{\prime\prime} \left\Vert C_\alpha
\right\Vert X^\prime, T^\prime \right\rangle = \int\nolimits_{\alpha}
\delta X\delta T \int dN\,\delta \bigl[\dot T-N\bigr] \exp
\left(i\int\nolimits^1_0 d\lambda N\left[\frac{M}{2}\left(\frac{\dot
X}{N}\right)^2 - V(X)\right]\right)\, . 
\label{seventhreeeight}
\end{equation}
There remains an integral over the paths in the $(X,T)$ configuration
space and a single integral over the constant value of $N$.
This path-integral involves a Lagrangian action that is different from
\eqref{seventhreethree} but becomes equivalent to it if the $\delta$-function
in \eqref{seventhreeeight} is used to eliminate the multiplier.

To continue, we consider the two possibilities for fine-grained histories
separately.  If the paths are restricted to move forward in $T$ then
$\dot T$ is positive.  As a consequence, if $T^{\prime\prime} >
T^\prime$, the unique value
\begin{equation}
N=T^{\prime\prime} - T^\prime \label{seventhreenine}
\end{equation}
contributes to the integration over $N$, and the unique path
\begin{equation}
T(\lambda) = T^\prime (1-\lambda) + T^{\prime\prime}\lambda
\label{seventhreeten}
\end{equation}
to the integration over the functions $T(\lambda)$.  The result is
\begin{equation}
\left\langle X^{\prime\prime}, T^{\prime\prime} \left\Vert C_\alpha
\right\Vert X^\prime, T^\prime\right\rangle = \theta \left(T^{\prime\prime}
- T^\prime\right) \int\nolimits_{\alpha} \delta X \exp
\bigl(iS[X(T)]\bigr) 
\label{seventhreeeleven}
\end{equation}
where $S$ is the deparametrized action \eqref{seventhreeone}.
The class operators thus coincide with those of the
non-relativistic theory described in Section V.  We write
\begin{equation}
\left\langle X^{\prime\prime}, T^{\prime\prime} \left\Vert C_\alpha
\right\Vert X^\prime, T^\prime\right\rangle = \theta
\left(T^{\prime\prime} - T^\prime\right) \left\langle X^{\prime\prime}
\left | C_\alpha\right | X^\prime\right\rangle 
\label{seventhreetwelve}
\end{equation}
understanding that the matrix element on the right refers to the  partition
of non-relativistic paths moving forward on the interval
$[T^\prime, T^{\prime\prime}]$
induced by the partition $c_\alpha$ of all paths. This was
defined in Section V
and we are
 using the notation of that section in which the dependence of
$\langle X^{\prime\prime}|C_\alpha | X^\prime\rangle$ on
$T^{\prime\prime}$ and $T^\prime$ has been suppressed.

The result when the fine-grained histories move both forward and
backward in time is different, but not very different. If the multiplier
integration is over a positive range then again only the unique value of
$N$ in \eqref{seventhreenine} and the unique path in \eqref{seventhreeten}
contribute and the result is \eqref{seventhreetwelve}.  If the multiplier
integration is over the whole range of $N$ then there is an additional
contribution from a unique negative $N$ and the same unique path when $T^\prime
> T^{\prime\prime}$.  One finds for the range $-\infty < N < \infty$
\begin{equation}
\left\langle X^{\prime\prime}, T^{\prime\prime} \left\Vert C_\alpha
\right\Vert X^\prime, T^\prime \right\rangle = \left\langle
X^{\prime\prime} \left| C_\alpha \right| X^\prime \right\rangle\, .
\label{seventhreethirteen}
\end{equation}

The important point about these results is that paths that move {\it both}
forward and backward in time do not contribute to the path-integrals
defining the class operators.  Partitions of all paths may therefore be
effectively regarded as partitions of paths that are single-valued in
time. Therefore, whether we take the
fine-grained histories to be all paths or just those single valued in
$T$, whether the multiplier is integrated over all $N$ or just positive
$N$, if $T^{\prime\prime} > T^\prime$, we recover the matrix elements of
the usual formulation of non-relativistic quantum mechanics.  As we
shall see, this is enough to ensure equivalence with that theory.

To complete the construction of the decoherence functional for
parametrized non-relativistic quantum mechanics according to
\eqref{sevenoneeight} and \eqref{sevenonenine} we must specify the product 
$\circ$ and
the space of wave functions representing initial and final conditions.  For
this it is important to consider the role of the constraints. In
the discussion of  gauge
theories in Section VI, we ensured that wave functions defined on the
configuration space of gauge potentials that represented states on a
spacelike surface and
depended only on the true physical degrees of freedom by using an operator
representation of the constraints to enforce the condition $(constraint)
\psi = 0$ [\cf \eqref{sixfivesix}] on the initial and final condition.
Enforcing this condition on the initial
and final conditions was enough to guarantee that it was satisfied on
all spacelike surfaces [\cf \eqref{sixfivethree}, \eqref{sixfivesix}]. In a
generalized quantum mechanics we do not necessarily have a notion of
``state on a spacelike surface'' and therefore of ``states depending
only on true physical degrees of freedom''. However, we achieve a
similar objective by enforcing the constraints as operator conditions on
the wave functions representing the initial and final conditions.  Then
when states on spacelike surfaces can be defined, either generally or in
the context of specific approximations and limits, we expect that these
will satisfy the constraints. Even where
states cannot be defined, we shall see that enforcing the constraints in
this way leads to important and attractive features for the resulting
generalized quantum mechanics.

The operator form of the constraint \eqref{seventhreefive} is
\begin{equation}
\left[ - i\frac{\partial}{\partial T} + h\left(-i\frac{\partial}{\partial
X}, X\right)\right] \psi (X,T) = 0 
\label{seventhreefourteen}
\end{equation}
which will be recognized as the Schr\"odinger equation.  If the initial
and final wave functions are required to satisfy \eqref{seventhreefourteen}, 
they cannot
be members of the Hilbert space ${\cal H}^Q = {\cal H}^{(X,T)}$ of
square integrable
wave functions on $(X,T)$-configuration space nor can we use the
inner product of that space
 as the product $\circ$ in \eqref{sevenoneeight}.  There are no
solutions of \eqref{seventhreefourteen} that lie in ${\cal H}^Q$ because, for
them
\begin{equation}
\int\nolimits^{+\infty}_{-\infty} dT \int\nolimits^{+\infty}_{-\infty} dX
|\psi (X,T) |^2 = \int\nolimits^{+\infty}_{-\infty} dT \cdot {\rm const.}
= \infty \label{seventhreefifteen}
\end{equation}
by the usual conservation of probability.
However, we can construct the decoherence functional using the familiar
Hilbert space ${\cal H}^X$ of square integrable functions of $X$ as
follows:  Choose two surfaces of constant
time $T^\prime$ and $T^{\prime\prime}$ respectively, with $T^{\prime\prime} >
T^\prime$, such that any coarse graining of interest does not restrict
the paths on these surfaces.  The $\circ$ product may be defined on such
constant time surfaces by
\begin{equation}
\phi (X, T) \circ \psi (X,T) = \int\nolimits_T dX\,\phi^* (X, T) \psi
(X, T)\, .
\label{seventhreefifteena}
\end{equation}
Thus, \eqref{sevenoneeight} is implemented as
\begin{equation}
\left\langle \phi_i\left| C_\alpha \right| \psi_j\right\rangle =
\int\nolimits_{T''} dX^{\prime\prime}
\int\nolimits_{T'} dX^\prime \phi^*_i (X^{\prime\prime}, T'')
 \left\langle X^{\prime\prime} T^{\prime\prime} \left\Vert
C_\alpha \right\Vert X^\prime T^\prime \right\rangle \psi_j (X^\prime,
T')\, .
\label{seventhreesixteen}
\end{equation}
These matrix elements are independent of $T^\prime$ and
$T^{\prime\prime}$ provided these surfaces lie outside the domain of $(X,
T)$ that is restricted by the coarse graining.  This follows because the
class operator matrix elements satisfy the Schr\"odinger equation [\cf
\eqref{seventhreefourteen}, \eqref{fiveoneten}], the initial and final wave
functions do likewise by assumption, and the $\circ$ product is preserved
by Schr\"odinger evolution.

With this choice, whether the class operators are given by
\eqref{seventhreetwelve} or \eqref{seventhreethirteen}, 
the decoherence functional for
parametrized non-relativistic quantum mechanics \eqref{sevenonenine}
 reduces to that of
non-relativistic quantum mechanics approached straightforwardly\break
\noindent [\cf
\eqref{fiveoneten}]. Coarse grainings may be regarded as coarse grainings of
paths moving forward in time because only those have non-vanishing
contributions to the class operators.  As described in Section IV.4,
 an equivalent Hamiltonian quantum
mechanics of states evolving unitarily and by reduction of the wave
packet may be derived for those coarse grainings which restrict the
paths only on successions of constant time surfaces.
 The trivial elevation of time to the status of a dynamical
variable has thus produced no change in non-relativistic
 quantum prediction.  This may seem to
be a round about way of approaching non-relativistic quantum mechanics
and indeed it is.  It is this model, however, that we shall follow in
constructing a generalized quantum mechanics of less trivial
reparametrization invariant theories.

\subsection{The Relativistic World Line --- Formulation with a
Preferred Time}

The most familiar example of a reparametrization invariant model is the
free relativistic particle whose classical dynamics are described by
either the action \eqref{sevenonetwo} or \eqref{sevenonethree}.  An elementary
calculation starting from either of these shows that the momenta $p_\alpha$
conjugate to the $x^\alpha$ satisfy the mass shell constraint.
\begin{equation}
p^2+ m^2 = 0\, . 
\label{sevenfourone}
\end{equation}
In the next two sections
 we shall construct two generalized quantum mechanical theories for
this model. These are distinguished primarily by different choices for
the set of fine-grained histories.

Identifying the fine-grained histories with
arbitrary curves in the four-dimensional configuration space of the
$\{x^\alpha\}$ is the most natural choice from the point of view of
Lorentz invariance.  However, from the
point of view of Hamiltonian quantum mechanics another choice is
possible.  This is to break Lorentz invariance, single out a preferred
Lorentz frame, and choose the {\it fine-grained histories} to be curves that
are {\it single-valued} in the time coordinate of that Lorentz case.
We shall consider this case first as it leads to the  usual Hamiltonian
formulation \cite{HK86}.

If the paths move forward in $t$ their {\it allowed coarse grainings} are
identical with those of non-relativistic quantum mechanics described in
Section V.  In
particular it is possible to coarse grain by regions of the spatial
coordinates, ${\bf x}$, on a
sequence of constant-$t$ surfaces.

To implement the general prescription for the class operators
\eqref{seventwonine}, first note that the Hamiltonian following from the
action \eqref{sevenonethree} is $H=(p^2+m^2)/(2m)$. The
canonical action \eqref{seventwoeight} is therefore
\begin{equation}
S\left[{p_\alpha}, {x^\alpha}, N \right]
= \int\nolimits^1_0
d\lambda\,\left[p\cdot{\dot x} - N (p^2 +
m^2)/(2m)\right]\, . 
\label{sevenfourtwo}
\end{equation}
Then note that, for paths that move forward in $t$, a convenient way to
fix the parametrization of the curves is to take $\lambda$ to be equal
to $t$ up to a scale, specifically to choose
\begin{equation}
\Phi = t-[t''\lambda + t'(1-\lambda)]\, . 
\label{sevenfourthree}
\end{equation}
The Faddeev-Popov determinant for this gauge condition is
\begin{equation}
\Delta_\Phi = |\{\Phi,H\}|=|p^0/m| 
\label{sevenfourthreea}
\end{equation}
where $\{,\}$ is the Poisson bracket\footnote{If the construction of
the determinant from the gauge fixing condition is not familiar
see Faddeev  \cite{Fad69} or \cite{HK86} in the specific case of the
relativistic particle.}.
With this choice of parametrization fixing condition, a unique path
\begin{equation}
t(\lambda) = t^{\prime\prime} \lambda + t^\prime (1-\lambda)
\label{sevenfourfour}
\end{equation}
contributes to the path-integral over $t(\lambda)$. The expression for
the class operators becomes
\begin{equation}
\left\langle x^{\prime\prime} \left\Vert C_\alpha\right\Vert
x^\prime\right\rangle = \int\nolimits_{\alpha} \delta p \delta
{\bf x}\delta N
\left(\prod\nolimits\left|\frac{p^0}{m}\right|\right)
\exp \left(i \int\nolimits^{t^{\prime\prime}}_{t^\prime}
dt\,\left[p\cdot (dx/dt) - N(p^2 +
m^2)/(2m)\right]\right)\, . 
\label{sevenfourfive}
\end{equation}
where, in a time-slicing implementation of the path integral analogous
to \eqref{fivetwothree},
 the product is of factors on each time-slice but the last.
Integrating the multiplier $N$ over the positive real axis corresponds
to the usual quantum theory of a positive frequency relativistic
particle.  To see this carry out the integration over $N(\lambda)$ on
each time slice to yield
\begin{equation}
\left\langle{\bf x}^{\prime\prime}, t^{\prime\prime} \left\Vert
C_\alpha\right\Vert {\bf x}^\prime, t^\prime\right\rangle =
\int\nolimits_{\alpha}\delta p \delta {\bf x}
\left[\prod\nolimits\left( \frac{-2ip^0}{p^2+m^2-i\epsilon}
\right)\right]{\exp
\left(i\int\nolimits^{t^{\prime\prime}}_{t^\prime} dt\, p \cdot
(dx/dt)\right)}\, .
\label{sevenfoursix}
\end{equation}
The integration over $p^0$ can be completed into a closed contour in the
upper half-plane and evaluated by the method of residues giving
\begin{equation}
\left\langle {\bf x}^{\prime\prime}, t^{\prime\prime} \left\Vert C_\alpha
\right\Vert {\bf x}^\prime, t^\prime\right\rangle =
\int\nolimits_{\alpha} \delta {\bf p} \delta {\bf x} \exp \left(i
\int\nolimits^{t^{\prime\prime}}_{t^\prime} dt\, \left[{\bf p} \cdot
d{\bf x}/dt - \sqrt{{\bf p}^2 + m^2}\right]\right)\ .
\label{sevenfourseven}
\end{equation}
This is just the phase space path-integral for a ``non-relativistic''
system with Hamiltonian
\begin{equation}
h\left({\bf p}, {\bf x}\right) = \sqrt{{\bf p}^2 + m^2} \ .
\label{sevenfoureight}
\end{equation}
The class operators thus reduce to the ones for the usual
single-particle theory of a free relativistic particle. For example,
the matrix elements $C_u$ defined by the sum over {\it all} paths
is the usual propagator between Newton-Wigner localized states
\cite{NW49}.\footnote{The path integral can, in fact, be done by
carrying out the integrals
over the ${\bf x}$'s to yield $\delta$-functions enforcing the
conservation of momentum and then using these to carry out all the
integrations over the momenta except the last.}
The choice of the ${\cal H}^{\bf x}$, the space of square integrable
wave functions on ${\bf x}$, for the space of initial and final wave
functions and its inner product $\circ$
 on surfaces of constant time for the
product $\circ$ in \eqref{sevenoneeight} gives
\begin{equation}
\left\langle\phi_i \left|C_\alpha \right| \psi_j \right\rangle =
\int\nolimits_{t''}
d^3 x^{\prime\prime} \int\nolimits_{t'}
d^3 x^\prime \phi^*_i ({\bf x}^{\prime\prime}, t'')
\left\langle{\bf x}^{\prime\prime}, t^{\prime\prime} \left\Vert C_\alpha
\right\Vert {\bf x}^\prime, t^\prime \right\rangle \psi_j ({\bf x}', t') \ .
\label{sevenfournine}
\end{equation}

This completes the correspondence with the usual Hamiltonian quantum
theory of a positive frequency free relativistic particle.  The
$\phi_i({\bf x}, t)$
 and $\psi_j({\bf x}, t)$ are Newton-Wigner wave functions.  If the
coarse grainings are restricted to alternatives on the surfaces of
constant preferred time, then the construction sketched in Section IV.4
can be used to define states on these surfaces.
These are
represented by Newton-Wigner wave functions in ${\cal H}^{\bf x}$ that
evolve
either unitarily with the Hamiltonian \eqref{sevenfoureight} or by reduction
of the wave packet. Hamiltonian quantum mechanics is thus recovered for
these coarse grainings.

The important lesson of this model is that by introducing a preferred
time in which the histories are single-valued we recover the usual
Hamiltonian form of quantum theory with its two laws of evolution.  We
shall now see that, when such a preferred time is not introduced, there is
no Hamiltonian formulation of the quantum mechanics of a relativistic
particle but there is a predictive generalized quantum mechanics.

\subsection{The Relativistic World Line ---
Formulation Without a Preferred Time}

\subsubsection{Fine-Grained Histories, Coarse Grainings, and
Decoherence Functional}

In this section we formulate a generalized quantum mechanics for a single
relativistic world line using a set of Lorentz invariant fine-grained
histories that to not single out a preferred time.
The most obvious Lorentz invariant set of fine-grained histories for a
single relativistic particle is the set of all curves in
spacetime. Such curves generally move both forward and backward in the
time of any Lorentz frame, perhaps intersecting a surface of constant
time many times. We shall now construct a sum-over-histories generalized
quantum mechanics of a single relativistic particle world line based on this
set of fine-grained histories.

It should be stressed that we do not mean the resulting theory to be
a realistic theory of relativistic particles such as protons and
electrons. That is supplied by quantum field theory. The theory that
we shall construct is of a different kind. It is a quantum theory of a
single world line. As we shall describe, when the single world line
interacts with an external potential, certain $S$-matrix elements of this
model coincide with the $S$-matrix elements of field theory. In general,
however, the theories are different because they deal with different
alternatives. We consider this generalized quantum mechanics of a single
world line, not as a theory of realistic elementary particles, but  rather as a
model for quantum cosmology which necessarily
is the quantum mechanics of a single universe.

The allowed coarse grainings of this generalized quantum mechanics are
partitions of the fine-grained
histories into Lorentz invariant and reparametrization invariant
classes, most generally by the values of Lorentz and reparametrization
invariant functionals. We illustrate with a few examples:

Partitions by the values of position at moments of the time of some
particular Lorentz frame are not possible because paths may cross a
constant time surface, not just at one place, but at an arbitrary number
of positions.  However, one can still partition the paths, say, by the
location of the particle's first passage of a given spacelike surface after the
the initial condition.  Partitions by whether paths cross or do not
cross a set of spacetime regions are possible.  In addition, the
existence of a reparametrization invariant proper time along a curve
$x(\lambda)$ between invariantly defined points $\lambda'$ and
$\lambda''$
\begin{equation}
\tau\bigl(\lambda^{\prime\prime}, \lambda^\prime, N(\lambda)\bigr] =
\int\nolimits^{\lambda^{\prime\prime}}_{\lambda^\prime} N(\lambda) d\lambda
\label{sevenfiveone}
\end{equation}
allows further kinds of coarse grainings.  For example, we could
partition the paths by the total proper time that elapses between the
initial and final condition or by the point in spacetime the particle
has reached a certain proper time  after the initial condition.
We shall illustrate the
calculation of the class operators for some of these coarse grainings
below.

The general form of the matrix elements defining the class operators
corresponding to an individual coarse-grained history is
\eqref{seventwonine} with the action \eqref{sevenfourtwo}.
Again the condition
\begin{equation}
\Phi = \dot N = 0 \label{sevenfivetwo}
\end{equation}
is convenient to fix the parametrization.  The only remaining choice is the
range of the multiplier
integration.  As we shall see the range $0$ to $\infty$ leads to the
closest correspondence with field theory.  The matrix elements of the
class operators are then
\begin{equation}
\left\langle x^{\prime\prime}\left\Vert C_\alpha\right\Vert
x^\prime\right\rangle = \int\nolimits_{\alpha} dN\,\delta x\delta p
\exp \left\{i\int\nolimits^1_0 d\lambda\,\left[p\cdot \dot x - N
(p^2 + m^2)/(2m)\right]\right\}\, . 
\label{sevenfivethree}
\end{equation}
where the integral is over the positive constant value of $N$  and over paths
in the class
$c_\alpha$.  The choice of positive $N$ is perhaps suggested by
the consequent value of
$C_u$ --- the integral over all paths between $x^\prime$ and
$x^{\prime\prime}$.  Rescaling the
parameter $\lambda$ to write $w=\lambda N$, the integral in
\eqref{sevenfivethree} for the matrix elements of $C_u$ can be written
\begin{equation}
\left\langle x^{\prime\prime} \left\Vert C_u\right\Vert x^\prime
\right\rangle =
\int\nolimits^\infty_0 dN\,\left\langle x^{\prime\prime}, N\bigl\Vert
x^\prime, 0\right\rangle 
\label{sevenfivefour}
\end{equation}
where the integrand is defined as
\begin{equation}
\left\langle x^{\prime\prime}, N\big\Vert x^\prime, 0 \right\rangle =
\int \delta x \delta p\, \exp\left\{i\,\int\nolimits^N_0 dw \left[p
\cdot dx/dw - (p^2 +m^2)/(2m)\right]\right\}\, .
\label{sevenfivefive}
\end{equation}
This has the form of the momentum-space path integral for the
propagator of a free non-relativistic particle in four-dimensions over a
time $N$. (Hence the choice of notation on the left hand side of
 \eqref{sevenfivefive}.)  Thus, either by recognizing this connection or
by explicit evaluation of the Gaussian functional integrals:
\begin{equation}
\left\langle x^{\prime\prime}, N\big\Vert x^\prime, 0\right\rangle = \int
\frac{d^4p}{(2\pi)^4}\, \exp\left\{i\left[-\frac{1}{2m}\bigl(p^2 + m^2\bigr)N 
+ p\cdot
\bigl(x^{\prime\prime} - x^\prime\bigr)\right]\right\}\, .
\label{sevenfivesix}
\end{equation}
It is then an elementary calculation to verify that as a consequence of
the positive multiplier range the matrix element \eqref{sevenfivefour} is,
up to a factor, just
the Feynman propagator
\begin{equation}
\left\langle x^{\prime\prime} \left\Vert C_u \right\Vert x^\prime
\right\rangle = -2mi\Delta_F \left(x^{\prime\prime} - x^\prime\right)\ .
\label{sevenfiveseven}
\end{equation}

To construct the decoherence functional, we must identify the space of
wave functions that supply the initial and final conditions and the
product $\circ$ in \eqref{sevenoneeight}.  
As in the case of the non-relativistic
particle discussed in the previous section, initial and final wave
functions that satisfy the constraint will ensure the closest
correspondence with the usual quantum mechanics of special-relativistic
systems.

In the case of the free relativistic particle the constraint,
eq.~\eqref{sevenfourone}, is the Klein-Gordon equation
\begin{equation}
(-\nabla^2 +m^2)\psi(x) = \left(\frac{\partial^2}{\partial t^2} -
\vec\nabla^2 +m^2\right) \psi (x) = 0\, . 
\label{sevenfiveeight}
\end{equation}
The Klein-Gordon equation has a conserved current and thus there
are no solutions in the Hilbert space ${\cal H}^x$ of square integrable
functions on four-dimensional spacetime. The norms of solutions diverge,
as in \eqref{seventhreefifteen}.  Therefore, the inner product of ${\cal H}^x$
cannot be used as the product $\circ$ in \eqref{sevenoneeight}.  However, the
Klein-Gordon product on a spacelike surface $\sigma$ can be used.  This
is
\begin{equation}
\phi(x) \circ \psi(x) = i\int_\sigma d\Sigma^\mu \phi^* (x)
 \buildrel\leftrightarrow \over{\nabla}_\mu \psi (x)
\label{sevenfivenine}
\end{equation}
where $d\Sigma^\mu$ is the surface area element of the surface $\sigma$.
The product is independent of $\sigma$ if $\phi(x)$ and $\psi(x)$
satisfy the constraint, \eqref{sevenfiveeight}.

Therefore, pick two non-intersecting spacelike surfaces $\sigma^\prime$
and $\sigma^{\prime\prime}$ and define
\begin{equation}
\left\langle\phi_i\left|C_\alpha\right|\psi_j\right\rangle = -
\int_{\sigma^{\prime\prime}} d\Sigma^{\prime\prime\mu}
\int_{\sigma^\prime} d\Sigma^{\prime\nu} \phi^*_i (x^{\prime\prime})
\buildrel\leftrightarrow \over{\nabla}^{\prime\prime}_\mu \left\langle
x^{\prime\prime}\left\Vert C_\alpha\right\Vert x^\prime\right\rangle
 \buildrel\leftrightarrow\over{\nabla}^\prime_\nu \psi_j
(x^\prime)\, . 
\label{sevenfiveten}
\end{equation}
The construction in eqs \eqref{sevenonenine} and \eqref{sevenoneten} yields a
decoherence function that satisfies all of the general requirements
\eqref{fouroneone i} -- \eqref{fouroneone iv} of Section IV.  
It, therefore, completes the specification of
a generalized quantum mechanics for the single, free, relativistic particle
world line which
does not single out a preferred time.

The construction \eqref{sevenfiveten} appears to depend on the choice of
surfaces $\sigma^\prime$ and $\sigma^{\prime\prime}$ but in fact is
largely independent of these choices for partitions that distinguish
paths only in some compact region of spacetime $R$.  Choose
$\sigma^{\prime\prime}$ to be to the future of $R$, and $\sigma^\prime$
to be a surface to it past that does not intersect $\sigma''$.
For points $x^{\prime\prime}$ located on
$\sigma^{\prime\prime}$ we can show
\begin{equation}
\left(-\nabla^2_{x^{\prime\prime}} + m^2\right) \left\langle
x^{\prime\prime}\left\Vert C_\alpha\right\Vert x^\prime \right\rangle =
0 \ .\label{sevenfiveeleven}
\end{equation}
The same relation holds for points $x^\prime$ on $\sigma^\prime$.  This
is immediate in the case when $c_\alpha$ is the class of all paths, $u$,
because then $\langle x^{\prime\prime} \Vert C_u\Vert x^\prime\rangle$
is the Feynman propagator [\cf \eqref{sevenfiveseven}]. We shall demonstrate
\eqref{sevenfiveeleven} more generally below, but first note a consequence.
Outside of $R$, \eqref{sevenfiveten} is of the form of two Klein-Gordon
products between two solutions of the Klein-Gordon equation.  The matrix
elements \eqref{sevenfiveten} are therefore independent of the choice of
the spacelike surfaces $\sigma^{\prime\prime}$ and $\sigma^\prime$ as
long as they do not intersect the region of coarse graining, $R$, or each
other.

Not only does the
Feynman propagator solve the Klein-Gordon
equation for $x^\prime\not= x^{\prime\prime}$, it is also composed just
of {\it positive frequency} solutions for $t^{\prime\prime} > t^\prime$.
As we shall show below, this also turns out to be a general property
of the matrix elements
$\langle x^{\prime\prime}\Vert C_\alpha\Vert x^\prime \rangle$ --- a
consequence of the positive multiplier range in \eqref{sevenfivethree}.
Positive frequency solutions {\it do} form a Hilbert space ${\cal H}^{(+)}$
with the inner product \eqref{sevenfivenine}.  The Klein-Gordon inner
product between positive and negative frequency solutions of
the constraints vanishes. Without losing generality we
may therefore write for the decoherence functional \eqref{sevenonenine}
\begin{equation}
D\left(\alpha^\prime, \alpha\right) = {\cal N}\sum\nolimits_{ij}
p^{\prime\prime}_i \left\langle \phi_i \left| C_{\alpha^\prime}\right|
\psi_i \right\rangle\,\left\langle \phi_i \left| C_\alpha \right | \psi_j
\right\rangle^* p^\prime_j 
\label{sevenfivetwelve}
\end{equation}
where the sums are over positive frequency solutions in ${\cal H}^{(+)}$.

The normalization factor in \eqref{sevenfivetwelve} is  given by 
\eqref{sevenoneten}. In
the present case of a {\it free}-relativistic particle it may be evaluated
explicitly using \eqref{sevenfiveseven}.  One finds
\begin{equation}
\bigl\langle\phi_i|C_u|\psi_j\bigr\rangle= 2m \left(\phi_i \circ
\psi_j\right)\ .
\label{sevenfivethirteen}
\end{equation}
The normalization factor is then
\begin{equation}
{\cal N}^{-1} = 4m^2 Tr\left(\rho_f\rho_i\right) 
\label{sevenfivefourteen}
\end{equation}
where the trace in ${\cal H}^{(+)}$ is over the density matrices
constructed from the initial and final wave functions and probabilities
in the usual Klein-Gordon sense.

In the case of a pure initial condition, represented by a single wave
function $\psi(x)$ and a final condition of indifference with respect to
final state, \eqref{sevenfivetwelve}
reduces to
\begin{equation}
D\left(\alpha^\prime, \alpha\right) = \bigl(4m^2\bigr)^{-1}
\sum\nolimits_i \left\langle
\phi_i \left |
C_{\alpha^\prime} \right| \psi \right\rangle\,\left\langle \phi_i \left|
C_\alpha \right| \psi \right\rangle^* 
\label{sevenfivefifteen}
\end{equation}
where the sum is over a complete set of states in ${\cal H}^{(+)}$.

We now return to sketch the demonstration that the $\langle
x^{\prime\prime}, N\Vert C_\alpha \Vert x^\prime, 0 \rangle$ are
positive frequency solutions of the Klein-Gordon equation,
\eqref{sevenfiveeleven}, when $t^{\prime\prime}>t^\prime$.  The key point is
that, in the $\dot N=0$ gauge, a partition of the paths restricted to a
spacetime region $R$ cannot restrict the constant value of $N$ because,
in that gauge, the constant value
is the overall proper time
between initial and final surface [\cf \eqref{sevenfiveone}]. For any $R$
this will depend on the paths outside $R$.
Thus, from \eqref{sevenfivethree} we can
write
\begin{equation}
\left\langle x^{\prime\prime} \left\Vert C_\alpha \right\Vert x^\prime
\right\rangle = \int\nolimits^\infty_0 dN\,\left\langle x^{\prime\prime}, N
\left\Vert C_\alpha \right\Vert x^\prime, 0 \right\rangle 
\label{sevenfivesixteen}
\end{equation}
where the integrand is the sum over all paths in the class
$c_\alpha$ that travel from $x^\prime$ to $x^{\prime\prime}$ in proper
time $N$.  Using the parameter $w=N\lambda$, this can be
written
\begin{equation}
\left\langle x^{\prime\prime}, N \left\Vert C_\alpha \right\Vert x, 0
\right\rangle = \int\nolimits_{\alpha} \delta x \delta p
\, \exp
\left\{i\int\nolimits^N_0 dw\, \left[ p \cdot (dx/dw) - (p^2 + m^2
)/(2m)\right]\right\}\, . 
\label{sevenfiveseventeen}
\end{equation}
This is of the form of an integral defining a non-relativistic
propagator over a time interval $N$.
As long as $x^{\prime\prime}$ is outside the region $R$
constrained by the partition it satisfies the ``Schr\"odinger
equation'':
\begin{equation}
\left[-i\frac{\partial}{\partial N} + \frac{1}{2m}\left(-\nabla_{x''}^2 +
m^2\right)
\right] \ \left\langle x^{\prime\prime}, N \left \Vert C_\alpha \right
\Vert x, 0 \right\rangle = 0 
\label{sevenfiveeighteen}
\end{equation}
with the boundary condition
\begin{equation}
\left\langle x^{\prime\prime}, 0 \left\Vert C_\alpha \right\Vert
x^\prime, 0 \right\rangle = \delta^{(4)} \left(x^{\prime\prime} -
x^\prime\right)\, . 
\label{sevenfivenineteen}
\end{equation}
Now operate with $(-\nabla_{x''}^2 + m^2)$ on both sides of
\eqref{sevenfivesixteen}.  Use \eqref{sevenfiveeighteen} 
to convert the integrand on
the left hand side to a total derivative in $N$.  Use
\eqref{sevenfivenineteen}
and wave packet spreading to evaluate the limits and conclude that
$\langle x'' \Vert C_\alpha \Vert x' \rangle$ satisfies the Klein-Gordon
equation,
\eqref{sevenfiveeleven}, when $x''$ is distinct from $x'$.

To show that the $\langle x'' \Vert C_\alpha \Vert x' \rangle$ are {\it
positive frequency} solutions of the Klein-Gordon equation we argue as
follows: The solutions to \eqref{sevenfiveeighteen} outside of $R$ can be
written
\begin{equation}
\left\langle x^{\prime\prime}, N \left\Vert C_\alpha \right\Vert x', 0
\right\rangle = \int\frac{d^4 p}{(2\pi)^4}\, e^{-iN\left(p^2+m^2\right)/2m}
e^{ip\cdot x^{\prime\prime}}
\Phi_\alpha \left(p, x^\prime\right) 
\label{sevenfivetwenty}
\end{equation}
for some $\Phi_\alpha(p)$.  Carry out the integral over $N$ over
the range $N=0$ to $N=+\infty$ in
\eqref{sevenfivesixteen} to yield the following representation
\begin{equation}
\left\langle x^{\prime\prime} \left\Vert C_\alpha \right\Vert
x^\prime\right\rangle = -2mi \int \frac{d^4 p}{(2\pi)^4} \ \frac{e^{ip\cdot
x''}}{p^2+m^2-i\epsilon} \Phi_\alpha \left(p, x^\prime\right)\, .
\label{sevenfivetwentyone}
\end{equation}
Since $t^{\prime\prime}$ can be made arbitrarily large, we expect that
the contour of $p^0$
integration can be completed in the upper-half plane.
The poles from the denominator contribute only positive frequency
solutions of the Klein-Gordon equation.  Singularities of $\Phi_\alpha$
can contribute nothing more since we already know that the left hand side of
\eqref{sevenfivetwentyone} satisfies that equation. The multiplier range
$N=0$ to $N=\infty$
therefore corresponds to positive frequency solutions of the
Klein-Gordon equation when $x^{\prime\prime}$ is to the future of $R$.

\subsubsection{Explicit Examples}

We next consider two examples of coarse grainings of the paths
of a relativistic particle between spacetime points $x^\prime$ and
$x^{\prime\prime}$ for which the matrix elements of the class operators
$\langle x^{\prime\prime} \left\Vert C_\alpha \right\Vert
x^{\prime}\rangle$ can be explicitly calculated.

The simplest example of a coarse graining of the paths between
$x^\prime$ and $x^{\prime\prime}$ is a partition by the
alternative values of $x$ they have reached at a given proper time $\tau$
after $x^\prime$, if they have not already reached $x^{\prime\prime}$.
This would not be an interesting partition for a theory of elementary
particles for we surely have no direct and independent access to the
proper time along an elementary particle's path.       However, the
analogous question of the proper time elapsed in a universe is
meaningful.  In addition these class operators  have the virtue of being
immediately and transparently calculable.

More specifically, this  partition of paths is defined as follows: Divide
four-dimensional
spacetime into an exhaustive set of exclusive regions,
$\{\Delta_\alpha\},\; \alpha = 1, 2, 3, \cdots$.  The partition consists of
the class $c_0$ of paths that pass between $x^\prime$ and
$x^{\prime\prime}$ in a proper time less than $\tau$, and the classes
$c_\alpha, \; \alpha=1, 2, \cdots$ of paths that are in region
$\Delta_\alpha$ a proper time $\tau$ after $x^\prime$.  Employing the
$\dot N = 0$ gauge, where $\tau = N\lambda$, and the notation of the
previous section, the integral over all paths in $c_0$ is
\begin{equation}
\left\langle x''\Vert C_0 \Vert x' \right\rangle =
\int\nolimits^\tau_0 dN \left\langle x'', N \Vert x', 0
\right\rangle\, . 
\label{sevenfivetwentytwo}
\end{equation}
The integrals over the paths in $c_\alpha$ are a sum over $x\in
\Delta_\alpha$ of a product of two factors.  The first is the integral
over all paths from $x'$ to $x$ in proper time $\tau$. The second is the
integral over paths from $x$ to $x''$ in any proper time greater than
$\tau$.
This product is
\begin{equation}
\int\nolimits^\infty_\tau dN \left\langle x^{\prime\prime}, N\left\Vert
x, \tau \right\rangle\,\left\langle x, \tau \right\Vert x^\prime,
0\right\rangle\, . 
\label{sevenfivetwentythree}
\end{equation}
Here, as follows from \eqref{sevenfivesix},
\begin{equation}
\left\langle x^{\prime\prime}, \tau^{\prime\prime} \big\Vert x^\prime,
\tau^\prime \right\rangle = -i\left[\frac{m}{2\pi i (\tau'' - \tau')}\right]^2
\,\exp\left\{i\left[\half m (\tau'' -\tau') +
\frac{m\left(x^{\prime\prime}-x^\prime\right)^2}
{2\left(\tau^{\prime\prime}-\tau^\prime\right)} \right]\right\}\, .
\label{sevenfivetwentyfour}
\end{equation}
Because of the $\tau$-translation invariance of \eqref{sevenfivetwentyfour},
the integral in
\eqref{sevenfivetwentythree} over all
paths that go from $x$ to $x^{\prime\prime}$ in a proper time greater
than $\tau$ is the same as the integral over all paths between $x$ and
$x^{\prime\prime}$.  Thus, using \eqref{sevenfiveseven},
\begin{equation}
\left\langle x^{\prime\prime} \left\Vert C_\alpha\right\Vert
x^\prime\right\rangle = -2mi\int\nolimits_{\Delta_\alpha} d^4 x\,\Delta_F
\left(x^{\prime\prime}-x\right)\, \left\langle x, \tau \big\Vert
x^\prime, 0\right\rangle
\label{sevenfivetwentyfive}
\end{equation}
where the second factor is given by \eqref{sevenfivetwentyfour}.
We note that, because we are dealing with a coarse graining that involves
the proper time from the initial slice $\sigma'$, it is not restricted
to  partitioning the paths in a compact
spacetime region $R$ and the resulting
class operators do not satisfy the constraint, \eqref{sevenfiveeleven}.

Another coarse graining that is easily calculable, although not easily
useful, is to partition the paths between $x^\prime$ and
$x^{\prime\prime}$ by the position,
$x$, of their first passage
through a given spacelike surface $\sigma$ after $x^\prime$.  Divide the
spacelike surface up into spatial regions $\{\Delta_\alpha\}$.  The path
integral over all paths in the class $c_\alpha$ whose first crossing of
$\sigma$ is in $\Delta_\alpha$ is the integral over $x\in
\Delta_\alpha$ of the product of two factors.\footnote{Analogously
to the calculation of Halliwell and Ortiz \cite{HO93}.} The first is the
integral over all paths from $x^\prime$ to $x$ that never cross
$\sigma$.  Denote this by $\Delta_{1\sigma} \left(x, x^\prime \right)$.
The second is the sum over all paths between $x$ and $x^{\prime\prime}$
that may cross $\sigma$ an arbitrary number of times.  This is the same
as the sum over all paths between $x$ and $x^{\prime\prime}$, that is,
it is a factor times  the Feynman propagator $\Delta_F(x^{\prime\prime}-x)$
[\cf \eqref{sevenfiveseven}].
Thus,
\begin{equation}
\left\langle x^{\prime\prime} \left\Vert C_\alpha \right\Vert x^\prime
\right\rangle = -2mi\int_{\Delta_\alpha} d \Sigma\,\Delta_F
\left(x^{\prime\prime}-x\right) \Delta_{1\sigma} \left(x,
x^\prime\right)\, . 
\label{sevenfivetwentyfivea}
\end{equation}
where $d\Sigma$ is the volume element in $\sigma$.

The propagator $\Delta_{1\sigma}$ is easily evaluated in the case that
$\sigma$ is the surface of constant time $t$ in a particular Lorentz
frame.  For then, if both sides of \eqref{sevenfivetwentyfivea} 
are summed over all
$\alpha$, we must recover on the left the sum over all paths between
$x^\prime$ and $x^{\prime\prime}$.  That is,
\begin{equation}
\Delta_F \left(t^{\prime\prime} - t^\prime, {\bf x}^{\prime\prime} - {\bf
x}^\prime\right) = \int d^3 x\,\Delta_F \left(t^{\prime\prime} - t, {\bf
x}^{\prime\prime} - {\bf x} \right)\, \Delta_{1t} \left(t, {\bf x};
t^\prime, {\bf x}^\prime\right)\, . 
\label{sevenfivetwentysix}
\end{equation}
This integral equation is easily inverted to find $\Delta_{1t}$.
It is
\begin{equation}
\Delta_{1t} \left(t, {\bf x}; t^\prime, {\bf x}^\prime \right) =
\int \frac{d^3p}{(2\pi)^3}\, e^{-i\omega_p\left(t-t^\prime\right)}
e^{i{\bf p}\cdot\left({\bf x} - {\bf x}^\prime \right)} 
\label{sevenfivetwentyseven}
\end{equation}
where $\omega_p = \sqrt{{\bf p}^2 + m^2}$. That is, $\Delta_{1t}$ is just
the propagator between Newton-Wigner localized states \cite{NW49}.

We shall consider a further example of an explicitly calculable coarse
graining defining four-momentum in connection with a discussion of the
constraints in Section VII.5.6.

\subsubsection{Connection with Field Theory}

For coarse grainings that define $S$-matrix elements, the quantum
mechanics of the relativistic world line that we have been discussing
yields $S$-matrix elements that coincide with those of usual field
theory.  To make that correspondence a non-trivial statement, let us
consider the relativistic particle interacting with a fixed external
electromagnetic field.  The action \eqref{sevenonethree} becomes
\begin{equation}
S_A\left[x^\alpha, N\right] = \int\nolimits^1_0 d\lambda \left\{\half m
\left[\frac{\left(\dot x\right)^2}{N} - N\right] + q \dot
x \cdot  A (x) \right\} 
\label{sevenfivetwentyeight}
\end{equation}
where $A_\alpha(x)$ is the potential of the external field and $q$ is the
particle's charge.  It is then a well established fact
\cite{Fey50} that the path-integral
\begin{equation}
\left\langle  x^{\prime\prime}\left\Vert C_u \right\Vert x^\prime
\right\rangle = \int \delta x\, \delta N\, \Delta_\Phi [Q, N]\,\delta
[\Phi(Q, N)]\,\exp \left(iS_A\left[x^\alpha, N\right]\right)
\label{sevenfivetwentynine}
\end{equation}
taken over all paths between $x^\prime$ and $x^{\prime\prime}$, with the
measure
induced from the Liouville measure in phase space, is just the two point
function for a scalar field in the external potential {\it provided} the
multiplier $N$ is integrated over a positive range.  That is
\begin{equation}
\left\langle x^{\prime\prime} \left\Vert C_u \right\Vert x^\prime
\right\rangle = 2m\bigl\langle 0_+ | {\bf T}
(\phi\left(x^{\prime\prime}\right) \phi
\left(x^\prime\right)) | 0_- \bigr\rangle /\bigl\langle 0_+ | 0_-
\bigr\rangle\, .
\label{sevenfivethirty}
\end{equation}
where $|0_+\rangle$ and $|0_-\rangle$ are the initial and
final vacuum states
and ${\bf T}$ denotes time ordering.  To derive this result
\cite{Fey50}, one expands both sides of \eqref{sevenfivethirty}
 in powers of the charge
$q$, and checks
the identity order by order in perturbation theory using
\eqref{sevenfiveseven} relating the free $\langle x^{\prime\prime}
\Vert C_u \Vert x^\prime \rangle$ and the Feynman propagator.
The positive range of
the lapse is necessary to ensure this equivalence.
Examining \eqref{sevenfiveten}  we see that if the surfaces
$\sigma^\prime$ and $\sigma^{\prime\prime}$
are taken to infinity, and the initial and final wave functions
$\psi_j (x)$ and $\phi_i(x)$ are
positive frequency solutions of the Klein-Gordon equation, then 
\eqref{sevenfiveten}
is just the usual
formula for a
one-particle to one-particle $S$-matrix element.  That is
\begin{equation}
(-2m)^{-1}
\left\langle \phi_i \left| C_u \right| \psi_j \right\rangle = S_{ij}
\label{sevenfivethirtyone}
\end{equation}
where $i$ and $j$ are one-particle states.  This connection with familiar
field theory is a strong motivation for choosing a positive multiplier
range to define the generalized quantum mechanics of a relativistic
particle.

While the sum over {\it all} paths generates a known matrix element in field
theory, it is not evident that there is any correspondence for other
partitions of the paths of a relativistic particle.
We have in \eqref{sevenfivethirtyone}, the connection
\begin{equation}
(2m)^{-1}
\left\langle x^{\prime\prime} \left\Vert C_u \right\Vert x^\prime
\right\rangle = \frac{\int\nolimits_u
\delta\phi\,\phi(x^{\prime\prime}) \phi (x^\prime)
\, e^{iS_A[\phi]}}{\int\nolimits_u \delta\phi\,e^{iS_A[\phi]}}
\label{sevenfivethirtytwo}
\end{equation}
where $S_A [\phi]$ is the action for a scalar field interacting with the
external electromagnetic field and the integral is over all fields with
suitable asymptotic boundary conditions.  But it is unlikely that
there is any restriction on the field integration that would reproduce
$\langle x^{\prime\prime} \Vert C_\alpha \Vert x^\prime \rangle$ for a
general coarse graining. Similarly, there are no evident partitions of
the paths of particles that will reproduce partitions by field values in
field theory.  Field theory and the present quantum mechanics
of a relativistic particle coincide for one important class of coarse
grainings but are probably distinct quantum theories because they are
concerned with different alternatives.

Of course, field theory specifies more general $S$-matrix elements than
the single-particle  ones of \eqref{sevenfivethirtyone}.  There
are pair creation amplitudes for example.  These too can be expressed as
integrals over paths as Feynman originally showed.  The amplitudes for
pair creation involve paths that connect two points on the final surface
$\sigma^{\prime\prime}$. However, for the analogy with
cosmology we want a quantum theory of a single world line --- the analog
of the history of a closed universe.
Generalized quantum mechanics allows us to
construct such a theory with the decoherence functional
\eqref{sevenonenine}.
However, the existence of pair creation in the corresponding field
theory means that the normalizing factor ${\cal N}$ for that
one-particle theory in \eqref{sevenonenine}
will be non trivial. In $S$-matrix terms, if there is a single one
particle state $\psi_i(x)$ that specifies
the initial condition, and a condition of
final indifference then
\begin{equation}
{\cal N}^{-1} = 4m^2\sum\nolimits_j\, S^\dagger_{ij} S_{ji} \ .
\label{sevenfivethirtythree}
\end{equation}
In field theory terms, this is $4m^2$ times the probability that the single
particle state represented by $\psi_i(x)$ persists in being a a single
particle state.  That is not unity because of the possibility of pair
creation.

\subsubsection{No Equivalent Hamiltonian Formulation}

The generalized quantum mechanics of a single relativistic world line
that we have constructed does not have an equivalent Hamiltonian
formulation in terms of states on spacelike surfaces in spacetime that
evolve unitarily or by state vector reduction that is valid for all
coarse grainings we have discussed.  This was already true for general
spacetime
coarse grainings in the case of the non-relativistic systems and the
free-relativistic particle with a preferred time discussed previously.
However, in these cases, the fact that the fine-grained histories are
single-valued in a preferred time permitted the construction of an
equivalent Hamiltonian
 formulation by the methods discussed in Section IV.4 for those
coarse grainings that distinguished positions on surfaces of the
preferred time.   For the theory of the relativistic world line without
a preferred time, there are no such coarse grainings and no
corresponding factorization as in Section IV.4 because the paths may cross
a given surface in spacetime an arbitrary number of
times.\footnote{One might imagine that one could construct a tower
of wave functions a given member of which would correspond to a specific
number of crossings at specified positions.  However, non-differentiable
paths dominate the sum over histories.  The expected number of crossings
and the amplitude for any finite number of crossings is zero.  Each
entry in the tower would therefore vanish. For explicit calculations in
the case of non-relativistic quantum mechanics,
see \cite{Har88a} and Yamada and Takagi \cite{YT91a}.}

\subsubsection{The Probability of the Constraint}

Classically, the four-momentum of a relativistic particle is constrained
to the mass shell, $p^2= - m^2$.  That same constraint is the starting
point for Dirac quantization of this system (see Section
VII.6). In the generalized quantum
mechanics of a relativistic world lines under discussion,
four-momentum may be defined  through
partitions of the paths by proper-time-of-flight through spacetime in
analogy
to the definition of three-momentum in Section V.4.2.
The question of whether the constraint is satisfied is
then the physical question of the probabilities for the various values
of $p$.    In the
following we shall show that the probability is zero for values of $p^2\not=
-m^2$ because the class operators
vanish for values of
$p^2 \not= -m^2$.

More specifically, to define the four-momentum, we shall consider
partitions of the paths between the initial and final points $x^\prime$
and $x^{\prime\prime}$ by the value of the spacetime displacement, $d$,
that the particle travels between a proper time $\tau_1$, after the
initial position $x'$, and a later proper time $\tau_2 = \tau_1 + T$.
Classically the four-momentum is
\begin{equation}
p= md/T\, . \label{sevenfivethirtyfour}
\end{equation}
Quantum mechanically we expect the same formula to define four-momentum
for suitable coarse grainings of spacetime position in the limit of very
large $T$ for the physical reasons described in Section V.4.2.

We therefore begin by partitioning the paths between $x^\prime$ and
$x^{\prime\prime}$ into the class of paths that makes this passage in a
total proper time less than $\tau_2 = \tau_1 + T$ and the class that
takes more proper time.  Clearly only the latter class is of interest in
defining four-momentum as described above.  We partition {\it this}
class by the positions in spacetime, $x_1$ and $x_2$, that the particle
has reached at proper times $\tau_1$ and $\tau_2 = \tau_1 + T$
respectively.  We then coarse grain this partition by the values of the
displacement $d$
between $x_1$, and $x_2$ specified to an accuracy so that $p$ defined
by \eqref{sevenfivethirtyfour} lies in a range $\widetilde\Delta$.
The resulting partition by values of $d$ is then a partition by the
corresponding value of $p$ through \eqref{sevenfivethirtyfour} in the very
large $T$ limit.

Working in
the $\dot N = 0$ gauge, where the constant value of $N$  is the
elapsed proper time, we can write for the class
operator $C_{\widetilde \Delta}$
\[
\bigl\langle x^{\prime\prime}\Vert C_{\widetilde\Delta} \Vert x^\prime
\bigr\rangle = \int^\infty_{T+\tau_1} d\tau \int d^4x_2
\int d^4x_1 e_{\widetilde\Delta}\left[m(x_2-x_1)/T\right]
\]
\begin{equation}
\times\,\left\langle x^{\prime\prime}, \tau \Vert x_2, \tau_1 +
T\right\rangle\,\left\langle x_2, \tau_1 + T \Vert x_1, \tau_1
\right\rangle\,\left\langle x_1, \tau_1 \Vert x^\prime, 0\right\rangle
\label{sevenfivethirtyfive}
\end{equation}
where $e_{\widetilde\Delta}(x)$ is the characteristic function for the
four-vector range $\widetilde\Delta$.  No elaborate calculation is
needed to evaluate \eqref{sevenfivethirtyfive}.  Except for the integration
over proper time it has essentially the same form as the corresponding
integral \eqref{fivefoureleven} in the non relativistic case.
Making use of the translation
invariance of the propagators in proper time and \eqref{sevenfiveseven},
 we can write this as
\[
\bigl\langle x^{\prime\prime}\Vert C_{\widetilde\Delta} \Vert x^\prime
\bigr\rangle = -2mi \int d^4x_2 \int d^4x_1 e_{\widetilde\Delta}
\left[m(x_2-x_1)/T\right] \Delta_F \left(x^{\prime\prime} - x_2 \right)
\]
\begin{equation}
\times\,\left\langle x_2, T\Vert x_1 0\right\rangle\,\left\langle x_1, \tau_1
\Vert x^\prime, 0 \right\rangle\, . 
\label{sevenfivethirtysix}
\end{equation}
Now let us adjoin initial and final wave functions according to
\eqref{sevenfiveten}, assuming that the initial and final surfaces
$\sigma^{\prime\prime}$ and $\sigma^\prime$ are surfaces of constant
time $t$.
The result is
\begin{equation}
\bigl\langle \phi_i | C_{\widetilde\Delta} | \psi_j \bigr\rangle = -2 mi
\ \int d^4x_2 \int d^4x_1 e_{\tilde\Delta}
\left[m(x_2 - x_1)/T\right] \Phi^*_i (x_2)
\left\langle x_2 T\Vert x_1 0 \right\rangle \Psi_j (x_1)
\label{sevenfivethirtyseven}
\end{equation}
where we have defined
\begin{subequations}
\label{sevenfivethirtyeight}
\begin{equation}
\Phi^*_i (x_2) = i\int_{t^{\prime\prime}} d^3 x^{\prime\prime}
\phi^*_i (x^{\prime\prime})\frac{\buildrel\leftrightarrow\over\partial}
{\partial t}\ \Delta_F \left(x^{\prime\prime}-x_2\right)
\label{sevenfivethirtyeight a}
\end{equation}
and
\begin{equation}
\Psi_j (x_1) = i \int_{t^\prime} d^3 x^\prime \left\langle x_1,
\tau_1 \Vert x^\prime, 0 \right\rangle
\ \frac{\buildrel\leftrightarrow\over\partial}{\partial t}\ \psi_j
(x^\prime)\, . 
\label{sevenfivethirtyeight b}
\end{equation}
\end{subequations}
In this expression, introduce the corresponding momentum space wave
functions
\begin{equation}
\Psi_j (x) = \int \frac{d^4 q}{(2\pi)^4}\ e^{ik\cdot x}
\widetilde\Psi_j (k) 
\label{sevenfivethirtynine}
\end{equation}
with a similar definition for $\widetilde\Phi_i (x)$.
Write
\begin{equation}
e_{\widetilde\Delta} (k) = \int_{\widetilde\Delta} d^4 p
\delta^{(4)} (k-p)\, . 
\label{sevenfiveforty}
\end{equation}
and incorporate the explicit representation of the propagator
\eqref{sevenfivetwentyfour}. All the integrals over positions and some of the
integrals over momenta may be then carried out with the following result:
\[
\bigl\langle \phi_i | C_{\widetilde\Delta} | \psi_j \bigr\rangle = 2m
\left(\frac{T}{m}\right)^4 \int_{\widetilde\Delta} d^4 p \int
\frac{d^4 k}{(2\pi)^4}\ \Phi^*_i (k) \Psi_j (k)\ \left(\frac{m}{2\pi i
T}\right)^2
\]
\begin{equation}
\times\ \exp \left[-\frac{iT}{2m}
 \left(k^2 + m^2\right) \right]\ \exp \left[\frac{iT}{2m}\, (p-k)^2
\right]\, . 
\label{sevenfivefortyone}
\end{equation}
Equation \eqref{sevenfivefortyone} has essentially the same form as
\eqref{fivefourtwelve} with the important exception of the additional factor
$\exp[-i(T/2m)(k^2 + m^2)]$.  This difference arises because of the
integration over the total proper time in \eqref{sevenfivethirtyfive}. The
difference is important because it is the presence of this factor that
enforces the constraint.

In the limit of very large proper-time-of-flight, $T$, the integrals in
\eqref{sevenfivefortyone} may be evaluated by the method of stationary
phase.  The second exponential factor enforces the equality of $p$ and
$k$.  The first exponential factor enforces the constraint.  After a
straightforward, but messy calculation, one finds the following:
(1)\ The class operator matrix elements vanish if $\widetilde\Delta$
does not intersect the mass shell $p^2 = -m^2$.  (2)\ If
$\widetilde\Delta$ does intersect the mass shell then
\begin{equation}
\bigl\langle\phi_i | C_{\widetilde\Delta} | \psi_j \bigr\rangle = m
\int_{\widetilde\Delta} \frac{d^3 p}{(2\pi)^3
2\omega_p}\ \widetilde\phi^*_i ({\bf p}) \widetilde\psi_j ({\bf p})
\label{sevenfivefortytwo}
\end{equation}
where ${\tilde\psi}_j ({\bf p})$ and $ {\tilde\phi}_i ({\bf p})$
are the momentum space
representatives of the positive frequency solutions to the Klein-Gordon
equation $\psi_j (x)$ and
$\phi_i (x)$ and $\omega_{\bf p} =
\sqrt{{\bf p}^2 + m^2}$. The integral is over three momenta that lie in
the range $\widetilde\Delta$.

The first part of this result, (1), means that the class operators
vanish for values of $p$ that do not satisfy $p^2 =
-m^2$.  It is in this physical, probabilistic sense that the constraints
are satisfied.  The second part, (2), shows that the partition by values of
$p$ that do satisfy the constraints is the same as would be defined by
projections on momentum in usual relativistic quantum mechanics up to an
over all factor of $m$ arising from the proper time integration
 that cancels in the construction of conditional
probabilities.

\subsection{Relation to Dirac Quantization}

In the preceding discussion we saw that, utilizing coarse grainings
that define the momentum, the constraint $p^2=-m^2$ is satisfied with
probability unity. Such constraints play a central role in Dirac
quantizations.
This section discusses the senses in which the present formulation of
the quantum mechanics of systems with a single reparametrization
invariance coincides and does not coincide with the ideas of Dirac
quantization.  We will give a general discussion in the framework used
Sections VII.1.~and VII.2.~that is not restricted to the two specific
 models we
have considered.

The starting points for Dirac quantization are wave functions that are
annihilated by operator versions of the constraints and operators
(``observables'') that commute with the constraints.  In the case of the
systems with a single reparametrization invariance we would write
\begin{equation}
H\bigl(P_i,Q^i\bigr) \Psi (Q) = 0
\label{sevensixone}
\end{equation}
and
\begin{equation}
\bigl[H\bigl(P_i, Q^i\bigr), {\cal O}\bigr] = 0\, .
\label{sevensixtwo}
\end{equation}
Of course, there is much more to Dirac quantization than just these two
equations.\footnote{As in the lectures of Ashtekar in this volume!}  For
example, an inner product between wave functions
satisfying \eqref{sevensixone} must be specified as well as the rules
for how the state vector is reduced (the ``second law of evolution'') when an
observable has been measured and the rules for
constructing the probabilities of [time (?)] sequences of such
measurements.
However, without entering into these issues, let us
ask whether there are natural ways in which \eqref{sevensixone} and
\eqref{sevensixtwo} are satisfied in the present framework.  In particular, we
analyze the question of whether the class operators commute with the
constraints and
whether branch wave functions corresponding to individual
coarse-grained histories are annihilated by the constraints.

We begin with the question of whether the class operators commute with
the constraints.  Consider a partition of the paths between $Q^\prime$
and $Q^{\prime\prime}$ into classes $\{c_\alpha\}$ defined by whether the
value of a reparametrization invariant functional $F[Q,N]$ lies in an
interval $\Delta_\alpha$ that is one of an exhaustive and exclusive
set of such intervals.  Following \eqref{fivefourfive} we write
\begin{equation}
\left\langle Q^{\prime\prime} \left\Vert C_\alpha \right\Vert Q^\prime
\right\rangle = \int^{+\infty}_{-\infty} d\mu\,\tilde e_\alpha
(\mu)\ \left\langle Q^{\prime\prime} \left\Vert C_\mu \right\Vert
Q^\prime \right\rangle\, .
\label{sevensixthree}
\end{equation}
Here $\tilde e_\mu$ is the Fourier transform for the characteristic
function
for the interval $\Delta_\alpha$ [\cf \eqref{fivefourfour}] and
\begin{equation}
\left\langle Q^{\prime\prime} \left\Vert C_\mu \right\Vert Q^\prime
\right\rangle = \int \delta P \delta Q \delta N\ \Delta_\Phi
[Q,N]\,\delta\,[\Phi(Q,N)] \,\exp\Bigl\{i\bigl(S[P,Q,N] + \mu F [Q,N]
\bigr)\Bigr\} 
\label{sevensixfour}
\end{equation}
where $S$ is as in \eqref{seventwoeight} and  the integrations over the $Q$'s,
$P$'s, and multiplier are unrestricted by any coarse graining.

Within the class of reparametrization-invariant coarse grainings, one
can distinguish those that partition the paths only by their behavior
inside a region of configuration space that is bounded away from the
surfaces $\sigma^\prime$ and $\sigma^{\prime\prime}$ on which the initial
and final conditions are specified. For the relativistic world line, the
partition by the position of first passage through a spacelike surface
$\sigma$ not intersecting $\sigma^\prime$ and
$\sigma^{\prime\prime}$ described in Section VII.5.2 is of this class.
The partitions by value of spacetime position a certain proper time
after $\sigma^\prime$ is not.  The subclass of partitions that
discriminate between paths only by their behavior inside a region $R$ is
important for two reasons: First, it is physically realistic when the
relativistic world line is a model for quantum cosmology. Our
observations restrict the history of the universe only in a limited
region of its configuration space. We certainly do not have direct
access to anything like the proper time from the initial conditions.
Second, the decoherence functional and its consequent probabilities are
independent of the choice of $\sigma^\prime$ and $\sigma^{\prime\prime}$
for coarse grainings that only discriminate between paths inside a
region $R$ provided $\sigma^\prime$ and $\sigma^{\prime\prime}$ lie
outside $R$.  This too is physically reasonable.  For these reasons we
shall focus exclusively on coarse grainings of this type in what
follows.

In Section VII.4.2.~we gave a demonstration that for the free
relativistic particle the matrix elements $\langle x^{\prime\prime}
\Vert C_\alpha \Vert x^\prime \rangle$ satisfied the Klein-Gordon
equation \eqref{sevenfiveeleven}, when $x^{\prime\prime}$ and $x^\prime$
were outside the region of spacetime restricted  by the coarse graining
and $x^{\prime\prime}\not = x^\prime$. We did that by working in the
gauge $\dot N = 0$ in which $N$ is the total proper time between $x^\prime$
and $x^{\prime\prime}$, deriving a ``Schr\"odinger equation'' in that
proper time for the functional integral over just the $p$'s and $x$'s,
and then carrying out the remaining integral over $N$.  The same
derivation can be carried out in this more general case supposing
$F[Q,N]$ depends only on the portions of the paths inside some region $R$ of
configuration space.  The result is
\begin{equation}
\left[-i\frac{\partial}{\partial N} + H \left(-i \frac{\delta}{\delta
Q''},
Q^{\prime\prime}\right)\right]\,\left\langle Q^{\prime\prime},
N\left\Vert C_\mu \right\Vert Q^\prime, 0 \right\rangle = 0
\label{sevensixfive}
\end{equation}
{\it provided } $Q^{\prime\prime}$ is outside of $R$ and $Q^{\prime\prime}
\not= Q^\prime$.  A similar ``Schr\"odinger equation'' holds in
$Q^\prime$.  Subtracting these, integrating over $N$, either from
$-\infty$ to $+\infty$ or from $0$ to $\infty$ one finds
\begin{equation}
\left\langle Q^{\prime\prime} \left\Vert \left[C_\alpha, H \right]
\right\Vert Q^\prime\right\rangle = 0,\quad Q^{\prime\prime}\notin
R,\, Q^\prime\notin R\, .
\label{sevensixsix}
\end{equation}
The restriction $Q^{\prime\prime}\not = Q^\prime$ is no longer necessary
because the $\delta$-functions analogous to \eqref{sevenfivenineteen} 
cancel in the
construction of the commutator.  It is possible to give more
sophisticated and careful derivations of this result\footnote{For
example, by using the methods of \cite{HH91} and an argument similar
to that used to demonstrate \eqref{sixfivefive}.} but since we are
about to describe a negative result inside $R$, we shall not pursue
them.

Inside $R$ the story is different. That can be seen most easily by
considering the particular class of reparametrization invariant
functionals
\begin{equation}
F[Q,N] = \int^1_0 d\lambda N(\lambda) f\bigl(Q(\lambda)\bigr)
\, .\label{sevensixseven}
\end{equation}
This would give an effective action in the exponent of \eqref{sevensixfour}
that has the same form as \eqref{seventwoeight} but with $H$ replaced by
$H+\mu f$.  The class operators for this reparametrization-invariant
coarse graining therefore do not commute with $H$ but with $H + \mu f$.
  Class operators therefore generally do not commute
with the constraints.

A natural candidate for a branch wave function of a pure initial
condition that correspond to an
individual coarse-grained history is
\begin{equation}
\Psi_\alpha (Q) \equiv \left\langle Q\bigl\Vert C_\alpha \bigr| \Psi
\right\rangle = \left\langle Q \left\Vert C_\alpha \right\Vert Q^\prime
\right\rangle \circ \Psi \left(Q^\prime\right)
\label{sevensixeight}
\end{equation}
where $\Psi(Q)$ is the wave function representing the initial condition.
The same argument that was used above in \eqref{sevensixfive}
to establish $H\langle
Q^{\prime\prime}\Vert C_\mu \Vert Q^\prime \rangle = 0$ when $Q^{\prime\prime}
\not= Q^\prime$ and both are outside $R$ also suffices to show that
\begin{equation}
H\Psi_\alpha (Q) = 0, \quad Q \notin R \ {\rm or}\ \sigma^\prime
\label{sevensixnine}
\end{equation}
where $\sigma^\prime$ is the surface on which the product $\circ$ is
constructed.  However we do not expect that \eqref{sevensixnine} will be
satisfied inside $R$ for general reparametrization-invariant coarse
grainings.  Branch wave functions are therefore not generally annihilated
by the constraints.

The commutation of operators representing ``observables'' with the
constraints and the annihilation of ``physical'' wave functions by them
are two of the starting points of Dirac quantization.  The reason for
departures from the natural analogs of these relations in the present
quantum mechanics of systems with a single reparametrization invariance
can be traced to the more general nature of the alternatives
that generalized quantum mechanics considers,
as we now describe.

First, the fact that certain class operators do not commute with the
constraint does not signal a breakdown of reparametrization invariance.
The alternatives are a reparametriza-\break
\noindent tion-invariant partition of the
paths and the construction of their class operators has been
reparametrization invariant throughout.  Commutation of operators
representing alternatives with operators representing the constraint
implied by reparametrization invariance is therefore not a necessary
condition for invariance in the present formulation.

However, the alternatives whose class operators do not commute with the
constraint are of a more general character than those normally
considered in Dirac quantization.  The alternatives of Dirac quantization
correspond to functions on phase space.  The closest analog in the
present sum-over-configuration-space-histories formulation would
probably be partitions by reparametrization-invariant functionals of
the $Q$'s alone independent of the multiplier. These can be shown to have
class operators that {\it do} commute with the constraint. [The effective
action in \eqref{sevensixfour} 
in these cases does not imply a modification of $H$ as do
partitions by functionals that depend on $N$, \eg \eqref{sevensixseven}].
Unfortunately, there are only trivial examples of
reparametrization-invariant functionals of the $Q$'s that are
independent of $N$.  However, it is possible to extend the present
configuration space sum-over-histories formulation of the relativistic
world line to one that allows for phase space alternatives
\cite{Harup}. There the class operators for partitions by
reparametrization-invariant functionals of phase space histories that do
not involve the multiplier commute with the constraints.  Those that
involve the multiplier generally do not as the example of
\eqref{sevensixseven} shows.

The failure of the natural analogs of the branch wave functions to be
annihilated by constraints can be similarly traced to the more general
nature of the alternatives.
In the Dirac quantization of reparametrization invariant theories, the
equation $H\Psi = 0$ plays the role of a dynamical evolution equation
like the Schr\"odinger equation of non-relativistic theory.  For the
relativistic particle $H\Psi = 0$ is the Klein-Gordon equation; for
parametrized non-relativistic quantum mechanics it {\it is}
 the Schr\"odinger equation 
 [\cf \eqref{seventhreefourteen}].  However, in the canonical
formulation of quantum mechanics there are {\it two} laws of evolution.
Unitary evolution by a dynamical equation and reduction of the state
vector.  In sum-over-histories quantum mechanics, these are unified in a
single path-integral description.
We would thus expect $H\Psi=0$ for those regions of configuration space
where paths are unrestricted by the coarse graining and nothing like a
``second law of evolution'' was operative.  That is exactly the content
of \eqref{sevensixnine}. Where the paths {\it are} restricted by coarse
graining we expect $H\Psi$ to continue to vanish if the class operators
commute with the constraints but not for the more general alternatives
whose class operators do not.

For a gauge theory the Dirac condition $(constraint) \Psi=0$ ensures
that wave functions of states on spacelike surface depend only on the
``true physical degrees of freedom''.  For the
reparametrization-invariant systems under discussion that idea is
captured in wave functions representing the initial and final
condition that satisfy the constraint.
However, in a theory where there is no natural construction
of a state on a spacelike surface, and therefore no natural quantum
mechanical notion of a ``degree of freedom'' on such a surface, it is
perhaps not surprising to find that $H\Psi\not=0$ for all branch wave
functions.

\section{General Relativity}
\setcounter{footnote}{0}

\subsection{General Relativity and Quantum Gravity}

We come, at last, to a generalized quantum mechanics for general
relativity --- a theory that exhibits both the reparametrization
invariance of the models discussed in the
preceding  section and gauge symmetries analogous to
those discussed in the section before that.  Classical general relativity is a
theory of spacetime geometry and a quantum theory of general relativity
assumes spacetime geometry as a fundamental dynamical variable.
It may be, as suggested by string theory or by the
non-perturbative canonical quantum gravity program, that qualitatively
different kinds of fundamental variables are needed to formulate a
successful quantum of gravity.  Spacetime geometry would then be a
particular type of coarse graining of these fundamental variables.  In
the face of such uncertainty about the fundamentals why consider a generalized
quantum
mechanics for spacetime at all?  There are at least three reasons:
\begin{itemize}
\item First, even if there is a more fundamental theory, it is
unlikely that it will involve a fixed background spacetime.  That theory
too will therefore require a generalization of quantum mechanics to deal
with the ``problem of time''.  A formal generalized quantum mechanics of
Einstein's theory can thus serve as a model for the kind of quantum
mechanics that will be needed and offer insight into
 the kinds of questions that can be
asked of it.
\item Second, and more importantly, in a more fundamental
theory it must
be possible to describe spacetime because we successfully employ this mode of
description for a wide range of phenomena here and now.  A quantum
theory of gravity must be able to predict, for example, the
probabilities that spacetime geometry on accessible scales conforms to
the classical Einstein equation.  Further, since
Einstein's theory is the unique low energy limit of any quantum theory
of gravity \cite{Des70, BD75}, we expect
that quantum gravitational phenomena will be
approximately described on the scales most easily accessible
to us by a quantum theory of spacetime based on Einstein's action
suitably cut off at very short distances.  We expect, for example, weak
gravitons to be adequately described in this way. In quantum cosmology
most predictions of low energy properties of the universe such as the
galaxy--galaxy correlation function are predicted using such a quantum
theory.
A generalized quantum mechanics of spacetime is therefore needed just
for this approximation and to pose this kind of question.
\item The third reason for exploring a generalized quantum mechanics
for general relativity is that Ashtekar and Smolin\footnote{As in
Ashtekar's lectures in this volume.}, Agishtein and Migdal
\cite{AM93}, DeWitt \cite{DeWup}, Hamber \cite{Ham93},
and others
could be right in their various ways
 in believing that Einstein's theory, or simple modifications of it,
 make sense
non-perturbatively.  In that case we want to be ready with an
understanding of how such a theory could be used to make predictions!
\end{itemize}

The role of a generalized quantum mechanics for general relativity can be
stated more precisely if we imagine a hierarchy of approximations.  At
the most fundamental level there is {\it the} fundamental theory with
its fine-grained histories, coarse grainings and decoherence functional,
$D_{\rm fundamental} (\alpha^\prime, \alpha)$. Some coarse-grained sets
$\{c_\alpha\}$
must describe alternative spacetime geometries at scales, say, above
the Planck length.  For these coarse grainings and for {\it certain}
initial conditions we expect
\begin{equation}
D_{\rm fundamental} (\alpha^\prime, \alpha) \cong D_{\rm quantum\ GR}
(\alpha^\prime, \alpha)
\label{eightone}
\end{equation}
where $D_{\rm quantum\ GR}(\alpha^\prime, \alpha)$ is the decoherence
functional
for a generalized quantum mechanics of spacetime based on Einstein's
action cut off at short distances if necessary.  It would be in this
sense that quantum general relativity could be an approximation to a
more fundamental theory --- not generally, but for certain coarse grainings
and certain initial conditions.

Further coarse-graining can define spacetime geometry on scales much
above the Planck length.  Two geometries lie in the same
coarse-grained history if they differ only in structures on scales well
beneath those accessible to us.
For such coarse-grained sets of histories $\{c_{\bar \alpha}\}$ and for {\it
some} initial conditions it may happen that
\begin{equation}
D_{\rm quantum\ GR} (\bar \alpha^\prime, \bar \alpha) \simeq D_{\rm classical
\ GR} (\bar \alpha^\prime, \bar \alpha)
\label{eighttwo}
\end{equation}
where $D_{\rm classical\ GR} (\bar \alpha^\prime,
\bar \alpha)$ is the decoherence
functional for classical general relativity as described in Section
IV.5.  This is the sense in which classical general relativity is the limit of
quantum general relativity or the sense in which classically behaving
spacetime is predicted by a theory of the initial condition.  We do not
expect \eqref{eighttwo} to hold for all coarse grainings.  Coarse grainings
that specify alternative values of spacetime geometry on Planck scales
or on any scale  in the Planck
epoch of the early universe
 are unlikely to work. Neither do we expect \eqref{eighttwo} to hold for
all initial conditions because the classical behavior of spacetime
geometry, like classical behavior generally, requires some restriction
on the initial condition.

In this section, therefore,  we describe the construction of a generalized
quantum mechanics of general relativity, and in particular, its
decoherence functional $D_{\rm
quantum\ GR} (\alpha^\prime, \alpha)$.  Our considerations will necessarily be
formal since we are far from knowing how to do sums over geometries in
most cases, but we shall try to make the constructions concrete in
discrete approximations to them.

\subsection{Fine-Grained Histories of Metric and Fields and their
Simplicial\\
 Approximation}

To construct a generalized quantum mechanics one must specify the
fine-grained histories, the allowed coarse grainings, and the
decoherence functional. We begin in this subsection with the
fine-grained histories for a quantum theory of general relativity.
As throughout most of
these lectures, we shall take the sum-over-histories point
of view in which there is a unique fine-grained set of histories.
The fine-grained histories of {\it classical} general relativity
 are manifolds endowed with
Lorentz signatured metrics and four-dimensional matter field
configurations satisfying the Einstein equation and matter field
equations. For quantum general relativity we therefore take the
fine-grained histories to be four-dimensional manifolds with arbitrary
Lorentz signatured metrics and matter field configurations.
 One of the advantages of generalized quantum mechanics
is that {\it different} topologies can be included in
the set of fine-grained histories and the quantum mechanics of topology
change investigated.\footnote{Classically the restriction of
geometries to be manifolds with metrics is the mathematical
implementation of the principle of equivalence. However, the
undecidability of the homeomorphism problem for four-manifolds may make
it natural to consider metrics on more general topological spaces than
manifolds as the fine-grained histories of the quantum theory. For
discussion see \cite{Har85b} and, for a
specific proposal see
Schleich and Witt \cite{SW93}. We will not discuss
sums over topologies in these lectures and restrict attention to a fixed
manifold.}  However, it is simplest to begin by fixing the
topology to the manifold ${\bf I} \times M^3$ where ${\bf I}$
is a finite interval of ${\bf
R}$ and $M^3$ is a closed three manifold and we shall do so throughout
the remaining sections. We are thus considering
spatially closed universes with two $M^3$ boundaries, $\partial
M^\prime$ and $\partial M^{\prime\prime}$.  This is the
simplest case compatible with non-singular Lorentz signatured
metrics with no closed timelike curves\footnote{As in the result
of Geroch \cite{Ger67}, see, however, Horowitz \cite{Hor91}.}
and the one most relevant for quantum cosmology.  It is also the case
that is closest in analogy with the reparametrization invariant models
just discussed in Section VII.  The ${\bf I}\ \times M^3$ geometry is
analogous to a particle path and its two boundaries to the endpoints of
the paths.
In analogy with field theory and particle quantum mechanics, we expect
important contributions to the functional integrals defining decoherence
functionals from metrics and field configurations which are
non-differentiable.

The fine-grained histories for our discussion are therefore Lorentz
signatured metrics and matter field configurations that are continuous,
but not generally differentiable,  on the fixed manifold $M= {\bf I}
\times ^3M$.
We write metrics as $g_{\alpha\beta} (x)$ or $g(x)$ and for the most
part we consider a single scalar matter field, $\phi(x)$, for
illustrative purposes. We denote by $h^\prime_{ij} ({\bf x})$ and
$\chi^\prime ({\bf x})$ the metric and matter field induced in
$\partial M^\prime$ and by $h^{\prime\prime}_{ij} ({\bf x})$ and
$\chi^{\prime\prime} ({\bf x})$ those induced in $\partial
M^{\prime\prime}$.

It is possible to construct a  generalized quantum mechanics by
restricting the fine-grained histories to special
subsets
metrics and field configurations on $M$.  For example, following Section
IV.5., we could consider classical general relativity as a generalized
quantum mechanics by restricting the fine-grained histories to be
solutions of the classical field equation.  Following the example of the
relativistic particle in Section VII.4 one could break general
covariance and introduce a preferred time variable.  The fine-grained
histories would then be restricted to those metrics that can be uniquely
foliated by this time variable, that is, metrics for which each value of
the variable labels a unique spacelike surface in the four-geometry.
However, the natural, generally covariant, choice of the set of
fine-grained histories is the class of {\it all} four-metrics and matter
field, configurations on $M$. This is the choice we shall use for the
generalized quantum mechanics to be constructed in these lectures.

In view of possible formal character of functional integrals over
metrics, it is useful to understand how to formulate a generalized
quantum mechanics for a cut-off version of general relativity.  We shall
return to discuss such a generalized quantum mechanics in Section VIII.7
but we describe its fine-grained histories here as a concrete aid to
thinking about the more formal case of continuous metrics.

\begin{figure}[t]
\begin{center}
\includegraphics[width=6in]{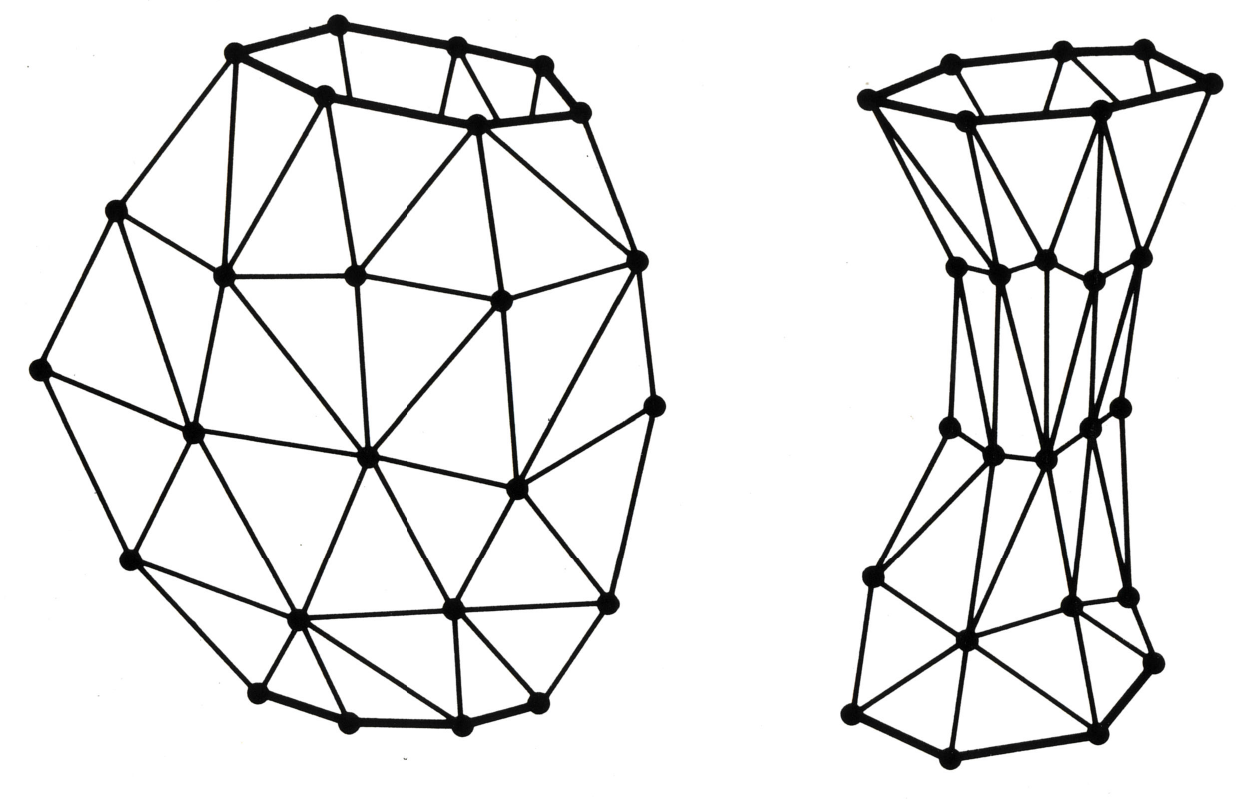}
\caption{Two-dimensional simplicial geometries.
Two-dimensional surfaces can be made up by joining together flat
triangles to form  simplicial manifold.  A geometry of the surface is
specified by an assignment of squared edge-lengths to the triangles.
The figure shows two different geometries obtained by a different assignment
of squared edge-lengths to the same simplicial manifold.  The
generalization of these ideas to four dimensions and Lorentz signature
gives the natural lattice version of general relativity --- the Regge
calculus.  In a sum-over-histories quantum theory of simplicial
spacetimes, sums over geometries are represented by integrals over the
squared edge-lengths.  Diffeomorphism invariant alternatives can be
defined by partitioning the space of allowed squared edge-lengths into
exhaustive sets of exclusive regions.  For example, one could partition
closed cosmological geometries into the class that has no simplicial
spacelike three surface greater than a certain volume and the class that
has at least one such surface.  In a given simplicial manifold it is
possible to enumerate all three surfaces and identify the regions in the
space of squared edge-lengths to which each class corresponds.}
\end{center}
\end{figure}

Simplicial manifolds provide the natural lattice version of general
relativity and the most direct route to implementing a generally
covariant cut-off.  A surface in two dimensions can be built up from
flat triangles as in a ``geodesic dome''.  The topology of the surface is
specified by the way the triangles are joined together.  A metric is
specified by giving the squared edge-lengths of each triangle and
a flat metric for its interior.  In this way various
two-dimensional simplicial geometries can be constructed (see Figure 12).
The situation is similar in four dimensions.  A geometry can be built up
from flat, four-dimensional simplices.  The topology of the simplicial
manifold is specified by the way the simplices are joined together.  A
Lorentz metric is specified by giving the values of the $n_1$ squared edge
lengths of the four-simplices, $s^i$,  and a Lorentz signatured flat
metric in their interiors.  For the edge-lengths to be compatible with a
Lorentz signatured flat metric, there must be
some restrictions on the $s^i$ analogous to the triangle inequalities
and $s^i$ will be negative if their edges define timelike directions.
Values of a scalar field $\phi^i$
 can be assigned to the $n_0$ vertices.  The space of fine-grained
histories for this simplicial approximation is then the domain of ${\bf
R}^{n_1}\times {\bf R}^{n_0}$ consistent with the analogs of the
triangle inequalities.  A particular fine-grained history is a point in
this space.

\subsection{Coarse Grainings of Spacetime}

Every assertion that we make about the universe corresponds to a
partition of its histories into those for which the assertion is true and
those for which it is false.  If we assert that the universe is nearly
homogeneous and isotropic on large scales at late times, we are utilizing a
partition
of the four-dimensional geometries into the class of those that
are nearly homogeneous and isotropic at late enough times and the class of
those that are not
and asserting that our universe lies in the former class.  Similarly, to
say that the spacetime
of the late universe behaves classically on accessible scales presumes
that we can divide the cosmological histories into those correlated by
Einstein's equations in accessible coarse grainings and those which
are not so correlated.

Even an assertion that refers to our own experience, such as the
assertion that spacetime is nearly flat in the neighborhood of
our sun, presumes a distinction of this form from the point of view of
cosmology.
To make the necessary partitions
we would first have to describe what we mean by
``our sun''.  If we were kidnaped by aliens in UFO's and set down again
on a planet, how would we tell if it is our own earth and how would we
tell if the star about which it orbits is our own sun?  We would, of
course, compare the planet of arrival to a description of the earth
recorded in our memories.
It is the remembered description that defines the physical situation that
we mean by ``our sun''.
Utilizing such a description, it is possible to
partition the histories into the classes that contain ``our
sun'' with a nearly flat
spacetime about it, the class that contains
``our sun'' with a highly curved spacetime, and the class
that does not contain ``our sun''.  This is a coarse graining of the
histories of the universe and a very coarse graining at that.

Thus, at a fundamental level every assertion about the
universe, from assertions about large scale structure to statement about
the everyday here and now, is the assertion that the history of the
universe lies in the coarse-grained class in which the assertion is true
and not in the class in which it is false.
An assertion which does not unambiguously correspond to such a partition
is not well defined.
Generalized quantum mechanics predicts the probabilities for such
alternative coarse-grained sets of histories.

Each of the examples of
coarse graining discussed above is manifestly
diffeomorphism invariant --- no mention of coordinates went into their
description.  The
{\it allowed coarse grainings} of this generalized quantum mechanics
are more generally
 partitions of the fine-grained histories of metrics and matter field
configurations into an exhaustive set of
exclusive, {\it diffeomorphism invariant} classes. We now describe some
further
examples of partitions of four-metrics into diffeomorphism invariant
classes.

A familiar question in quantum cosmology is ``What are the probabilities
of the possible maximum volumes the universe may reach in the course of
its history?''
The answer is of use, for example, in determining whether it is probable
that a closed universe will be nearly spatially flat and exist for a long
time -- two features observed of our universe.
We can state this question precisely utilizing
a coarse graining that divides all four metrics into two
diffeomorphism invariant classes $c_0$ and $c_1$ as follows:
\begin{itemize}
\item[$c_0$:] The class of metrics for which all spacelike
three-surfaces have volumes less than a fiducial volume $V_0$.
\item[$c_1$:] The class of metrics for each of which there is
at least one three-surface with a
volume larger than $V_0$.
\end{itemize}
\noindent This is a manifestly exhaustive set of exclusive
 diffeomorphism invariant
alternatives. If it decoheres, the probability of $c_0$ is what we
mean by the probability the the universe has a maximum
volume\footnote{Note
that we cannot usefully turn this around and ask whether the
universe has a minimum volume which is less than $V_0$. That is because
a general Lorentzian four-geometry will contain three-surfaces of
arbitrarily small volume with segments that are close to
 null.~~The question can be asked whether the universe has a spacelike
three-surface with volume less than a fixed $V_0$, but the answer will
be ``yes'' with probability one.} not   greater than $V_0$.

The following example illustrates that care must be used to choose
coarse grainings that are genuine partitions of the set of fine-grained
histories.  It is sometimes suggested that one way of resolving the
problem of time is to use some property of a three-surface, say, the
total volume as a time variable.  One could then
define alternatives at a given value of total volume, say alternative
possibilities for the rest of the three-geometry on that surface.
However, this is not a genuine partition of the fine-grained histories
because a given four-geometry may contain arbitrarily many
three-surfaces of a given volume  each with {\it different}
three-geometries.
This is the geometrical analog of paths which forward and backward in
time, intersecting a surface of constant time more than once, which was
discussed in the case of the reparametrization invariant models
of Section VII.

The analogy with systems like the relativistic particle
 may be made somewhat more precise by
utilizing a particular gauge and representing  four-dimensional histories
as  curves in the superspace of three-dimensional geometries and
matter field configurations (Figure 13).  The analog of spacetime in the
case of the relativistic particle is superspace in the case of spacetime
geometry. Fine-grained histories of the relativistic particle are
curves in spacetime.  Fine-grained histories of spacetime geometry are
curves in superspace. The analog of a surface of constant
time in spacetime in the case of the relativistic particle
would be a surface in superspace. For any surface one chooses in
superspace there are fine-grained histories --- spacetimes ---
that correspond to curves
that intersect it an arbitrarily large number of times as was the case
in the quantum mechanics of a single relativistic world line discussed
in Section VII.  Partitions by the location in superspace that a curve
crosses a surface in superspace are therefore not possible.  In this
sense there is no property of three-geometry that can play the usual
role of  time in this generalized quantum mechanics.

\begin{figure}[t]
\begin{center}
\includegraphics[width=6in]{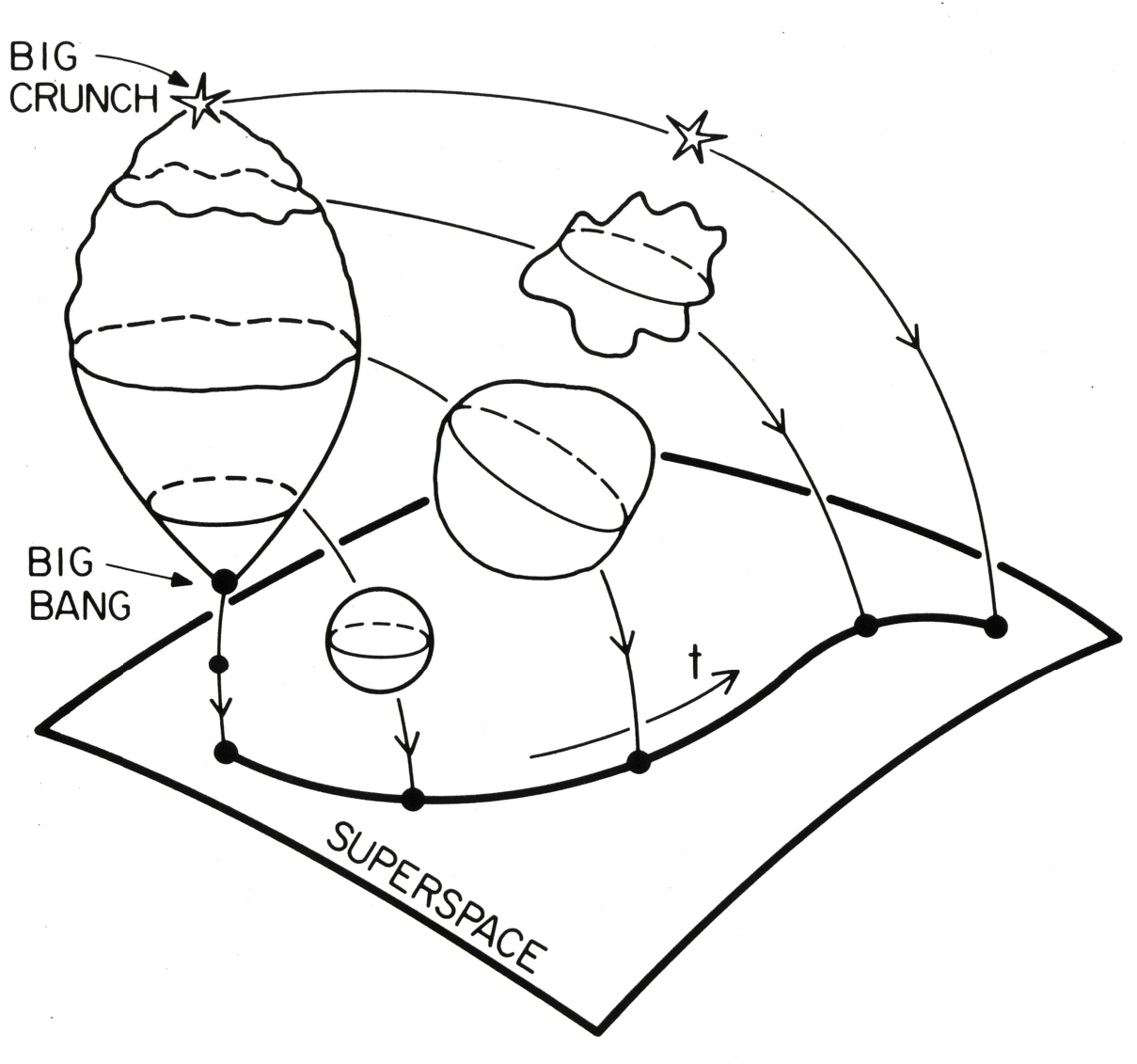}
\caption{Superspace.  A cosmological
history is a four-dimensional cosmological spacetime with matter fields
upon it.  A two-dimensional representation of such a history is shown in
the upper left of this
figure proceeding from a big bang to a big crunch.  In the Gaussian
gauge of
$ds^2 = -dt^2 + h_{ij} (x^k, t) dx^i dx^j$
 a cosmological history can be thought of as a parametrized
succession of
three-dimensional geometries and spatial matter field configuration.
Superspace is
the space of such three-dimensional geometries and matter field
configurations.
A ``point'' in superspace is a {\sl particular} three-geometry and spatial
matter
field
configuration.  The succession of three-geometries and matter fields
that
make up a four-geometry and field history, therefore, trace out a path
in
superspace.}
\end{center}
\end{figure}

Allowed coarse grainings involving the geometries of spacelike surfaces
can be constructed as follows: Define a range $R$ of three-geometries
--- a region in superspace --- by a set of restrictions that are
invariant under three-dimensional diffeomorphisms.  For example, we
might consider the region $R$ in which the total volume lies in a range
$\Delta_v$ the integrated square of the three-dimensional Riemann tensor
lies in another range $\Delta_{{\rm Riem}^2}$, and the average value of
the scalar field lies in yet another range $\Delta_{\bar\phi}$.  The
fine-grained histories can be partitioned into the following two
classes:
{}~~(1) the class of all histories that have no three surface in the
region $R$, and ~~(2) the class of all histories that have {\it at
least} one three surface in the region $R$.  Such partitions are the
analog of the partitions by a spacetime region discussed in Section
V.3.2 for a non-relativistic particle and in Section VII.5.2 for a
relativistic particle.  By partitioning the paths according to their
behavior with respect to many such regions of superspace a rich variety
of coarse grainings analogous to the time sequences of non-relativistic
quantum mechanics can be built up.

However, coarse grainings are not restricted just to those that distinguish
 the geometries of spacelike
surfaces.  For example, we could consider coarse-grainings by values of
the proper four-volume in between spacelike surfaces or the values of
the proper time on curves that connect spacelike surfaces.

The assumption that the fine-grained histories are continuous but not
necessarily differentiable, means that some partitions of classical
differentiable histories become vacuous like the
partitions by finite values of the derivatives of the paths
of a single particle discussed in Section V.4.2.
Quantities like momenta, or the extrinsic curvature of surfaces, can still
be defined utilizing a spacetime description, but only approximately.
We shall return to what we might
mean by a sum-over-non-differentiable
geometries in connection with simplicial approximations below.

The {\it general}\ notion of coarse graining is by ranges of values of
diffeomorphism invariant functionals of four-geometry and matter field
configurations. These are especially easy to illustrate in the
simplicial approximation
described in Section VIII.2. Consider a fixed simplicial net as
illustrated in Figure 12 and suppose that the fields and squared edge
lengths are fixed on the two boundaries.  The fine-grained histories are
then specified by the values of the interior squared edge-lengths,
$s^i$, and the field values, $\phi^i$, at the interior vertices.
The
general notion of coarse graining is by ranges of values of functions
$F_A(s^i, \phi^i)$, $A=1, 2, \cdots$
of the interior squared edge-lengths and field values
that are invariant under any symmetry group of lattice that is a remnant
of diffeomorphism invariance.

The main concluding point about the coarse grainings discussed here
is that they
supply a much larger set of diffeomorphism invariant alternatives for
quantum cosmology than those conventionally contemplated on spacelike
surfaces.  Within this larger class are the coarse grainings
that are directly accessible and easily interpretable by us.

\subsection{The Decoherence Functional for General Relativity}

In this section we shall describe a construction for the decoherence
functional of general relativity.  The essential ideas of the
construction have already been illustrated in the gauge-invariant and
reparametrization invariant models previously studied.  As we mentioned in
Section VI.1., general relativity exhibits both kinds of symmetry.  It
is necessary only to spell out the details of how the ideas illustrated
in the models are combined.

\subsubsection{Actions, Invariance, Constraints}

The action for general relativity is a sum of the gravitational action
for the metric and an action for the matter field
\begin{equation}
S[g, \phi] = S_E [g] + S_M[g,\phi]\, . 
\label{eightthree}
\end{equation}
 For illustrative purposes,  we shall assume for the matter a
scalar field with the action
\begin{equation}
S_M [g, \phi] = -\half \int_M d^4 x (-g)^\half \left[(\nabla\phi)^2 +
V(\phi)\right] 
\label{eightfour}
\end{equation}
for some potential $V(\phi)$.  The action for Einstein's theory that is
appropriate when the three-metric, $h_{ij}$, is fixed on the boundaries of
$M$ is
\begin{equation}
\ell^2 S_E [g,\phi] = \int_M d^4 x (-g)^\half (R-2\Lambda) +
2\int_{\partial M} d^3 x h^\half K 
\label{eightfive}
\end{equation}
where $\ell = (16\pi G)^\half$ is $4\pi^\half$ times the Planck
length.
In the first integral, $R$ is the scalar curvature, $\Lambda$
is the cosmological constant and the integration
range is the whole of the manifold
 $M$. The
surface term is necessary to compensate for the second derivatives in the
scalar curvature so as
to make the action additive on spacetime regions.  It
is an integral over each boundary three-surface in which $h_{ij}$ is the
metric induced by $g_{\alpha\beta}$ in the surface.  The quantity $K$ is
the trace of the extrinsic curvature of the surface , $K_{ij}$, defined
as the projection into the surface of the
derivative $-\nabla_\alpha n_\beta$ where $n_\alpha$ is the
normal to the surface.

The canonical form of the action will be useful in constructing the
functional integrals that define the decoherence functional because it
is in phase-space that the measure for these integrals is most easily
defined.  We rapidly recall the canonical formulation.\footnote{For
more details, see \cite{ADM62, Kuc74, HRT76}.}
The first step is to write the action
\eqref{eightfive} in $3+1$ form using the $3+1$ decomposition of the metric
with respect to a foliating family of spacelike surfaces that was
discussed in Section VI.1. We assume that two members of the foliating
family coincide with the boundary surfaces $\partial M'$ and $\partial
M''$.  In terms of the lapse,
shift, induced three-metric and extrinsic curvature of the constant $t$
surfaces, the action $S_E$ takes the simple form
\begin{equation}
\ell^2 S_E \bigl[N, N^i, h_{ij}\bigr] = \int_M dt\, d^3 x\, h^\half N
\ \left[K_{ij}
K^{ij} - K^2 - (2\Lambda -^3R)\right]\, . 
\label{eightsix}
\end{equation}
Here, $^3R$ is the scalar curvature of the foliating surfaces and
$K_{ij}$ is their extrinsic curvature. Explicitly
\begin{equation}
K_{ij} = (2N)^{-1} \left[-\dot h_{ij} + 2D_{(i}N_{j)} \right]
\label{eightseven}
\end{equation}
where $D_i$ is the derivative in the three-dimensional
constant-$t$ surfaces and the dot
denotes a derivative with respect to $t$. The momenta conjugate to the
$h_{ij}$ may be calculated straightforwardly from the action
\eqref{eightsix}
and are
\begin{equation}
\ell^2\pi^{ij} = -h^\half \left[K^{ij} - h^{ij} K\right]\, .
\label{eighteight}
\end{equation}
The action \eqref{eightfive} may then be re\"expressed in canonical form as
\begin{equation}
S_E \bigl[N,N^i, \pi^{ij}, h_{ij} \bigr] = \int\nolimits_M dt\, d^3 x
\left[\pi^{ij} \dot h_{ij} - NH (\pi^{ij}, h_{ij}) - N^i H_i
(\pi^{ij}, h_{ij})\right]\, , 
\label{eightnine}
\end{equation}
where the $H$ and $H_i$ are defined as follows:
\begin{subequations}
\label{eightten}
\begin{eqnarray}
H & = & \ell^2 G_{ijkl}\, \pi^{ij} \pi^{kl} + \ell^{-2} h^\half (2\Lambda
-^3R)\, , \label{eightten a}\\
H_i & = & -2D^j \pi_{ij}\, , \label{eightten b}
\end{eqnarray}
\end{subequations}
with the DeWitt supermetric $G_{ijkl}$ being defined by
\begin{equation}
G_{ijkl} = \half h^{-\half} \left(h_{ik} h_{jl} + h_{il}h_{jk} - h_{ij}
h_{kl} \right) \, . 
\label{eighttwelve}
\end{equation}
The evident symmetry with which the $(N, N_i)$ and $(H, H_i)$ enter
\eqref{eightnine} makes it useful to introduce the notation
\begin{equation}
N^0=N\ ,\ H_0=H \label{eightthirteen}
\end{equation}
so that the canonical action can be rewritten compactly as
\begin{equation}
S_E \bigl[N^\alpha, \pi^{ij}, h_{ij}\bigr] = \int\nolimits_M dt\, d^3 x
\ \left[\pi^{ij} \dot h_{ij} - N^\alpha H_\alpha\right]\, .
\label{eightfourteen}
\end{equation}

The action for the matter field may be expressed in
a canonical form similar to \eqref{eightnine}.  In analogy to $h_{ij} ({\bf
x})$, we write $\chi ({\bf x})$ for the value of the field on a
constant-$t$ surface and $\pi_\chi ({\bf x})$ for its conjugate
momentum.  For a scalar field
$\chi ({\bf x}, t) = \phi({\bf x}, t)$.
The total action, $S_E + S_M$, takes the form
\begin{equation}
S\bigl[N^\alpha, \pi^{ij}, \pi_\chi, h_{ij}, \chi\bigr] = \int_M dt\, d^3
x \ \left[\pi^{ij} \dot h_{ij} + \pi_\chi \dot \chi - N^\alpha {\cal
H}_\alpha \bigl(\pi^{ij}, \pi_\chi, h_{ij}, \chi \bigr)\right]\, .
\label{eightfifteen}
\end{equation}
Here, ${\cal H}_0$ and ${\cal H}_i$ are functions of the canonical
coordinates and momenta defined by
\begin{subequations}
\label{eightsixteen}
\begin{eqnarray}
{\cal H}_0 & = & H+ h^\half T_{nn}\, , \label{eightsixteen a}\\
{\cal H}_i & = & H_i + h^\half T_{ni}\, , \label{eightsixteen b}
\end{eqnarray}
\end{subequations}
where $T_{\alpha\beta}$ is the stress energy tensor of the scalar field
expressed as a function of $\pi_\chi,  \chi$, and $h_{ij}$, an index
$n$ indicating that it is projected onto the normal, $n^\alpha$,
 of the constant-$t$
surfaces, \viz $T_{ni} = n_\alpha T^\alpha_i,\quad T_{nn} = n_\alpha
T^{\alpha\beta} n_\beta$.

The absence of any term in \eqref{eightfifteen} that is just a function
of the coordinates and momenta and not proportional to lapse or shift
is a signal of diffeomorphism invariance as we shall shortly see.  Very
little of the subsequent argument will depend on the specific forms of
${\cal H}_0$ and ${\cal H}_i$ beyond the fact that they are at most
quadratic in the momenta.  Almost everything we shall need follows from
the form \eqref{eightfifteen}.

The diffeomorphism invariance of general relativity implies four
constraints between the canonical coordinates $(h_{ij}, \chi)$ and
their conjugate momenta $(\pi^{ij}, \pi_\chi)$ as the general argument
in Section VII.2.~shows.  With the action in the form \eqref{eightfifteen},
they are not difficult to find.  They are the equations that result from
extremizing \eqref{eightfifteen} with respect to lapse and shift:
\begin{equation}
{\cal H}_\mu \bigl(\pi^{ij}({\bf x}), \pi_{\chi}({\bf x}), h_{ij}({\bf x}),
\chi({\bf x})  \bigr) = 0\, .
\label{eighteighteen}
\end{equation}
These four relations among the canonical coordinates and momenta
 are constraints
that must be satisfied by any initial data for Einstein's
equation.

Dynamical equations in canonical form result from varying the action
with respect to the canonical coordinates.  For example, by varying
with respect to $\pi^{ij}$ and $h_{ij}$ one finds

\begin{eqnarray}
\dot h_{ij}({\bf x}) & = & N^\mu \left(\partial {\cal H}_\mu ({\bf x})
/\partial\pi^{ij}({\bf x})\right)\, ,
\label{eightnineteen}\\
\dot \pi^{ij}({\bf x}) & = & -N^\mu
\left(\partial {\cal H}_\mu ({\bf x})/\partial h_{ij}({\bf x})
\right)\, , \label{eighttwenty}
\end{eqnarray}
and similar equations for the matter field and its momentum.  The
equations of motion may be written compactly by introducing the
contraction
\begin{equation}
{\cal H} (N) = \int_t d^3 x N^\mu ({\bf x}) {\cal H}_\mu ({\bf x})
\label{eighttwentyone}
\end{equation}
and the Poisson bracket $\{~,~\}$ with conventions such that $\{q^A,
p_B\} = \delta^A_B$. Then,
\begin{subequations}
\label{eighttwentytwo}
\begin{eqnarray}
\dot h_{ij} ({\bf x}) & = &  \{h_{ij} ({\bf x}), {\cal H}(N)\}\, ,
\label{eighttwentytwo a}\\
\dot\pi^{ij} ({\bf x}) & = & \{\pi^{ij} ({\bf x}), {\cal H} (N)\}\, ,
\label{eighttwentytwo b}
\end{eqnarray}
\end{subequations}
and similar equations for the matter degrees of freedom.  Thus the
constraints generate dynamics by specifying how the canonical
coordinates change between two surfaces connected by lapse $N$ and
shift $N^i$ (Figure 11).  The constraints \eqref{eighteighteen} together with
the dynamical equations \eqref{eighttwentytwo} are the Einstein
equation written in canonical form.

The constraints of classical general relativity are closed under the
Poisson bracket operation.  That is, with all quantities evaluated on a
common constant-$t$ surface
\begin{equation}
\left\{{\cal H}_\mu \left({\bf x}^\prime\right), {\cal H}_\nu\left({\bf
x}^{\prime\prime}\right)\right\}
 = \int_t d^3 x^{\prime\prime\,\prime}
\ U^\gamma_{\mu\nu} \left({\bf x}^\prime, {\bf x}^{\prime\prime},
{\bf x}^{\prime\prime\,\prime}\right) {\cal H}_\gamma \left({\bf
x}^{\prime\prime\,\prime}\right)\, . 
\label{eighttwentythree}
\end{equation}
The structure functions, $U^\gamma_{\mu\nu}$, involve various
$\delta$-functions, derivatives and the metric $h_{ij}$.  Their explicit
form, which will not be necessary for us, can be found in many standard
references (\eg \cite{Kuc74}, p.~250).

Closure of the constraints
 under the Poisson bracket operation is necessary for
consistency.  Otherwise the Poisson bracket of two constraints would
represent new and different constraints on the canonical coordinates.
However, because the structure functions depend on the coordinates
(specifically the $h_{ij}$), the relations
\eqref{eighttwentythree} do not define an algebra (although they are often
referred to informally as such). In particular,  they do not define the
algebra of four-dimensional
diffeomorphisms that were the origin of the constraints and that
fact has
important consequences for the canonical theory.

A straightforward calculation shows that the action \eqref{eightfifteen} is
invariant under the following canonical transformation generated by an
infinitesimal vector $\epsilon^\alpha ({\bf x})$:
\begin{subequations}
\label{eighttwentyfour}
\begin{eqnarray}
\delta h_{ij} ({\bf x}) & = & \left\{h_{ij} ({\bf x}), {\cal H}
(\epsilon)\right\}\, , \label{eighttwentyfour a}\\
\delta\pi^{ij}({\bf x}) & = & \{\pi^{ij} ({\bf x}), {\cal H}(\epsilon)
\}\, , \label{eighttwentyfour b}\\
\delta \chi ({\bf x}) & = & \{\chi ({\bf x}), {\cal H} (\epsilon)
\}\, , \label{eighttwentyfour c} \\
\delta\pi_\chi ({\bf x}) & = & \left\{\pi_\chi ({\bf x}), {\cal H}
(\epsilon) \right\}\, , \label{eighttwentyfour d}
\end{eqnarray}
together with the related transformation of the lapse and shift:
\begin{equation}
\delta N^\alpha ({\bf x})  =  \dot\epsilon^\alpha ({\bf x}) - \int_t
d^3x^\prime
\int\nolimits_t d^3 x^{\prime\prime} U^\alpha_{\beta\gamma} \left({\bf
x}^\prime, {\bf x}^{\prime\prime}, {\bf x}\right) N^\beta
\left({\bf x}^\prime \right) \epsilon^\gamma \left({\bf
x}^{\prime\prime} \right)\, , 
\label{eighttwentyfour e}
\end{equation}
\end{subequations}
where all functions are evaluated
 on a common constant-$t$ surface. These are the
generalizations of the symmetries \eqref{seventwoeighta} in the case of the
relativistic world line. The action
\eqref{eightfifteen} is invariant under the transformations
\eqref{eighttwentyfour}
 {\it provided}, as stressed by Teitelboim
\cite {Tei83a}, that the normal component of $\epsilon^\mu$,
vanishes on all components of the boundary of $M$.

The infinitesimal symmetry \eqref{eighttwentyfour} of the canonical action is
closely connected with diffeomorphism invariance but does not coincide
with it \cite{BK72, LW90}.  Indeed, the two symmetries
act on different
spaces.  The canonical symmetry acts on the space of extended
phase-space histories, while diffeomorphisms act
on the space of
four-dimensional metrics and field configurations.  Under an
infinitesimal diffeomorphism generated by a vector field $\xi^\mu
(x)$, the metric and matter field change by
\begin{subequations}
\label{eighttwentyfive}
\begin{eqnarray}
\delta g_{\alpha\beta} (x) & = & 2\nabla_{(\alpha}\xi_{\beta)} (x)\, , 
\label{eighttwentyfive a}\\
\delta\phi & = & \xi^\alpha (x) \nabla_\alpha\phi (x)\, . 
\label{eighttwentyfive b}
\end{eqnarray}
\end{subequations}
Equations \eqref{eighttwentyfive} coincide with the transformations of 
three-metric
and lapse given by
\eqref{eighttwentyfour a} if the
components of $\epsilon^\mu$ are identified with the normal component
and the projection of $\xi^\mu$ into the surface  as follows
\begin{equation}
\xi^0 = \epsilon^0/N  ,\,\,\quad \xi^i = \epsilon^i -N^i \epsilon^0/N \, .
\label{eighttwentysix}
\end{equation}
{\it provided} that the equations of motion relating the time
derivatives of canonical
 coordinates to momenta are satisfied.  The
infinitesimal canonical symmetry \eqref{eighttwentyfour} thus coincides
with diffeomorphism invariance only
when certain (not all) of the equations of motion are satisfied.
However, for theories with constraints that are at most quadratic in the
momenta these equations of motion {\it are} effectively satisfied in
path integral constructions at least for gauge conditions that do not
restrict the momenta.  Gaussian integrals over the momenta
effectively replace the $\pi$'s by the right combinations of $\dot q$'s,
and integrals over exponents linear in the momenta lead to
$\delta$-functions that enforce the relevant relation exactly.
It's important to keep in mind, however,
that the relation \eqref{eighttwentysix} holds only when both $\xi^\mu$
and $\epsilon^\mu$ are infinitesimal and will fail, for example, near
$N=0$. The quantities
$\epsilon^\mu$ must therefore be {\it further} restricted to coincide
with diffeomorphisms than just invariance of the canonical action under
\eqref{eighttwentyfour} would require. Further restrictions are needed to
ensure that the resulting $\xi^\mu$ correspond to one-to-one mappings
of the manifold into itself. For these reasons, the symmetries generated
by \eqref{eighttwentyfour} are a {\it larger} set than  
the diffeomorphisms which
they include \cite{BK72, LW90}. For general relativity, therefore, we
may use invariance
under the infinitesimal canonical symmetry to ensure invariance of the
measure under infinitesimal diffeomorphisms.

\subsubsection{Class Operators}

The construction of the class operators corresponding to the coarse
grainings discussed in VIII.3 follows that for gauge theories and models
with a single reparametrization invariance. As discussed in VIII.2 the
fine-grained histories are metrics and matter fields on the manifold
$M$ bounded by the two boundaries $\partial M^\prime$ and $\partial
M^{\prime\prime}$.  The Hilbert space in which the class operators act is
therefore formally the space ${\cal H}^{(h,\chi)}$ of square integrable
function{\it als} of three-metrics and matter field configurations on these
boundary surfaces. We therefore define
\[
\bigl\langle h^{\prime\prime}_{ij}, \chi^{\prime\prime}
\left\|C_\alpha\right\| h^\prime_{ij}, \chi^\prime\bigr\rangle = \int_\alpha
\delta\pi\delta h \delta \pi_\chi \delta\chi \delta N
\]
\begin{equation}
\times \Delta_\Phi \bigl[h_{ij}, \pi^{ij},  \chi, \pi_\chi,
N^\gamma\bigr]\ \delta
\bigl[\Phi^\beta\bigl[h_{ij}, \chi,  N^{\gamma}\bigr]\bigr]
\exp \bigl\{iS\bigl[N^\gamma, \pi^{ij}, \pi_\chi, h_{ij}, \chi
\bigr]\bigr\}
\label{eighttwentyseven}
\end{equation}
where $S$ is the canonical action of \eqref{eightfifteen} and the integral is
over all metrics $g_{\mu \nu}(x) = (N^\beta({\bf x}, t), h_{ij}({\bf x}, t))$
and field configurations $\phi(x) = \chi ({\bf x}, t)$ that lie in the
diffeomorphism invariant class $c_\alpha$.  A few words are of course in
order about the rest of \eqref{eighttwentyseven} and about the attitude
we shall adopt towards such formal expressions.  $\Phi^\beta$ stands for
four conditions that fix the four-dimensional symmetry
\eqref{eighttwentyfour}
and $\Delta_\Phi$
is the associated ``Faddeev-Popov factor''.\footnote{General
relativity, viewed as a constrained Hamiltonian system,
displays a rich and interesting canonical
structure that is reflected in the
construction of its phase-space path integrals. These are perhaps most
accurately dealt with by using the BRST-invariant constructions of
Batalin, Fradkin, and
Vilkovisky
However, in a subject where
it is unclear whether the basic integrals even exist it does not seem
appropriate to devote a great deal of attention to technical issues.
For this reason, we have not made use of BRST-BFV techniques
in these lectures in the hopes of not obscuring the argument. The author
believes that the path integrals we do use could be described in this
more precise language without essential difficulty. The standard
references are Fradkin and Vilkovisky
\cite{FV75, FV77} and Batalin and Vilkovisky \cite{BV77}. For
a lucid review see Henneaux \cite{Hen85}.}
These conditions are
assumed to leave the momenta unrestricted so they may be formally
integrated out.  The important remainder of \eqref{eighttwentyseven} is
the ``measure''.  This is assumed to be the canonical (Liouville) measure in
the canonical coordinates $(h_{ij} ({\bf x}), \chi ({\bf x}))$ and
their conjugate momenta $(\pi^{ij} ({\bf x}), \pi_\chi ({\bf x}))$. This
measure is formally invariant under infinitesimal canonical
transformations generated through Poisson brackets.  In particular it is
invariant\footnote{To see specifically that the Liouville
measure is invariant under
canonical transformations, one has only to calculate the Jacobian of the
transformation.  For infinitesimal transformations this is unity plus
the trace of a matrix.  This trace vanishes because of the antisymmetry
of the Poisson brackets.}
 under transformations (VIII.4.20a--d) that include
infinitesimal diffeomorphisms in the sense discussed in Section
VIII.4.1.

The only remaining choice needed to specify the class operators is the
range of integration of the multipliers.  We integrate the shifts,
$N^i(x)$, over ${\bf R}$ at each point $x$.  We integrate the lapse,
$N(x)$, over a positive range for each point $x$. This is a
diffeomorphism invariant range because the $3+1$ decomposition of the
metric depends only on $N^2$ [\cf \eqref{sixoneone}].  
All metrics are therefore
represented as $N$ ranges over positive values.  A positive range is not,
however, invariant under the larger group of transformations (VIII.4.20a--d)
that leave the canonical action and measure invariant.

With a positive lapse range,
if the symmetry fixing conditions $\Phi^\beta$ are
 chosen to be independent of
$\pi^{ij}$, these momenta can be formally integrated out of the matrix
elements of
the class operators since the action is quadratic in the $\pi^{ij}$ and
the partition $\{c_\alpha\}$ does not restrict them.\footnote{The
integrations over the momenta are not necessarily simple Gaussians
because the factor $\Delta_\Phi$ in \eqref{eighttwentyseven} may depend
on the momenta even when the gauge fixing functions are independent of
them. However, in relativity, where the constraints are at most quadratic
in the momenta, that dependence is typically at most polynomial in the
momenta. Integrals of polynomials times Gaussians differ by integrals of
pure Gaussians only by prefactors in front of a common exponential,
which in the present case is just the Lagrangian form of the action. We
have assumed all the prefactors have been absorbed into the measure
in (VIII.4.24). For more details on this type of technical point
see Fradkin and Vilkovisky \cite{FV77}. Thanks are due to A. Barvinsky for
a discussion of this issue.}
The result is a path integral in Lagrangian form
\begin{equation}
\left\langle h^{\prime\prime}, \chi^{\prime\prime} \left\| C_\alpha \right\|
h^\prime, \chi^\prime\right\rangle = \int_\alpha \delta h\delta \phi
\delta N
\Delta _\Phi \left[h, \phi, N\right]\, \delta\, \left[\Phi
\left[h, \phi, N\right]\right]
\exp \left\{iS\left[N, h, \phi\right]\right\}.
\label{eighttwentyeight}
\end{equation}
Here we have compressed the notation of \eqref{eighttwentyseven} even further
by omitting indices on vectors and tensors.
The ``measure'' is that induced by the Liouville measure on phase
space.\footnote{For further discussion of the
induced measure and its precise form,
see especially Fradkin and Vilkovisky \cite{FV73}.}  The action is the
usual Lagrangian action for general relativity \eqref{eightsix} 
coupled to matter.

The choice of a positive range for the lapse $N$ was advocated by
Teitelboim \cite{Tei83d} in his pioneering study of canonical path
integrals for general relativity and has a number of compelling
arguments in its favor.  First, as we saw in Section VII, in the case of
a relativistic particle interacting with an external potential, the
choice of positive multiplier range reproduces the usual $S$-matrix
elements of the corresponding field theory.  Second, and perhaps more
persuasively, the choice of a positive range for $N$ corresponds in
four-dimensional, geometrical terms to a direct implementation of
Feynman's sum-over-histories principles for quantum mechanics
\cite{HH91}.  To see this, note that in the $3+1$ form of the action
\eqref{eightsix}, $\sqrt{-g}$ is represented as $Nh^{1/2}$.  The integral
\eqref{eighttwentyeight} over
a positive range for $N$ can therefore be reexpressed as
\begin{equation}
\left\langle h^{\prime\prime}, \chi^{\prime\prime} \left\| C_\alpha
\right \| h^\prime, \chi^\prime \right\rangle = \int_\alpha \delta g\,
\delta\phi\, \Delta_\Phi [g,\phi]\,\delta[\Phi[g,\phi]]
\ \exp(iS[g,\phi])
\label{eighttwentynine}
\end{equation}
where $g$ and $\phi$ denote the four-dimensional metric and matter field
configuration respectively.  Reversal of the sign of $N$ in
\eqref{eightsix}
 changes the sign of the action.  A sum over both positive and negative
lapse therefore corresponds, not to sum over geometries weighted by
$\exp(iS)$, but rather by $\cos(S)$. This choice would define a
distinct
generalized quantum mechanics, but positive lapse and
\eqref{eighttwentynine} are closer to Feynman's original principle.

\subsubsection{Adjoining Initial and Final Conditions}

The rest of the construction of the decoherence functional for a quantum
theory of spacetime parallels that for theories with a single
reparametrization invariance discussed in Section VII.  Initial and
final conditions are represented by wave functions that satisfy the
constraints  on the superspace of
three-metrics and spatial matter field configurations.
For example, the initial condition might be represented by
a family of wave functions $\{\Psi_j[h_{ik} ({\bf x}), \chi ({\bf x})]\}$
that each satisfy
\begin{equation}
{\cal H}_\mu \left[\widehat\pi^{ik} ({\bf x}), \widehat\pi_\chi ({\bf x}),
h_{ik}({\bf x}), \chi ({\bf x})\right] \Psi_j \left[h_{ik}({\bf x}),
\chi ({\bf x})\right] = 0\, .
\label{eightfourtwentyeight}
\end{equation}
Here, we take $\widehat\pi^{ij} ({\bf x}) = -i\delta/\delta h_{ij} ({\bf x})$,
$\widehat\pi_{\chi} ({\bf x}) = -i\delta/\delta\chi({\bf x})$ and the ${\cal
H}_\mu$ are operators constructed from these quantities and the
three-metric
and scalar field that represent the classical constraints
\eqref{eighteighteen}. 
Simply writing these equations down should not obscure the
fact that there are serious problems to be faced with giving them a
precise meaning.  For instance, eq.~\eqref{eightfourtwentyeight}
 is not just four
equations but four functional differential equations for each point on
the manifold $M^3$.  The formal products of operators that occur in
${\cal H}_\mu$ are singular and must be regulated \cite{TW87}. Even given a
regularization there is the delicate question of finding an operator
ordering such that the constraints obey the ``algebra'' expected from
the classical algebra of Poisson brackets \eqref{eighttwentythree}.
We do not solve these problems here.

The next step in
constructing the decoherence functional is to attach the wave functions
satisfying \eqref{eightfourtwentyeight}
representing initial and final conditions to the class operator matrix
elements in analogy with (VII.1.8) for reparametrization invariant
theories.  We write
\begin{equation}
\left\langle \Phi_i \left|C_\alpha\right|\Psi_j\right\rangle = \Phi_i
\left[h^{\prime\prime}, \chi^{\prime\prime}\right]\, \circ\, \left\langle
h^{\prime\prime}, \chi^{\prime\prime}\left| C_\alpha\right| h^\prime,
\chi^\prime \right\rangle\, \circ\, \Psi_j \left[h^\prime, \chi^\prime
\right] \label{eightfourtwentynine}
\end{equation}
where $\circ$ is a Hermitian inner product between functionals on superspace
although not necessarily a positive definite one.  We shall return to a
discussion of candidates for this product in a moment, but first we
complete the construction of the decoherence functional.  Specify a set
of initial wave functions $\{\Psi_j [h, \chi]\}$ together with their
probabilities $\{p^\prime_j\}$.  Specify a set of final wave functions
$\{\Phi_i[h, \chi]\}$ together with their probabilities
$\{p^{\prime\prime}_i\}$. Construct
\begin{equation}
D\left(\alpha^\prime, \alpha\right) = {\cal N} \sum_{ij}
p^{\prime\prime}_i \left\langle \Phi_i \left| C_{\alpha^\prime} \right|
\Psi_j \right\rangle\ \left\langle \Phi_i \left| C_\alpha \right| \Psi_j
\right\rangle^* p^\prime_j\, . 
\label{eightfourthirty}
\end{equation}
With an appropriate choice for the constant ${\cal N}$ this satisfies
the requirements (i)--(iv) of \eqref{fouroneone} for a decoherence functional
of a generalized quantum mechanics.  It is Hermitian with
positive diagonal elements whether or not the product $\circ$ is positive.
The linearity of the sums over histories that define the class operators
$C_\alpha$ ensures the consistency with the principle of superposition.
Normalization fixes ${\cal N}$ as
\begin{equation}
{\cal N}^{-1} = \sum_{ij} p^{\prime\prime}_i \left|\left\langle \Phi_i
\left| C_u \right| \Psi_j \right\rangle\right|^2 p^\prime_j
\label{eightfourthirtyone}
\end{equation}
where $C_u$ is defined by the unpartitioned sum over {\it all} histories
in \eqref{eighttwentynine}.  The decoherence functional
\eqref{eightfourthirty}
is thus a natural basis for defining
decoherence and probabilities in a generalized quantum mechanics of
coarse-grained histories of spacetime geometry and matter fields.

There remains the specification of the inner product $\circ$
and the specification of initial and final conditions in particular
quantum cosmologies.  We consider the product in the rest of this
subsection and particular initial and final conditions in the next.

A positive, Hermitian, covariant, inner product between wave functions
on superspace that are annihilated by the constraints has been sought
in the Dirac approach to the quantization of general relativity for
nearly the past forty years.
The problem is
with the positivity.  Squaring and integrating over all of superspace
does not provide a suitable inner product because, like the case of the
relativistic particle, the constraints of general relativity imply a
conserved current in superspace \cite{DeW67}. This conserved current
means that wave functions that satisfy the constraints are not
necessarily
normalizable when squared and integrated over all of superspace with a
measure that makes the operators representing the constraints Hermitian
 [\cf the
discussion following VII.3.15].  There {\it is} an analog of the conserved
Klein-Gordon product on surfaces in superspace.  It is usually
called the DeWitt product and we shall exhibit it shortly.  However,
like the Klein-Gordon product, the DeWitt product is not generally
positive and therefore cannot serve as the basis for an inner product
defining a Hilbert space in which the norm of a state vector is related
to probability.

In free field theory in flat background spacetimes, the Klein-Gordon
inner product is positive on positive frequencies solutions of the
constraint. The existence of timelike Killing fields for the underlying
flat spacetime allows a notion of positive frequency to be consistently
specified over the whole of spacetime.  Time translation
invariance means positive frequency solutions of the Klein-Gordon
equation at one time remain positive frequency solutions at all times.  A
single particle Hilbert space can thus be constructed for a free
relativistic particle.  This free particle construction does not extend
to particles interacting with a potential and neither is it available in
general relativity for there are no Killing fields in superspace
\cite{Kuc81}.

It may be that a deeper investigation into the constraints of general
relativity will reveal a positive, Hermitian, covariant, inner product
on solutions to the constraints.  That is the aim of
some.\footnote{As
described in the lectures of Ashtekar in this volume.}
If found, it could be used to construct a decoherence
functional for a quantum theory of spacetime via \eqref{eightfourthirty} and
\eqref{eightfourtwentynine}.  Here, however, we shall follow a different
route. This is to note that in the present framework the wave functions
that satisfy the constraints and specify the initial and final
conditions do not have a direct probability interpretation. That is
provided by the decoherence functional.  The spaces of wave functions
specifying the initial and final conditions therefore do not need a
Hilbert space structure. We are therefore free to take a non-positive
product for $\circ$ and still have positive probabilities for decoherent
sets of coarse-grained histories.  The DeWitt product naturally suggests
itself and in the following we spell out what it is and what
the consequences of using it are.

By ``superspace'', ${\cal M}$, we mean the space of
three-metrics $h_{ij}({\bf x})$ and spatial matter field configurations
$\chi({\bf x})$ on a spacelike surface of topology $M^3$. ${\cal M}$ is
the product of ${\cal M}_h$, the space of three-metrics, and ${\cal
M}_\chi$, the space of spatial matter field configurations.  ${\cal M}_h$
may be thought of as the product of the six-dimensional space of metric
coefficients $h_{ij}({\bf x})$ at each point ${\bf x}$ of $M^3$.
Similarly, ${\cal M}_\chi$ may be thought of as the product of the
one-dimensional space of field values $\chi ({\bf x})$ at each point
$\bf x$ of $M^3$.  The formal cardinality of ${\cal M}$ is therefore
$\infty^{3(6+1)}$ where $\infty$ denotes cardinality of the real line.

The DeWitt metric $G_{ijkl} ({\bf x})$ was introduced in
\eqref{eighttwelve}
and provides an inner product on the six-dimensional space of three-metrics
at a point ${\bf x}$.  To find an explicit expression, one can think of
a correspondence between the six dimensions and the six possible
symmetric pairs of indices $i$ and $j$, but it is easier to write
expressions directly in terms of the usual three-dimensional tensor
indices.  Thus, for example, the inverse of $G_{ijkl}$ is defined by
\begin{equation}
\bar G^{ijkl} G_{klmn} = \half \left(\delta^i_m \delta^j_n + \delta^i_n
\delta^j_m\right) 
\label{eightfourthirtytwo}
\end{equation}
and is
\begin{equation}
\bar G^{ijkl} = \half h^\half \left(h^{ij} h^{kl} + h^{il} h^{jk} -
2h^{ij} h^{kl}\right)\, . 
\label{eightfourthirtythree}
\end{equation}
The inner product between two vectors $\delta h^1_{ij} ({\bf x})$ and
$\delta h^2_{ij} ({\bf x})$ tangent to superspace at ${\bf x}$ is then
\begin{equation}
\bar G^{ijkl} ({\bf x}) \, \delta h^1_{ij} ({\bf x}) \delta h^2_{ij}
({\bf x})\, . 
\label{eightfourthirtyfour}
\end{equation}
The inner product on the whole of ${\cal M}_h$ is the sum of these inner
products over positions~${\bf x}$,
\begin{equation}
(\delta h^1, \delta h^2) = \int_{M^3} d^3 x\ \bar G^{ijkl}
({\bf x}) \delta h^1_{ij} ({\bf x}) \delta h^2_{ij} ({\bf x})\, .
\label{eightfourthirtyfive}
\end{equation}
In a similar way on ${\cal M}_\chi$ we can put
\begin{equation}
(\delta\chi^1, \delta\chi^2) = \int_{M^3} d^3 x h^\half\, \chi^1({\bf
x}) \chi^2 ({\bf x})\, . 
\label{eightfourthirtysix}
\end{equation}
Thus superspace, ${\cal M}$, acquires a metric structure.

The DeWitt metric is not positive definite.  Of six orthogonal directions at
a point, one will be timelike and five will be spacelike.
  Conformal deformations of the metric,
\begin{equation}
\delta h_{ij} ({\bf x}) = \delta \lambda ({\bf x}) h_{ij} ({\bf x})
\label{eightfourthirtyseven}
\end{equation}
are timelike, for instance.  We can therefore define a notion of a
``spacelike surface'' $\sigma$ in ${\cal M}$.
For example, we might fix the value
of the determinant of the three-metric, $h({\bf x})$, at each point.  The
DeWitt metric provides a notion of volume element $d\Sigma_{ij}({\bf
x})$ in such a surface at each ${\bf x}$.  Using this the DeWitt product
between wave functionals on superspace $\Psi^1[h,\chi]$ and
$\Psi^2[h,\chi]$ can be defined formally as
\begin{equation}
\Psi^1 \circ \Psi^2 = iZ\int_\sigma \Psi^{1*}[h_{ij}({\bf x}), \chi({\bf
x})] \left[\prod\limits_{\bf y} \left(d\chi ({\bf y}) \,
d\Sigma_{kl} ({\bf y})\right)
\frac{\buildrel\leftrightarrow\over\delta}{\delta h_{kl} ({\bf y})}\right]
\Psi^2 \left[h_{ij} ({\bf x}), \chi ({\bf x})\right].
\label{eightfourthirtyeight}
\end{equation}
A constant factor $Z$ has been included in \eqref{eightfourthirtyeight} to
absorb divergences arising from the fact that wave functionals
satisfying the constraints are constant on orbits of the
diffeomorphisms of $M^3$ in superspace.
This constant will cancel in the construction of probabilities
 if the DeWitt
product is used to construct the decoherence functional as described
above.  Alternatively, the product could be defined with suitable gauge
fixing machinery.\footnote{For more on this factoring out of
three-dimensional diffeomorphisms see H\'aj\'\i\v cek and Kucha\v r
\cite{HK90} and Barvinsky \cite{Bar93}.}

The DeWitt product defined by \eqref{eightfourthirtyeight} 
is the formal analog
in  superspace endowed  with the DeWitt metric of the Klein-Gordon
product in  spacetime with the Minkowski metric.  Like
the Klein-Gordon inner product it is not positive.
Like the Klein-Gordon product, the DeWitt product
is formally independent of the surface $\sigma$
provided $\Psi^1$ and $\Psi^2$ are solutions of the Wheeler-DeWitt
equation.  In the construction of the decoherence functional the wave
functions specifying the initial and final conditions are {\it assumed}
to satisfy the Wheeler-DeWitt equation.  The class operator matrix
elements satisfy the same equation in each argument in the neighborhood
of surfaces $\sigma^\prime$ and $\sigma^{\prime\prime}$ that are outside
the restrictions of the coarse graining.  Thus, at a formal level,
surface independence for the spacetime decoherence functional is
achieved in the same way that it is for the relativistic world line.

An important difference between this generalized quantum mechanics of
spacetime and that for the relativistic world line in flat spacetime
described in Section VII concerns the space of wave functionals describing
initial and final conditions.  In the case of the relativistic particle,
the choice of the range $N>0$ meant that the matrix elements $\langle
x^{\prime\prime} \| C_\alpha \| x^\prime\rangle$ contained only positive
frequencies in their time variation with respect to any direction defined
by a timelike Killing vector of flat spacetime.  As a consequence, the
wave functions describing initial and final conditions could be
restricted to the linear space of
 positive frequency solutions of the Klein-Gordon equations
without loss of generality.  The Klein-Gordon product is positive for
such positive frequency solutions making that space a Hilbert space.  The
quantum mechanics of spacetime described here retains the positive range
for the lapse integration.  However, this does not correspond to a
notion of positive frequency for solutions of the Wheeler-DeWitt
equation because there are generally no Killing vectors on superspace.
The space of initial or final wave functions endowed with the DeWitt
product is not, therefore, a Hilbert space.

\subsection{Discussion --- The Problem of Time}

The specification of the decoherence functional \eqref{eightfourthirty}
completes the formulation of a generalized sum-over-histories quantum
mechanics for spacetime geometry suitable for application to cosmology.
Fine-grained histories are manifolds,
metrics and matter field configurations.
Sets of alternative
coarse-grained histories are diffeomorphism-invariant partitions of
these.
The decoherence functional defines a notion of interference between
coarse-grained histories that is consistent with the principle of
superposition. Given initial and final conditions, this
decoherence functional can be used to determine which sets of
coarse-grained histories of the universe can be assigned consistent
probabilities, and what those probabilities are, according to the
principles of generalized quantum mechanics described in Section IV.

This is a fully four-dimensional formulation of a quantum mechanics of
spacetime.
Fine-grained histories are {\it four}-dimensional metrics and field
configurations.  {\it Four}-dimensional
alternatives are defined by partitions
of these fine-grained histories into classes that are invariant under
{\it four}-dimensional diffeomorphisms.  Dynamics is specified in the
decoherence functional by sums over four-dimensional histories involving
a {\it four}-dimensional diffeomorphism invariant action and measure.

This is a generally
covariant formulation of the quantum mechanics of spacetime.  No
additional ingredients beyond the metric and field configurations were
needed to specify either fine- or coarse-grained histories.  In
particular no preferred sets of spacelike surfaces in superspace or
spacetime were singled out in
the construction of the decoherence functional.  We have a quantum
mechanics of spacetime that is free from the problem of time.

Can this four-dimensional sum-over-histories quantum mechanics be
reformulated as a quantum mechanics of states on spacelike surfaces in
superspace and
their unitary evolution by a Hamiltonian or by state vector reduction?
It seems unlikely.  The standard reconstruction of Hamiltonian quantum
mechanics from a sum-over-histories formulation involves identifying a
family of surfaces in the space of coordinates
which each history intersects once and only once.  (See the discussion
in Section IV.4).  For gravity this would mean a set of surfaces in
superspace that each geometrical history intersects once and only once.
That would define a quantity that would uniquely label a set of
spacelike surfaces in every possible cosmological four-geometry.  While
there may be such quantities for certain classical spacetimes satisfying the
Einstein equation \cite{MT80}, there are none for a general
four-dimensional cosmological geometry.  A general cosmological
geometry, for example, could have arbitrarily many surfaces of a given
three-volume or trace of the extrinsic curvature. As in the case of the
theory of a relativistic particle without a preferred time, we are
unlikely to be able to formulate this generalized sum-over-histories
quantum mechanics in terms of states on  spacelike surfaces.  There is
no preferred time with which to do so.

We should probably stress that
the use of wave functions to specify initial and final conditions or the
use of functional integrals to define them is not to be construed as a
definition of a notion of state on a spacelike surface.  In the present
framework, these wave functions generally have no direct probability
interpretation.  Rather, they are part of the specification of the
decoherence functional which determines the probabilities of decoherent
{\it spacetime} alternatives as we have described.

Of course, were general covariance broken at the quantum mechanical
level so that the fine-grained histories were restricted to those in
which some superspace quantity uniquely labeled a foliating
family of spacelike surfaces in every possible spacetime, then it would
be still possible to construct a generalized quantum mechanics according
to the principles we have described.  It would have as its starting
point the more restricted set of histories which were foliable by the
quantity involved. Its construction would be analogous to the
formulation of the quantum mechanics of a relativistic particle with the
preferred time of a particular Lorentz frame that was discussed in
Section VII.4.  As there,
 an equivalent formulation in terms of states on the
corresponding surfaces in superspace would be expected.  It is important to
note that such restrictions imply a definite physical prediction.  To
restrict the fine-grained histories, for example, to a set where a type
of surface of a given three-volume occurs once and only once is to
predict that once that volume occurs there is zero probability for it
ever to occur again.

A generalization of Hamiltonian quantum mechanics, such as that of this
section, which dispenses with the familiar
notion of ``state on a spacelike surface'' has the heavy obligation to
show how it is recovered again in a suitable limit.  We shall discuss
this question in Section IX.  There we shall argue that, in those
limiting situations where spacetime behaves classically, we recover from
this generalized quantum mechanics of spacetime geometry and matter
fields an approximate quantum mechanics of matter fields in which the
preferred time necessary for a formulation in terms of states is
supplied by the background classical geometry.

\subsection{Discussion -- Constraints}

Are the constraints satisfied in this generalized quantum mechanics for
general relativity?  In the cases of electromagnetism and the relativistic
world line, we were able to give two distinct meanings to the question of
whether
the constraints were satisfied. The first was to partition the histories
by the values of the constraints and ask whether the probability was
unity that they were satisfied. The second was to ask whether
class operators commuted with the constraints and whether branch
wave functions were annihilated by them. In this subsection we offer
some thoughts on these questions in the quantum mechanics of
 general relativity we have constructed.

In the case of electromagnetism and the relativistic world line the
constraints restricted the values of certain combinations of the
momenta. The restrictions were $\pi^L({\bf x})=0$ in the case of
electromagnetism and $p^2=-m^2$ in the case of the relativistic world
line. We were able to give meaning to a partition of the histories by
the values of $\pi^L({\bf x})$ and $p^2$  by defining the momenta
as partitions by ``displacements in flight'' in the limit of very
long intervals of time. We found vanishing probability for values
of the momenta that did not satisfy the constraints. In this physical
sense, the
theories could be said to imply the constraints with probability unity.

To assess the probability that the constraints are satisfied in the
present quantum mechanics of spacetime, we must first exhibit a
diffeomorphism-invariant partition of metrics and field configurations
into a class where the constraints are satisfied and a class where they
are not. This is a more difficult problem than exhibiting similar
partitions in the cases of gauge theories or the relativistic world line
for two reasons: First, the constraints are not combinations of the momenta
alone when written in the form of
\eqref{eightsixteen} and \eqref{eightten},
 so that identifying the spacetime metrics in which they are satisfied
approximately is
not a question resolved as straightforwardly as with the ``time of
flight'' constructions in the simpler examples. (Remember the
fine-grained histories are not generally differentiable!)
Second, the
partition must include a diffeomorphism invariant specification of the
spacelike surfaces on which the constraints are to be investigated.  We
cannot, for example, usefully partition the fine-grained histories into
the class in which the constraints are defined and satisfied on {\it
every} spacelike surface and the class in which they are not.  If a
geometry satisfies the constraints on {\it every} spacelike then it
solves the Einstein equation {\cite{Kuc81b}.
A partition into classical histories and
non-classical ones {\it is} diffeomorphism invariant but also
 trivial in quantum
mechanics.  Rather, it is necessary to investigate the constraints on
some {\it specific} family of spacelike surfaces.  One could perhaps
imagine, in analogy with the relativistic world line, specifying such a
family using distances along suitable curves from $\partial
M^\prime$. However, such partitions are not likely to be of much use in
practical quantum cosmology.  We shall not pursue them further here.

We can more readily investigate the questions
of whether class operators commute with operator versions of the
constraints in the ``Hilbert space'' of functionals of three-metrics and
whether branch wave functionals corresponding to individual histories
in a coarse-grained set are annihilated by operator forms of the
constraints. For simplicity, we confine the discussion to the case
of pure gravity.

We first must draw a distinction between the momentum
constraints $H_i=0$ and the Hamiltonian constraint $H=0$ in the
notation of \eqref{eightten}. In the 3+1 decomposition through which they
are defined, the momentum constraints generate three-dimensional
diffeomorphisms in the sense that
\begin{equation}
h_{ij}+\xi^k \{h_{ij},H_k\} = h_{ij}+D_i\xi_j + D_j\xi_i\, .
\label{eighteightyone}
\end{equation}
The Hamiltonian constraint, on the other hand, generates changes more
analogous to reparametrization transformations.

For the reparametrization-invariant relativistic world line discussed in
Section VII.6, class
operators neither generally commuted with the constraint nor were branch
wave functions annihilated by them.
We can hardly expect more for
the Hamiltonian constraint in general relativity for similar reasons.
However, the momentum constraints generate three-dimensional
diffeomorphisms that are the analogs of spatial gauge transformations in
electromagnetism. The same argument that showed that, when defined with
a certain class of gauge-fixing conditions,  the class operators
corresponding to gauge invariant partitions in electromagnetism
commuted with the $\pi^L({\bf x})$ can be generalized to show a similar
result for the momentum constraints in general relativity (although
we shall not give the details here).
A notion of a branch wave functional may be defined by
\begin{equation}
\Psi_\alpha[h_{ij}]=\langle h_{ij}\|C_\alpha|\Psi\rangle
=\langle h_{ij} \|C_\alpha\|h'_{ij}\rangle\circ\Psi[h'_{ij}]\, .
\label{eighteightyfour}
\end{equation}
Then, {\it provided} that the surface $\sigma'$ on which the $\circ$ product
is calculated is itself defined by a three-dimensional diffeomorphism
invariant condition, and {\it provided} the matrix elements of $C_\alpha$
are defined with invariant gauge-fixing conditions as described above,
one can show formally that
\begin{equation}
H_i({\bf x})\Psi_\alpha[h_{ij}]=0\, . 
\label{eighteightyfive}
\end{equation}
Only in such a circumscribed way can have we been able to make limited contact
with the ideas of Dirac quantization.

\subsection{Simplicial Models}

In the absence of any conclusive evidence that its defining
functional integrals converge, the generalized quantum mechanics for
spacetime described in the preceding three subsections must be regarded
as a formal construction for the moment.  Whether the Einstein action can be
used as the
starting point for a complete, finite, manageable quantum theory of
gravity in which the ingredients of the above framework can be given
concrete meaning is at best an open question.  Therefore, to investigate
the decoherence and calculate the probabilities of the alternative
histories of our universe that might be confronted with observation, we must
either find the correct quantum theory of gravity or retain the Einstein
action but turn to finite models in which its ultraviolet divergences
have been cut off.  This second approach will be useful if, for a
realistic initial condition, the predictions of very low energy
phenomena, such as the probabilities of various galaxy-galaxy correlation
functions at the present epoch, are insensitive to this cut-off.
This subsection describes (very briefly) a class of such finite
models based on
the simplicial approximation to smooth geometries and the methods of the
Regge calculus\footnote{ The original paper is Regge \cite{Reg61}. For
a review and bibliography see Williams and Tuckey \cite{WT92}.
For a lucid introduction to the Regge calculus see the lectures by
F.~David in this volume. }.

As mentioned in Section VIII.2, a simplicial four-manifold can be
constructed by joining together four-simplices --- the four-dimensional
analogs of triangles in two-dimensions.  A metric on such a simplicial
manifold is specified by assigning definite values to the squared
lengths of the edges and a flat metric consistent with these values to
the interior of the simplices.  Both Lorentzian and Euclidean geometries
can be represented in this way, the signature in each simplex being
determined by the values of its squared edge-lengths.
Euclidean geometries have all positive squared edge-lengths that satisfy
the higher dimensional analogs of the triangle inequalities.
Lorentzian geometries may have some negative (timelike) squared
edge-lengths and are restricted by analogous inequalities.  Thus,
geometry is represented discretely and finitely.
Matter field configurations can also be represented discretely, for
example, in the case of a scalar field by specifying the value of the
field at each vertex.

In four-dimensions, the curvature of a simplicial geometry is
concentrated on the triangles in the same way that curvature in a
two-dimensional simplicial surface is concentrated at the vertices.
The deficit angle, $\theta$, is a
measure of the curvature. In two
dimensions, the deficit angle of a vertex is the difference between
$2\pi$ and the
sum of the interior angles between edges meeting at that vertex. In
four dimensions, the deficit angle of a triangle is $2\pi$ minus the sum
of the dihedral angles between the three-simplices that meet in that
triangle.  A flat geometry has vanishing deficit angles.

Einstein's action \eqref{eightfive} has a beautifully simple, geometrical
expression for a simplicial geometry.  It is most
straightforwardly stated for a Euclidean geometry.  The form for a
Lorentzian geometry, can be found by analytic continuation of the
squared edge-lengths to the values that specify a Lorentzian signatured
geometry.
The Euclidean action is \cite{Reg61, HS81}
\begin{equation}
\ell^2 I = - \sum_{\scriptstyle{\rm interior}\atop \scriptstyle{\rm
triangles}} 2A\theta + \ \sum\limits_{ \scriptstyle{\rm
four-}\atop \scriptstyle{\rm simplices}} 2\Lambda V_4
         -  \sum_{\scriptstyle{\rm boundary}\atop\scriptstyle{\rm
triangles}} 2A\psi\, .
\label{eightsixone}
\end{equation}
The first two terms correspond to the scalar curvature and cosmological
constant terms in \eqref{eightfive}.  Here $A$ is the area of a triangle,
$\theta$ is its deficit angle, and $V_4$ is the volume of a
four-simplex.  The last term is the boundary term.  Again $A$ is the
area of a triangle in the boundary and $\psi$ is $\pi$ minus the sum of
the dihedral angles between the three simplices that intersect in a
boundary triangle. Each of the quantities that enters into the action
can be expressed in terms of the squared edge-lengths by standard
geometrical formulae for areas, volumes, angles, etc.\footnote{See
\eg \cite{Har85a} for explicit and practical details.} The Regge action $I$
thus becomes a function of the squared edge-lengths specifying a
simplicial geometry.

We now describe how to construct a generalized quantum mechanics for
simplicial geometries on a fixed simplicial manifold. For simplicity we
consider pure gravity with no matter.
The fine-grained histories of the model are the Lorentz signatured
simplicial geometries. An individual fine-grained history is specified
by giving all the squared edge-lengths $\{s^i\}$ of the simplicial net.
A fine-grained history is thus a point in the space of squared
edge-lengths ${\cal S}$ in the region ${\cal S}_L$ corresponding to
Lorentz signature.  In general, two different assignments of
edge-lengths will correspond to two distinct geometries. (An exception
is flat space where different assignments {\it can} correspond to the
same flat geometry.) In general, therefore, there is no
diffeomorphism symmetry of the action \eqref{eightsixone}.
Integrating over distinct values of the
$\{s^i\}$ therefore corresponds to summing over distinct geometries.

The set of fine-grained histories  may be coarse-grained by
values of functions of the squared edge-lengths. A partition of ${\cal
S}_L$ into a set of exclusive  regions $\{c_\alpha\}$ is an example.
To define the corresponding
class operators we consider histories on
a fixed simplicial manifold $M$ with two boundaries $\partial M^\prime$ and
$\partial M^{\prime\prime}$ such as that illustrated in two dimensions
in Figure 12. Let $\{t^{\prime i}\}$ and $\{t^{\prime\prime i}\}$ respectively
be the
squared lengths of the edges in these boundaries.  We
define
\begin{equation}
\langle t^{\prime\prime i} \| C_\alpha \| t^{\prime i}
\rangle = \int_{{\cal S}^{\rm int}_L} d\mu_{\rm int} (s^i)
e_\alpha (s^i)\, \exp [iS(s^i)]\, .
\label{eightsixtwo}
\end{equation}
The multiple integration is over all interior edge-lengths keeping
$\{t^{\prime i}\}$ and $\{t^{\prime\prime i}\}$ fixed.  The
characteristic function $e_\alpha$
 is $1$ when the $s^i$ lie in the region
$c_\alpha$ and is zero otherwise.  The action $S(s^i)$ is $i$ times the
$I(s^i)$ of
\eqref{eightsixone} consistently continued to ${\cal S}_L$. The quantity
 $\mu_{\rm int}
(s^i)$ is an appropriate measure on the space of squared edge-lengths
which we shall not specify further in this discussion.

The boundary $\partial M^\prime$ is a closed simplicial three-manifold
made up of three-simplices.  The space, ${\cal T}^\prime$, of squared
edge-lengths $\{t^{\prime i}\}$ consistent with Euclidean signatured
three-geometries is a simplicial analog of superspace.  Wave functions
describing initial and final conditions are functions on ${\cal
T}^\prime$. There is similar space ${\cal T}^{\prime\prime}$ for the
boundary $\partial M^{\prime\prime}$.

To define the analog of the DeWitt metric on ${\cal T}^\prime$ we note that
each simplicial geometry in ${\cal T}^\prime$ corresponds to a class of
three-metrics in superspace that is invariant under three-dimensional
diffeomorphisms.  The DeWitt metric on ${\cal T}'$ may be identified with
the DeWitt product between a representative of  these three-metrics
\begin{equation}
G^\prime_{mn} \left(t^{\prime p}\right)\, \delta t^m \delta t^n = G^{ijkl}
\left(h_{rs}\right)\, \delta h_{ij} \delta h_{kl}\, .
\label{eightsixthree}
\end{equation}
In this equation, Latin indices range over all edges in $\partial
M^\prime$ on the left hand side and over the three spatial dimensions on the
right.  On the right, $h_{rs}$ is a three-metric representing the geometry
specified by $\{t^p\}$ and $\delta h_{ij}$ is a perturbation in  that
metric induced by $\{\delta t^m\}$.  Lund and Regge \cite{LR74, PW86}
have given a simple formula for $G'_{ij}$.  It is
\begin{equation}
G^\prime_{mn} = -\sum_{\scriptstyle{\rm three}\atop \scriptstyle{\rm
simplices}} \frac{1}{V_3}\ \frac{\partial V^2_3}{\partial t^m \partial
t^n} \label{eightsixfour}
\end{equation}
where $V_3$ is the volume of a three-simplex expressed as a function of
its squared edge-lengths and the sum is over all of them in $\partial
M^\prime$.  $G^\prime_{ij}$ is a metric on ${\cal T}'$ There is a
similar construction for ${\cal T}^{\prime\prime}$.

Given a spacelike surface $\sigma$ in ${\cal T}$, the DeWitt product
between wave functions $\Psi^1 (t^k)$ and $\Psi^2 (t^k)$ is defined by
\begin{equation}
\Psi^1 \circ \Psi^2 = i\int_\sigma d\Sigma^i\ \Psi^{1*} (t^k)
\buildrel\leftrightarrow\over \nabla_i \Psi^2 (t^k)\, ,
\label{eightsixfive}
\end{equation}
that is, by the usual Klein-Gordon product on the space ${\cal T}^\prime$
endowed with the metric $G^\prime_{ij}$.

We can now construct the decoherence functional for the simplicial model
following the discussion in Section VIII.4.3.  For sets of wave
functions $\{\Phi_i (t^k)\}$ and $\{\Psi_i (t^k)\}$
representing final and initial conditions respectively, we define
\begin{equation}
\left\langle\Phi_i\left| C_\alpha \right| \Psi_j \right\rangle = \Phi_i
\left(t^{\prime\prime m}\right) \circ \left\langle t^{\prime\prime m}
\left\| C_\alpha \right\| t^{\prime n} \right\rangle \circ \Psi
\left(t^{\prime n}\right) 
\label{eightsixsix}
\end{equation}
where the products are to be taken over initial and final surfaces
$\sigma^\prime$ and $\sigma^{\prime\prime}$ in ${\cal T}^\prime$ and
${\cal T}^{\prime\prime}$ respectively.  As in \eqref{eightfourthirty} the
decoherence
functional is
\begin{equation}
D\left(\alpha^\prime, \alpha\right) = {\cal N}\sum\limits_{ij}
p^{\prime\prime}_i
\left\langle\Phi_i \left| C_{\alpha^\prime} \right| \Psi_j \right\rangle
\ \left\langle\Phi_i \left| C_\alpha \right | \Psi_j \right\rangle^*
p^\prime_j\, . 
\label{eightsixseven}
\end{equation}
Normalization fixes  $\cal N$ as
\begin{equation}
{\cal N}^{-1} = \sum_{ij} p^{\prime\prime}_i \left|\left\langle \Phi_i
\left| C_u \right| \Psi_j \right\rangle\right|^2 p^\prime_j
\label{eightsixsevena}
\end{equation}
where the integral defining $\langle\Phi_i|C_u|\Psi_j\rangle$ is defined
by \eqref{eightsixfive} and \eqref{eightsixtwo} with $e_\alpha=1$.
This construction will not be independent of the surfaces $\sigma^\prime$
and $\sigma^{\prime\prime}$ unless further conditions are put on the
wave functions $\Psi_j(t^m)$ and $\Phi_i(t^m)$.  The values and
derivatives of $\Psi_j(t^m)$ on two different surfaces in ${\cal
T}^\prime$, for example, should be related in such a way that the change
in $\langle\Phi_i | C_\alpha | \Psi_j \rangle$ is zero or becomes small
in the limit of increasingly fine simplicial subdivisions of the
manifold $M$.  That is the limit in which, were the theory well behaved,
we would expect to recover the continuum behavior described formally in
Section VIII.4.  In particular, in the continuum the matrix elements
\eqref{eightsixsix} are formally independent of surface when the wave
functions satisfy the Wheeler-DeWitt equation.  A precise form of the
analogous conditions for the simplicial model is not known at the time
of writing.\footnote{For possible direction, see the discussion of
constraints in the continuum time Regge calculus formalism in Piran and
Williams  \cite{PW86} and Friedman and Jack \cite{FJ86}.}

Whether or not the continuum limit of the simplicial model exists and
whether or not its construction is independent of the surfaces
$\sigma^\prime$ and $\sigma^{\prime\prime}$, the constructions sketched
above define a generalized
 quantum theory of simplicial spacetime that is consistent
with the principles of Section IV. It, therefore, is
a tractable model with which to test the decoherence of model spacetime
coarse grainings and the predictions of particular theories of the
initial and final conditions of our universe.

\subsection{Initial and Final Conditions in Quantum
Cosmology}

The quantum mechanics of cosmological spacetimes described in this
section can be used to calculate the probabilistic predictions of
particular theories of the initial and final conditions of our universe.
Cosmologically interesting coarse-grained alternatives include whether
or not spacetime geometry behaves classically in the later universe on
scales above the Planck scale, whether or not the universe is
homogeneous and isotropic on large scales, alternative values of the
fluctuations that produced the large scale structure, alternative values
of the present age, etc., etc.  Precisely defined, each of these sets of
alternatives corresponds to a diffeomorphism invariant partition of
spacetime geometry and matter fields for which
a decoherence functional can be calculated, given a cut off theory of
quantum gravity  and a specification of the initial and final
conditions.  The discussion of particular theories of the initial and
final conditions,
their virtues and failings, lies outside the scope of these lectures,
but it is perhaps appropriate to offer a few speculations on their
nature.

Theories of the initial condition of the universe have been much
discussed and there are many candidates.\footnote{For a review see
Halliwell \cite{Hal91}.} Typically a single ``wave function of the universe''
is specified for the set $\{\Psi_j\}$ described above.  An example, not
chosen independently of the prejudices of the author, is the ``no
boundary'' wave function \cite{HH83}.  The no-boundary wave function
is the cosmological analog of the ground state wave functions of quantum
mechanics and field theory.  The analogy is not to a state which is the
lowest eigenstate of a Hamiltonian.  As we have mentioned, for closed
cosmological spacetimes there is no preferred notion of time, therefore
no preferred notion of energy, therefore no covariant notion of
Hamiltonian and no covariant notion of the ground state of a
Hamiltonian.  However, in theories that have a well-defined notion of
time and a corresponding Hamiltonian, the ground state wave function
which is the lowest eigenstate of that Hamiltonian may be alternatively
expressed as a functional
integral over Euclidean histories with suitable boundary conditions.  It is
this
construction that covariantly generalizes to give the ``no boundary''
wave function of the universe.  More explicitly, the no boundary wave
function, in its simplest version, is defined as an integral over
metrics and fields on a compact manifold $M$ with a single boundary
$\partial M$, of the form
\begin{equation}
\Psi \left[h, \chi\right] = \int_{\cal C} \delta g\,\delta\phi\ \Delta_\Phi
[g,\phi]\delta[\Phi[g,\phi]]\exp(-I[g,\phi])\ .
\label{eightsevenone}
\end{equation}
Here, the integral is over four-dimensional metrics $g$ and fields
$\phi$ on $M$ that match the arguments of the wave function on the
single boundary $\partial M$ along some appropriate contour $\cal C$.  The
action $I$ is the Euclidean Einstein
action for gravity coupled to matter.  Of course, much remains to be
specified in making a schematic form like \eqref{eightsevenone} concrete.  In
particular, the manifold $M$ (or the class of manifolds to be summed over if
many are allowed), the measure, and the contour of integration $\cal C$. The
latter must be complex because the integral would diverge along a purely
real contour, the Euclidean Einstein action being unbounded below.
Various possibilities have been discussed for these, but in view of the
remaining ambiguities it might be more accurate
to speak of various possible no boundary proposals
corresponding to different choices of the contour \cite{HH90, HL89}.
If these choices are
made so that they are invariant under the symmetry \eqref{eighttwentyfour}
generated by the
constraints then \eqref{eightsevenone} is an integral representation of a
wave function that formally satisfies an operator form of the
constraints. (See \eg \cite{HH91}).
The wave function $\Psi$ defined on superspace by \eqref{eightsevenone} is
thus a possible candidate for a theory of the initial condition in the
predictive formalism we have described.  Observations reveal the
early universe to have been remarkably simple so this cosmological analog
of the ground state is a plausible candidate for the initial condition of
our universe.

The use of a functional integral over complex metrics to define a wave
function representing the initial condition for cosmology is not to be
interpreted to mean
 that probabilities are assigned to complex values of the
metric of spacetime.  In the present framework the wave function has no
direct probabilistic interpretation.  Rather, it is an input to the
construction of the decoherence functional which determines the
probabilities for decoherent coarse-grainings of real, Lorentzian,
cosmological geometries.  We shall see in Section IX, however, that
certain predictions for the classical behavior of spacetime can be
extracted directly from initial wave functions in domains of superspace
where they have semiclassical form.

In contrast to the initial condition, the final condition of the
universe has received little discussion.  Yet, in the time-neutral
formulation of quantum mechanics used here (Section IV.6) the
specification of a final condition is just as necessary as is the
initial one.  As described in Section IV.6,
available evidence is consistent with a special
condition like the no-boundary proposal
 at one end of the histories and a condition analogous to the
condition of indifference with respect to final state used in the usual
formulations of quantum mechanics.
What is the analog of a condition of final indifference in a generalized
quantum mechanics that does not possess a notion of state on a spacelike
surface?  What sets of final wave functions $\{\Phi_i\}$ should be
summed over in \eqref{eightfourthirty} and what are the probabilities
$\{p^{\prime\prime}_i\}$? This is a subject for further
research.\footnote{For one idea see Sorkin \cite{Sor89}.  There
are several others.}

\section{Semiclassical Predictions}
\setcounter{footnote}{0}

\subsection{The Semiclassical Regime}

Extracting the predictions of a theory of the initial condition of the
universe for observations today is the
central application of the generalized quantum mechanics
developed in the preceding section to the subject
of quantum cosmology.
To find these predictions one must calculate
which present alternatives decohere and use the resulting joint
probabilities to search for
conditional probabilities sufficiently near unity.  These are the
definite predictions with which the theory of the initial condition can be
tested.

By and large, even for specific alternatives of interest, nothing like
this program has been carried out in detail for any of the proposed
theories of the initial condition.
Earlier work has, for the most
part, focused on predictions of the most likely classical spacetimes and
matter field configurations that
the late universe will exhibit. Clearly
these are the predictions most directly testable by observations of the
large scale structure of the present universe.  Practical prescriptions
have been developed
for extracting predictions of classical histories from a wave function
encapsulating the theory of the initial condition in
analogy to those that are used for interpreting WKB wave functions
in non-relativistic quantum mechanics.  Typically, these
prescriptions posit that,  in regimes where the wave function describing the
initial condition has the semiclassical form of a slowly varying prefactor
 times $\exp[i$(classical
action)], it can be interpreted as predicting the ensemble of classical
histories that correspond to the classical action with a likelihood
measured by the size of the prefactor.  The decoherence of the
alternative classical histories of spacetime is implicitly assumed.

However, fundamentally, the prediction of classical behavior in quantum
mechanics is not a matter of a separately posited rule; it is a matter
of the probabilities of histories.  A system exhibits classical behavior
when, in a suitably coarse-grained decoherent set of histories, the
probability is high only for those histories correlated by deterministic
laws.\footnote{For more extensive discussions of classical
behavior from the point of view of the quantum mechanics of histories
see \cite{GH90a, GH93}.} Practical prescriptions
for the extraction of
classical predictions from the form of a wave function must therefore be
justified in terms of the probabilities of such sets of histories.  In this
section
we take some important steps in the direction of  justifying
the rules for semiclassical prediction that have been employed in quantum
cosmology, using the quantum mechanics of histories of geometry
developed in the preceding section.

To show that a wave function of semiclassical form predicts classical
histories in generalized quantum mechanics it is necessary to do three things:
First, one must
exhibit a coarse graining in which classical histories correlated by
deterministic laws can
be distinguished from non-classical ones not so correlated.
Second, one should show that this set of histories decoheres as
a consequence of the initial wave function. Third, one should show
that the  probability is high only for histories correlated by
deterministic laws and calculate the relative probabilities for
the different histories that exhibit these correlations.
In Section III.4 we carried out such a combined analysis of the
decoherence of histories and probabilities of deterministic
correlations for non-relativistic systems using a model
class of coarse grainings. Consideration of both is important because
both contribute to establishing the
requirements on the coarse graining necessary for classical behavior.  Coarse
graining is needed for decoherence and further coarse graining is needed
to achieve classical predictability in the presence of the noise that
typical mechanisms of decoherence produce.

However, at the time of writing, no calculations of the decoherence of
coarse-grained histories of spacetime geometry have been carried out
using the generalized quantum framework presented
here.\footnote{Although suggestive calculations have been carried
out using not unrelated ideas by Zeh \cite{Zeh86, Zeh88},
Kiefer \cite{Kie87}, Fukuyama and Morikawa \cite{FM89}, Halliwell
\cite{Hal89}, and Padmanabhan \cite{Pad89}.}
 We shall therefore
investigate a more limited question.  We shall {\it assume} that wave
functions of semiclassical form lead to the {\it decoherence} of suitably
coarse-grained sets
of histories of spacetime, but {\it demonstrate} how these can be
classically correlated in time.  We begin with the analogous
demonstration in non-relativistic quantum mechanics.

\subsection{The Semiclassical Approximation to the Quantum Mechanics of a
Non-Relativistic Particle}

Let us recall how the semiclassical approximation works in
non-relativistic particle
quantum mechanics. A set of coarse-grained histories for the particle
could be defined by giving exhaustive sets of exclusive intervals
$\{\Delta^k_{\alpha_k}\}$ at  various times $\{t_k\}$ . We shall {\it assume }
that the
decoherence of such a set has been accomplished
 by the interaction of the particle with a larger system,
as in a measurement situation.
(See the discussion in Section II.6.) We can then focus on the probabilities
of correlations
in a particular history which
is described by particular sequence of intervals $\Delta_1, \cdots, \Delta_n$
at times $t_1, \cdots, t_n$ (dropping the superscripts on the $\Delta$'s to
simplify the notation).

 Suppose we are given an initial wave
function $\psi(q_0)$ at $t=0$. From \eqref{threeoneeighteen},
the the class operator corresponding to the coarse grained history in
which the particle passes through
the position intervals
$\Delta_1,\cdots,\Delta_n$ at times $t_1, \cdots,  t_n$
has the matrix elements
\begin{equation}
\langle q_f, t_f|P_{\Delta_n}(t_n)\cdots P_{\Delta_1}(t_1)|\psi\rangle=\int
dq_0\int\nolimits_{\alpha}\delta q\ e^{iS[q(\tau)]/\hbar}
\psi(q_0)\, .
\label{nineone}
\end{equation}
The sum is over the class $c_\alpha$ of all
paths that start at $q_0$ at $t=0$, pass through
the intervals
$\Delta_1, \cdots, \Delta_n$ at the appointed times, and wind up at $q_f$
at time
$t_f$ (see Fig. 14).

\begin{figure}[t]
\begin{center}
\includegraphics[width=4in]{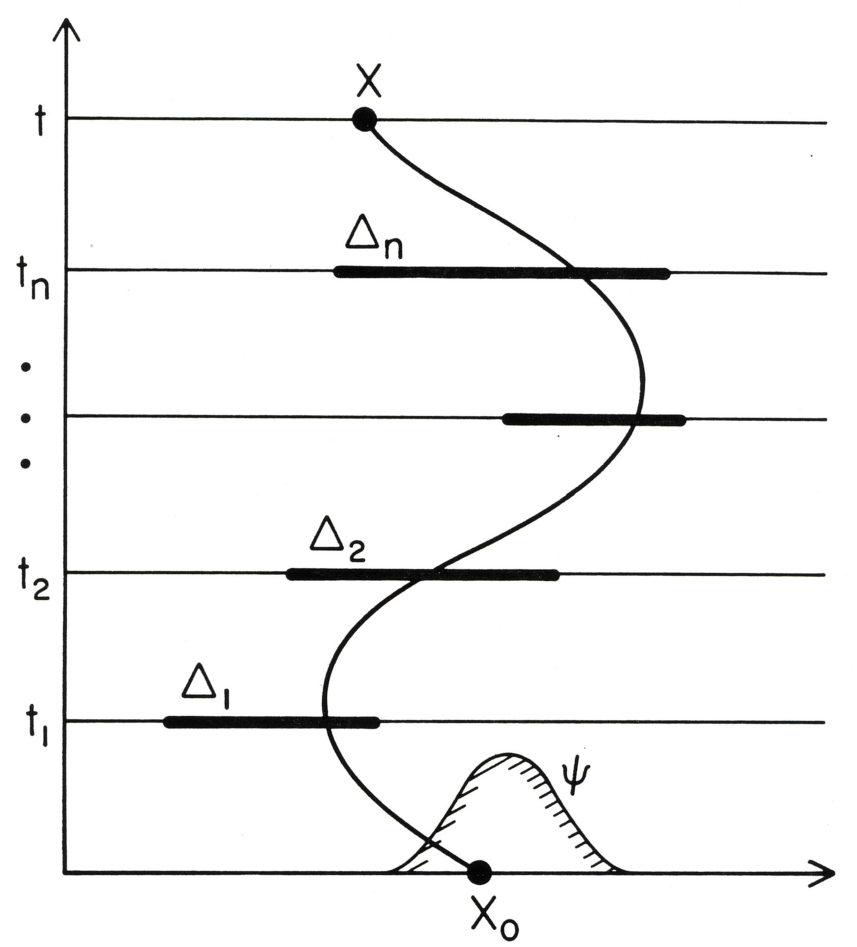}
\caption{The semiclassical approximation
to the quantum
mechanics of a non-relativistic particle.  Suppose at time $t=0$ the
particle
is in a state described is a wave function $\psi(q_0)$.  Its subsequent
evolution exhibits classical correlations in time if successive
determinations
of position are correlated according to classical laws, that is, if the
amplitude for non-classically correlated positions is near zero.  The
existence of such classical correlations is, therefore, a property not
only of
the initial condition but also the {\sl coarse graining} used to analyze
the
subsequent motion.  Classical correlations are properties of
coarse-grained
{\sl sets} of histories of the particle.  The amplitude for the particle
to
pass through intervals $\Delta_1, \Delta_2, \cdots, \Delta_n$ at times
$t_1, \cdots, t_n$ and arrive at $q_f$ at $t_f$ is the sum of exp ($iS$)
over all paths to
$q(t)$ that pass through the intervals, weighted by the initial wave
function.
For suitably spaced intervals in time, suitably large intervals
$\Delta_k$, and
suitable initial wave function $\psi$, this sum may be well approximated
by the
method of stationary phase.  In that case, only when the intervals
$\Delta_k$
are aligned about a classical path will there be a significant
contribution to
this sum.  Classical correlations are thus recovered.  How many
classical paths
contribute depends on the initial condition $\psi(q_0)$.  If, as
illustrated
here, it is a wave packet whose center follows a particular classical
history
then only that particular path will contribute significantly.  By
contrast, if
$\psi$  is  proportional to exp $[iS(q_0)]$ for
some classical action $S(q_0)$, then all classical paths that satisfy
$m\dot q
=\partial S/\partial q$ will contribute.  Then the prediction is of an
{\sl
ensemble} of classical histories, each one correlated according to the
classical
equations of motion.}
\end{center}
\end{figure}

Classical correlations are predicted when the path integral in
\eqref{nineone}
can be done by the
method of stationary phase.  For then, only when $\Delta_1, \cdots,
\Delta_n$
are lined up
so that a classical path from some $q_0$ to $q_f$ passes through them  will the
amplitude \eqref{nineone} be non-vanishing.

Whether a stationary phase approximation is appropriate for the path
integral in \eqref{nineone}
depends on the intervals $\Delta_1, \cdots, \Delta_n$, the times $t_1,
\cdots,
 t_n$, and
the initial wave function $\psi(q_0)$.  The $\Delta_1, \cdots,
\Delta_n$ must be large
enough and the times $t_1,\cdots, t_n$ separated enough to permit the
destructive interference of the non-classical paths by
which the stationary phase approximation operates.  The $\Delta$'s must
be small enough that a unique classical path passes through them.
But, in addition to these requirements on the coarse graining, the
initial $\psi(q_0)$ must be
right as well.  There are a number of standard forms for $\psi(q_0)$ for
which
the stationary phase
approximation can be seen to be valid.  For example, if
$\psi(q_0)$
describes a
wave packet with position and momentum defined to an accuracy consistent
with
the uncertainty principle, and the time intervals between the $t_k$
are short compared with the
time
over which it spreads, and the $\Delta_k$ are greater than its
initial
width,
then only a single path will contribute significantly to the integral
--- that
classical path with the initial position and momentum of the wave
packet.
Another case leading to the validity of the stationary phase
approximation is when  $\psi(q_0)$
corresponds to {\sl two} initially separated wave packets.  Then, {\sl
two
different} classical paths contribute to the stationary phase
approximation to \eqref{nineone}
corresponding to the two sets of initial data.  A unique classical
trajectory
is not predicted but rather  one of two possible classical evolutions
each with some probability determined by $\psi(q_0)$.

In general, therefore, a detailed examination of the initial wave
function $\psi(q_0)$ is
needed to determine if it predicts classical correlations in a
suitably coarse-grained set of histories.
However, there is a simple case when the requirements can be seen to be
satisfied.
This is when the Schr\"odinger evolution of the
wave function $\psi(q_0)$ is
well
approximated by forms like
\begin{equation}
\psi(q,t) \approx A(q,t)\ e^{\pm iS(q,t)/\hbar}
\label{ninetwo}
\end{equation}
where $A(q,t)$ is a real slowly varying function of
$q$ and $S/\hbar$ is a real, rapidly varying function of $q$.  
Eq.~\eqref{ninetwo}
thus  separates $\psi(q_0)\equiv \psi(q_0, 0)$ into a
slowly varying  prefactor and a rapidly varying exponential.  It follows
from the Schr\"odinger equation in these circumstances that $S$ is a
classical
action approximately satisfying the Hamilton-Jacobi equation
\begin{equation}
-\frac{\partial S}{\partial t} + H\left(\frac{\partial S}{\partial q},
q\right) = 0\, ,
\label{ninethree}
\end{equation}
where $H$ is the Hamiltonian:
\begin{equation}
H=  \frac{p^2}{2M} + V(q)\, .
\label{ninethreea}
\end{equation}
The form \eqref{ninetwo} 
is general enough to include the familiar WKB case when
$\psi(q,t)$ is an energy eigenfunction and $S(q)$ and $A(q)$ are
independent of time with $S(q)$ satisfying $H(\partial S/\partial q, q)
= E$.

The forms \eqref{ninetwo} are called 
{\sl semiclassical approximations}. When the
semiclassical approximation \eqref{ninetwo} is
inserted in \eqref{nineone}, the functional integral over paths $q(t)$
 and the integral
over $q_0$ are   integrals
of a slowly varying prefactor
times
a rapidly varying exponential.  This is immediately of the form for which
the
stationary phase approximation will be valid for suitably large
intervals\footnote{For  different perspectives on how much coarse graining
is necessary for
classical behavior to be predicted in the semiclassical approximation
 see e.g. Habib and Laflamme \cite{HL90} and \cite{GH93}.}
$\Delta_1,
\cdots, \Delta_n$ and times $t_1,\ \cdots,\ t_n$.  The exponent of the
integral is
\begin{equation}
S\big[q(\tau); q_f, q_0\big) + S(q_0, 0)\, .
\label{ninethreeb}
\end{equation}
Here, $S[q(\tau); q_f, q_0)$ is the action {\it functional} for paths
between $q_0$ and $q_f$ while $S(q_0, 0)$ is the classical action {\it
function} specifying the initial semiclassical wave function.  Extremization
of the exponent \eqref{ninethreeb}
with respect to the paths $q(t)$ and the  value of $q_0$ give the values
that dominate the integral \eqref{nineone}.  Extremizing with respect to
$q(t)$ keeping $q_0$ and $q_f$ fixed means the dominant paths satisfy the
classical equations of motion.  Extremizing with respect to $q_0$ gives
the initial momentum of a path in terms of $S(q_0, 0)$:
\begin{equation}
p_0 = \partial S/\partial q_0\, . 
\label{ninethreec}
\end{equation}
Like the two wave
packet
example above,   a unique
classical trajectory is not predicted.  The wave function \eqref{ninetwo}
 is not
peaked
about some {\sl particular }
initial data.  In fact, since $A(q_0, 0)$ varies slowly, it treats
many $q_0$'s equally.  Thus, for suitable subsequent intervals $\Delta_k$ and
times, $t_1,
\cdots, t_n$ a semiclassical
wave function predicts not     just one  classical trajectory, nor all of
them, but
just those for which the initial coordinates and  momenta are related
by \eqref{ninethreec}  for the particular classical action $S$
defined by the initial wave function.
A wave function of semiclassical form thus predicts an ensemble of classical
trajectories, each
differing from the other by the constant needed to integrate
\eqref{ninethreec}.

The prefactor $A$ is also of significance. Its square,  $|A(q_0,0)|^2$, is
the probability of an
initial $q_0$.  Given that subsequent values of $q$ are correlated by
the
classical trajectory with this initial $q_0$ and the initial momentum
\eqref{ninethreec},
$|A(q_0,0)|^2$ may be thought of as the probability of a particular
classical trajectory crossing the surface $t=0$. The
order $\hbar$ implication of the Schr\"odinger equation is that
\begin{equation}
\frac{\partial |A|^2}{\partial t} + \vec \bigtriangledown\ .
\left(|A|^2\frac{\vec\bigtriangledown S}{M}\right) =0
\label{ninesix}
\end{equation}
so that the probability density $|A|^2$ is conserved along the
trajectories.

\subsection{The Semiclassical Approximation for the Relativistic Particle}

The argument that wave functions of semiclassical form imply classical
correlations in time for suitably coarse-grained sets of histories
extends straightforwardly from non-relativistic quantum mechanics to the
quantum mechanics of spacetime.  As a warm-up for the latter problem,
however, we begin by considering a system with a single reparametrization
invariance, specifically the free relativistic world line in flat
spacetime with paths
that move forward and backward in time.  The
generalized  quantum mechanics for this system was developed in Section
VII.

We suppose that the initial condition for the relativistic particle is
supplied by a pure state whose Klein-Gordon wave function $\psi(x)$  is
well approximated, in some region of spacetime, by the semiclassical
form
\begin{equation}
\psi(x) \approx A(x) e^{iS(x)/\hbar} 
\label{ninethreeone}
\end{equation}
where $A(x)$ is a slowly varying prefactor while the exponential varies
rapidly. In this approximation, the most
rapidly varying part of the Klein-Gordon equation implies the
Hamilton-Jacobi equation for $S(x)$:
\begin{equation}
\left(\nabla S\right)^2 + m^2 = 0\, , 
\label{ninethreetwo}
\end{equation}
which shows that $S(x)$ is a classical action. The conservation of
Klein-Gordon current in this approximation of rapid variation of $S$ gives
\begin{equation}
\nabla \cdot \bigl(\bigl|A\bigr|^2 \nabla S\bigr) = 0\, ,
\label{ninethreethree}
\end{equation}
which shows that $|A|^2$ is conserved along the integral curves of the
classical action $S(x)$.

Unlike the case of non-relativistic quantum mechanics, the wave function
$\psi(x)$ does not have a direct probability interpretation in the
generalized quantum mechanics developed in Section VII.
 Rather, it supplies the initial condition for
the decoherence functional from which the probabilities for decoherent
partitions of the particle's
paths are determined.  Despite this, we
shall see that probabilities for those coarse grainings that define sets
of histories that behave semiclassically can be extracted simply from
the form of \eqref{ninethreeone}.

We consider a coarse graining of the paths of the relativistic particle
into classes $c_\alpha$ that distinguish classical from non-classical
behavior.  An example would be coarse graining the paths $x(\lambda)$
by their behavior with respect to a division of spacetime into cells.
Classical paths will go only through certain sequences of cells
consistent with obeying the classical equation of motion.  In the case
of a free relativistic particle, the classical paths go through cells
connected by straight lines in spacetime intersecting each cell once and
only once.  Non-classical paths connect cells in other ways.

We assume the decoherence of such a coarse-grained set of histories,
$\{c_\alpha\}$, either by themselves or through interaction with a
larger system as in a measurement situation.  The relevant class
operator matrix element for a
particular coarse-grained history $c_\alpha$ is [\cf VII.4.19)]
\begin{equation}
\langle x'' \|C_\alpha|\psi\rangle =
i\int\nolimits_{\sigma^\prime} d\Sigma^\nu\left\langle x'' \left\| C_\alpha
\right\| x^\prime \right\rangle
{\buildrel\leftrightarrow\over{\nabla'}}_\nu \psi\left(x^\prime\right)
\label{ninethreefour}
\end{equation}
where the matrix elements of the class operator are defined by the path
integral \eqref{seventwonine} and the integral is taken over a spacelike surface
$\sigma^\prime$ to the past of any restriction by the coarse graining.

We further suppose, as a consequence of the semiclassical form
\eqref{ninethreeone} and the nature of the coarse grainings, that the integral
over $x^\prime$ and the path integral defining the matrix elements of
$C_\alpha$ can be done by the method of stationary phase.  Inserting
\eqref{ninethreeone} in \eqref{ninethreefour}, using the gauge
$\dot N=0$ in \eqref{seventwoeleven},
we find an integral that is proportional to an exponential of the
following combination:
\begin{equation}
S\left[x(\lambda), N; x'', x^\prime\right) + S \left( x^\prime\right)
\label{ninethreefive}
\end{equation}
where we have indicated the dependence of the action on the
endpoints explicitly.  In the stationary phase approximation, only paths that
extremize this combination with respect to the variables integrated over
in \eqref{ninethreefour}
contribute to the integral.  Extremization with respect to
$x(\lambda)$ and $N$ yield the classical equations of motion.  The
variable $x^\prime$ is to be extremized in the surface
$\sigma^\prime$ in which it is integrated.  This gives the connection
between the components of the momentum of the classical path in the
surface and the tangential derivatives of $S(x)$. The remaining
component is determined by the constraint \eqref{ninethreetwo} and we can
therefore write initially
\begin{equation}
p'=\nabla_{x'} S\, .
\label{ninethreesix}
\end{equation}
Thus, with these assumptions, a wave function of the semiclassical form
\eqref{ninethreeone} predicts that suitable coarse grainings will define an
ensemble of histories correlated in time by the classical equations of
motion having any of the possible initial positions on the initial surface
$\sigma^\prime$ with an initial momentum  determined \eqref{ninethreesix}.
To calculate the probability of a particular history
we only have to calculate the probability of the position that it
crosses $\sigma^\prime$.  These are determined by the diagonal elements of
the decoherence functional \eqref{sevenfivefifteen}.

For simplicity, let us, assume that the surfaces $\sigma^\prime$ and
$\sigma^{\prime\prime}$ are surfaces of constant time, $t^\prime$ and
$t^{\prime\prime}$ respectively.  We calculate the probability that the
classical path passes through a particular spatial region $\alpha$ of
the surface $\sigma^\prime$ having volume $\Delta_\alpha$ centered about
position ${\bf x}_\alpha$.  According to \eqref{sevenfivefifteen} this is
\begin{equation}
p(\alpha) = {\cal N} \int \frac {d^3p^{\prime\prime}}{(2\pi\hbar)^3
2\omega_{p''}} \left|\left\langle
\phi_{{\bf p}^{\prime\prime}}\right| C_\alpha \left | \psi \right\rangle
\right|^2.
\label{ninethreeseven}
\end{equation}
Here, we have assumed  a final condition of indifference with respect to
final states implemented by a sum over a
complete set of positive frequency momentum
eigenstates having Klein-Gordon wave functions
\begin{equation}
\phi_{\bf p} (x) =  \hbar^\half \exp
\left[i\left(-\omega_p t + {\bf p} \cdot {\bf x}\right) \right /\hbar]
\label{ninethreeeight}
\end{equation}
where, as usual, $\omega_p = \sqrt{{\bf p}^{~2} + m^2}$. The matrix elements
of $C_\alpha$ are
\begin{equation}
\left\langle\phi_{{\bf p}^{~\prime\prime}}\left| C_\alpha\right| \psi
\right\rangle = - \int_{{\bf R}^3} d^3x^{\prime\prime} \int_\alpha
d^3x^\prime\ \phi^*_{{\bf p}^{~\prime\prime}} (x^{\prime\prime})
\frac{\buildrel\leftrightarrow\over\partial}{\partial t^{\prime\prime}}
\Delta_F \left(x^{\prime\prime} - x^\prime\right)
\frac{\buildrel\leftrightarrow\over \partial}{\partial t^\prime}
\psi(x^\prime)\, . 
\label{ninethreenine}
\end{equation}
Noting that
\begin{equation}
\rho_f \left(x'', x^\prime\right) \equiv  \int \frac{d^3 p}{(2\pi\hbar)^3
2\omega_p}\,\phi_{\bf p}
\left(x^{\prime\prime}\right)\,\phi^*_{\bf p} \left(x^\prime\right) =
-i \Delta_F\left(x^{\prime\prime} - x^\prime\right)\, ,
\label{ninethreeten}
\end{equation}
and composing Feynman propagators where appropriate, we can write for the
probability \eqref{ninethreeseven}
\begin{equation}
p(\alpha) = {\cal N} \int_\alpha d^3 x^{\prime\prime} \int_\alpha d^3
x^\prime\ \psi^*\left(x^{\prime\prime}\right)
\ \frac{\buildrel\leftrightarrow\over\partial}{\partial t^{\prime\prime}}
\Delta_F \left(x^{\prime\prime} - x^\prime\right)
\ \frac{\buildrel\leftrightarrow\over\partial}{\partial t^\prime} \psi
\left(x^\prime\right) 
\label{ninethreeeleven}
\end{equation}
where $x^{\prime\prime}$ and $x^\prime$ both lie on the constant time
spacelike surface $\sigma^\prime$.  This can be  evaluated as follows:
Insert the semiclassical form \eqref{ninethreeone} in
\eqref{ninethreeeleven} and
note that the slowly varying prefactor $A(x)$ can be pulled outside all
integrations and evaluated at the center ${\bf x}_\alpha$ of the interval
$\Delta_\alpha$.  To
evaluate the remainder of the integral insert the standard integral
representation for $\Delta_F$ following from \eqref{ninethreeten} and
\eqref{ninethreeeight} into \eqref{ninethreeeleven}.  Note that, if the
characteristic size of the region $\Delta_\alpha$ is large compared to the
Compton wave length $\hbar/m$, the various factors of $\omega_p$ may
be replaced by $(\partial S/\partial t^\prime)$, when the latter is
positive like $\omega_p$,  because the integrations over
${\bf x}^{\prime\prime}$  and
${\bf x}^{\prime}$ approximately enforce the connection
\eqref{ninethreesix}.  If
$\partial S/\partial t^\prime$ is negative the integral is zero in these
approximations.  Carrying out the remaining integrations one finds
\begin{equation}
p(\alpha) = {\cal N}\left| A\left(t^\prime, {\bf x}_\alpha\right)\right|^2
\Delta_\alpha
\, \theta\left(\partial S/\partial t^\prime\right)\, \partial \left[S\left(t',
{\bf x}_\alpha\right)/\hbar\right]/\partial t^\prime\, .
\label{ninethreetwelve}
\end{equation}
The normalization factor, ${\cal N}$ can be determined in this
approximation by requiring the probabilities to be normalized,
$\Sigma_\alpha p(\alpha)=1$.

The restriction to coarse grainings that distinguish classical paths
only up to errors in position larger than the Compton wavelength may be
understood in another way.  The exact notion of localization for the
relativistic particle is provided by the Newton-Wigner position operator
\cite{NW49}.  A state localized in the Newton-Wigner sense does not
have a localized Klein-Gordon wave function, rather one spread out over
coordinate intervals of order $\hbar/m$. Throughout, we have been
discussing coarse grainings defined in terms of the coordinates of
spacetime by which the fine-grained histories are defined.  We
should, therefore, not expect to obtain a notion of classical position
that is defined more accurately than the Compton wavelength. We do
not.

The result \eqref{ninethreetwelve} is not a surprise.  The conservation of
Klein-Gordon current leads to the conservation of $|A|^2\nabla S$ in the
semiclassical approximation [\cf \eqref{ninethreethree}].  In view of the
connection
\eqref{ninethreesix} between $\nabla S$ and four-velocity,
this can be interpreted as the conservation of the
density $|A|^2$ along classical trajectories.  We, therefore, naturally
are led to think of
\begin{equation}
|A|^2 \nabla_\mu S d\Sigma^\mu\, , 
\label{ninethreethirteen}
\end{equation}
when positive, as the relative probability that the classical
trajectories cross an element of spacelike hypersurface $d\Sigma^\mu$.
This is the rule that was advocated by many authors for a probability
interpretation of semiclassical wave functions of reparametrization
invariant systems especially clearly and completely by Vilenkin
\cite{Vil88}.  Here, we have {\it derived} this rule from a more fundamental
probability interpretation through which the limitations of the
approximation can be explored.

It should be stressed that we have exhibited no readily applicable rule
for determining which coarse grainings lead to classical correlations.
For example, we expect the semiclassical approximation \eqref{ninethreeone}
to be valid only in some region of spacetime.  Coarse grainings that
distinguish between paths outside this region cannot be expected to
exhibit classical correlations.  In particular, in evaluating the
integral over $x^\prime$ that led to \eqref{ninethreesix} we, in effect,
assumed that the semiclassical form \eqref{ninethreeone} was valid over the
whole of the surface $\sigma^\prime$.  If that is not true a more
delicate argument with possibly more stringent requirements on the
coarse graining may be needed to exhibit classical correlations in time.
The important point is that the generalized quantum mechanics for a
single relativistic particle gives us a precise meaning for the
probabilities of decoherent sets of coarse-grained histories in which
various approximation schemes can be analyzed and their limitations
explored.

\subsection{The Approximation of Field Theory in Semiclassical
Spacetime}

Any generalization of quantum mechanics that is proposed to deal with
the problem of time in quantum gravity must reproduce the usual Hamiltonian
quantum
mechanics of matter fields in a fixed background spacetime for those
coarse-grained histories in which spacetime geometry behaves
classically.  To discuss this question, a more refined type of
semiclassical approximation is needed than the kind we have discussed
for non-relativistic systems or the relativistic particle. In these, all
variables  behave classically.  To discuss the recovery of
quantum field theory in classical spacetime we need to treat the matter
field variables fully quantum mechanically in situations where geometry
behaves approximately classically.  Such approximations
are familiar from
other areas of physics.  In the Born-Oppenheimer approximation to
molecular dynamics, for example, the motion of the nuclei is treated
classically while the dynamics of the electrons is treated quantum
mechanically.

In ordinary quantum mechanics, wave functions that are products of a
rapidly oscillating function of semiclassical form like \eqref{ninetwo} in some
variables times a more slowly varying function of the remaining ones
lead to classical behavior of the former
 and quantum behavior of the latter.  Typically there is a
scale that governs the separation into rapidly and slowly varying parts.
In the case of the Born-Oppenheimer approximation it is the ratio of the
mass of the nuclei to that of the electron. The ratio of the Planck mass
to characteristic particle energies will be the important ratio in
approximations where spacetime geometry behaves classically but matter
behaves quantum mechanically.

The initial condition for cosmology  is represented by wave function(s)
on superspace that
solve operator versions of the constraints.  Procedures for constructing
wave functions of various semiclassical forms that approximately
satisfy the constraints have been widely discussed in the literature and
we shall only briefly review them here.  For details and references to
the original literature the reader can consult the papers of Halliwell
\cite {Hal87} and Padmanabhan and Singh \cite{PS90}.
Many different semiclassical forms are possible depending
on what variables the
rapidly and slowly varying parts of the wave function depend on.  To
illustrate with a simple case we  consider wave functions of the form
\begin{equation}
\Psi\left[h_{ij} ({\bf x}), \chi({\bf x})\right] = A[h_{ij}({\bf
x})]
\exp(\pm
iS_0[h_{ij}({\bf x})])\psi\left[h_{ij}({\bf x}), \chi({\bf
x})\right]\, ,
\label{ninefourone}
\end{equation}
where $A$ and $\psi$ are slowly varying functionals of $h_{ij}({\bf
x})$
and
$S_0[h_{ij} ({\bf x})]$ is a real classical action for gravity
alone and we have reverted to units where $\hbar=1$ for the remainder of
this section. The action $S_0[h_{ij}({\bf x})]$ satisfies the
classical Hamilton-Jacobi equations \cite{Per62, Ger69} that arise
from the constraints of general relativity
 $H(\pi^{ij}, h_{ij})=0$ and $H_i(\pi^{ij},
h_{ij})=0$ when the momentum $\pi^{ij}({\bf x})$ conjugate to
$h_{ij}({\bf x})$ is related to $S_0$ by
\begin{equation}
\pi^{ij}({\bf x}) = \frac{\delta S_0}{\delta h_{ij}({\bf x})}\, .
\label{ninefourtwo}
\end{equation}
Explicitly [\cf (VIII.4.8)] these constraints are:
\begin{subequations}
\label{ninefourthree}
\begin{equation}
\ell^2 G_{ijk\ell}({\bf x})\pi^{ij}({\bf x})\pi^{k\ell}({\bf x}) +
\ell^{-2}
h^{\half}({\bf x})\left(2\Lambda-^3R({\bf x})\right) = 0\, .
\label{ninefourthree a}
\end{equation}
\begin{equation}
D_j\pi^{ij}({\bf x}) =0\, .
\label{ninefourthree b}
\end{equation}
\end{subequations}
The gradient \eqref{ninefourtwo}
 defines a vector field on superspace and its integral
curves
are the classical spacetimes that give rise to the action $S_0$.  For
example,
if we work in the gauge where four-metrics have the form
\begin{equation}
ds^2 = -d\tau^2 + h_{ij}(\tau,{\bf x}) dx^i dx^j\, ,
\label{ninefourfour}
\end{equation}
then eq.~\eqref{ninefourtwo} becomes
\begin{equation}
\half \frac{dh_{ij}}{d\tau} = G_{ijk\ell}\frac{\delta S_0}{\delta
h_{k\ell}}\, .
\label{ninefourfive}
\end{equation}
Integrating \eqref{ninefourfive},  we recover a four-metric
\eqref{ninefourfour}
 that satisfies the
Einstein
equation.
The values of $\psi$ along such an integral curve define $\psi$ as a
function of
$\tau$
\begin{equation}
\psi = \psi\left[h_{ij}(\tau,{\bf x}), \chi({\bf x})\right] = \psi
\left[\tau, \chi({\bf x})\right]\, .
\label{ninefoursix}
\end{equation}

The wave function $\Psi[h_{ij} ({\bf x}),\chi ({\bf x})]$ must satisfy
the operator form of the constraints \eqref{eighteighteen}  that implement the
underlying
gravitational dynamics.  The three momentum constraints, ${\cal
H}_i\Psi=0$,
guarantee that $\Psi$ is independent of the choice of coordinates in
the spacelike surface.  The Hamiltonian
constraint may be written out formally as
\begin{equation}
{\cal H}_0({\bf x}) \Psi = \left[-\ell^2\nabla^2_{\bf x} + \ell^{-2}
h^{\half} (2\Lambda-^3R) + h^{\half} \hat T_{nn} (\chi, - i\delta/\delta
\chi)
\right]\Psi=0\, .
\label{ninefourseven}
\end{equation}
Here,
\begin{equation}
\nabla^2_{\bf x} = G_{ijk\ell}({\bf x})\frac{\delta^2}{\delta
h_{ij}({\bf x})
\delta h_{k\ell}({\bf x})} + \left(\begin{array}{l}
                                   {\rm linear\ derivative}\\
                                    {\rm terms\ depending}\\
                                    {\rm on\ factor\ ordering}
                                         \end{array} \right)
\label{ninefoureight}
\end{equation}
and $\hat T_{nn}$ is the stress-energy of the matter field projected
into the
spacelike surface (the Hamiltonian density)  expressed as a function of
the matter field $\chi({\bf x})$ and the operator 
$-i\delta/\delta\chi({\bf x})$
corresponding to its conjugate momentum.  This operator form of the
Hamiltonian constraint is
called the
Wheeler-DeWitt equation \cite{DeW67, Whe68}.  The implications of
the Wheeler-DeWitt equation \eqref{ninefourseven} 
for that part of the semiclassical
approximation
that varies slowly with three-metric may be found by inserting the
approximation \eqref{ninefourone} into
\eqref{ninefourseven}, using the Hamilton-Jacobi equation
\eqref{ninefourthree},
and neglecting second
derivatives of terms varying   slowly with respect to the three-metric.
The result is an equation for $A\psi$ that can be organized in the
following form:
\begin{equation}
-i\psi\left[(\nabla^2_{\bf x} S_0)A + 2G_{ijk\ell} \frac{\delta
S_0}{\delta h_{ij}}\frac{\delta A}{\delta h_{k\ell}}\right]
+ A\left[-2iG_{ijk\ell}\frac{\delta S_0}{\delta h_{ij}}
\frac{\delta\psi}{\delta h_{k\ell}} + \ell^{-2}
h^\half \hat T_{nn}\psi\right]=0\, .
\label{ninefournine}
\end{equation}
We now impose the condition that the two terms in \eqref{ninefournine}
vanish separately.  This defines a decomposition of the slowly varying
part,
$A\psi$, into $A$ and $\psi$.

The condition on the $\psi$ resulting from \eqref{ninefournine}
may be rewritten using
\eqref{ninefourfive}
and \eqref{ninefourseven} as
\begin{equation}
i\frac{\partial\psi}{\partial\tau} = h^\half \hat T_{nn} \left(\chi,
-i \frac{\delta}{\delta\chi}\right)\psi\, .
\label{ninefourten}
\end{equation}
This is the Schr\"odinger equation in the field representation for a
quantum
matter field $\chi$ executing dynamics in a background geometry of the
form \eqref{ninefourfour}.

The condition on $A$ arising from \eqref{ninefournine} implies the following
relation
\begin{equation}
G_{ijk\ell} \frac{\delta}{\delta h_{ij}} \left(|A|^2\frac{\delta
S_0}{\delta h_{k\ell}}\right) = 0\, .
\label{ninefoureleven}
\end{equation}
This is the equation of conservation of the current $|A|^2(\delta
S_0/\delta h_{ij})$ in superspace.  It is the analog of the similar
relation \eqref{ninesix} in non-relativistic
quantum mechanics and \eqref{ninethreethree} in the quantum mechanics of the
relativistic particle.  Indeed, in view of \eqref{ninefourfive},
this is just the statement
that the ``density in superspace'', $|A|^2$, is conserved along
classical trajectories, the integral curves of \eqref{ninefourfive}.

Many other semiclassical approximations are possible besides the one
based on
the form \eqref{ninethreeone}.
For example, an approximate form
in which both spacetime and some matter variables behave classically would
involve
an action defining the rapidly
varying part of
the wave function which depended on both kinds of variables.  One can
consider
ensembles of classical geometries driven
by expectation values of matter fields in which the constraints
\eqref{ninefourthree}
contain
such terms as sources.  Systematic approaches to obtaining such
approximate wave functions by expanding the solutions to the
Wheeler-DeWitt
equation in powers of the inverse Planck length have been extensively
discussed.
Indeed, it is essential
to consider approximations with both matter and geometry behaving
classically
since  the late universe
is certainly not a solution of the vacuum Einstein equation.
Superpositions of semiclassical forms like those of \eqref{ninefourone}, such
as those which arise from the ``no boundary'' proposal of the initial
condition \cite{HH83}, may also be considered.  Provided that there is
no interference between the branches arising from distinct semiclassical
forms, the probability of a coarse-grained history is just the sum of
contributions from each.
The common feature of all these semiclassical
approximations is the separation of the wave
function into superposition of pieces each having
a part rapidly varying in certain variables governed by  a
classical action and a more slowly varying part.  There are different
approximations depending on what variables are distinguished in this
way.

We now sketch a derivation of how an initial condition of the form
\eqref{ninefourone}
 can imply the classical behavior of geometry in suitable coarse
grainings and the familiar quantum mechanics of matter
fields in the resulting background classical spacetimes.  We shall give only
the broad outlines of a demonstration making many assumptions that must be
made precise and justified to complete it.

We assume that we have a coarse graining of geometry that distinguishes
classical from non-classical behavior.  That is, we assume that the
four-dimensional metrics that are the fine-grained histories of geometry
are partitioned into classes $\{c_\alpha\}$ such that some of the
classes can be said to exhibit the classical correlations implied by
Einstein's equation to a sufficient accuracy while the rest do not.  We
let the index $\gamma$ range over the subset of the $\alpha$
corresponding to  possible classical
histories so that $\{c_\gamma\}$ is the set of possible coarse-grained
classical histories.  Each of the classes $c_\gamma$ may be further
partitioned by the behavior of the matter field into a finer set of
classes $\{c_{\gamma\beta}\}$.  The classes $c_{\gamma\beta}$ of
physical interest will typically be highly branch dependent as described
in Section III.1.1, that is, the
partitions of the matter field of interest will depend on the classical
spacetime geometry $\gamma$.  We thus have a division of the
fine-grained histories into non-classical geometries and various
classical geometries with different behaviors for the matter field in
those classical spacetimes.  We
denote the coarse-grained classes by $\{c_{\alpha\beta}\}$ understanding
that for the non-classical alternatives for the geometry there is but a
single alternative $\beta$ for the matter --- all possible field
histories. Our central assumption is that the geometrical alternatives
decohere, that is that the decoherence functional is approximately
diagonal in the alternatives $\alpha$ (which include the alternative
classical histories, $\gamma$).

The decoherence functional \eqref{eightfourthirty}
 is constructed from amplitudes of
the form
\begin{equation}
\left\langle h^{\prime\prime}, \chi^{\prime\prime}\left\|
C_{\alpha\beta} \right\| h^\prime, \chi^\prime \right\rangle \circ \Psi
\left[h^\prime, \chi^\prime\right]
\label{ninefourtwelve}
\end{equation}
where we are using a compressed notation in which indices (including
coordinate labels) have been suppressed.  The class operator matrix
elements are [{\it cf.} \eqref{eighttwentynine}]
\begin{equation}
\left\langle h^{\prime\prime}, \chi^{\prime\prime} \left\|
C_{\alpha\beta} \right\| h^\prime, \chi^\prime \right\rangle =
\int_{\alpha\beta} \delta g \delta \phi\, \Delta_\Phi[g,\phi]\delta[\Phi[
g,\phi]]\exp \left\{i\left(S_E[g] +
S_M [g, \phi]\right)\right\}\, . 
\label{ninefourthirteen}
\end{equation}

Including the integral over $h^\prime$ and $\chi^\prime$ involved in the
$\circ$
product, the amplitude \eqref{ninefourtwelve} is defined by a functional
integral over metrics and matter fields including their values on the
initial surface $\sigma^\prime$.  We now assume that the coarse
graining is such that, for wave functions of the semiclassical form
\eqref{ninefourone}, the integral over {\it metrics} can be carried out by the
method of stationary phase.  Significant contributions come only from
the extrema of the exponent
\begin{equation}
S_E\left[g; h^{\prime\prime}, h^\prime\right] + S_0 \left[h^\prime\right]
\label{ninefourfourteen}
\end{equation}
with respect to $g$ and $h^\prime$.  Eq.~\eqref{ninefourfourteen} is
extremized with respect to $g$ by solutions of the Einstein equations
with no matter sources.  Eq.~\eqref{ninefourfourteen} is an extremum with
respect to $h^\prime$ when the initial momenta of these classical
solutions is connected to $S_0$ by the classical relation
\eqref{ninefourtwo}. In this approximation, therefore, amplitude for
non-classical behavior of the geometry is zero;  classical spacetime is
predicted.

We assume that the coarse graining defining the classical classes
$\{c_{\gamma\beta}\}$ is fine enough that an essentially unique geometry
(up to the accuracy of the coarse graining) provides the extremum
 between $\sigma^{\prime\prime}$ and
$\sigma^\prime$ and dominate the sum over metrics in the corresponding
amplitudes \eqref{ninefourtwelve}.  Denote by $g_\gamma$ a metric representing
this solution of the Einstein equation
that satisfies the gauge conditions $\Phi^\alpha[g]=0$.
Denote by $\sigma^{\prime\prime}_\gamma$ and $\sigma^\prime_\gamma$
respectively the hypersurfaces in the classical spacetime that
respectively correspond to the surfaces $\sigma^{\prime\prime}$ and
$\sigma^\prime$ in superspace.  Taking account of the semiclassical form
\eqref{ninefourone}, the amplitude \eqref{ninefourtwelve} may be written
\begin{equation}
A_\gamma F_\gamma \int \delta\chi^\prime \left\langle\chi^{\prime\prime},
\sigma^{\prime\prime}_\gamma \left\| C_{\gamma\beta}\right\| \chi^\prime,
\sigma^\prime_\gamma \right\rangle\,\psi \left[\chi^\prime,
\sigma^\prime_\gamma\right]\, .
\label{ninefourfifteen}
\end{equation}
where
\begin{equation}
\left\langle\chi^{\prime\prime},\sigma^{\prime\prime}_\gamma \left\|
C_{\gamma\beta} \right\| \chi^\prime,\sigma^\prime_\gamma\right\rangle
= \int_{\gamma\beta} \delta\phi\,\exp\left(iS_M\left[g_\gamma,
\phi\right]\right)\, .
\label{ninefoursixteen}
\end{equation}
These expressions were arrived at as follows: The slowly varying factor
$A[h_{ij}]$ in \eqref{ninefourone} was evaluated at the value of $h_{ij}$
corresponding to the classical solution $g_\gamma$, pulled out of the
integral and written $A_\gamma$.  The functional integral over fields in
the class $c_{\gamma\beta}$ occurs in \eqref{ninefoursixteen}.  It is an
integral over all fields that are in the class $c_{\gamma\beta}$ and
match the values $\chi^\prime$ and $\chi^{\prime\prime}$ on the
hypersurfaces $\sigma^\prime_\gamma$ and $\sigma^{\prime\prime}_\gamma$
respectively. The remaining factors arising from the stationary
phase approximation to the integral over metrics are lumped together in
$F_\gamma$.

Assuming that measure induced from \eqref{eighttwentyseven} has an
appropriate form, the matrix elements \eqref{ninefoursixteen}
define the class operators of a
matter field theory in the background spacetime $g_\gamma$. The
composition with the wave function $\psi[\chi^\prime,
\sigma^\prime_\gamma]$ is the usual inner product between states of
definite field on a hypersurface $\sigma^\prime_\gamma$.  Now assume
that the wave functions $\{\Phi_i(h^{\prime\prime},
\chi^{\prime\prime})\}$ specifying the final condition factor into products of
functions of $h^{\prime\prime}$ and functions of $\chi^{\prime\prime}$.
When we construct the full decoherence functional from the amplitudes
\eqref{ninefourfifteen} we find for the only non-vanishing values
\begin{equation}
D\left(\gamma^\prime, \beta^\prime; \gamma, \beta\right) \cong
\delta_{\gamma^\prime\gamma} \left|A_\gamma\right|^2 {\cal F}_\gamma
D^M_\gamma \left(\beta^\prime, \beta\right)
\label{ninefourseventeen}
\end{equation}
where $D^M_\gamma \left(\beta^\prime, \beta\right)$ is the decoherence
functional for matter field alternatives $\{c_{\gamma\beta}\}$ in the
fixed background spacetime $g_\gamma$. The factor ${\cal F}_\gamma$
represents the combination of the factors $F_\gamma$ and the final
conditions on geometry.

Eq.~\eqref{ninefourseventeen} shows the sense in which the generalized quantum
mechanics of spacetime and matter fields reproduces field theory in
curved spacetime when the geometry behaves classically. The decoherence
and probabilities of matter alternatives are governed by the field
theory in curved spacetime decoherence functional
$D^M_\gamma(\beta^\prime, \beta)$ in each classical spacetime
$g_\gamma$. The probabilities of the different possible classical
geometries themselves are given by $|A_\gamma|^2 {\cal F}_\gamma$.  The
conservation of current \eqref{ninefoureleven} makes it plausible that with
suitable final conditions ${\cal F}_\gamma$ will be the ``velocity''
$\delta S_0/\delta h_{ij}$.  However, a more careful analysis of the
final conditions and the stationary phase approximation would be needed
to conclude such a result.

\subsection{Rules for Semiclassical Prediction and the
Emergence of Time}

While not complete, the discussion in this section points to two
conclusions: First, the usual rules for extracting the semiclassical
predictions of a wave function of the universe can be made precise and
justified in the generalized sum-over-histories quantum mechanics of
cosmological spacetimes.  A wave function specifying an initial
condition does not generally have a direct probability interpretation in
this framework.  It is an input to the calculation of the probabilities
of partitions of cosmological four-geometries and matter field
configurations into decoherent classes.  However, wave functions of the
semiclassical form \eqref{ninefourone} imply that suitably coarse-grained,
decoherent sets of histories will, with high probability, exhibit the
correlations of classical spacetime.  For each initial three-metric
there is such a classical spacetime.  It can be found by integrating the
Einstein equation with the initial data
\begin{equation}
h_{ij}({\bf x}), \,\,\,\,\pi^{ij}({\bf x}) = 
\delta S_0/\delta h_{ij} ({\bf x})\, .
\label{ninefiveone}
\end{equation}
Thus, an initial condition represented by a wave function that is
approximately of the semiclassical form \eqref{ninefourone} may be said to
predict the ensemble of classical spacetimes with the initial data of
\eqref{ninefiveone}.  Not all classical spacetimes are predicted for that
would correspond to all data $(h_{ij}, \pi^{ij})$
consistent with the constraints.  Rather, only classical spacetimes
corresponding to the initial data of the particular form of
the initial wave function through \eqref{ninefiveone} are predicted.  The
relative probabilities of these classical spacetimes are proportional to
$|A[h_{ij}]|^2$ as \eqref{ninefourseventeen} shows.

The utility of a general framework for prediction is not simply to justify
the rules for semiclassical prediction that were
posited on the basis of analogy with non-relativistic quantum mechanics.
The more general framework allows us to analyze the deviations from
these rules.  It permits classical behavior to be precisely defined in
terms of the probabilities for histories.  It permits an analysis of
what level of coarse graining is necessary for a classical description.
It allows us to understand quantitatively how close the initial wave
function has to come to a semiclassical form to predict classical
histories.  It allows us to calculate the probabilities for deviations
from classical behavior and to analyze when the semiclassical rules
break down.  It permits the calculation of probabilities for highly
non-classical alternatives.  Most importantly, it allows us to analyze
which sets of alternative coarse-grained sets of histories of the
universe decohere.

The second conclusion which the discussion of this section points to
concerns the problem of time in quantum gravity.  The sum-over-histories
generalized quantum mechanics we have been describing is in fully
four-dimensional form and does not require the specification of a
preferred family of spacelike surfaces.  Yet we have seen in
eq.~\eqref{ninefourseventeen} how for coarse grainings that exhibit the
correlations of classical geometry, the decoherence functional can
reduce
to the decoherence functional $D^M_\gamma (\beta^\prime, \beta)$ for
field theory in a curved spacetime, $g_\gamma$.  That theory {\it does} have an
equivalent Hamiltonian formulation in terms of states on any family of
spacelike surfaces that foliate the background spacetime, $g_\gamma$.  Its
construction follows the discussion in Section IV.4. Assuming that the
measure induced from the Liouville construction in
\eqref{eighttwentyseven} is appropriate, the resulting states evolve
unitarily when unrestricted by the coarse graining.  States and
unitarity may thus be recovered in quantum theory, not generally, but in
approximations in which spacetime geometry behaves classically.
 The consistency of
Hamiltonian formulation on different foliating families of spacelike
surfaces is guaranteed by their equivalence with the sum-over-histories
formulation and traceable ultimately to the causal structure supplied
by the background geometry $g_\gamma$.

For example, let us consider how the {\it time ordering} of the
alternatives in the class operators of Hamiltonian quantum mechanics
[{\it cf.} \eqref{fourtwo}] emerges from the generalized quantum mechanics of
spacetime which has no preferred time and therefore {\it a fortiori} no
notion of time ordering. Making essential use of the possibility of branch
dependent
partitions, consider a coarse graining of the matter field histories by
ranges of field averages over spatial regions on a succession of
non-intersecting spacelike surfaces $\sigma_{\gamma1}, \cdots,
\sigma_{\gamma n}$ of the background geometry $g_\gamma$. The class
operators defined by \eqref{ninefoursixteen} will be given by matrix elements
of projection operators that are {\it time-ordered} with respect to the
causal structure of the background geometry. That is because functional
integrals defining matrix elements of products of operators
automatically time-order them [\cf \eqref{fivethreeeleven}].  Arrows of time,
such as the second law of thermodynamics, then can arise from asymmetries
between the initial and final conditions on the matter fields as
described in Section IV.7.  It is in such ways that a preferred notion
of time enters quantum mechanics when there is a classical background
spacetime to supply it.

\section{Summation}
\setcounter{footnote}{0}
These lectures have developed generalized  quantum frameworks for
non-relativistic quantum mechanics, field theory, and a single relativistic
world line in which quantum theory is put into fully spacetime form
both with respect to dynamics and alternatives. These frameworks
motivate the proposal of Section VIII for a quantum framework for cosmology
incorporating a quantum dynamics of spacetime geometry.  The three basic
elements of a generalized quantum theory are compared for these
frameworks in the table above. 

\begin{figure}[t]
\centerline{\includegraphics[width=8.0in]{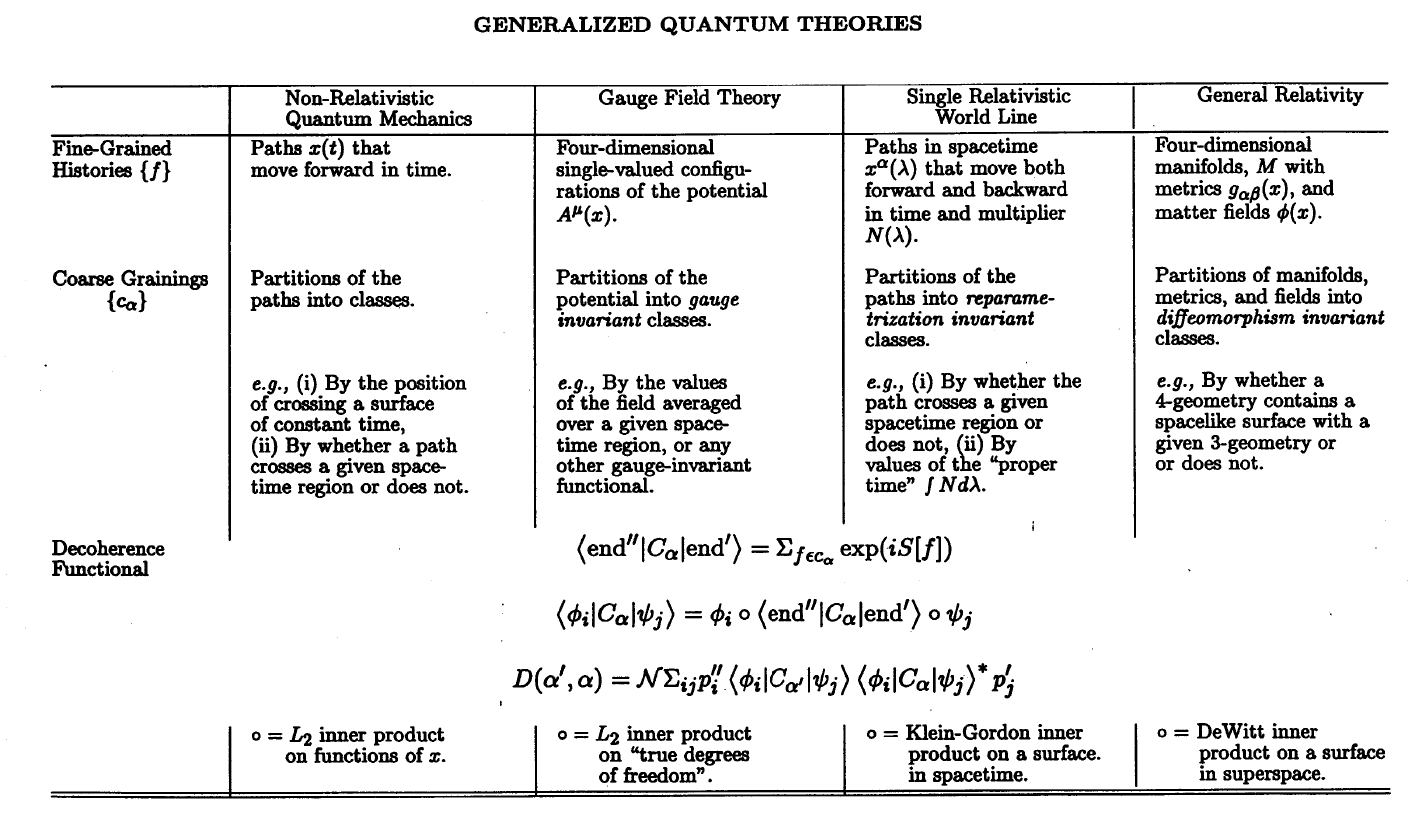}}
\end{figure}

\begin{itemize}
\item Quantum mechanics is formulated for a closed system ---
the universe.  Decoherence rather than measurement distinguishes those
alternatives which
 may consistently be assigned probabilities from those  which may
not.  The framework may thus be applied to make predictions of
alternatives of interest to cosmology in the very early universe or on
very large distance scales which are not part of any measurement
situation.

\item The sum-over-histories approach to quantum mechanics is
used to formulate the quantum mechanics of cosmology in fully spacetime
form.  Dynamics is expressed in terms of sums over fine-grained
histories that are four-dimensional manifolds, metrics, and matter field
configurations.  Alternatives are defined by partitions
(coarse-grainings) of these four-dimensional, fine-grained histories into
exhaustive sets of exclusive diffeomorphism invariant classes.  The
analogs of ``unitary evolution'' and ``reduction of the wave packet''
are given a unified sum-over-histories expression.
The formulation is manifestly four-dimensionally
diffeomorphism invariant.

\item The alternatives to which this quantum theory assigns
probabilities, if they decohere, are at once more general and more
restricted than the ``observables'' that are often considered in other
formulations.  Four-dimensional
diffeomorphism invariant alternatives {\it on a spacelike
surface}, for example, usually
are restricted to classical constants of the motion in the sense that
they commute with the constraints.  The
present formulation considers the much larger, more realistic, and more
accessible class of diffeomorphism invariant {\it spacetime}
alternatives.  However, in its present form the theory considers only
alternatives describable in spacetime form as partitions of the unique
fine-grained set of histories of the sum-over-histories formulation.
Alternatives analogous to all the Hermitian observables of
transformation theory are considered approximately by expressing them in
spacetime form.  A spacetime description is adequate for our experience
and for cosmology.  It remains to be seen whether it is fundamental, as
assumed here, or whether the theory can be extended to an even richer
class of alternatives.

\item The generalized quantum mechanics of spacetime is free
from the ``problem of time''.  No preferred family of spacelike surfaces
was needed either to define the fine-grained histories or
quantum evolution or the alternatives for which probabilities are
predicted.  These were specified directly in four-dimensional,
geometrical, terms. This does not mean that the notion of time has been
eliminated from this framework, for this is a quantum theory of
space{\it time}!  But this generalized quantum framework for spacetime
neither requires nor specifies a preferred family of spacelike surfaces.

\item Familiar Hamiltonian quantum mechanics of matter
fields, with its preferred time(s), is an approximation to this
generalized quantum mechanics of spacetime.  The approximation is
appropriate for decoherent coarse-grainings that specify coarse-grained
geometries that are correlated classically with high probability.  The
classical geometries that summarize these correlations supply the notion
of time for an approximate Hamiltonian quantum mechanics of matter
fields.  Such classical behavior of geometry is an emergent feature of
the boundary conditions in cosmology. Having generalized Hamiltonian
quantum mechanics to deal with quantum spacetime, we recover known
physics in a suitable limit.

\item A significant advantage of the sum-over-histories
formulation of quantum mechanics is that the classical limit may be
analyzed directly.  That is especially important in quantum cosmology
where we expect that most predictions of particular theories of the
initial condition that can be confronted with observation will be
semiclassical in nature.  A system behaves classically when, in a
suitably coarse-grained decoherent set of histories, the probability is
high for histories correlated by deterministic laws.  These
probabilities are supplied by this generalized quantum framework.  The
wave function that specifies the initial condition does not have a
direct probabilistic interpretation in this framework.  However,
assuming their decoherence, the probabilities for histories can be used
to provide a justification for the familiar rules that have been used to
extract semiclassical predictions directly from wave functions of semiclassical
form.

\item A lattice version of this generalized quantum mechanics
can be constructed using the methods of the Regge calculus to consider
fine-grained histories that are four-dimensional simplicial geometries.
Such quantum models are a natural cut-off version of general relativity.
They supply a finite and tractable arena in which to examine the low
energy, large scale predictions of specific proposals for initial
condition and with which to test the sensitivity of these predictions to
the nature of quantum gravity at smaller scales.

\item This sum-over-histories formulation of the quantum
mechanics of cosmological spacetimes is a {\it generalization} of
familiar quantum mechanics that neither utilizes states on spacelike
surfaces nor even permits their construction in general.  It is
therefore different from the usual versions of Dirac or ADM quantum
mechanics
which are formulated in terms of states on a spacelike surface.
Constraints do not play a primary role in constructing quantum dynamics.
States satisfying the constraints are used to specify the initial and
final conditions of a quantum cosmology but it is only in this sense that
``true physical degrees of freedom'' are defined. However, should a
preferred time be discovered in classical general relativity nothing
necessarily needs to be changed in this formulation of the quantum
mechanics of spacetime as long as that preferred structure is
expressible in terms of the metric.  Further, should experiment show
that quantum theory breaks general covariance by singling out a
preferred family of spacelike surfaces not distinguished by the
classical theory it is still possible to construct a generalized quantum
mechanics on the principles described here, by suitably restricting the
set of fine-grained histories.

\end{itemize}

This short list of attractive features does not mean that the
generalized quantum mechanics of spacetime that we have described
 is correct.  That determination is,
in principle, a matter for experiment and observation.  Of course, we
are unlikely to have such experimental checks any time in the near
future.  As far as quantum cosmology is concerned, the main result of
these investigations is to show that the rules for semiclassical
prediction that are commonly employed
can be
put on a firmer probabilistic footing in a generalized quantum
framework that does not require a preferred notion of time or or a
definition of
measurement.

Beyond theories of the initial condition, it is possible that these
ideas may be useful in formulating a complete and manageable quantum
theory of gravity which must necessarily predict the quantum behavior of
spacetime geometry in a suitable limit.  Thus, while we have learned little
about a correct quantum theory of gravity in these lectures, we may have
learned something of how to formulate questions to ask of it.

\acknowledgments

The author is grateful to Bernard Julia for much encouragement in the
preparation of these lectures and much patience in their completion. In
many places the lectures reflect the discussions that the
author had with the students and other lecturers
at the school. Thanks are due to B.~Julia and J.~Zinn-Justin for the
stimulating and hospitable atmosphere at Les Houches.
 Over a longer period of time
the author has benefited with conversations from physicists too numerous
to mention on the subjects of these lectures. Special thanks should
be made to M.~Gell-Mann whose joint work with the author on the quantum
mechanics of closed systems forms the basis of much of this material, to
J. Halliwell for many discussions on quantum cosmology, to
K. Kucha\v r for many discussions on the problem of time in quantum
gravity, and to C. Teitelboim and R. Sorkin for conversations on the
sum-over-histories formulation of quantum mechanics. Thanks are due to
A.~Barvinsky, J.~Halliwell, C.~Isham, K.~Kucha\v r,
 R.~Laflamme, D.~Page, R.~Sorkin, and R.~Tate
for critical readings of the manuscript. Their constructive criticisms
are reflected throughout although they are not, of course, responsible
for the remaining errors nor do they necessarily endorse every point of
view expressed here. The preparation of
these lectures as well as the work they describe was supported in part
by the US National Science Foundation under grant PHY-90-08502.

\section*{\bf Notation and Conventions}
\setcounter{footnote}{0}

For the most part we follow the conventions of Misner, Thorne, and
Wheeler \cite{MTW70} with respect to signature, curvature, and
indices.  In particular:

\noindent{\sl Signature} --- $(-, +, +, +)$ for Lorentzian spacetimes.

\noindent{\sl Indices} --- Greek indices range over spacetime from 0 to 3.
Latin indices range over space from 1 to 3. Indices on tensors are often
suppressed where convenient.

\noindent{\sl Units} --- In Sections VI-VIII we use units in which $\hbar =
c = 1$. In Section IX we include $\hbar$ explicitly but set $c=1$;
 The length $\ell$ is $\ell = (16\pi G)^\half = 1.15 \times
10^{-32}$cm which is $(4\pi)^\half$ times the Planck length.

\noindent{\sl Coordinates and Momenta} --- The four coordinates of
spacetime $\{x^\alpha\}$ are frequently abbreviated just as $x$.
Similarly, conjugate momenta $\{p_\alpha\}$ are abbreviated as $p$.
Spatial coordinates $\{x^i\}$ are written ${\bf x}$ and spatial
momenta $\{p_i\}$ as ${\bf p}$.  Thus $p\cdot x = p_\alpha x^\alpha$
and ${\bf p}\cdot {\bf x} = p_i x^i$. Similarly, configuration space
coordinates $\{q^i\}$ are written as $q$, conjugate momenta $\{p_i\}$
as $p$, and $p\cdot q = p_i q^i$.

\noindent{\sl Vectors} --- Four-vectors $a^\alpha, b^\alpha, \cdots$ are
written $a, b, c \cdots$ and their inner products as $a\cdot b$, etc.
Three-vectors are written as $\vec a, \vec b, \vec c \cdots$ and their
inner products as $\vec a\cdot \vec b$, etc.  Thus, in the case of
displacement vectors and their conjugate momenta we use ${\bf p} \cdot
{\bf x} = \vec p \cdot \vec x$ interchangeably.

\noindent{\sl Covariant Derivatives} --- $\nabla_\alpha$ denotes a spacetime
covariant derivative and $D_i$ a spatial one. $\nabla^2 = \nabla_\alpha
\nabla^\alpha$. In flat space $\nabla f$ is $\nabla_\alpha f$ and $\vec
\nabla f$ is the usual three-dimensional gradient.

\noindent{\sl Traces and Determinants} --- Traces of second rank tensors
$K_{\alpha\beta}$ are written as $K=K^\alpha_\alpha$ except when the
tensor is the metric in which case $g$ is the determinant of
$g_{\alpha\beta}$ and $h$ the determinant of spatial metric $h_{ij}$;

\noindent{\sl Extrinsic Curvatures} --- If $n_\alpha$ is the unit normal to
a spacelike hypersurface in a Lorentzian spacetime, we define its
extrinsic curvature to be
\[
K_{ij} = - \nabla_i\, n_j\, .
\]
{\sl Intrinsic Curvatures} --- Intrinsic curvatures are defined so that the
scalar curvature of a sphere is positive.

\noindent{\sl Momentum Space Normalization} --- We use Lorentz invariant
normalization for momentum states of a relativistic particle and include
factors of $2\pi$ and $\hbar$  as follows:
\[
\left\langle{\bf p}~^{\prime\prime} \big| {\bf p}~^\prime\right\rangle =
(2\pi \hbar)^3
(2\omega_p)\, \delta^{(3)}\left({\bf p}~^{\prime\prime} - {\bf
p}~^\prime\right)
\]
where $\omega_p = \sqrt{{\bf p}~^2 + m^2}$.  Similarly in the
non-relativistic case
\[
\left\langle{\bf p}~^{\prime\prime} \big| {\bf  p}~^\prime\right\rangle =
(2\pi \hbar)^3 \delta^{(3)} \left({\bf p}~^{\prime\prime} - {\bf
p}~^\prime\right)\, .
\]
This convention means that sums over momenta occur as
$d^3p/[(2\omega_p)(2\pi\hbar)^3]$ or as $d^3p/(2\pi\hbar)^3$ respectively.

\noindent{\sl Klein-Gordon Inner Product} ---
\[
i \int\nolimits_t d^3 x \phi^* (x) \frac{\buildrel
\leftrightarrow\over\partial}{\delta t}\ \psi (x) = i \int\nolimits_t d^3 x
\left[\phi^* (x)\,\frac{\partial\psi(x)}{\partial t}\ -\
\frac{\partial\phi^*(x)}{\partial t}\ \psi (x)\right]\,  .
\]

\noindent{\sl The Feynman Propagator} ---
\[
\Delta_F (x) = \hbar^2 \int \frac{d^4p}{(2\pi\hbar)^4}\ \ \frac{e^{ip\cdot
x/\hbar}}{p^2+m^2-i\epsilon} \, .
\]

\end{document}